\pdfoutput=1

\documentclass[iop,apj,numberedappendix,appendixfloats]{emulateapj}

\usepackage{amsmath}
\usepackage{amssymb}
\usepackage{xspace}
\usepackage{url}
\usepackage{datetime}
\usepackage[backref, breaklinks, colorlinks, urlcolor=magenta, citecolor=blue, linkcolor=red]{hyperref}

\usepackage[all]{hypcap} 

\slugcomment{Accepted for publication in The Astrophysical Journal}

\shorttitle{Ionized gas properties of star-forming galaxies at $z\simeq3.3$}
\shortauthors{Onodera et al.}

\newcommand{\kms}{\ensuremath{\,\text{km}~\text{s}^{-1}}}

\newcommand{\oiii}{\ensuremath{\text{[\ion{O}{3}]}}\xspace}
\newcommand{\oiiitot}{\ensuremath{\text{[\ion{O}{3}]}\lambda\lambda4959,5007}\xspace}
\newcommand{\oiiione}{\ensuremath{\text{[\ion{O}{3}]}\lambda4959}\xspace}
\newcommand{\oiiitwo}{\ensuremath{\text{[\ion{O}{3}]}\lambda5007}\xspace}
\newcommand{\oiiidirect}{\ensuremath{\text{[\ion{O}{3}]}\lambda4363}\xspace}
\newcommand{\oii}{\ensuremath{\text{[\ion{O}{2}]}\lambda3727}\xspace}
\newcommand{\oiitot}{\ensuremath{\text{[\ion{O}{2}]}\lambda\lambda3726,3729}\xspace}
\newcommand{\oiione}{\ensuremath{\text{[\ion{O}{2}]}\lambda3726}\xspace}
\newcommand{\oiitwo}{\ensuremath{\text{[\ion{O}{2}]}\lambda3729}\xspace}

\newcommand{\neiiione}{\ensuremath{\text{[\ion{Ne}{3}]}\lambda3869}\xspace}
\newcommand{\neiiitwo}{\ensuremath{\text{[\ion{Ne}{3}]}\lambda3969}\xspace}

\newcommand{\niitwo}{\ensuremath{\text{[\ion{N}{2}]}\lambda6583}\xspace}
\newcommand{\nii}{\ensuremath{\text{[\ion{N}{2}]}}\xspace}

\newcommand{\height}{\ensuremath{\text{H}8}\xspace}
\newcommand{\hepsilon}{\ensuremath{\text{H}\epsilon}\xspace}
\newcommand{\hdelta}{\ensuremath{\text{H}\delta}\xspace}
\newcommand{\hgamma}{\ensuremath{\text{H}\gamma}\xspace}
\newcommand{\hbeta}{\ensuremath{\text{H}\beta}\xspace}
\newcommand{\halpha}{\ensuremath{\text{H}\alpha}\xspace}
\newcommand{\lya}{\ensuremath{\text{Ly}\alpha}\xspace}

\newcommand{\uvbeta}{\ensuremath{\beta_\text{UV}}\xspace}
\newcommand{\ohmetal}{\ensuremath{12+\log(\text{O/H})}\xspace}

\newcommand{\zmsfr}{\ensuremath{Z(M_\star,\text{SFR})}\xspace}

\newcommand{\autorefsec}[1]{\hyperref[#1]{Section~\ref*{#1}}}

\providecommand\natexlab[1]{#1}
\providecommand\JournalTitle[1]{#1}

\catcode`\@=11
\def\gsim{\ifmmode{\mathrel{\mathpalette\@versim>}}
    \else{$\mathrel{\mathpalette\@versim>}$}\fi}
\def\lsim{\ifmmode{\mathrel{\mathpalette\@versim<}}
    \else{$\mathrel{\mathpalette\@versim<}$}\fi}
\def\@versim#1#2{\lower 2.9truept \vbox{\baselineskip 0pt \lineskip
    0.5truept \ialign{$\m@th#1\hfil##\hfil$\crcr#2\crcr\sim\crcr}}}
\catcode`\@=12

\begin{document}


\title{ISM excitation and metallicity of star-forming galaxies at $z \simeq 3.3$ from near-IR spectroscopy}


\author{
  M. Onodera\altaffilmark{1}, 
  C.~M. Carollo\altaffilmark{1},
  S. Lilly\altaffilmark{1},
  A. Renzini\altaffilmark{2},
  N. Arimoto\altaffilmark{3,4},
  P. Capak\altaffilmark{5,6},
  E. Daddi\altaffilmark{7},
  N. Scoville\altaffilmark{6},
  S. Tacchella\altaffilmark{1},
  S. Tatehora\altaffilmark{4},
  G. Zamorani\altaffilmark{8},
}

\affil{$^{1}$Institute for Astronomy, ETH Z\"urich, Wolfgang-Pauli-Strasse 27, 8093 Z\"urich, Switzerland}
\affil{$^{2}$INAF-Osservatorio Astronomico di Padova, Vicolo dell'Osservatorio 5, I-35122, Padova, Italy}
\affil{$^{3}$Subaru Telescope, National Astronomical Observatory of Japan, 650 North A'ohoku Place, Hilo, Hawaii 96720, USA}
\affil{$^{4}$Graduate University for Advanced Studies, 2-21-1 Osawa, Mitaka, Tokyo, Japan}
\affil{$^{5}$Infrared Processing and Analysis Center (IPAC), 1200 East California Boulevard, Pasadena, California 91125, USA}
\affil{$^{6}$California Institute of Technology, 1200 East California Boulevard, Pasadena, California 91125, USA}
\affil{$^{7}$CEA, Laboratoire AIM-CNRS-Universit\'e Paris Diderot, Irfu/SAp, Orme des Merisiers, F-91191 Gif-sur-Yvette, France}
\affil{$^{8}$INAF-Osservatorio Astronomico di Bologna, Via Ranzani 1, I-40127 Bologna, Italy}

\email{monodera@phys.ethz.ch}



\begin{abstract}
  We study the relationship between stellar mass, star formation rate (SFR),
  ionization state, and gas-phase metallicity for a sample of 41 normal star-forming galaxies
  at $3 \lesssim z \lesssim 3.7$.
  The gas-phase oxygen abundance, ionization parameter, and electron density of ionized gas
  are derived from rest-frame optical strong emission lines
  measured on near-infrared spectra obtained with Keck/MOSFIRE.
  We remove the effect of these strong emission lines
  in the broad-band fluxes to compute stellar masses 
  via spectral energy distribution fitting, 
  while the SFR is derived from the dust-corrected ultraviolet luminosity.
  The ionization parameter is weakly correlated with the specific SFR,
  but otherwise the ionization parameter and electron density do not correlate with 
  other global galaxy properties such as stellar mass, SFR, and metallicity. 
  The mass--metallicity relation (MZR) at $z\simeq3.3$
  shows lower metallicity by $\simeq 0.7$~dex than that at $z=0$
  at the same stellar mass. 
  Our sample shows an offset by $\simeq 0.3$~dex from the locally defined mass--metallicity--SFR relation,
  indicating that simply extrapolating such relation to higher redshift
  may predict an incorrect evolution of MZR. 
  Furthermore, within the uncertainties we find no SFR--metallicity correlation, 
  suggesting a less important role of SFR in controlling the metallicity at high redshift. 
  We finally investigate the redshift evolution of the MZR
  by using the model by \citet{lilly:2013}, 
  finding that the observed evolution from $z=0$ to $z\simeq3.3$
  can be accounted for by the model assuming
  a weak redshift evolution of the star formation efficiency. 
\end{abstract}


\section{Introduction}

It is well established that, 
up to redshift $z \simeq 6$,
the star formation rate (SFR) of star-forming galaxies (SFGs)
tightly correlates with their stellar mass ($M_\star$),
producing a so-called star-forming main-sequence (MS)
where the specific SFR ($\text{sSFR}\equiv \text{SFR}/M_\star$) depends only weakly on stellar mass
\citep[e.g.,][]{brinchmann:2004, daddi:2007:sfr, elbaz:2007, noeske:2007:ms, pannella:2009, pannella:2015, %
  magdis:2010, peng:2010, karim:2011, rodighiero:2011, rodighiero:2014, whitaker:2012:ms, %
  whitaker:2014:ms, kashino:2013, gonzalez:2014, speagle:2014, steinhardt:2014, renzini:2015:ms}. 
While the normalization of the MS increases by a factor of $\gtrsim20$ from $z=0$ to $z=2$, 
the scatter ($\simeq 0.3$ dex) and slope of the MS are almost independent of redshift over the entire $0 \lesssim z \lesssim 4$ period
\citep[e.g.,][]{salmi:2012,speagle:2014,schreiber:2015}.
  At $z\simeq2$,
  the fraction of starbursts defined as SFGs with more than 4 times higher SFR than that of the MS
  is observationally constrained to be $\sim 2$\%{}
  of total star-forming population
  and they account for $\sim 10$\%{} of the cosmic SFR density \citep{rodighiero:2011},
and these fractions appear to be constant up to $z\simeq4$ \citep{schreiber:2015}.
This indicates that galaxies spend most of their star-forming phases on the MS
due to the interplay of gas inflow, star formation, and gas outflow \citep[e.g.,][]{tacchella:2015:ms}.

The evolution of galaxies on the MS
can be attributed to the evolution of baryon  accretion
into dark matter haloes \citep[e.g.,][]{dutton:2010,dave:2012,dekel:2013b,lilly:2013,forbes:2014,sparre:2015}.
Cosmological hydrodynamical simulations are able to reproduce  the confinement of SFGs on the MS, 
with  galaxies fluctuating  within the observed SFR scatter of the MS in response to fluctuations in the accretion rate
\citep[][]{tacchella:2015:ms}.
The simulations can also naturally reproduce the observed trend of the so-called ``inside-out quenching''
\citep[e.g.,][]{morishita:2015,nelson:2015,tacchella:2015:profile, tacchella:2015:science}: in the early phase the 
stellar mass profile grows in a self-similar manner, i.e., no or little radial dependence of the sSFR.  Then the sSFR in the central part starts being suppressed, 
once the central density of the bulge reaches a critical value of $\gtrsim 10^{10}M_\odot\,\text{kpc}^{-2}$, at the expenses of having consumed most of the gas,
while star formation and mass growth still continue  in the disk \citep{tacchella:2015:data}.
Eventually, the suppression of star formation takes place in the entire galaxy (\citealt{carollo:2014}; \citealt{tacchella:2015:science}; Lilly \&{} Carollo, in preparation). 
This process appears to be rapid at early cosmic times, i.e, $z \gtrsim 2$, with a timescale of $<1$ Gyr for massive ($M \gtrsim 10^{11}M_\odot$) galaxies
\citep[e.g.,][]{onodera:2010:pbzk, onodera:2012, onodera:2015, vandesande:2013, barro:2015, belli:2015},
while the timescale becomes longer for less massive  galaxies at later cosmic times, i.e, $z \lesssim 2$
\citep[e.g.,][]{thomas:2005,thomas:2010,mcdermid:2015,tacchella:2015:science}.
Therefore, less massive SFGs will stay on the MS longer before the cessation of star formation,
increasing the quenched galaxy population at later epochs \citep{carollo:2013}.

Star formation enriches the gas  metal content via supernova explosions and stellar mass loss,
and a tight correlation between galaxy stellar mass and gas-phase metallicity ($Z$) in SFGs
has been known since 1970s \citep[e.g.,][]{lequeux:1979}. The stellar mass--gas-phase metallicity relation (MZR) is now well established for galaxies  in the local Universe \citep{tremonti:2004}.
Studies of the MZR at high redshifts  find systematically lower gas-phase metallicities than in their local counterparts  at a given stellar mass, by up to $\simeq 0.8$ dex at $z \simeq 3$
\citep[e.g.,][see also \citealt{hayashi:2009,onodera:2010:sbzk}]{carollo:2001, lilly:2003, kobulnicky:2004, savaglio:2005, erb:2006:metallicity,
  maiolino:2008, mannucci:2009, yoshikawa:2010, yabe:2012, yabe:2014, henry:2013,
  cullen:2014, maier:2014, masters:2014, steidel:2014, troncoso:2014, zahid:2014, wuyts:2014, sanders:2015:metal}.

The MS and the MZR are likely to be closely related to  each other. 
There is indeed some evidence that the metallicity depends also on SFR 
as a second parameter in the MZR,  leading to a  a \zmsfr relation \citep{ellison:2008, mannucci:2010, laralopez:2010}.
These authors  found that local SFGs lie on a thin 2 dimensional (2D) surface
in the 3D $M$--$Z$--SFR
space, with  a  strikingly small scatter, only 0.05 dex, in metallicity.
This surface, dubbed the  ``fundamental metallicity relation'' (FMR),
has been suggested to accommodate SFGs up to $z \simeq 2.5$ \citep[][but see \citealt{wuyts:2014}]{mannucci:2010, belli:2013}, 
even as the typical SFR at a given stellar mass is higher  by a factor of $\sim 20$.

The presence of SFR as a second parameter in the MZR follows quite  naturally 
if star formation in galaxies is regulated by their gas reservoir \citep{bouche:2010,dave:2012,krumholz:2012,dayal:2013,lilly:2013,forbes:2014}.
A redshift-independent \zmsfr relation is theoretically predicted
if the parameters of the regulator,
i.e., the mass-loading of the wind outflow and the star-formation efficiency,
are roughly constant in time.
Such a gas-regulated model also links the decline of the sSFR in galaxies to the decline of the
specific accretion rate of dark matter haloes
\citep[with an offset of about a factor of two;][]{lilly:2013}.
In this gas-regulation model the gas consumption timescale,
that is observationally estimated to be $\lsim 1$ Gyr for MS galaxies
\citep[e.g.,][]{daddi:2010:co, genzel:2010}, 
is shorter than the mass increase timescale, i.e., $\text{sSFR}^{-1}$,
so that a quasi-steady-state is maintained between inflow, star formation and outflow.
However, at very high redshifts, $z\gtrsim3$,  the gas-regulator may break down as  a number of timescales tend to converge,  namely the gas depletion
timescale, the halo dynamical time, and the mass increase timescale of galaxies.
An important diagnostics of this break down would may come from  changes in the \zmsfr relation.

Some indication that the fundamental \zmsfr relation breaks down at $z>2.5$ has indeed been reported \citep{mannucci:2010, troncoso:2014}.
However, the samples studied so far at $z>2.5$ do not include the most massive galaxies, i.e., $M \simeq 10^{11}M_\odot$
despite the claim that the evolution of the \zmsfr relation may be  more prominent at high stellar mass regime \citep[][but see \citealt{zahid:2014}]{troncoso:2014}. 
Also, this  depends on how the local \zmsfr relation is extrapolated into regions of $(M_\star,\text{SFR})$ 
that are poorly populated at low redshifts \citep[i.e., high stellar mass and high SFR;][]{maier:2014}.

Therefore, firmly establishing whether,  and with which functional form,  the  \zmsfr relation extends beyond  $z\sim3$
could provide an important clue on the regulation of star formation in galaxies at such earlier epochs.
In order to properly assess the existence and form of a \zmsfr relation at $z \gtrsim 3$, 
it is essential to measure gas metallicities for a large number of galaxies
spanning a broad range of stellar mass and SFR, including also a fair number of the most massive objects.

In this study, we use the Multi-Object Spectrograph for Infra-Red Exploration \citep[MOSFIRE;][]{mclean:2010:mosfire,mclean:2012:mosfire}
on the Keck I telescope 
to obtain rest-frame optical emission lines that enable us to measure the gas-phase oxygen abundance, \ohmetal,
for a sample of SFGs at $3\lesssim z \lesssim 3.7$.

Compared to the previous study  of \citet{troncoso:2014}, our sample and data offer several advantages:
galaxies are selected in a single field, i.e., the COSMOS field,
enabling us to use homogeneous multi-band photometry; 
our sample includes more objects with measured metallicity and 
extends to $M_\star \simeq 10^{11}\,M_\odot$;
resolving the  \oiitot doublet allows  us to estimate the electron density of the ionized gas.
Using the measured metallicity and wealth of multi-wavelength dataset,
we then study the \zmsfr relation at the critical epoch to test the gas-regulator model of star formation in galaxies.

In \autorefsec{sec:data}, we introduce our data used in this study.
Basic measurements of emission lines, AGN contamination, SED fitting, SFR, and spectral stacking procedure
are presented in \autorefsec{sec:measurement}, 
and measurements of the ionized gas properties are reported in \autorefsec{sec:gasprop}.
In \autorefsec{sec:discussion}, we discuss the relation between the ionized gas properties
and other galaxy properties such as stellar mass and SFR.
We summarize our results in \autorefsec{sec:summary}.

Throughout the analysis, we adopt a $\Lambda$-dominated cold dark matter ($\Lambda$CDM) cosmology
with cosmological parameters of
$H_0=70\kms\,\mathrm{Mpc}^{-1}$, $\Omega_{\rm m}=0.3$, and $\Omega_\Lambda=0.7$
and AB magnitude system \citep{oke:1983}.
Unless explicitly stated, we refer ``metallicity'' and ``$Z$'' with
the same meaning as ``gas-phase oxygen abundance'' or ``\ohmetal''
and we use a base 10 for logarithm.

\section{Data}
\label{sec:data}

\subsection{Sample selection}

Our primary goal of the project is to measure metallicity at $z \gtrsim 3$ using strong rest-frame optical emission lines
such as \oiitot, \hbeta, and \oiiitot.
At $3<z<3.8$, \oiitot are observed in \textit{H}-band and \hbeta and \oiiitot can be accessible in \textit{K}-band.

Our primary sample has been extracted from zCOSMOS-Deep redshift catalog \citep[][and in preparation]{lilly:2007}.
We have first selected objects in the redshift range $3<z_\text{spec}<3.8$
which are spectroscopically classified as galaxies (i.e., neither stars nor broad-line AGN)
and the redshifts are reliable with confidence classes (CC) in the range of $2.5 \le \text{CC} \le 4.5$
which includes some insecure spectroscopic redshifts but which agree well with photometric reshifts \citep[see Table 1 in][]{lilly:2009}.
Then we have given higher priority for those with the expected line positions of \oii, \hbeta, and \oiiitwo
being away more than 2 MOSFIRE resolution element (i.e., $R\sim3600$) from the nearest OH line.
Since it is expected that spectroscopic redshifts at $z\sim3$ in the optical spectra have been determined primarily by the presence of \lya{},
and a velocity offset in \lya relative to the systemic velocity is also expected \citep{erb:2014,finkelstein:2011,hashimoto:2013,mclinden:2011},
we gave lower priority for those with some degree of OH contamination instead of fully excluding them from the sample.

The number density of the zCOSMOS-Deep selected objects is $\lesssim 5$ per MOSFIRE FoV,
so we filled the remaining multiplex with various classes of galaxies in the 2014 run which will be presented elsewhere.
On the other hand, in the 2015 run, star-forming galaxies at $3<z_\text{phot}<3.8$ have been selected
from the 30 band COSMOS photometric redshift catalog on UltraVISTA photometry \citep{mccracken:2012,ilbert:2013}.
We have further selected objects with predicted \hbeta flux of $>5\times 10^{-18}\,\text{erg}\,\text{s}^{-1}\,\text{cm}^{-2}$
scaled from SFR in the catalog to \halpha flux by using a conversion of \citet{kennicutt:2012},
assuming an intrinsic \halpha/\hbeta ratio of 2.86,
and applying dust extinction by using $E(B-V)_\text{star}$ from the catalog with the Calzetti extinction law \citep{calzetti:2000}.
Conservatively, we adopt $E(B-V)_\text{gas}=E(B-V)_\text{star}/0.44$ extinction in \hbeta following the original recipe of \citet{calzetti:2000},
though some studies suggest the factor appears to be closer to unity at high redshifts \citep[e.g.,][]{kashino:2013,pannella:2015}.
A MOSFIRE FoV typically contains $\sim25$ these photo-$z$ objects, which is enough to fill the multiplex.

Finally, we observed 54 objects, one of which was observed in both observing runs,
with 43 of them having robust spectroscopic identification by one or more emission lines in the MOSFIRE spectra.
\autoref{tab:objprop} summarizes our sample of 43 objects with detected emission lines.

\capstartfalse
\begin{deluxetable*}{cccccccccc}
  \tablecaption{Properties of the galaxy sample and the observervation log.\label{tab:objprop}}
  \tablehead{
    \colhead{ID} &
    \colhead{R.A.} &
    \colhead{Decl.} &
    \colhead{$z_\text{phot}$} &
    \colhead{$z_\text{zCOSMOS}$} &
    \colhead{Run} &
    \colhead{$t_{\text{exp},\mathit{H}}$} &
    \colhead{$t_{\text{exp},\mathit{K}}$} &
    \colhead{\textit{B}$_\text{AB}$} &
    \colhead{\textit{Ks}$_\text{AB}$\tablenotemark{a}}
    \\
    \colhead{}       &
    \colhead{(deg)}  &
    \colhead{(deg)}  &
    \colhead{}       &
    \colhead{}       &
    \colhead{}       &
    \colhead{(min)}  &
    \colhead{(min)}  &
    \colhead{(mag)}  &
    \colhead{(mag)}
    \\
    \colhead{(1)} &
    \colhead{(2)} &
    \colhead{(3)} &
    \colhead{(4)} &
    \colhead{(5)} &
    \colhead{(6)} &
    \colhead{(7)} &
    \colhead{(8)} &
    \colhead{(9)} &
    \colhead{(10)}
  }
\startdata
434625 & 150.09385 & 2.421970 & 3.155 &  3.0717 & 2014 & 28 & 30 & 25.29 & 25.38 \\
413136 & 150.10600 & 2.436190 & 2.838 &  3.1911 & 2014 & 28 & 30 & 24.91 & 22.99 \\
413646 & 150.08025 & 2.469180 & 3.132 &  3.0389 & 2014 & 28 & 30 & 25.23 & 23.70 \\
413453 & 150.13365 & 2.457430 & 3.121 &  3.1892 & 2014 & 28 & 30 & 25.07 & 22.24 \\
434585 & 149.84702 & 2.373020 & 3.435 &  3.3557 & 2014 & 28 & 30 & 24.66 & 22.82 \\
434571 & 149.83784 & 2.362050 & 3.142 &  3.1292 & 2014 & 28 & 30 & 25.77 & 24.36 \\
413391 & 149.78424 & 2.452890 & 3.469 &  3.3626 & 2014 & 28 & 30 & 24.78 & 22.44 \\
427122 & 150.07483 & 2.032280 & 2.917 &  3.0454 & 2014 & 28 & 30 & 25.14 & 24.64 \\
434148 & 150.03551 & 1.965090 & 3.144 &  3.1457 & 2014 & 28 & 30 & 25.13 & 23.63 \\
434082 & 150.35009 & 1.904380 & 3.449 &  3.4750 & 2014 & 28 & 42 & 24.34 & 22.65 \\
434126 & 150.37178 & 1.947200 & 3.427 &  3.2715 & 2014 & 28 & 42 & 25.35 & 23.76 \\
434139 & 150.38895 & 1.956150 & 3.108 &  3.0355 & 2014 & 28 & 42 & 24.86 & 23.74 \\
434145 & 150.40089 & 1.963460 & 3.442 &  3.4208 & 2014 & 28 & 42 & 25.35 & 23.05 \\
434242 & 150.29841 & 2.075330 & 3.271 &  3.2376 & 2014 & 28 & 30 & 25.74 & 24.04 \\
406390 & 150.29878 & 2.068820 & 3.149 &  3.0487 & 2014 & 28 & 30 & 25.05 & 23.41 \\
406444 & 150.33032 & 2.072270 & 3.400 &  3.3060 & 2014 & 28 & 30 & 24.45 & 21.71 \\
434227 & 150.32680 & 2.053940 & 3.355 &  3.2136 & 2014 & 28 & 30 & 24.71 & 22.57 \\
191932 & 150.27605 & 2.299900 & 3.414 & \nodata & 2015 & 56 & 72 & 26.03 & 23.41 \\
434547 & 150.27003 & 2.333980 & 3.261 &  3.1875 & 2015 & 56 & 72 & 25.49 & 23.54 \\
192129 & 150.30078 & 2.300540 & 3.457 &  3.5002 & 2015 & 56 & 72 & 24.83 & 22.89 \\
193914 & 150.31923 & 2.306690 & 3.023 & \nodata & 2015 & 56 & 72 & 24.65 & 23.43 \\
212863 & 150.29286 & 2.367090 & 3.372 & \nodata & 2015 & 56 & 72 & 26.14 & 23.69 \\
195044 & 150.32696 & 2.310780 & 3.185 &  3.1149 & 2015 & 56 & 72 & 25.14 & 23.38 \\
208115 & 150.31325 & 2.351720 & 3.661 & \nodata & 2015 & 56 & 72 & 25.71 & 22.52 \\
200355 & 150.33103 & 2.327790 & 3.476 & \nodata & 2015 & 56 & 72 & 25.87 & 23.48 \\
214339 & 150.31607 & 2.372240 & 3.392 & \nodata & 2015 & 56 & 72 & 25.94 & 22.44 \\
411078 & 150.35142 & 2.322420 & 3.106 &  3.1104 & 2015 & 56 & 72 & 23.84 & 22.44 \\
212298 & 150.34268 & 2.365390 & 3.107 &  3.1024 & 2015 & 56 & 72 & 25.26 & 22.79 \\
412808 & 149.83413 & 2.416690 & 3.274 &  3.3050 & 2015 & 72 & 84 & 25.05 & 24.13 \\
223954 & 149.83188 & 2.404150 & 3.436 & \nodata & 2015 & 72 & 84 & 25.62 & 23.75 \\
220771 & 149.83628 & 2.393190 & 3.473 & \nodata & 2015 & 72 & 84 & 26.63 & 24.64 \\
219315 & 149.83952 & 2.388460 & 3.240 & \nodata & 2015 & 72 & 84 & 25.72 & 23.90 \\
434618 & 149.89213 & 2.414710 & 3.289 &  3.2819 & 2015 & 72 & 84 & 25.18 & 23.24 \\
221039 & 149.86938 & 2.394510 & 3.069 & \nodata & 2015 & 72 & 84 & 25.00 & 23.05 \\
215511 & 149.84826 & 2.376170 & 3.424 & \nodata & 2015 & 72 & 84 & 26.33 & 23.92 \\
217597 & 149.86534 & 2.382770 & 3.415 & \nodata & 2015 & 72 & 84 & 26.76 & 24.06 \\
211934 & 149.84725 & 2.364040 & 3.773 & \nodata & 2015 & 72 & 84 & 28.08 & 24.44 \\
434579 & 149.86179 & 2.366931 & 3.145 &  3.1889 & 2015 & 72 & 84 & 26.16 & 23.87 \\
217753 & 149.89451 & 2.383700 & 3.438 & \nodata & 2015 & 72 & 84 & 26.51 & 23.06 \\
218783 & 149.92082 & 2.387060 & 3.504 & \nodata & 2015 & 72 & 84 & 25.57 & 22.80 \\
217090 & 149.91827 & 2.381250 & 3.669 & \nodata & 2015 & 72 & 84 & 26.21 & 24.41 \\
210037 & 149.90451 & 2.357800 & 3.579 & \nodata & 2015 & 72 & 84 & 27.06 & 24.33 \\
208681 & 149.90551 & 2.353990 & 3.342 &  3.2671 & 2015 & 72 & 84 & 24.20 & 22.17
\enddata
\tablecomments{
(1) Object ID;
(2) Right ascention;
(3) Declination;
(4) Photometric redsfhit from \citet{ilbert:2013};
(5) Spectroscopic redshift from zCOSMOS-Deep;
(6) Observing run;
(7) Exposure time in \textit{H}-band;
(8) Exposure time in \textit{Ks}-band;
(9) \textit{B}-band total magnitude;
(10) \textit{Ks}-band total magnitude.
}
\tablenotetext{a}{\textit{Ks} magnitudes are not corrected for emission lines.}
\end{deluxetable*}
\capstarttrue

\subsection{Observation and data reduction}

Observation has been carried out on 20--22 Jan 2014 and 15 Jan 2015 using MOSFIRE on Keck-I \citep{mclean:2010:mosfire,mclean:2012:mosfire}.
We used $1''$ and $0\farcs7$ slit width in 2014 and 2015 runs, respectively,
which provides instrumental resolution of $R\simeq2500$ and $3600$, respectively. 
We observed in \textit{J}, \textit{H}, and \textit{K} gratings in the 2014 run
because there are some lower redshift objects for which \oii falls into \textit{J}-band, 
while only \textit{H} and \textit{K} gratings are used in the 2015 run.
Following the exposure recommendation, we used 120 s, 120 s, and 180 s per exposure
in \textit{J}, \textit{H}, and \textit{K} band, respectively.
Exposure was done in a sequence of either AB or ABBA dithering with a distance between A and B positions of $2\farcs5$.
The Observing run and total exposure time for each object in \textit{H} and \textit{K} bands are listed in \autoref{tab:objprop}.

We observed a couple of white dwarfs and A0V stars per night, using them as standard stars for flux calibration.

Data were reduced with the MOSFIRE data reduction pipeline version 1.1\footnote{\url{https://keck-datareductionpipelines.github.io/MosfireDRP}}
for the science frames and its \texttt{Longslit} branch to make the standard star reduction consistent with the science frames.
The pipeline performs flat-fielding, wavelength calibration, sky subtraction, rectification and coaddition of each exposure, 
and produces rectified two-dimensional (2D) spectra with associated noise as well as exposure maps.
One-dimensional (1D) spectra were extracted with $0\farcs7$--$1''$ aperture depending on the spatial extent of detected emission lines
to maximize the signal-to-noise ratio (S/N).
Corresponding 1D noise spectra were also extracted from 2D noise spectra by using the same aperture and summing them up in quadrature. 
Flux calibration was carried out using the standard star closest in time to the corresponding science exposure.
At the same time with the telluric correction, we also carried out absolute flux calibration 
by scaling the observed standard star spectra to 2MASS magnitudes correcting for the slit loss for a point source.

\section{Basic Measurement}
\label{sec:measurement}

In this section, we describe the measurements of emission line fluxes, stellar population properties,
dust extinction, and SFR.
We also present two AGN in our sample, and spectral stacking in bins of stellar mass and SFR.

Among the 54 observed objects at $3 \lesssim z \lesssim 3.8$,
43 show clear detection of emission lines which allow us to measure the line properties,
while 11 objects show either non-detection or very faint spectral features with too low S/N,
so that we are not able to claim a detection.
Among the 11 objects with non-detection, 2 objects were selected from the zCOSMOS-Deep catalog.
One of them was in a slit which did not work properly,
and the other is a filler object with $\text{CC}=2.1$
and its photometric redshift $z_\text{phot}=0.53$
differs substantially from the spectroscopic one.
The rest of the objects with non-detection of emission lines were photometrically selected.
We speculate that the main reasons of non-detections could be either wrong photometric redshifts
or wrong emission line flux predictions propagated from the best-fit SED parameters. 
Also, in some of these objects emission lines may have been obliterated by the OH air glow.
In the following analysis, we focus on the 43 objects with detected emission lines.

\subsection{Emission line measurement}

The emission lines are fit in two steps.
In the first step, \oiione, \oiitwo, \hbeta, \oiiione, and \oiiitwo are fit simultaneously. 
We assume that each emission line can be described by a simple Gaussian redshifted by the same amount 
with a common velocity dispersion $\sigma_\text{vel}$ on a constant continuum component.
Therefore, free parameters of the first fitting step are redshift, velocity dispersion,
flux of each emission line, and continuum component.
The continuum is assumed to be a constant within each band,
i.e., \oiione and \oiitwo have the same continuum in \textit{H}-band
and so do \hbeta, \oiiione, and \oiiitwo in \textit{K}-band.
In the second step, \neiiione, \height, \neiiitwo+\hepsilon, \hdelta, and \hgamma are fit 
individually by fixing the redshift and line width derived in the first path
and leaving the line flux and constant continuum as free parameters.
We used \texttt{MPFIT}\footnote{\url{http://www.physics.wisc.edu/~craigm/idl/fitting.html}} \citep{markwardt:2009} for the fitting.
During the procedure, we put constraints on $\sigma_\text{vel}$ and emission line fluxes
to be larger than the instrumental resolution and to be positive, respectively.

To obtain each parameter and the associated uncertainty,
we carried out a Monte Carlo simulation
by perturbing each pixel value of the spectra with
the associated noise multiplied by a normally distributed random number.
Mean and standard deviation of the measured distribution of $10^3$ realizations 
are adopted for each parameter and its $1\sigma$ uncertainty ($\sigma_\text{MC}$), respectively. 
The best-fit spectra and observed 1-dimensional spectra are shown in \autoref{fig:spec_begin}.
We computed another $1\sigma$ error ($\sigma_\text{direct}$) for each emission line flux
directly from the associated noise spectrum
by integrating $\pm 2\sigma_\text{vel}$ from the line center in quadrature.
In the case of $F_\text{MC}/\sigma_\text{direct}>3$, where $F_\text{MC}$ is flux measured by the Monte Carlo simulation,
we claim the detection of the emission line
and use $\sigma_\text{MC}$ as the corresponding $1\sigma$ error.
Otherwise, we adopt $3\sigma_\text{direct}$ as the $3\sigma$ upper limit. 

Measured \hbeta fluxes were corrected for underlying stellar absorption
by assuming $\text{EW}(\hbeta)=2$ \AA{} \citep{nakamura:2004}.
To compute EW, the continuum flux was estimated from the  total \textit{Ks}-band magnitude. 
The resulting correction factor is up to $\simeq 20\%$ with a median of $3\%$.

\autoref{tab:lineflux} lists measured redshifts, reddening uncorrected emission line fluxes,
and correction factors applied to stellar \hbeta absorption.

\capstartfalse
\begin{deluxetable*}{cccccccccc}
    \tablecaption{Emission line measurements.\label{tab:lineflux}}
    \tablehead{
      \colhead{ID} &
      \colhead{$z_\mathrm{MOSFIRE}$} &
      \colhead{$F(\oii)$} &
      \colhead{$F(\neiiione)$} &
      \colhead{$F(\hbeta)$\tablenotemark{a}} &
      \colhead{$F(\oiiione)$} &
      \colhead{$F(\oiiitwo)$} &
      \colhead{$f_\text{corr}(\hbeta)$} &
      \colhead{$f_{\text{em},\mathit{H}}$} &
      \colhead{$f_{\text{em},\mathit{K}}$} 
      \\
      \colhead{(1)} &
      \colhead{(2)} &
      \colhead{(3)} &
      \colhead{(4)} &
      \colhead{(5)} &
      \colhead{(6)} &
      \colhead{(7)} &
      \colhead{(8)} &
      \colhead{(9)} &
      \colhead{(10)}
    }
\startdata
434625  &  3.07948  &  $2.41 \pm  0.27$  &  $< 0.88        $  &  $<1.40         $  &  $1.70 \pm  0.18$  &  $5.28 \pm  0.20$  &  1.00   &  0.07  &  1.00  \\
413136  &  3.19037  &  $< 3.56        $  &  $< 1.21        $  &  $1.17 \pm  0.31$  &  $< 0.71        $  &  $1.84 \pm  0.31$  &  1.10   &  0.00  &  0.08  \\
413646  &  3.04045  &  $< 6.75        $  &  $< 2.31        $  &  \nodata           &  $< 3.61        $  &  $3.95 \pm  0.31$  &  1.00   &  0.00  &  0.21  \\
413453  &  3.18824  &  $< 4.84        $  &  $< 3.05        $  &  $< 2.64        $  &  $2.89 \pm  0.31$  &  $10.27 \pm 0.35$  &  1.00   &  0.00  &  0.14  \\
434585  &  3.36345  &  $2.94 \pm  1.23$  &  $< 3.69        $  &  $1.56 \pm  0.46$  &  $< 1.14        $  &  $3.77 \pm  0.96$  &  1.09   &  0.07  &  0.12  \\
434571  &  3.12729  &  $< 3.57        $  &  $< 4.23        $  &  \nodata           &  $1.24 \pm  0.11$  &  $3.03 \pm  0.19$  &  1.00   &  0.00  &  0.30  \\
413391  &  3.36544  &  $3.16 \pm  0.47$  &  $< 3.71        $  &  $0.91 \pm  0.26$  &  $2.22 \pm  0.24$  &  $4.84 \pm  0.69$  &  1.22   &  0.04  &  0.09  \\
427122  &  3.04157  &  $< 2.24        $  &  $< 0.47        $  &  $0.44 \pm  0.13$  &  $< 0.66        $  &  $1.90 \pm  0.14$  &  1.04   &  0.00  &  0.29  \\
434148  &  3.14780  &  $1.29 \pm  0.24$  &  $< 0.30        $  &  $< 0.51        $  &  $< 0.61        $  &  $2.26 \pm  0.10$  &  1.00   &  0.04  &  0.11  \\
434082  &  3.46860  &  $5.84 \pm  0.38$  &  \nodata           &  $4.33 \pm  0.33$  &  $6.97 \pm  0.41$  &  $20.96 \pm 0.51$  &  1.02   &  0.08  &  0.49  \\
434126  &  3.27209  &  $2.90 \pm  0.40$  &  $< 0.87        $  &  $1.19 \pm  0.23$  &  $2.14 \pm  0.34$  &  $7.93 \pm  0.36$  &  1.03   &  0.10  &  0.50  \\
434139  &  3.03623  &  $< 7.31        $  &  $< 0.79        $  &  $< 1.83        $  &  $< 3.41        $  &  $6.25 \pm  0.42$  &  1.02   &  0.00  &  0.35  \\
434145  &  3.42400  &  $5.15 \pm  0.77$  &  $< 4.47        $  &  $< 3.23        $  &  $3.65 \pm  0.48$  &  $11.91 \pm 0.48$  &  1.02   &  0.09  &  0.35  \\
434242  &  3.23290  &  $1.44 \pm  0.34$  &  $< 1.44        $  &  $1.37 \pm  0.34$  &  $2.16 \pm  0.18$  &  $6.98 \pm  0.33$  &  1.01   &  0.28  &  0.59  \\
406390  &  3.05272  &  $2.16 \pm  0.57$  &  $< 0.56        $  &  $< 2.31        $  &  $< 1.39        $  &  $4.35 \pm  0.48$  &  1.00   &  0.06  &  0.18  \\
406444  &  3.30355  &  $5.15 \pm  0.80$  &  $1.27 \pm  0.27$  &  $2.86 \pm  0.34$  &  $4.27 \pm  0.38$  &  $9.66 \pm  0.73$  &  1.13   &  0.01  &  0.10  \\
434227  &  3.21501  &  $< 2.54        $  &  $< 2.40        $  &  $< 2.06        $  &  $< 2.50        $  &  $5.56 \pm  1.05$  &  1.00   &  0.00  &  0.10  \\
191932  &  3.18246  &  $1.50 \pm  0.12$  &  $< 0.52        $  &  $< 0.57        $  &  $0.92 \pm  0.10$  &  $2.69 \pm  0.11$  &  1.14   &  0.05  &  0.11  \\
434547  &  3.19167  &  $1.33 \pm  0.10$  &  $< 0.23        $  &  $0.40 \pm  0.05$  &  $0.90 \pm  0.05$  &  $2.81 \pm  0.07$  &  1.16   &  0.04  &  0.14  \\
192129  &  3.49466  &  $1.82 \pm  0.14$  &  \nodata           &  $1.25 \pm  0.12$  &  $1.44 \pm  0.14$  &  $3.61 \pm  0.19$  &  1.10   &  0.04  &  0.12  \\
193914  &  2.96871  &  $1.76 \pm  0.28$  &  $0.52 \pm  0.06$  &  \nodata           &  $2.01 \pm  0.11$  &  $6.45 \pm  0.09$  &  1.00   &  0.04  &  0.27  \\
212863  &  3.29183  &  $1.83 \pm  0.08$  &  $< 0.25        $  &  $1.29 \pm  0.11$  &  $2.05 \pm  0.10$  &  $5.46 \pm  0.13$  &  1.03   &  0.25  &  0.34  \\
195044  &  3.11091  &  $1.42 \pm  0.21$  &  $< 0.15        $  &  $0.30 \pm  0.11$  &  $0.26 \pm  0.08$  &  $0.76 \pm  0.10$  &  1.00   &  0.03  &  0.04  \\
208115  &  3.63568  &  $1.49 \pm  0.13$  &  $3.70 \pm  0.10$  &  $6.78 \pm  0.16$  &  $17.99 \pm 0.25$  &  $53.21 \pm 0.30$  &  1.00   &  0.06  &  1.00  \\
200355  &  3.56796  &  $1.71 \pm  0.10$  &  \nodata           &  $1.34 \pm  0.15$  &  $2.94 \pm  0.17$  &  $7.47 \pm  0.27$  &  1.04   &  0.07  &  0.37  \\
214339  &  3.60885  &  $1.59 \pm  0.20$  &  $< 0.43        $  &  $0.83 \pm  0.18$  &  $1.34 \pm  0.25$  &  $3.98 \pm  0.31$  &  1.27   &  0.04  &  0.08  \\
411078  &  3.11280  &  $6.07 \pm  0.14$  &  $1.99 \pm  0.07$  &  $4.23 \pm  0.20$  &  $10.26 \pm 0.10$  &  $29.80 \pm 0.13$  &  1.02   &  0.10  &  0.55  \\
212298  &  3.10781  &  $3.97 \pm  0.20$  &  $0.65 \pm  0.10$  &  $2.25 \pm  0.11$  &  $2.54 \pm  0.09$  &  $7.18 \pm  0.24$  &  1.05   &  0.04  &  0.21  \\
412808  &  3.30614  &  $1.48 \pm  0.09$  &  $0.57 \pm  0.06$  &  $1.04 \pm  0.08$  &  $2.25 \pm  0.11$  &  $7.63 \pm  0.13$  &  1.01   &  0.06  &  0.67  \\
223954  &  3.37076  &  $1.72 \pm  0.11$  &  $< 0.61        $  &  $0.59 \pm  0.13$  &  $1.30 \pm  0.12$  &  $3.38 \pm  0.14$  &  1.09   &  0.06  &  0.21  \\
220771  &  3.35871  &  $0.77 \pm  0.10$  &  $0.20 \pm  0.04$  &  $0.37 \pm  0.08$  &  $< 0.24        $  &  $1.14 \pm  0.10$  &  1.06   &  0.79  &  0.18  \\
219315  &  3.36253  &  $0.96 \pm  0.09$  &  $0.21 \pm  0.05$  &  $< 0.23        $  &  $0.75 \pm  0.12$  &  $1.61 \pm  0.10$  &  1.00   &  0.05  &  0.10  \\
434618  &  3.28463  &  $2.45 \pm  0.09$  &  $< 0.26        $  &  $0.98 \pm  0.08$  &  $1.72 \pm  0.14$  &  $5.18 \pm  0.09$  &  1.08   &  0.06  &  0.21  \\
221039  &  3.05515  &  $4.67 \pm  0.15$  &  $0.66 \pm  0.05$  &  $1.73 \pm  0.08$  &  $3.00 \pm  0.24$  &  $8.49 \pm  0.10$  &  1.05   &  0.10  &  0.29  \\
215511  &  3.36345  &  $2.44 \pm  0.17$  &  $0.40 \pm  0.08$  &  $1.12 \pm  0.13$  &  $2.32 \pm  0.16$  &  $5.76 \pm  0.20$  &  1.03   &  0.10  &  0.43  \\
217597  &  3.28377  &  $1.69 \pm  0.09$  &  $0.24 \pm  0.04$  &  $0.52 \pm  0.10$  &  $1.39 \pm  0.18$  &  $3.77 \pm  0.12$  &  1.06   &  0.09  &  0.31  \\
211934  &  3.35538  &  $1.32 \pm  0.28$  &  $< 0.09        $  &  $0.53 \pm  0.16$  &  $0.43 \pm  0.11$  &  $1.63 \pm  0.16$  &  1.05   &  0.23  &  0.22  \\
434579  &  3.18653  &  $0.82 \pm  0.09$  &  $< 0.34        $  &  $0.23 \pm  0.07$  &  $0.54 \pm  0.08$  &  $1.28 \pm  0.09$  &  1.22   &  0.05  &  0.09  \\
217753  &  3.25408  &  $2.38 \pm  0.45$  &  $< 0.54        $  &  $1.26 \pm  0.21$  &  $< 0.80        $  &  $1.82 \pm  0.18$  &  1.09   &  0.06  &  0.08  \\
218783  &  3.29703  &  $1.91 \pm  0.16$  &  $< 0.30        $  &  $0.75 \pm  0.10$  &  $0.87 \pm  0.12$  &  $2.53 \pm  0.22$  &  1.19   &  0.05  &  0.07  \\
217090  &  3.69253  &  $1.69 \pm  0.16$  &  \nodata           &  $1.65 \pm  0.19$  &  $3.25 \pm  0.21$  &  $9.10 \pm  0.36$  &  1.00   &  0.17  &  1.00  \\
210037  &  3.69083  &  $2.07 \pm  0.43$  &  \nodata           &  $< 0.61        $  &  $1.10 \pm  0.35$  &  $4.10 \pm  0.77$  &  1.00   &  0.48  &  0.39  \\
208681  &  3.26734  &  $3.58 \pm  0.30$  &  $< 0.91        $  &  $2.34 \pm  0.16$  &  $1.04 \pm  0.17$  &  $3.31 \pm  0.19$  &  1.11   &  0.05  &  0.07
\enddata
\tablecomments{
  (1) Object ID;
  (2) Spectroscopic redshift measured from MOSFIRE spectra. The associated $1\sigma$ error is typicall an order of $\lesssim 10^{-4}$;
  (3) \oii (i.e., $\oiione + \oiitwo$) flux;
  (4) \neiiione flux; (5) \hbeta flux; (6) \oiiione flux; (7) \oiiitwo flux;
  (8) Correction factor for \hbeta absorption assuming 2 \AA{} in the equivalent width;
  (9) and (10) fraction of emission line contribution in \textit{H}- and \textit{K}-band, respectively.
  All fluxes are in unit of $10^{-17}$ erg s$^{-1}$ cm$^{-2}$, and they are not corrected for dust extinction.
  Quoted upper limits are $3\sigma$ upper limit.
}
\tablenotetext{a}{
  Fuxes are not corrected for the underlying stellar absorption.
  To correct it, one need to multipy $F(\hbeta)$ by $f_\text{corr}(\hbeta)$.
}
\end{deluxetable*}
\capstarttrue

\begin{figure*}
  \centerline{
    \includegraphics[width=0.9\linewidth]{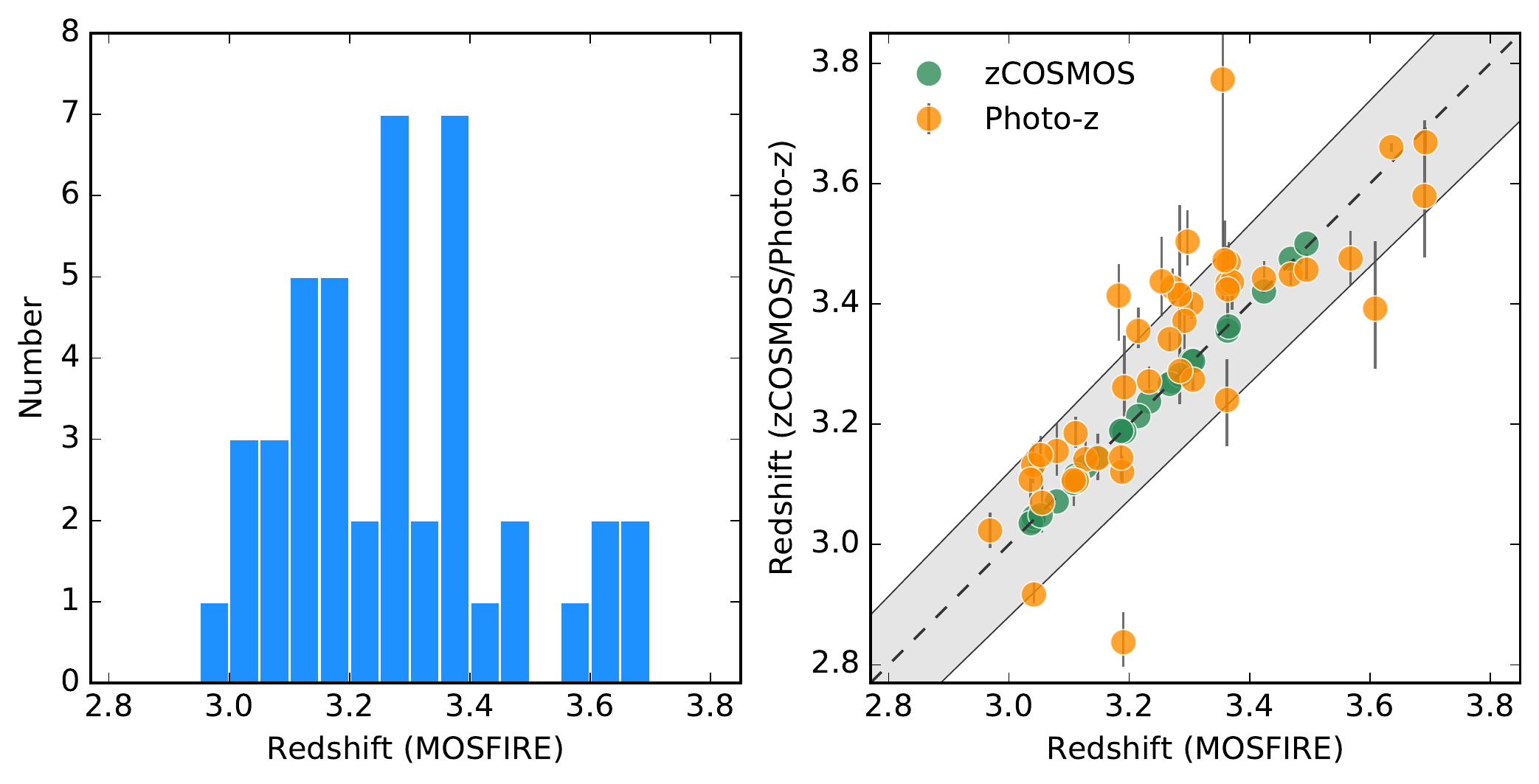}
  }
  \caption{%
    \textit{Left}: MOSFIRE redshift distribution of the sample.
    \textit{Right}: Comparison of spectroscopic redshifts measured in MOSFIRE spectra with those from zCOSMOS-Deep and
    photometric redshift from \citet{ilbert:2013}. 
  \label{fig:zcomp}}
\end{figure*}

In \autoref{fig:zcomp}, the MOSFIRE spectroscopic redshifts are compared with those from zCOSMOS-Deep or with photometric redshifts.
Agreement between MOSFIRE and zCOSMOS-Deep spectroscopic redshifts is
almost perfect with a standard deviation of 0.004 in $(z_\text{MOSFIRE}-z_\text{zCOSMOS})$.
Photometric redshifts also agree well with the spectroscopic ones with no catastrophic failure. 
The normalized median absolute deviation of $(z_\text{MOSFIRE}-z_\text{phot})/(1+z_\text{MOSFIRE})$
is 0.027 which is consistent with what reported in \citet{ilbert:2013} for high redshift galaxies.
The redshifts of our sample are in a range of $2.97<z<3.69$, with a median of 3.27.

\subsection{Note on AGN contamination}

There are two AGN candidates in the sample which are excluded in the following analysis as nuclear activity
is not the main focus of this paper. 
The object, 413453, has an X-ray counterpart detected in
the \textit{Chandra}-COSMOS survey \citep[CID 3636 in][]{civano:2011}
as well as in the more recent \textit{Chandra} COSMOS Legacy Survey \citep{marchesi:2016,civano:2016}. 
Although the object is not classified as a broad-line AGN in the zCOSMOS-Deep catalog,
it actually shows \ion{C}{4}$\lambda1549$ emission in the optical spectra.
Therefore, the \oiii and \hbeta emission lines are likely to be contaminated by nuclear activity.

Object 208115 is not detected in X-ray
but it shows very strong \neiiione emission with $\neiiione/\oii\simeq2.5$ and $\oiiidirect/\hgamma\simeq0.7$,
which appears to be due to AGN rather than star formation.
The median values of $\oiiidirect/\hgamma$ for the local star-forming galaxies and AGN sample presented by \citet{shirazi:2012}
are $0.13$ and $0.39$, respectively (M.~Shirazi, private communication;
see also \citealt{francis:1991,vandenberk:2001}).

\subsection{Stellar population properties}

In this section, we  carry out the spectral energy distribution (SED) fitting to the broad-band photometry.
We started from the PSF-homogenized UltraVISTA photometry catalog \citep{mccracken:2012}\footnote{\url{http://terapix.iap.fr/article.php?id_article=844}}.
Since our objects are in a narrow redshift range and have small angular extent  (\autoref{fig:hstimg}),
we adopted  \textit{uBVrizYJHK}  aperture magnitudes measured within $2''$ apertures.
These magnitudes were first corrected for the Galactic extinction based on
the calibration of dust map by \citet{schalafly:2011} and a extinction curve by \citet{fitzpatrick:1999} with $R_V=3.1$.
Then the aperture correction to the total magnitude was made by applying an average difference between
aperture and \texttt{AUTO} magnitudes across the used photometric bands.
Finally, 0.369 mag was subtracted from \textit{B}-band magnitude
as instructed in the \texttt{README} file of the catalog%
\footnote{
  This offset was introduced to convert the photometric zero-point in \citet{capak:2007}
  derived by using spectrophotometric standard stars to
  that based on sources with flat spectrum such as moderate redshift galaxies
  (\citealp{ilbert:2009,ilbert:2013}; Laigle et al., in preparation).
  The former zero-point suffered from the uncertainty
  caused by the combination of sharp Balmer absorption features of the calibration stars
  and the location of the blue edge of the filter which is sensitive to
  temperature and humidity, while the latter is less sensitive to exact knowledge of the bandpass.
}. %
Since one of the objects (434579) does not have a counterpart in the UltraVISTA catalog,
but found in the previous CFHT/WIRC \textit{K}-selected catalog \citep{mccracken:2010},
we adopted the photometry from the latter.
Optical-NIR photometry was then matched with \textit{Spitzer}/IRAC photometry in 4 channels 
by the S-COSMOS survey \citep{sanders:2007} with a search radius of 1 arcsec. 
We used $1\farcs9$ aperture flux and corrected it for total flux by using conversion factors listed in the \texttt{README} file of the catalog.

\subsubsection{Correction for emission line contributions in broad-band magnitudes}

Given the detection of strong emission lines in our MOSFIRE spectra,
a significant contribution from them to the broad band flux is expected \citep[e.g.,][]{schaerer:2013,stark:2013}.
We have estimated the contribution by comparing the broad-band magnitudes and emission line fluxes.
The contribution of \oiitot to \textit{H}-band flux ranges from 
1.2 to 80\%, with a median of 5.7\%{} and 8 objects contributing $>10\%$.
In the \textit{K}-band, \hbeta and \oiiitot contribute from 6 to $\sim 100\%$ with a median of 21\%{} and 6 objects showing $>50\%$.
In the case of the very high contributions, close to 100\%, the continuum broad-band magnitudes are  close to the detection limit.

Looking at the optical spectra from zCOSMOS-Deep, we have also estimated the contribution of \lya to \textit{V}-band magnitudes.
There are 20 objects with zCOSMOS-Deep spectra and 7 of them shows \lya in emission, which contributes 
 less than 10\%{} of the broad-band flux.

Another strong rest-frame optical emission line,
\halpha, does not contribute any of broad-band fluxes considered here,
since it is located in the gap between the \textit{K}-band and the  IRAC 3.6 \micron\ band
at the redshift of our sample.

We have corrected \textit{H}- and \textit{K}-band magnitudes for the emission lines,
while no correction has been applied for \lya
given its  minor contribution to the broad-band flux.

\subsubsection{SED fitting}
\label{sec:sedfitting}

\begin{figure*}
  \centerline{
    \includegraphics[width=\linewidth]{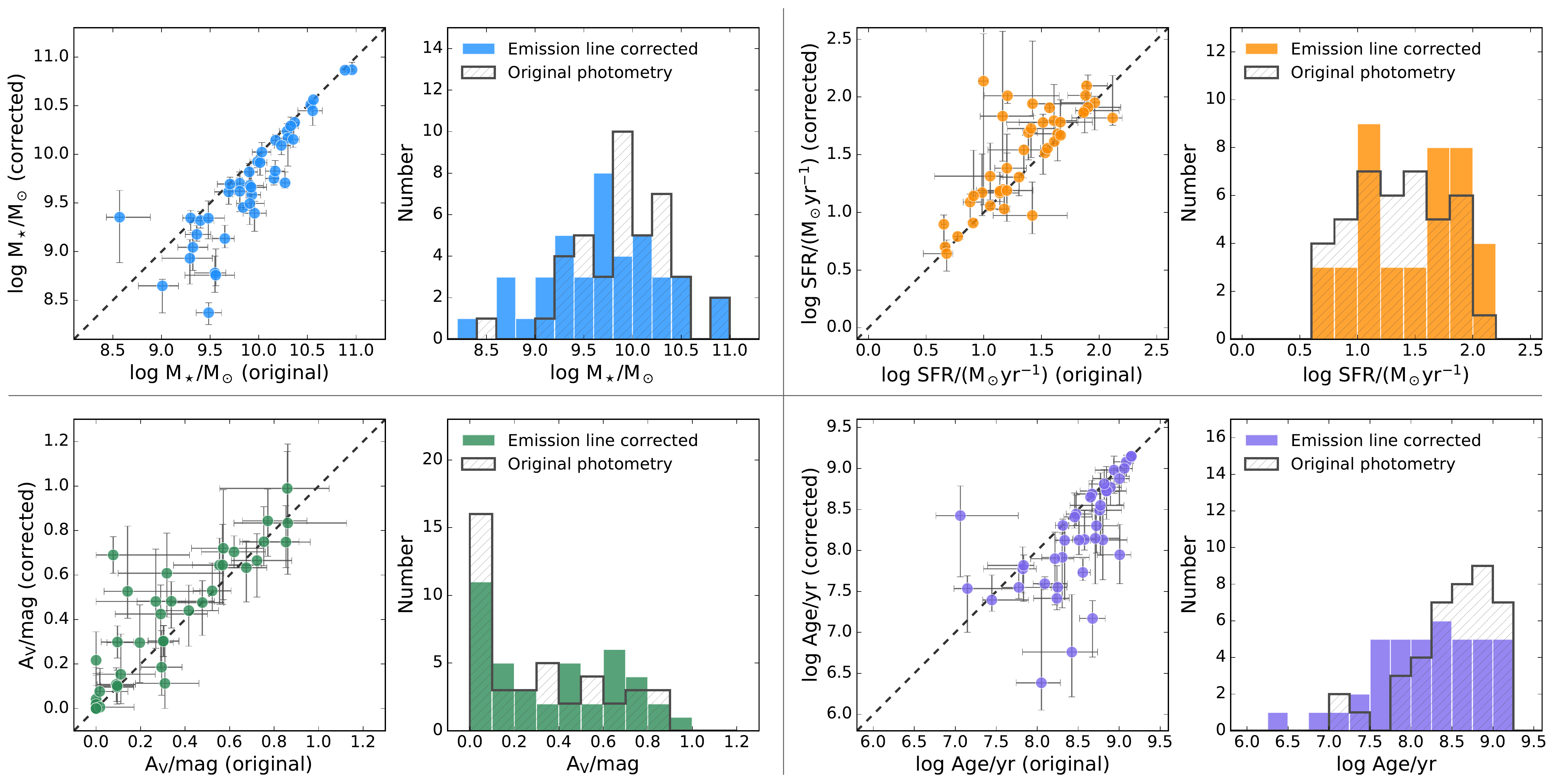}
  }
  \caption{%
    At each quadrant, a pair of panels shows
    the comparison of outputs from SED fitting with and without the corrections for strong emission line contributions
    to the broad-band fluxes (\textit{Left})
    and the distribution of each SED parameter (\textit{Right}).
    Hatched and color-filled histograms show those derived based on photometry with and without the correction
    for emission line contributions, respectively.
    The following parameters are shown:
    \textit{Top left}: Stellar masses.
    \textit{Top right}: SFR.
    \textit{Bottom left}: Attenuation at \textit{V}-band, $A_V$.
    \textit{Bottom right}: Age since the onset of star formation.
    \label{fig:sedprop}
  }
\end{figure*}

SED fitting was carried out for the emission line corrected photometry 
by using the  template SED-fitting code \texttt{ZEBRA+}
which is an updated version of photometric redshift code \texttt{ZEBRA}\footnote{\url{http://www.astro.ethz.ch/carollo/research/zebra.html}}
\citep{feldmann:2006:zebra}.  Redshifts were set to the spectroscopic ones.
As templates, we used composite stellar population models generated from the simple stellar population models
of \citet{bruzual:2003} with a Chabrier initial mass function \citep[IMF;][]{chabrier:2003}. 
We employed an exponentially declining star formation history (SFH), $\propto \exp(-t/\tau)$,
with $\log \tau/\text{yr}=8$--$11$ with steps of 0.1 dex.
Ages range in $\log \text{age/yr}=6$--$9.5$ with steps of 0.1 dex
where the upper limit of the age is chosen to be an approximate age of the universe at $z=3$. 
Metallicities of $0.2Z_\odot$, $0.4Z_\odot$, and $Z_\odot$ were used.
We also allowed dust extinction with $E(B-V)=0$--$0.8$ mag with steps of 0.05 mag 
following the Calzetti extinction curve \citep{calzetti:2000}.
The median values of stellar mass\footnote{Sum of living stars and remnants.}, SFR, $\tau$, age, $A_\mathit{V}$, and metallicity and
corresponding 68\%{} confidence intervals derived by marginalizing the likelihood distribution
were returned as the output.
\autoref{fig:bestsed_begin} in \autoref{sec:bestsed}
shows the best-fit template and emission line corrected observed photometry. 

  An exponentially declining SFH may not be the best  approximation for SFGs at high redshift
  \citep[e.g.,][]{renzini:2009,maraston:2010,reddy:2012}. Although the age derived here is formally meant to be the time elapsed since the onset of star formation,
  it  should actually be regarded as the time interval before the present  during which the bulk of stars were formed.
  Indeed, 85\%{} of the sample show $\text{age} / \tau < 2$, which means that the fitting procedure returns a nearly constant SFR.
  However, in our main analysis of MZR we will use only the stellar mass  among the outputs of the SED fitting, as the stellar mass is quite stable against the choice of SFH. For example,  employing a constant  or delayed-exponential SFHs
  would change the stellar mass by  only  $0.1$ dex. On the other hand, the SFR and dust extinction will be derived only from the rest-frame UV properties.   

\autoref{fig:sedprop} compares various outputs from the SED fitting based on the original photometry
with those derived from emission line corrected photometry, and their distributions.
Emission line corrected stellar masses ($M_\star^\text{corr}$) are generally lower than
those from the original photometry ($M_\star^\text{original}$)
and the effect is more prominent in less massive galaxies.
The median difference in stellar mass is $\log M_\star^\text{corr} - \log M_\star^\text{original} = -0.13$~dex.
On the other hand, the median differences in SFR, $A_\mathit{V}$, and age of stellar populations
are $-0.03$, $-0.02$, and $0.12$ dex, respectively.

\begin{figure}
  \centerline{
    \includegraphics[width=\linewidth]{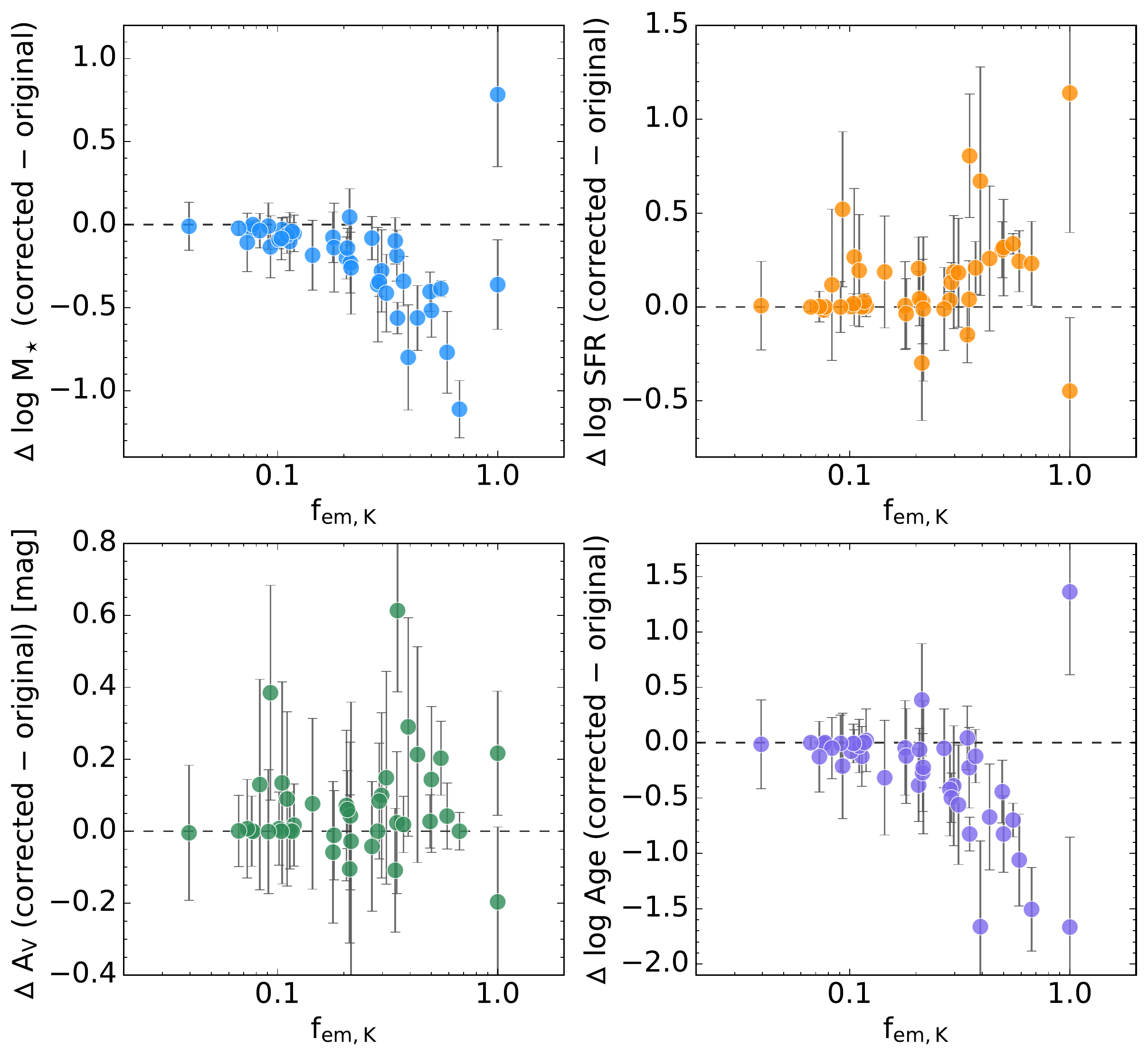}
  }
  \caption{%
    Difference in the best-fit SED fitting parameters (stellar mass, SFR, $A_\mathit{V}$, and age in clockwise from top left panel)
    between those derived from the original and emission line corrected photometry 
    as a function of the fraction of emission line fluxes in \textit{K}-band. 
    \label{fig:delta_param}
  }
\end{figure}

\autoref{fig:delta_param} shows the difference in the best-fit parameters
as a function of the fraction of emission line fluxes in the \textit{K}-band.
Stellar mass and age
show decreasing trends in the difference with increasing emission line contributions. 
SFRs are also affected by large emission line contributions,
but the trend appears weaker than those on stellar mass and age. 
On the other hand, no clear trend can be seen in $A_\mathit{V}$.
This can be understood as longer wavelength bands are more sensitive to the stellar mass and age,
while SFR and dust extinction are primarily captured by the rest-frame UV part of the SED,
where emission line contribution is not important.

  There are two  objects, 434625 and 217090,
  which have a $\sim 100\%$ contribution from emission lines in the \textit{K} band,
  and they do not seem to follow the general trend of the rest of our sample in \autoref{fig:delta_param}.
  These two objects are the only ones without detection both in the \textit{K} band (after emission-line correction) and in any of the IRAC bands.
  At $z\simeq 3.3$, only observing at wavelengths on and beyond the \textit{K} band one can capture the rest-frame wavelength of $>4000$~\AA{} 
  which is essential to  properly estimate the mass-to-light ratios of galaxies.
  This is likely the main reason why the two objects do not follow the trends,
  especially in stellar mass and age.

All this  indicates the importance of a proper assessment of emission line contribution
in near-infrared bands at $z>3$ \citep[e.g.,][]{schaerer:2013,stark:2013}.
In the following analysis, we adopt $M_\star^\text{corr}$ for the stellar masses of galaxies in our sample.

\capstartfalse
\begin{deluxetable*}{crccccc}
  \tablecaption{Stellar mass, dust extinction, and star formation rate.\label{tab:sedprop}}
  \tablehead{
    \colhead{ID} &
    \colhead{$\log M_*$} &
    \colhead{$A_\mathit{V}$ (SED)} &
    \colhead{$A_\mathit{V}$ (UV)} &
    \colhead{$\log \text{SFR}$ (SED)} &
    \colhead{$\log \text{SFR}$ (UV)} &
    \colhead{$\log \text{SFR}$ (\hbeta)} \\
    \colhead{} &
    \colhead{($M_\odot$)} &
    \colhead{(mag)} &
    \colhead{(mag)} &
    \colhead{($M_\odot\,\text{yr}^{-1}$)} &
    \colhead{($M_\odot\,\text{yr}^{-1}$)} &
    \colhead{($M_\odot\,\text{yr}^{-1}$)}
  }
  \startdata
434625 & $9.35_{- 0.47}^{+ 0.28}$ & $0.11_{-0.11}^{+0.19}$ & $0.33 \pm 0.14$ & $0.97_{-0.16}^{+0.29}$ & $1.09 \pm 0.06$ & $<1.40$ \\
413136 & $10.51_{- 0.04}^{+ 0.03}$ & $0.30_{-0.07}^{+0.07}$ & $0.84 \pm 0.10$ & $1.52_{-0.02}^{+0.02}$ & $1.86 \pm 0.04$ & $1.63 \pm 0.12$ \\
413646 & $ 9.35_{- 0.13}^{+ 0.08}$ & $0.75_{-0.12}^{+0.16}$ & $0.43 \pm 0.12$ & $1.82_{-0.06}^{+0.37}$ & $1.32 \pm 0.05$ & \nodata \\
413453 & $10.91_{- 0.02}^{+ 0.02}$ & $0.30_{-0.07}^{+0.07}$ & $0.83 \pm 0.09$ & $1.46_{-0.02}^{+0.02}$ & $1.80 \pm 0.04$ & $<1.94$ \\
434585 & $10.24_{- 0.08}^{+ 0.08}$ & $0.11_{-0.08}^{+0.08}$ & $0.38 \pm 0.07$ & $1.56_{-0.04}^{+0.03}$ & $1.71 \pm 0.03$ & $1.59 \pm 0.13$ \\
434571 & $ 9.04_{- 0.24}^{+ 0.15}$ & $0.30_{-0.18}^{+0.17}$ & $0.08 \pm 0.17$ & $1.17_{-0.20}^{+0.33}$ & $0.80 \pm 0.08$ & \nodata \\
413391 & $10.17_{- 0.29}^{+ 0.07}$ & $0.53_{-0.12}^{+0.29}$ & $0.54 \pm 0.08$ & $1.94_{-0.10}^{+0.54}$ & $1.88 \pm 0.04$ & $1.48 \pm 0.13$ \\
427122 & $ 8.93_{- 0.26}^{+ 0.19}$ & $0.00_{-0.00}^{+0.00}$ & $0.28 \pm 0.20$ & $0.70_{-0.04}^{+0.03}$ & $0.90 \pm 0.09$ & $0.88 \pm 0.15$ \\
434148 & $ 9.70_{- 0.13}^{+ 0.12}$ & $0.10_{-0.08}^{+0.08}$ & $0.02 \pm 0.14$ & $1.06_{-0.02}^{+0.02}$ & $0.92 \pm 0.06$ & $<0.84$ \\
434082 & $ 9.75_{- 0.07}^{+ 0.13}$ & $0.03_{-0.03}^{+0.12}$ & $0.35 \pm 0.07$ & $1.69_{-0.24}^{+0.03}$ & $1.74 \pm 0.03$ & $2.03 \pm 0.04$ \\
434126 & $ 9.13_{- 0.09}^{+ 0.14}$ & $0.48_{-0.12}^{+0.09}$ & $0.41 \pm 0.12$ & $1.73_{-0.25}^{+0.05}$ & $1.39 \pm 0.05$ & $1.44 \pm 0.10$ \\
434139 & $ 9.18_{- 0.07}^{+ 0.09}$ & $0.44_{-0.16}^{+0.11}$ & $0.11 \pm 0.12$ & $1.68_{-0.23}^{+0.05}$ & $1.09 \pm 0.05$ & $<1.41$ \\
434145 & $ 9.71_{- 0.06}^{+ 0.11}$ & $0.69_{-0.08}^{+0.08}$ & $0.69 \pm 0.11$ & $2.01_{-0.07}^{+0.02}$ & $1.77 \pm 0.05$ & $<2.04$ \\
434242 & $ 8.78_{- 0.20}^{+ 0.17}$ & $0.04_{-0.04}^{+0.14}$ & $0.00 \pm 0.22$ & $0.90_{-0.24}^{+0.08}$ & $0.66 \pm 0.10$ & $1.30 \pm 0.14$ \\
406390 & $ 9.61_{- 0.13}^{+ 0.10}$ & $0.66_{-0.16}^{+0.12}$ & $0.61 \pm 0.11$ & $1.88_{-0.19}^{+0.09}$ & $1.62 \pm 0.05$ & $<1.73$ \\
406444 & $10.87_{- 0.02}^{+ 0.08}$ & $0.10_{-0.07}^{+0.07}$ & $0.84 \pm 0.05$ & $1.61_{-0.02}^{+0.02}$ & $2.26 \pm 0.02$ & $2.07 \pm 0.06$ \\
434227 & $10.14_{- 0.06}^{+ 0.06}$ & $0.42_{-0.25}^{+0.13}$ & $0.64 \pm 0.11$ & $1.78_{-0.45}^{+0.07}$ & $1.86 \pm 0.05$ & $<1.75$ \\
191932 & $ 9.93_{- 0.06}^{+ 0.10}$ & $0.64_{-0.18}^{+0.12}$ & $0.48 \pm 0.16$ & $1.54_{-0.24}^{+0.03}$ & $1.24 \pm 0.07$ & $<1.16$ \\
434547 & $ 9.62_{- 0.21}^{+ 0.14}$ & $0.65_{-0.18}^{+0.18}$ & $0.55 \pm 0.12$ & $1.79_{-0.21}^{+0.31}$ & $1.49 \pm 0.06$ & $1.05 \pm 0.08$ \\
192129 & $10.33_{- 0.03}^{+ 0.05}$ & $0.00_{-0.00}^{+0.00}$ & $0.19 \pm 0.10$ & $1.20_{-0.04}^{+0.03}$ & $1.37 \pm 0.04$ & $1.46 \pm 0.06$ \\
193914 & $ 9.32_{- 0.08}^{+ 0.09}$ & $0.63_{-0.15}^{+0.12}$ & $0.28 \pm 0.11$ & $1.95_{-0.24}^{+0.06}$ & $1.30 \pm 0.05$ & \nodata \\
212863 & $ 9.91_{- 0.13}^{+ 0.10}$ & $0.19_{-0.13}^{+0.16}$ & $0.56 \pm 0.17$ & $1.03_{-0.03}^{+0.19}$ & $1.31 \pm 0.08$ & $1.55 \pm 0.08$ \\
195044 & $ 9.69_{- 0.13}^{+ 0.07}$ & $0.75_{-0.11}^{+0.16}$ & $0.55 \pm 0.10$ & $1.91_{-0.05}^{+0.28}$ & $1.56 \pm 0.05$ & $0.84 \pm 0.16$ \\
208115 & $10.86_{- 0.02}^{+ 0.02}$ & $0.00_{-0.00}^{+0.00}$ & $0.01 \pm 0.12$ & $1.11_{-0.02}^{+0.02}$ & $1.12 \pm 0.05$ & $2.11 \pm 0.05$ \\
200355 & $ 9.83_{- 0.12}^{+ 0.11}$ & $0.02_{-0.02}^{+0.14}$ & $0.45 \pm 0.15$ & $1.09_{-0.17}^{+0.07}$ & $1.35 \pm 0.07$ & $1.60 \pm 0.08$ \\
214339 & $10.56_{- 0.02}^{+ 0.02}$ & $0.00_{-0.00}^{+0.00}$ & $1.16 \pm 0.17$ & $0.91_{-0.02}^{+0.02}$ & $2.19 \pm 0.08$ & $1.81 \pm 0.12$ \\
411078 & $ 9.45_{- 0.02}^{+ 0.02}$ & $0.30_{-0.07}^{+0.07}$ & $0.00 \pm 0.06$ & $1.91_{-0.04}^{+0.02}$ & $1.38 \pm 0.03$ & $1.75 \pm 0.03$ \\
212298 & $10.15_{- 0.08}^{+ 0.11}$ & $0.84_{-0.16}^{+0.14}$ & $0.81 \pm 0.10$ & $2.10_{-0.21}^{+0.04}$ & $1.83 \pm 0.04$ & $1.85 \pm 0.04$ \\
412808 & $ 8.37_{- 0.12}^{+ 0.10}$ & $0.00_{-0.00}^{+0.10}$ & $0.08 \pm 0.14$ & $1.14_{-0.05}^{+0.40}$ & $0.99 \pm 0.06$ & $1.24 \pm 0.07$ \\
223954 & $ 9.68_{- 0.13}^{+ 0.12}$ & $0.15_{-0.13}^{+0.18}$ & $0.24 \pm 0.15$ & $1.17_{-0.07}^{+0.18}$ & $1.16 \pm 0.07$ & $1.11 \pm 0.11$ \\
220771 & $ 9.34_{- 0.21}^{+ 0.18}$ & $0.01_{-0.01}^{+0.17}$ & $0.00 \pm 0.33$ & $0.64_{-0.15}^{+0.12}$ & $0.56 \pm 0.15$ & $0.79 \pm 0.16$ \\
219315 & $ 9.82_{- 0.09}^{+ 0.09}$ & $0.00_{-0.00}^{+0.00}$ & $0.00 \pm 0.20$ & $0.79_{-0.04}^{+0.03}$ & $0.80 \pm 0.09$ & $<0.56$ \\
434618 & $10.09_{- 0.09}^{+ 0.08}$ & $0.08_{-0.08}^{+0.09}$ & $0.32 \pm 0.13$ & $1.18_{-0.16}^{+0.04}$ & $1.28 \pm 0.06$ & $1.34 \pm 0.06$ \\
221039 & $ 9.58_{- 0.04}^{+ 0.05}$ & $0.70_{-0.07}^{+0.07}$ & $0.41 \pm 0.08$ & $2.01_{-0.02}^{+0.02}$ & $1.51 \pm 0.03$ & $1.54 \pm 0.04$ \\
215511 & $ 9.39_{- 0.18}^{+ 0.12}$ & $0.48_{-0.21}^{+0.24}$ & $0.32 \pm 0.17$ & $1.31_{-0.21}^{+0.29}$ & $1.03 \pm 0.08$ & $1.40 \pm 0.08$ \\
217597 & $ 9.50_{- 0.22}^{+ 0.15}$ & $0.72_{-0.23}^{+0.26}$ & $0.50 \pm 0.23$ & $1.38_{-0.21}^{+0.30}$ & $1.05 \pm 0.10$ & $1.14 \pm 0.13$ \\
211934 & $ 9.66_{- 0.24}^{+ 0.20}$ & $0.83_{-0.23}^{+0.35}$ & $1.61 \pm 0.52$ & $1.19_{-0.27}^{+0.32}$ & $1.82 \pm 0.23$ & $1.67 \pm 0.24$ \\
434579 & $10.02_{- 0.10}^{+ 0.10}$ & $0.48_{-0.15}^{+0.10}$ & $0.36 \pm 0.39$ & $1.31_{-0.16}^{+0.03}$ & $1.05 \pm 0.17$ & $0.74 \pm 0.20$ \\
217753 & $10.29_{- 0.07}^{+ 0.09}$ & $0.99_{-0.15}^{+0.17}$ & $0.80 \pm 0.16$ & $1.78_{-0.12}^{+0.19}$ & $1.56 \pm 0.07$ & $1.66 \pm 0.09$ \\
218783 & $10.45_{- 0.15}^{+ 0.10}$ & $0.53_{-0.09}^{+0.12}$ & $0.72 \pm 0.12$ & $1.67_{-0.02}^{+0.13}$ & $1.75 \pm 0.05$ & $1.45 \pm 0.07$ \\
217090 & $ 8.65_{- 0.28}^{+ 0.07}$ & $0.22_{-0.22}^{+0.13}$ & $0.00 \pm 0.13$ & $2.14_{-1.07}^{+0.41}$ & $0.95 \pm 0.06$ & $1.51 \pm 0.07$ \\
210037 & $ 8.76_{- 0.11}^{+ 0.27}$ & $0.61_{-0.20}^{+0.18}$ & $0.77 \pm 0.20$ & $1.83_{-0.39}^{+0.74}$ & $1.50 \pm 0.09$ & $<1.43$ \\
208681 & $10.86_{- 0.02}^{+ 0.02}$ & $0.30_{-0.07}^{+0.07}$ & $0.35 \pm 0.05$ & $1.86_{-0.02}^{+0.02}$ & $1.75 \pm 0.02$ & $1.74 \pm 0.04$
\end{deluxetable*}
\capstarttrue

\subsection{UV  slope and dust extinction}

We computed the UV 
spectral slope $\uvbeta$ defined as $f_\lambda \propto \lambda^{\beta_\text{UV}}$
by fitting a linear function to the observed broad-band magnitudes from the  \textit{r} to the \textit{J} band
which cover the rest-frame wavelength range $1400 \lesssim \lambda \lesssim 2800$ \AA{} at $z=3.27$.
  We used the rest-frame wavelengths corresponding to the Galaxy Evolution Explorer (\textit{GALEX}) far-UV (FUV; $\lambda_c \simeq 1530$ \AA)
  and near-UV (NUV; $\lambda_c \simeq 2300$ \AA) bands
  to derive $\beta_\text{UV}$ photometrically, following the recipe used in \citet{pannella:2015}. 
  For all objects in the sample, this choice of passbands  encloses
  the rest-frame UV wavelengths with which UV slopes are determined.
Following \citet{nordon:2013}, magnitude errors are further weighted by
$1+|\lambda_\text{filter}-\lambda_\text{FUV}|/\lambda_\text{FUV}$, 
where $\lambda_\text{filter}$ and $\lambda_\text{FUV}$ are the rest-frame central wavelengths of each photometric band
and that of \textit{GALEX} FUV. 
By using the best-fit relation, \uvbeta was derived  as
\begin{align}
  \uvbeta & = \frac{\log(f_\text{$\lambda$,FUV}/f_\text{$\lambda$,NUV})}{\log(\lambda_\text{FUV}/\lambda_\text{NUV})} \\
  & = -0.4\frac{(m_\text{FUV}-m_\text{NUV})}{\log(\lambda_\text{FUV}/\lambda_\text{NUV})} - 2,
\end{align}
where $m_\text{FUV}$ and $m_\text{NUV}$ are the AB magnitudes in the \textit{GALEX} FUV and NUV bands, respectively
\citep[][]{nordon:2013,pannella:2015}.
Then assuming the Calzetti extinction law \citep{calzetti:2000,pannella:2015},
\uvbeta was converted to dust attenuation at $\lambda=1530$ \AA\  with 
\begin{equation}
  A_\text{UV} = 4.85 + 2.31\uvbeta,
\end{equation}
which assumes an intrinsic (un-extincted) slope $ \uvbeta(A_{\rm V}=0)=-2.10$  \citep{calzetti:2000}.

\begin{figure}
  \centerline{
    \includegraphics[width=0.95\linewidth]{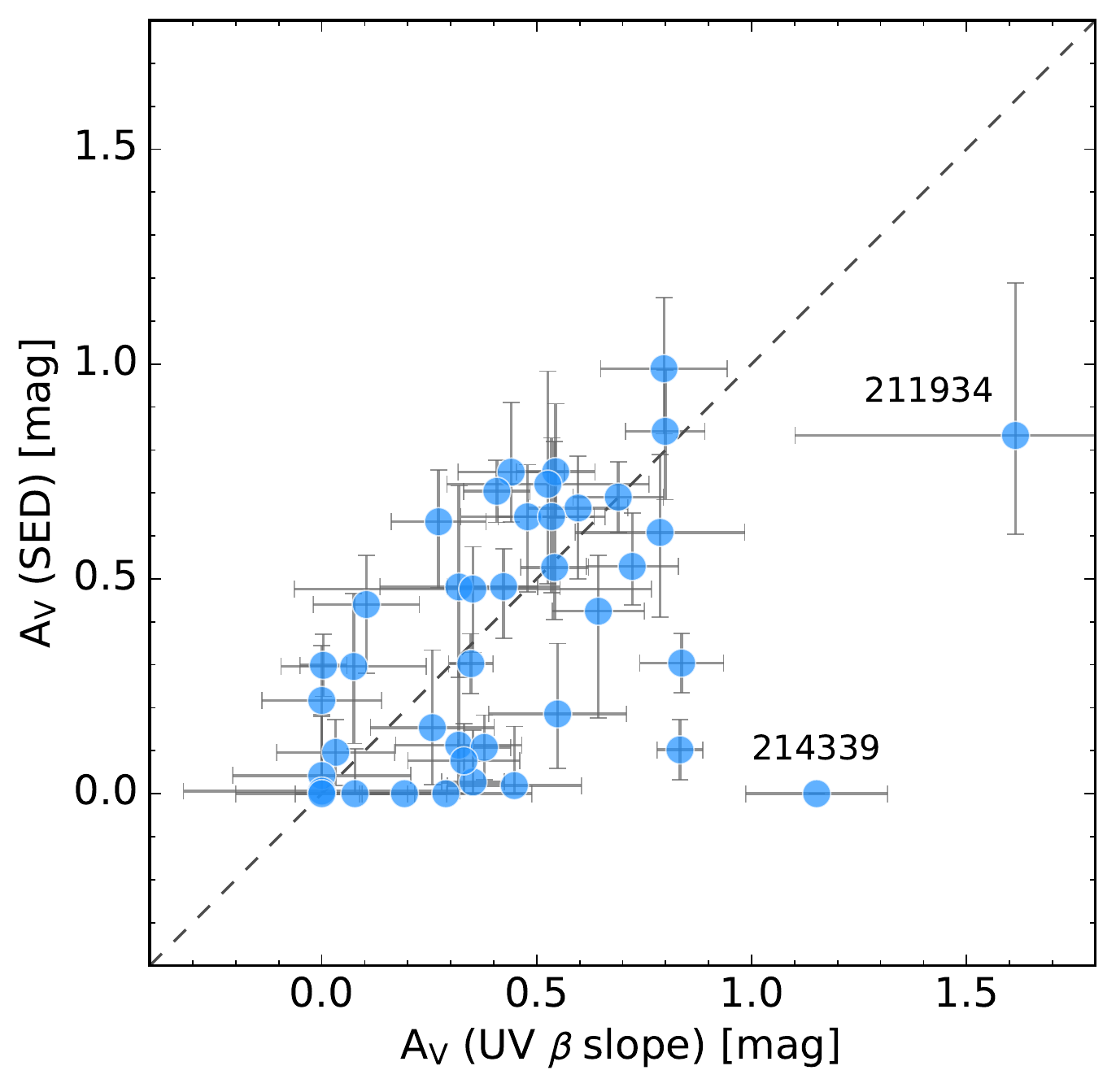}
  }
  \caption{%
    Comparison of the attenuation at \textit{V}-band estimated from SED fitting with
    that converted from $A_\text{UV}$ derived with UV $\beta$-slope and the Calzetti extinction law. 
    Dashed lines correspond to the one-to-one relation. 
    \label{fig:comp_uv_sed}
  }
\end{figure}

In \autoref{fig:comp_uv_sed}, the values of $A_\textit{V}$ from UV and SED fitting are compared, having converted 
$A_\text{UV}$ to $A_\mathit{V}$ by assuming  the Calzetti extinction law. 
There appears to be no tight correlation between the two estimates,
perhaps because dust extinction degenerates with other parameters
such as SFH and age in broad-band SED fitting \citep[e.g.,][]{kodama:1999,michalowski:2014}.
  Two outliers, 214339 and 211934, as indicated in \autoref{fig:comp_uv_sed},
  turn out to be the faintest galaxies in our sample. 
  Indeed, the broad-band SED of these objects shown in \autorefsec{sec:bestsed}
  appears to be quite noisy,
  which makes a robust estimate of \uvbeta difficult.

\begin{figure}[h]
  \centerline{
    \includegraphics[width=0.95\linewidth]{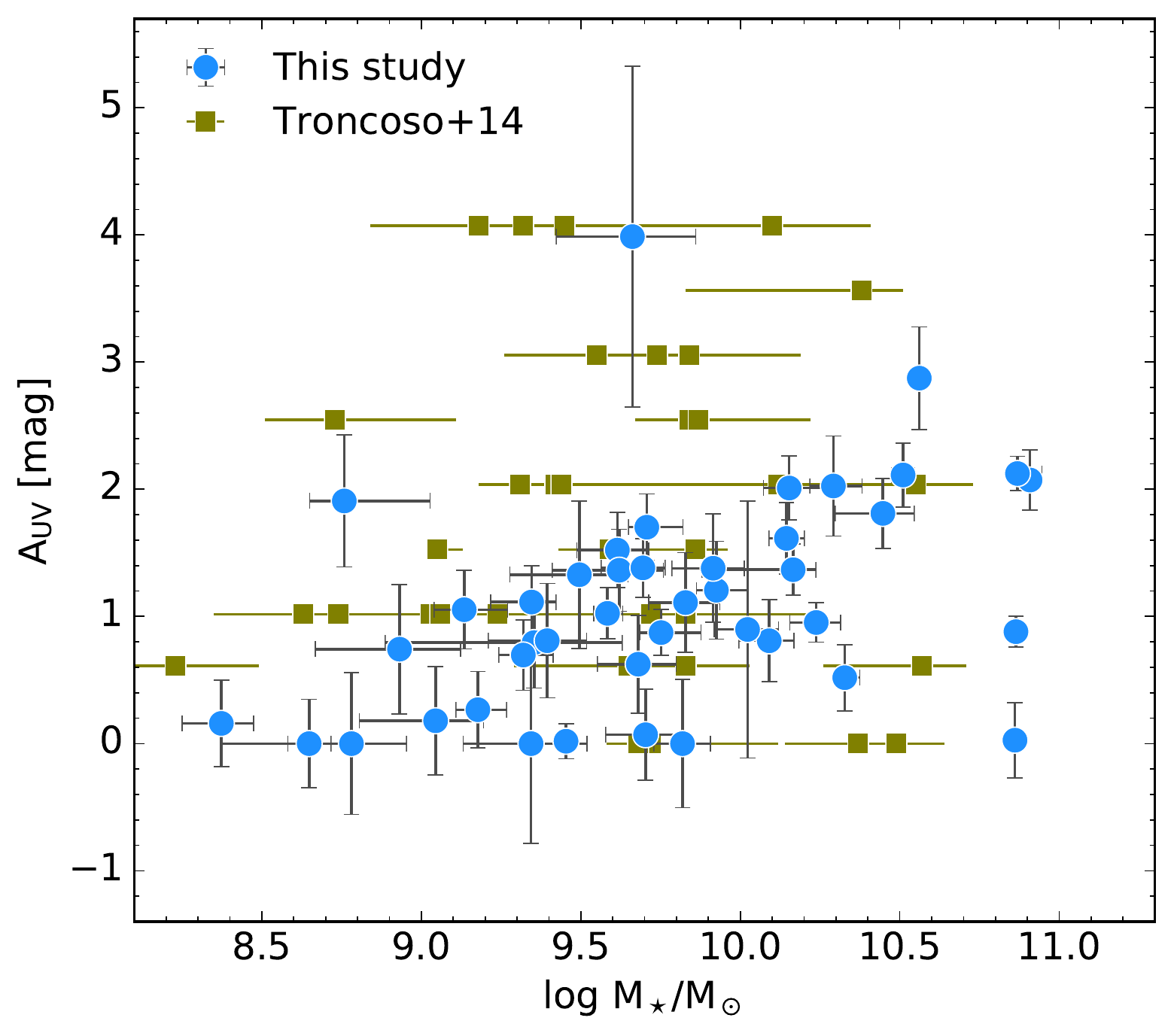}
  }
  \caption{
    $A_\text{UV}$ vs. stellar mass.
    Our sample is shown in blue circles and those taken from \citet{troncoso:2014} are shown with green squares.
    UV attenuation for \citet{troncoso:2014} are converted from $E(B-V)$ from broad-band SED fitting by assuming the Calzetti law. 
    \label{fig:mass_dust_uv}
  }
\end{figure}

\autoref{fig:mass_dust_uv} shows $A_\text{UV}$  vs. stellar mass.
There is a trend with more massive galaxies being  more dust attenuated.
The sample of $z\simeq 3.4$ SFGs from \citet{troncoso:2014} is also shown,
with  on average a higher  dust extinction compared to our sample.
Note that in their study extinction was derived from broad-band SED fitting,
while ours was based on \uvbeta slope.
\citet{troncoso:2014} specifically selected Lyman break galaxies (LBGs)  with redshifts from near-IR spectroscopy,
while we selected objects partly based on the availability of spectroscopic redshifts, which automatically
puts a limit on $B_\text{AB}=25$~mag for those galaxies culled from the zCOSMOS-Deep sample.
Moreover, targets were selected based on the predicted \hbeta flux of $>5\times10^{-18}\,\text{erg}\,\text{s}^{-1}\,\text{cm}^{-2}$.
Though this flux limit does not seem to be too high, the combination of these two criteria 
could result in selecting bluer, less dust attenuated objects
compared to the standard LBG selection employed by \citet{troncoso:2014}.

  Another noticeable feature of \autoref{fig:comp_uv_sed}
  is that there is an appreciable number of objects ($\sim 20\%$)
  showing $A_\textit{V}$ close or equal to zero,
  which seems somewhat at odds with the relatively high SFR of galaxies in our sample.
  The adopted calibration of the \uvbeta--$A_\text{UV}$ relation in Equation (3) was established 
  at $z=0$ but  the relation could be different at high redshift for a number of reasons,
  such as different intrinsic $\beta$ slope
  due to stellar population properties, IMF, SFH
  \citep[e.g.,][]{reddy:2010,wilkins:2011},
  or different extinction curve than one assumed here \citep[e.g.,][]{reddy:2015}.
  \citet{castellano:2014} investigated the UV dust extinction for a sample of
  LBGs at $z\simeq3$, obtaining an intrinsic  slope of $\uvbeta=-2.67$
  which is significantly steeper than 
  widely used values of $\uvbeta=-2.10$ \citep{calzetti:2000} and $\uvbeta=-2.23$ \citep{meurer:1999}.
  Indeed, for our galaxies with $A_\textit{V}$ close or equal to zero the measured $\uvbeta$ is close to or slightly steeper than $-2.10$, either because of measurement errors or because the intrinsic slope is actually steeper than $-2.10$.

\subsection{Star-formation rate}

\subsubsection{UV luminosity}

We adopt the calibration of the UV luminosity-based SFR from \citet{daddi:2004:bzk}
after converting from Salpeter IMF to Chabrier IMF by applying an offset of 0.23 dex:
\begin{equation}
  \text{SFR}_\text{UV}(M_\odot\, \text{yr}^{-1}) = 6.65 \times 10^{-29} L_\text{UV,corr} \; (\text{erg}\,\text{s}^{-1}\,\text{Hz}^{-1}), 
\end{equation}
where $L_\text{UV,corr}$ is extinction corrected UV luminosity derived using
\begin{equation}
  L_\text{UV,corr} = L_\text{UV,obs} \times 10^{0.4 A_\text{UV}}.
\end{equation}
Here, the UV luminosity is measured at rest-frame 1500 \AA{}. 

\begin{figure*}
  \centerline{
    \includegraphics[width=0.9\linewidth]{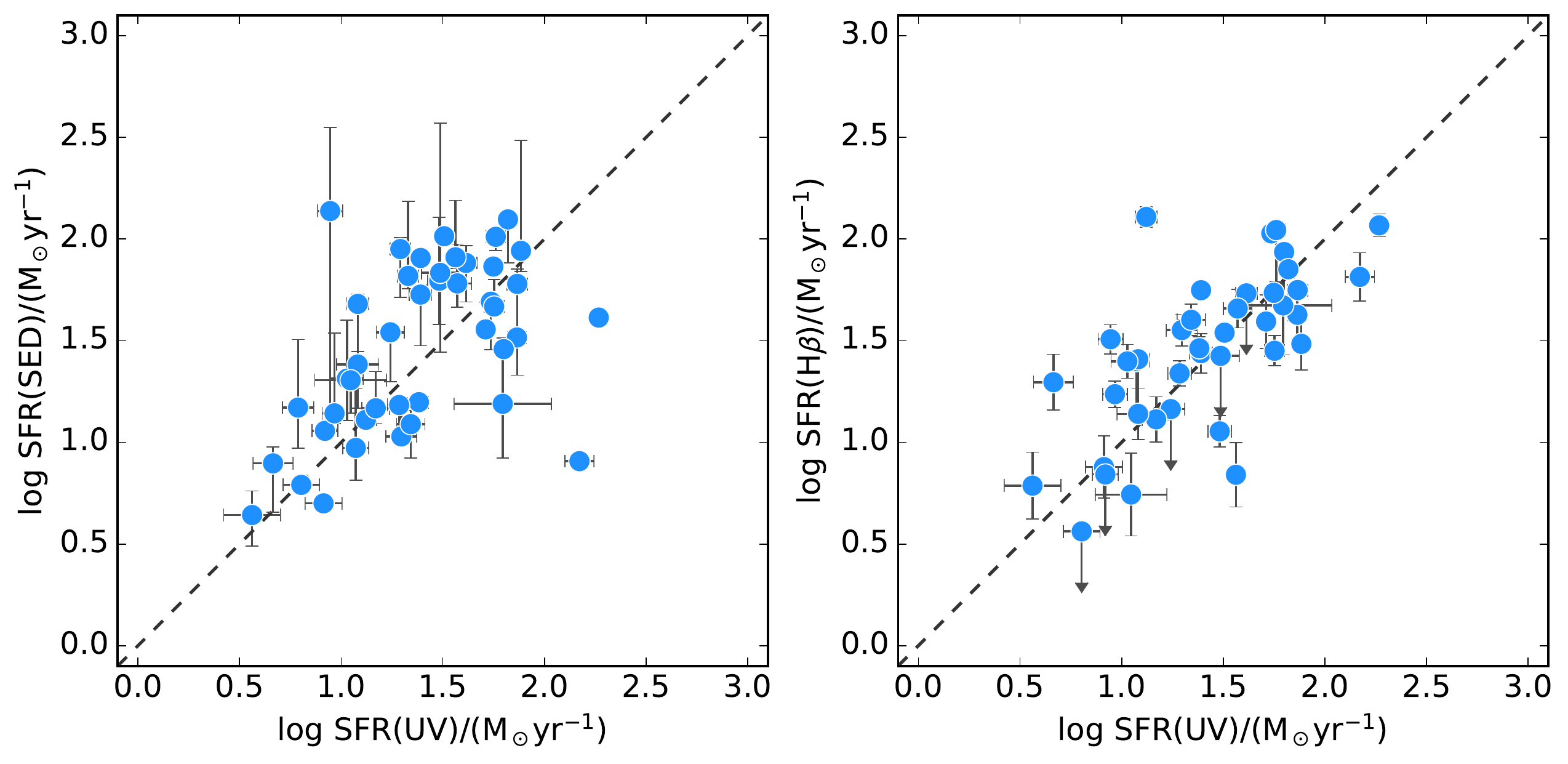}
  }
  \caption{%
    Comparison of SFR derived by UV luminosity with that derived from broad-band SED fitting (\textit{Left})
    and \hbeta{} (\textit{Right}),
    \hbeta{} SFRs are corrected for dust extinction from UV $\beta$-slope
    and assuming $E(B-V)_\text{neb}=E(B-V)_\text{star}$. 
    Dashed lines show the one-to-one relation.
    \label{fig:comp_sfr}
  }
\end{figure*}

In the left panel of \autoref{fig:comp_sfr}, we compare the UV-based SFRs with those from SED fitting, showing that the two measurements agree well with each other.

\begin{figure}
  \centerline{
    \includegraphics[width=0.95\linewidth]{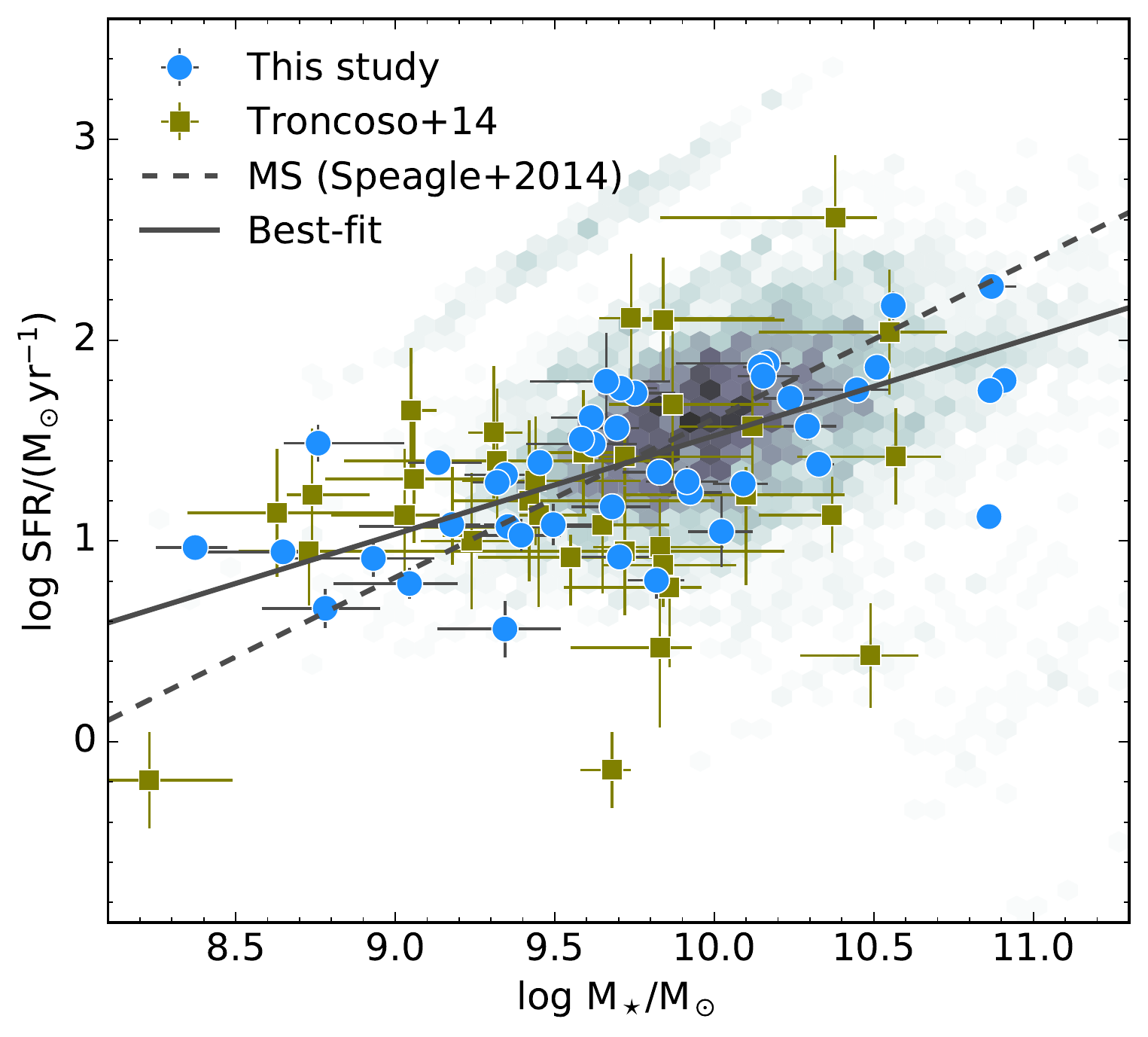}
  }
  \caption{
    SFR--stellar mass relation for galaxies at $3<z<3.8$.  Blue circles are shows UV-based SFR
    of our sample and background pixels show SFR from SED fitting of photo-$z$ selected galaxies at $3<z<3.8$ \citep{ilbert:2013}.
    Green squares are galaxies at $3<z<5$ by \citet{troncoso:2014}.
    SFR of their objects are estimated from \hbeta luminosity. 
    Dashed line indicate the star-forming main sequence at $z=3.27$ \citep{speagle:2014}.
    \label{fig:mass_sfr_uv}
  }
\end{figure}

\autoref{fig:mass_sfr_uv} shows SFR as a function of stellar mass for $z \simeq 3.3$ SFGs. 
Our sample covers about 2.5 dex in stellar mass from $\log M_\star/M_\odot\lesssim 8.5$ to $\simeq 11$,
essentially following the MS at $z=3.27$ derived by \citet{speagle:2014}
which also traces the parent sample of $z\simeq3.3$ SFGs shown with gray-scale.
We also derived the best-fit relation between stellar mass and SFR for our sample, finding
\begin{equation}
  \begin{split}
    & \log \text{SFR}/(M_\odot\,\text{yr}^{-1}) \\
    & = (1.52 \pm 0.05) + (0.49 \pm 0.08)\times(\log M/M_\odot-10),
  \end{split}
\end{equation}
shown with the solid line.
Because of the sample selection based either on the availability of spectroscopic redshift or
  on the predicted \hbeta flux, our sample is likely to be biased against especially low mass and low SFR
  galaxies.  We believe this is  the main reason why we obtained a flatter MS slope compared to
  that of the parent sample and that of the  \citet{speagle:2014} MS.
  We will use Equation (6)  only to separate the sample into bins of $M_\star$ and sSFR
  to create composite spectra (see \autorefsec{sec:stack}).

For comparison, overplotted in \autoref{fig:mass_sfr_uv} is the sample of $z\simeq3.4$ SFGs from \citet{troncoso:2014}.
Their sample shows an even  flatter slope than ours. 
Again, this could be due to their sample selection based on the LBG technique,
which prefers galaxies with blue SED and show a flat SFR--stellar mass relation \citep[e.g.,][]{erb:2006:survey}. 
Note that their SFRs are based on \hbeta{} luminosity with extinction correction
assuming an extra extinction in nebular component, i.e.,
$E(B-V)_\text{neb}=E(B-V)_\text{star}/0.44$, while UV-based SFRs are used for our sample.
Thus, both our best-fit and \citet{troncoso:2014} distributions
appear to be flatter than the canonical MS \citep{speagle:2014},
which can result  from  a bias favoring low-mass SFGs with above average sSFR and disfavoring high-mass galaxies  with high  dust extinction.
Relatively low dust extinction is indeed obtained for our sample in the analysis above.
In particular, we might miss high metallicity objects due to this possible bias,
if the correlation between metallicity and dust extinction claimed
at an intermediate redshift $z\simeq1.6$ \citep{zahid:2014}
still holds at $z\sim 3.3$.

\subsubsection{\hbeta luminosity}

We converted \hbeta luminosities to the \halpha luminosities,
assuming an intrinsic \halpha/\hbeta ratio of 2.86 \citep[Case B recombination with $T=10^4$~K and $n_e=10^2$~cm$^{-3}$; ][]{osterbrock:agnagn}. 
Then the SFR based on \halpha was computed following  \citet{kennicutt:2012},
\begin{equation}
  \log \text{SFR}_{\halpha} (M_\odot\,\text{yr}^{-1}) = \log L_\text{\halpha} (\text{erg}\,\text{s}^{-1}) - 41.27.
\end{equation}
For the objects in which \hbeta is not detected with more than $3\sigma$, we adopted $3\sigma$ upper limits.

In the original recipe by \citet{calzetti:2000}, calibrated with local UV-luminous starbursts,
nebular emission lines are attenuated more than stellar light following $E(B-V)_\text{neb}=E(B-V)_\text{star}/0.44$.
However, there have been various  claims in recent years indicating that at high redshift these two components may suffer similar attenuations
\citep[e.g.,][]{erb:2006:survey,kashino:2013,pannella:2015,puglisi:2016}. 
Therefore, we assumed the same amount of attenuation for the stellar light as for emission lines,
which actually provides a good agreement between UV-based and \hbeta-based SFRs
as shown in the left panel of \autoref{fig:comp_sfr}.

In addition to the relation between nebular emission lines and stellar continuum extinction, 
there is an uncertainty in the choice of the extinction curve
for nebular emission lines.
Related to the original derivation of the \citet{calzetti:2000} relation,
the Milky Way extinction curve \citep{cardelli:1989} or
small Magellanic cloud (SMC) extinction curve \citep[][]{gordon:2003}
is often preferred \citep[e.g., ][]{steidel:2014}.
In the estimate of \hbeta-based SFRs  
we use the  \citet{calzetti:2000} extinction curve.
Note that for our range of $A_\mathit{V}$, the change between \citet{calzetti:2000} and
\citet{cardelli:1989} extinction curves is very minor:
at most 2\%{} for the \hbeta flux and 0.03 dex for $\log(\oiii/\oii)$.

\subsection{Composite spectra}
\label{sec:stack}

In the next sections we will compare various properties of the ionized gas
with global properties of galaxies, such as stellar mass, SFR, and sSFR.
Besides  carrying out such comparison for  individual galaxies,
we will also compare average properties. 
For this purpose, we created composite spectra in bins of stellar mass and SFR. 
The sample,  excluding the two  AGN candidates, was split in three bins in stellar mass,
$\log M/M_\odot < 9.5$, $9.5 < \log M/M_\odot < 10.0$, and $\log M/M_\odot > 10.0$,
and two  bins in SFR, above and below the best-fit MS shown as the dashed-line in \autoref{fig:mass_sfr_uv},
i.e., $\Delta_\text{MS}<0$ and $\Delta_\text{MS}>0$ where $\Delta_\text{MS} \equiv \text{sSFR}/\text{sSFR}_\text{best-fit}$.

First, spectra of each object in both \textit{J}- and \textit{H}-band were normalized by total \oiiitwo flux and
registered by a linear interpolation to the rest-frame wavelength grid of 0.25\AA{} interval
which is slightly finer than the spectral resolution for the highest redshift object of our sample.
We also normalized the corresponding noise spectra by total \oiiitwo flux and 
registered to the identical rest-frame wavelength grid but interpolated in quadrature. 
Then composite spectra were constructed by taking an average at each wavelength pixel
weighted by the inverse variance.
We constructed the associated noise spectra via the standard error propagation
from the individual noise spectra.

Emission line fluxes are measured by fitting a Gaussian to each emission line
by assuming a common velocity shift and a velocity dispersion.
We fit simultaneously
\oiitot, \hbeta, \oiiitot, and \neiiione
with continuum described by a second order polynomial.
We computed the flux values and their $1\sigma$ errors by means of a Monte Carlo simulation with $10^3$ realizations.

\begin{figure*}
  \centerline{
    \includegraphics[width=\linewidth]{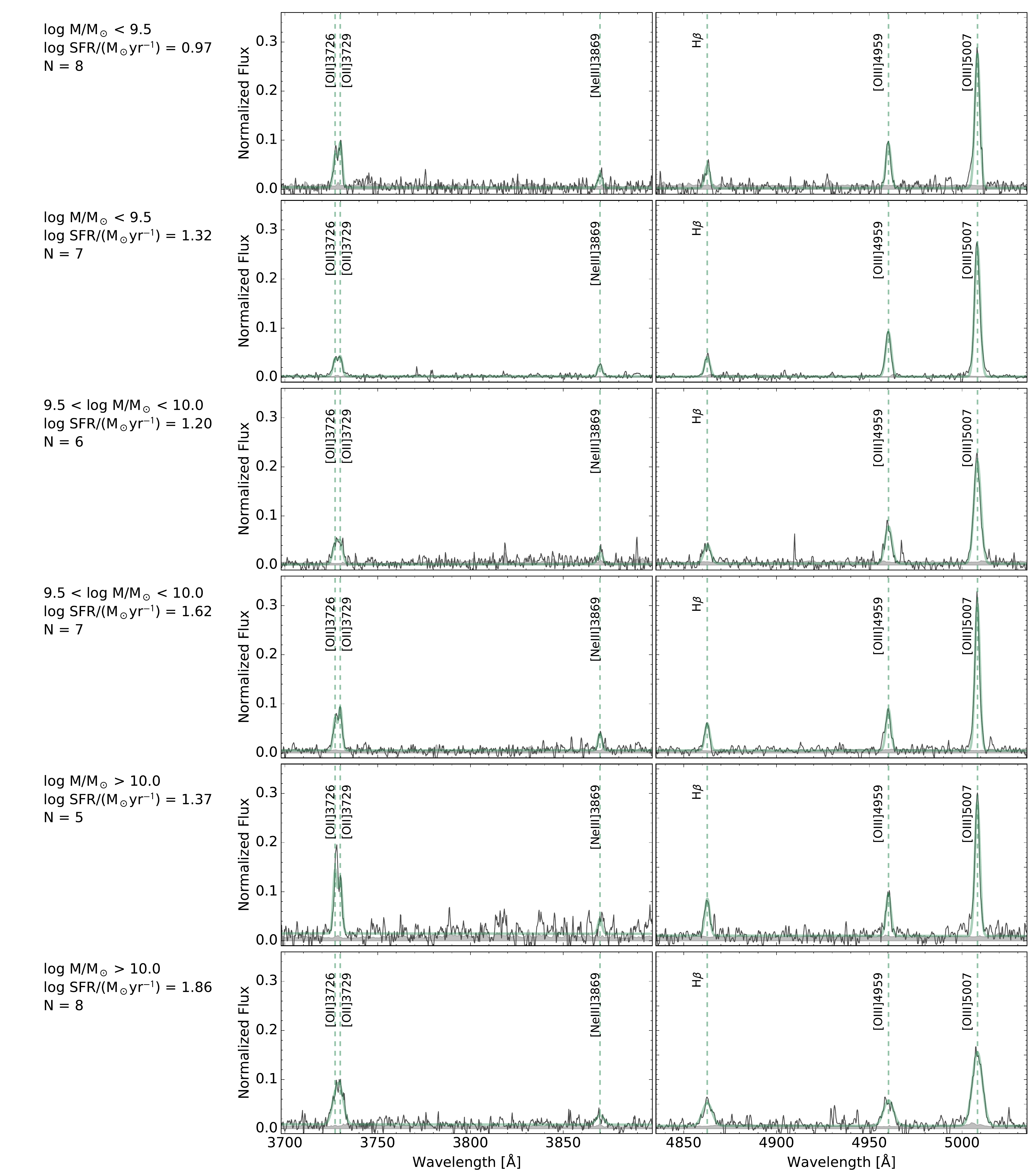}
  }
  \caption{
    Composite spectra in bins of stellar mass and SFR above and below the best-fit "main sequence" (\autoref{fig:mass_sfr_uv}).
    The range of stellar mass, median SFR, and number of objects stacked in each bin is indicated at each row.
   The  panels show the observed composite spectra (black) with associated $1\sigma$ noise (gray)
    and the best-fit Gaussians for the emission lines (green). 
    For each mass bin, the upper/lower panels show the composite spectra of galaxies below/above the adopted main sequence.
    \label{fig:stackspec_bin}}
\end{figure*}

The resulting stacked spectra are shown in \autoref{fig:stackspec_bin}.
We adopt median stellar mass, SFR, and $A_\mathit{V}$ whenever they are used in the subsequent analysis (see \autoref{tab:stack}).

\capstartfalse
\begin{deluxetable*}{ccccccc}
    \tablecaption{Properties of stacked spectra.\label{tab:stack}}
    \tablehead{
      \colhead{$N_\text{obj}$} &
      \colhead{$\log M_\star$} &
      \colhead{$\log\text{SFR}$} &
      \colhead{$A_\mathit{V}$} &
      \colhead{$\log q$} &
      \colhead{$\log n_e$} &
      \colhead{\ohmetal}
      \\
      \colhead{} & 
      \colhead{($M_\star/M_\odot$)} &
      \colhead{($M_\odot\,\text{yr}^{-1}$)} &
      \colhead{(mag)} &
      \colhead{(cm s$^{-1}$)} &
      \colhead{(cm$^{-3}$)} &
      \colhead{}
      \\
      \colhead{(1)} & 
      \colhead{(2)} & 
      \colhead{(3)} & 
      \colhead{(4)} & 
      \colhead{(5)} & 
      \colhead{(6)} & 
      \colhead{(7)}
    }
    \startdata
    \multicolumn{7}{c}{Above the best-fit MS} \\  [0.2em]
    \hline
8 &  9.26 & 0.97 & 0.20 & $7.77_{-0.01}^{+0.01}$ & $1.92_{-0.31}^{+0.21}$ & $8.15_{-0.10}^{+0.09}$ \\
6 &  9.82 & 1.20 & 0.34 & $7.86_{-0.01}^{+0.01}$ & $2.51_{-0.10}^{+0.09}$ & $8.04_{-0.12}^{+0.11}$ \\
5 & 10.29 & 1.37 & 0.35 & $7.69_{-0.01}^{+0.01}$ & $2.86_{-0.06}^{+0.05}$ & $8.38_{-0.09}^{+0.08}$ \\ 
    \hline
    \multicolumn{7}{c}{Below the best-fit MS} \\ [0.2em]
    \hline
7 &  9.13 & 1.32 & 0.28 & $7.95_{-0.00}^{+0.01}$ & $2.45_{-0.06}^{+0.06}$ & $7.97_{-0.10}^{+0.10}$ \\
7 &  9.66 & 1.62 & 0.55 & $7.76_{-0.01}^{+0.01}$ & $2.01_{-0.18}^{+0.15}$ & $8.15_{-0.10}^{+0.09}$ \\
8 & 10.34 & 1.86 & 0.76 & $7.74_{-0.01}^{+0.01}$ & $1.92_{-0.41}^{+0.25}$ & $8.41_{-0.09}^{+0.08}$
\enddata
\tablecomments{
  (1) Number of objects in the bin;
  (2) Median stellar mass;
  (3) Median SFR;
  (4) Median $A_\mathit{V}$;
  (5) Ionization parameter;
  (6) Electron density;
  (7) Gas-phase oxygen abundance.
}
\end{deluxetable*}
\capstarttrue

\section{Measurements of ionized gas proprieties}
\label{sec:gasprop}

In this section, we derive physical properties of ionized gas,
namely gas-phase oxygen abundance, ionization parameter, and electron density,
of our sample.
The obtained values are listed in \autoref{tab:stack} and \autoref{tab:metal}
for the stacked spectra and individual objects, respectively.

\subsection{Gas-phase oxygen abundance}
\label{sec:metal}

\begin{figure*}
  \centerline{
    \includegraphics[width=\linewidth]{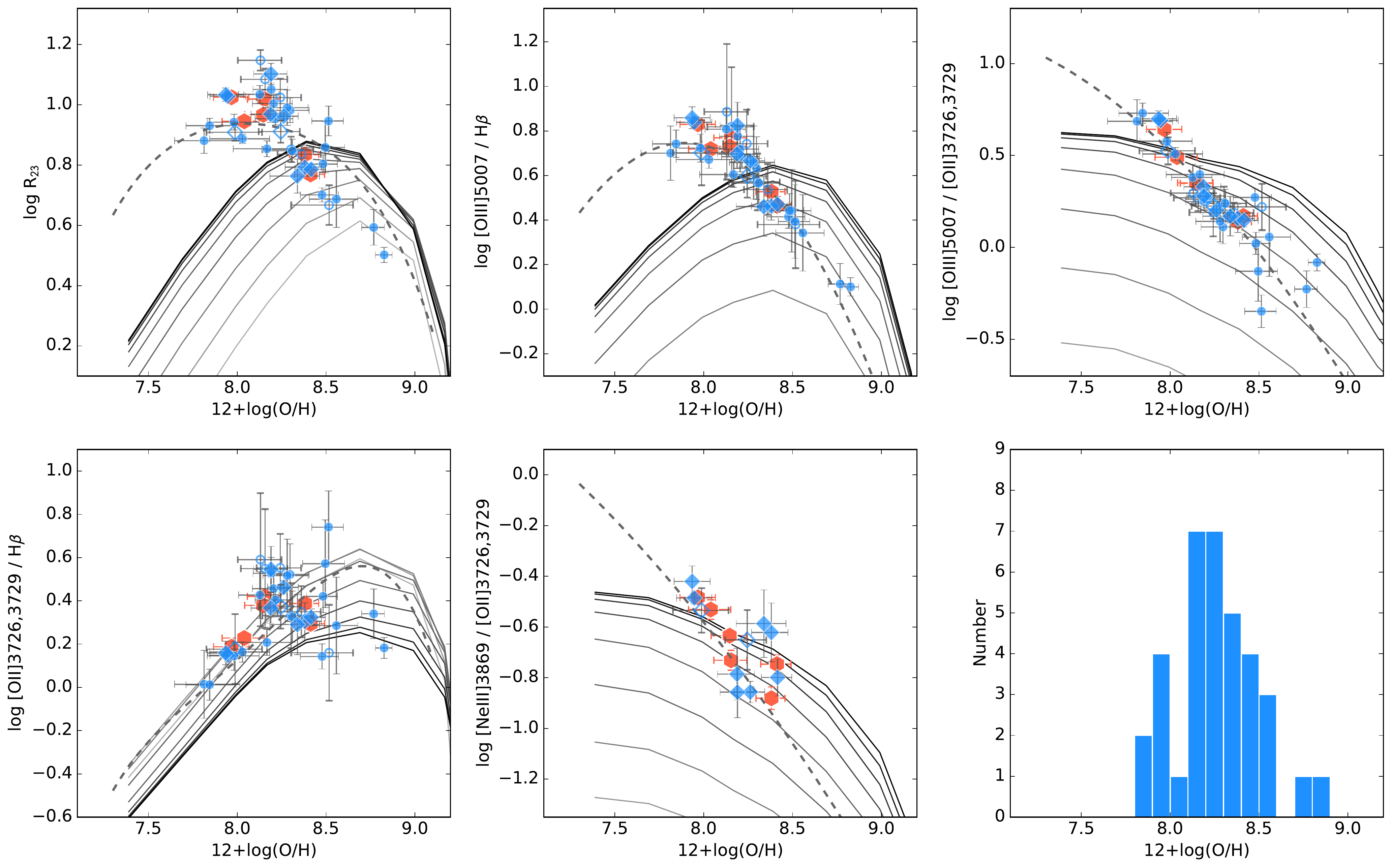}
  }
  \caption{
    Best-fit metallicity vs observed line ratios for our sample.
    Diamonds show objects with detected \oiitot, \oiii, and \neiiione,
    while circles are those with detected \oiitot and \oiii.
    Filled and open symbols show those with and without \hbeta detection, respectively. 
    In case of undetected \hbeta, its flux is supplemented from UV SFR.
    Orange open pentagon shows the values obtained from the stacking of all non-AGN objects.
    Solid lines are photo-ionization models by \citet{dopita:2013} with $\kappa=20$ for the $\kappa$-distribution of electron energies.
    Gray scale of each line
    indicates $\log q = 6.5, 6.75,  7.0, 7.25,  7.5, 7.75,  8.0, 8.25,  8.5$
    from faint to thick,
    where $q$ is ionization parameter defined as the ratio of the ionizing photon flux
    passing through a unit area and the local number density of hydrogen atoms.
    Dashed lines show a empirical calibration by \citet{maiolino:2008}
    that we adopted in this study. 
    Bottom left panel is a histogram of \ohmetal. 
  \label{fig:metalmeasure}}
\end{figure*}

The primary indicator for gas phase metallicity, \ohmetal, in this study is
$R_{23} \equiv (\oiione + \oiitwo + \oiiione + \oiiitwo)/\hbeta$ \citep{pagel:1979, kobulnicky:2003}.
For the metallicity measurement, we adopt the empirical calibration by \citet{maiolino:2008}
in which low metallicity regime ($\ohmetal \lesssim 8.3$) is directly calibrated by the electron temperature method. 
At this metallicity scale, the theoretical calibration has been known to have difficulties in reproducing the  observed emission line ratios,  e.g, \citet{kewley:2002}.
This appears to still exist in the recent photoionization models by \citet{dopita:2013} as shown in \autoref{fig:metalmeasure}, 
i.e., calibration lines by \citet{maiolino:2008} show higher ratios at low metallicity than those from \citet{dopita:2013}.
The line ratios from the stacked spectra (\autoref{fig:stackspec_bin}) are shown as hexagon symbols in \autoref{fig:metalmeasure}, 
 clearly showing that at low metallicities the photoionization models by \citet{dopita:2013} cannot account for  all five line ratios simultaneously.

For our metallicity estimates, following \citet{maiolino:2008} we used the five extinction-corrected line ratios shown in \autoref{fig:metalmeasure}, 
namely $R_{23}$, $\oiiitwo/\hbeta$, $\oiiitwo/\oiitot$, $\oiitot/\hbeta$, and $\neiiione/\oiitot$,
for the metallicity estimate. 
For the stacked spectra, we corrected for dust extinction with the median $A_\mathit{V}$ derived from the \uvbeta slope.

For the metallicity analysis of individual galaxies we removed seven objects without $3\sigma$ detection in $\oiione+\oiitwo$ and the two AGN candidates (one of two is also \oii non-detection). 
This leaves us with 35 objects.
In the case that either  $\oiiione$ or $\oiiitwo$ are undetected at the $3\sigma$ level,
the other [\ion{O}{3}] flux is complemented by assuming an intrinsic line ratio of $\oiiitwo/\oiiione=3$.
For eight  objects without $>3\sigma$ \hbeta detection, 
we used the \hbeta flux estimated from the UV-based SFR,  given the relatively tight  correlation
between the SFRs from the two estimators as shown in \autoref{fig:comp_sfr}.
\neiiione is detected for 10 out of the 35 objects. 

The gas-phase oxygen metallicity was then derived with the maximum likelihood method, 
 first computing
\begin{equation}
  \chi^2 = \sum_i \frac{\left(\log I_{i,\text{M08}} - \log I_{i,\text{obs}} \right)^2}{\sigma_{i,\text{obs}}^2+\sigma_{i,\text{rms}}^2}, 
\end{equation}
where $I_{i,\text{M08}}$ and $I_{i,\text{obs}}$ are $i$th line ratio from \citet{maiolino:2008}
calibration at a given \ohmetal and the one from observed spectra.
Further, $\sigma_{i,\text{obs}}$ and $\sigma_{i,\text{rms}}$ are the errors in the observed line ratio and
intrinsic scatter measured as in \citet{jones:2015} for $z = 0.8$ galaxies. 
Then, we translated $\chi^2$ to the likelihood distribution using $\mathcal{L}\propto \exp(-\chi^2/2)$. 
The metallicity and its confidence interval are then defined as the median and the 16 to 84 percentiles of the probability distribution, respectively.

Note that the \citet{maiolino:2008} calibration adopted in this study
implicitly assumes the ionization parameter at a given metallicity to be  independent of redshift, 
as it is derived for local SFGs.
Recent studies of gas-phase metallicity of high redshift SFGs show 
an  offset  in the BPT diagram  with  higher \oiii/\hbeta ratio at a given \nii/\halpha ratio,
which can be ascribed either to an elevated ionization parameters
in high redshift galaxies \citep[e.g.,][]{kewley:2013}, or to  an enhanced N/O  abundance ratio at a fixed O/H ratio
\citep[e.g.][]{masters:2014,sanders:2015:ionization,yabe:2015,cullen:2014, nakajima:2014}.
Moreover, \citet{sanders:2015:ionization} have argued
that there is no significant change in ionization parameter at a fixed metallicity
from $z \sim 0$ to $z \sim 2.3$.
Unfortunately, the \niitwo/\halpha ratio that is used to break the degeneracy
of metallicity estimates from   $R_{23}$ cannot be obtained from the ground for the redshift of our galaxies, but will become possible with 
the  \textit{James Webb Space Telescope}.
\citet{maier:2015} suggest that [\ion{O}{3}]$\lambda5007$/H$\beta$ ratio can be used to break
the degeneracy,  with  galaxies with $\log \oiiitwo/\hbeta > 0.26$ having  $\ohmetal < 8.6$ on the \citet{kewley:2002} metallicity scale.
Indeed, our sample with $\ohmetal < 8.6$ always have $\log \oiiitwo/\hbeta > 0.26$,
confirming the \citet{maier:2015} result.

\subsection{Ionization parameter}
\label{sec:logq}

The ionization parameter, $q$, is defined as $q \equiv Q_{\text{H}^{0}}/4\pi R_s^2 n$,
where $Q_{\text{H}^{0}}$ is the flux of ionizing photons above the Lyman limit,
$R_s$ is the Str{\"o}mgren radius, and $n$ is the local number density of hydrogen atoms \citep[e.g.,][]{kewley:2002}.
The ionization parameter can be derived through a $q$-sensitive line ratio, $O_{32}\equiv\oiiitot/\oiitot$ \citep[e.g.,][]{mcgaugh:1991}. 
Here, we adopt a metallicity dependent [\ion{O}{3}]/[\ion{O}{2}]--$q$ relation of \citet{kewley:2002}
parameterized by \citet{kobulnicky:2004} as
\begin{equation}
  \begin{split}
  \log q = & \Big\{32.81-1.153y^2 \\
  & +\left[\ohmetal\right]\left(-3.396-0.025y+0.1444y^2\right)\Big\}\\
  & \times \Big\{4.63-0.3119y-0.163y^2 \\
  & +\left[\ohmetal\right]\left(-0.48+0.0271y+0.02037y^2\right)\Big\}^{-1},
  \end{split}
\end{equation}
where $y=\log O_{32}$. The results are reported in Table 5.

Note that the ionization parameters derived in this way are not fully self-consistent,
as we used the \citet{maiolino:2008} calibration for gas-phase metallicity (see \autorefsec{sec:metal})
which implicitly assumed ionization parameters of normal star-forming galaxies in the local Universe.
Note also that in this and the following analysis all emission line fluxes have been corrected
for dust extinction using $E(B-V)_\text{star}$ derived based on \uvbeta slope 
by assuming the Calzetti extinction curve and $E(B-V)_\text{neb} = E(B-V)_\text{star}$.

\capstartfalse
\begin{deluxetable}{cccc}
  \tablecaption{Ionizing gas properties.\label{tab:metal}}
  \tablehead{
    \colhead{ID} &
    \colhead{$\log q$} &
    \colhead{$\log n_e$} &
    \colhead{$12+\log(\text{O/H})$} \\
    \colhead{} &
    \colhead{(cm s$^{-1}$)} &
    \colhead{(cm$^{-3}$)} &
    \colhead{} 
  }
\startdata
434625 &  $7.72_{-0.06}^{+0.06}$ &  \nodata & $8.13_{-0.13}^{+0.12}$ \\
413136 & $>7.42$ &  \nodata & \nodata \\
413646 & $>7.40$ &  \nodata & \nodata \\
413453 &  \nodata &  \nodata & \nodata \\
434585 &  $7.71_{-0.19}^{+0.18}$ & $>3.23$ & $8.56_{-0.15}^{+0.12}$ \\
434571 & $>7.50$ &  \nodata & \nodata \\
413391 &  $7.67_{-0.08}^{+0.07}$ &  $2.65_{-0.44}^{+0.29}$ & $8.30_{-0.12}^{+0.11}$ \\
427122 & $>7.50$ &  \nodata & \nodata \\
434148 &  $7.74_{-0.10}^{+0.09}$ & \nodata & $8.30_{-0.16}^{+0.13}$ \\
434082 &  $7.86_{-0.06}^{+0.06}$ &  $2.11_{-0.47}^{+0.24}$ & $8.03_{-0.18}^{+0.15}$ \\
434126 &  $7.77_{-0.07}^{+0.07}$ &  $3.01_{-0.41}^{+0.29}$ & $8.13_{-0.12}^{+0.11}$ \\
434139 & $>7.49$ &  \nodata & \nodata \\
434145 &  $7.71_{-0.07}^{+0.07}$ &  \nodata & $8.16_{-0.14}^{+0.12}$ \\
434242 &  $7.93_{-0.10}^{+0.11}$ & $<2.51$ & $7.81_{-0.17}^{+0.20}$ \\
406390 &  $7.83_{-0.16}^{+0.13}$ & $>2.90$ & $8.52_{-0.21}^{+0.14}$ \\
406444 &  $7.74_{-0.08}^{+0.07}$ & $<2.77$ & $8.38_{-0.11}^{+0.09}$ \\
434227 & $>7.87$ &  \nodata & \nodata \\
191932 &  $7.73_{-0.07}^{+0.06}$ &  $2.33_{-0.39}^{+0.28}$ & $8.36_{-0.13}^{+0.11}$ \\
434547 &  $7.71_{-0.05}^{+0.05}$ &  $2.76_{-0.20}^{+0.17}$ & $8.19_{-0.10}^{+0.09}$ \\
192129 &  $7.87_{-0.08}^{+0.06}$ &  $2.71_{-0.26}^{+0.19}$ & $8.48_{-0.12}^{+0.09}$ \\
193914 &  $7.85_{-0.08}^{+0.07}$ &  \nodata & $7.99_{-0.16}^{+0.15}$ \\
212863 &  $7.83_{-0.08}^{+0.06}$ &  $2.77_{-0.13}^{+0.11}$ & $8.17_{-0.19}^{+0.14}$ \\
195044 &  $7.38_{-0.07}^{+0.07}$ & \nodata & $8.51_{-0.09}^{+0.08}$ \\
208115 &  \nodata &  \nodata & \nodata \\
200355 &  $7.91_{-0.06}^{+0.06}$ &  \nodata & $7.98_{-0.15}^{+0.14}$ \\
214339 &  $7.75_{-0.08}^{+0.08}$ &  $2.69_{-0.46}^{+0.26}$ & $8.31_{-0.15}^{+0.12}$ \\
411078 &  $7.98_{-0.03}^{+0.03}$ &  $2.60_{-0.07}^{+0.07}$ & $7.95_{-0.10}^{+0.10}$ \\
212298 &  $7.72_{-0.05}^{+0.05}$ &  \nodata & $8.42_{-0.09}^{+0.08}$ \\
412808 &  $7.97_{-0.05}^{+0.05}$ &  \nodata & $7.94_{-0.10}^{+0.10}$ \\
223954 &  $7.74_{-0.06}^{+0.05}$ &  $2.12_{-0.38}^{+0.25}$ & $8.20_{-0.12}^{+0.11}$ \\
220771 &  $7.70_{-0.10}^{+0.10}$ &  $2.81_{-0.42}^{+0.29}$ & $8.34_{-0.15}^{+0.12}$ \\
219315 &  $7.74_{-0.07}^{+0.07}$ &  $2.71_{-0.31}^{+0.23}$ & $8.24_{-0.12}^{+0.11}$ \\
434618 &  $7.74_{-0.05}^{+0.05}$ &  $2.45_{-0.17}^{+0.14}$ & $8.21_{-0.11}^{+0.10}$ \\
221039 &  $7.70_{-0.04}^{+0.04}$ &  $2.54_{-0.12}^{+0.10}$ & $8.26_{-0.09}^{+0.08}$ \\
215511 &  $7.79_{-0.06}^{+0.05}$ &  $1.96_{-0.39}^{+0.30}$ & $8.19_{-0.11}^{+0.10}$ \\
217597 &  $7.74_{-0.06}^{+0.06}$ &  $2.52_{-0.23}^{+0.16}$ & $8.19_{-0.10}^{+0.09}$ \\
211934 &  $7.52_{-0.13}^{+0.13}$ &  \nodata & $8.50_{-0.13}^{+0.11}$ \\
434579 &  $7.68_{-0.10}^{+0.10}$ & $<2.78$ & $8.28_{-0.14}^{+0.12}$ \\
217753 &  $7.60_{-0.10}^{+0.10}$ &  $2.96_{-0.42}^{+0.37}$ & $8.77_{-0.07}^{+0.06}$ \\
218783 &  $7.65_{-0.06}^{+0.06}$ & \nodata & $8.48_{-0.09}^{+0.08}$ \\
217090 &  $7.99_{-0.06}^{+0.06}$ &  $2.38_{-0.43}^{+0.27}$ & $7.84_{-0.13}^{+0.15}$ \\
210037 &  $7.67_{-0.11}^{+0.10}$ &  $2.48_{-0.47}^{+0.30}$ & $8.24_{-0.13}^{+0.12}$ \\
208681 &  $7.77_{-0.06}^{+0.06}$ &  $3.32_{-0.26}^{+0.23}$ & $8.83_{-0.05}^{+0.05}$
\enddata
\end{deluxetable}
\capstarttrue

\subsection{Electron density}
\label{sec:ne}

The relatively high spectral resolution of MOSFIRE allows us to resolve
the \oiitot doublet as seen in individual spectra.
The ratio of the two [\ion{O}{2}] lines is sensitive to electron density \citep{osterbrock:agnagn} and
we used the \texttt{PyNeb}\footnote{\url{http://www.iac.es/proyecto/PyNeb/}} package \citep{pyneb}
to compute the electron density $n_e$ of the line-emitting regions of our galaxies.
We assumed an electron temperature of $T_e=10^4$~K.
At low to intermediate redshift, $T_e$ is indeed observed to be $\sim (1$--$2)\times10^4$~K
via the direct measurements of the [\ion{O}{3}]$\lambda4363$ line
\citep[e.g.,][]{izotov:2006,nagao:2006,andrews:2013,ly:2014,jones:2015}.
Assuming  a higher $T_e$, e.g., $2\times10^4$~K, the resulting $n_e$ will become systematically higher by $\sim 0.15$ dex.

When one of the [\ion{O}{2}] lines is not detected at the $3\sigma$ level,
either upper or lower limits of $n_e$ are  derived from the $3\sigma$ flux limit of the line.
In the case that both of [\ion{O}{2}] lines are detected,
we have carried out a Monte Carlo simulation with 500 realizations
by perturbing the measured line ratios with the associated $1\sigma$ uncertainties. 
The median and 16 and 84 percentiles of the resulting distribution have been taken as $n_e$
and $1\sigma$ confidence interval, respectively.

\section{Discussion}
\label{sec:discussion}

\subsection{Relation between ionized gas and galaxy properties}

\begin{figure}
  \centerline{
    \includegraphics[width=0.95\linewidth]{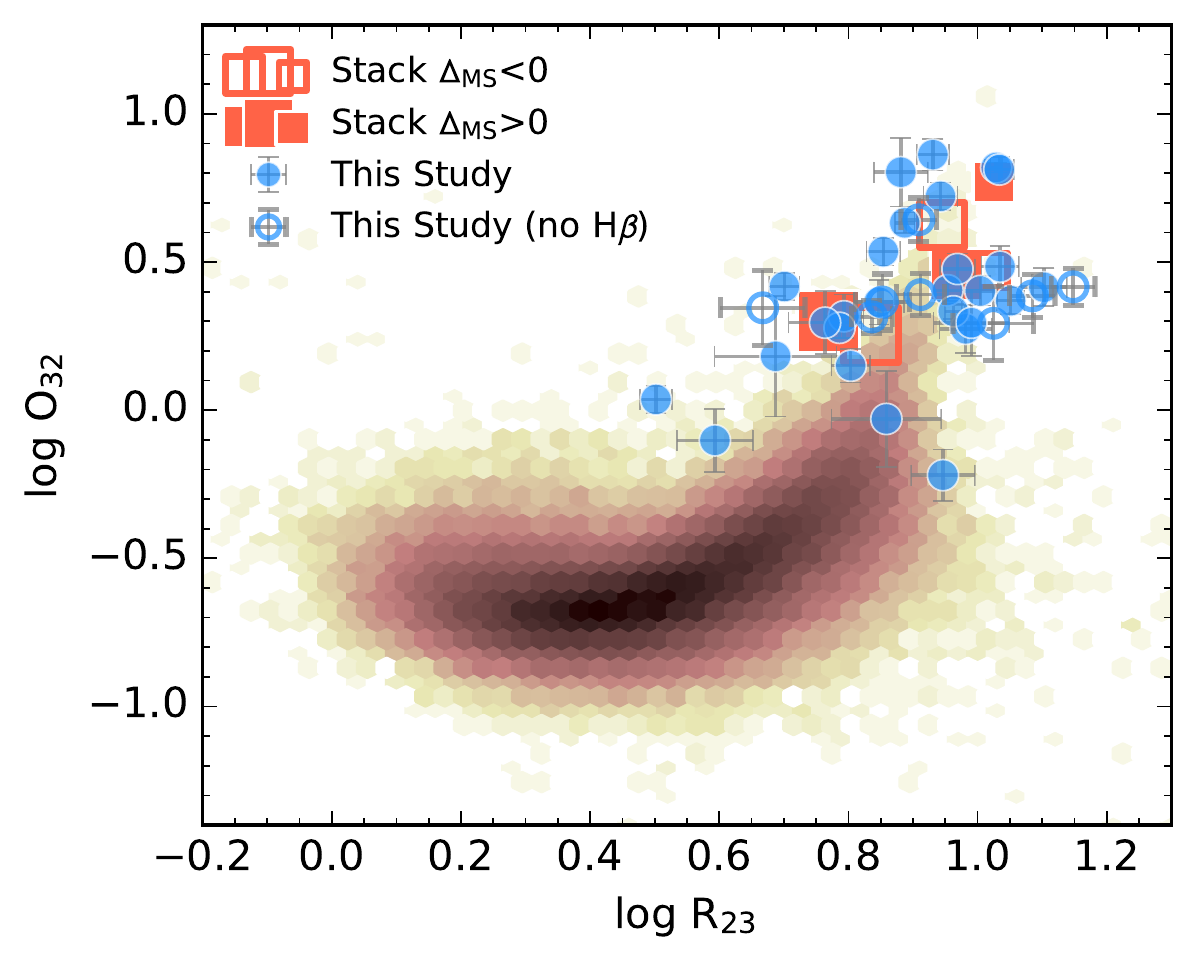}
  }
  \caption{
    $O_{32}$ vs $R_{23}$.
    Individual galaxies from our sample are shown in blue circles with filled and open ones corresponding to
    those with and without \hbeta detection, respectively.  \hbeta fluxes of objects without \hbeta detection
    are estimated from UV-based SFR. 
    For the stacked data shown in orange squares,
    open and filled squares represent $\Delta_\text{MS}<0$ and $\Delta_\text{MS}>0$, respectively,
    with the size proportional to the median stellar mass of each bin. 
    Since error bars for the stacked points are smaller than the size of the symbols,
    they are not shown here.
    Background pixels show the distribution of local SFGs selected based on the BPT diagram 
    drawn from the SDSS line measurement catalog of \citet{oh:2011}.
    \label{fig:o32r23}
  }
\end{figure}

\autoref{fig:o32r23} shows the relation between $R_{23}$ and $O_{32}$.
The local, $z \simeq 0$ galaxy sample shown in the background was selected from the OSSY catalog \citep{oh:2011}
as star-forming based on the BPT diagram \citep{baldwin:1991, kewley:2006}
by requiring all four emission lines (\hbeta, \oiiitwo, \halpha, and \niitwo)
as well as \oii with $\text{S/N}>3$.
The higher line ratios of $z\sim 3.3$ galaxies relative to the local SFGs indicate that
our galaxies have higher ionization parameter,  on average \citep{nakajima:2014,shirazi:2014}. 
Our $z=3.3$ galaxies  lie along the tail of the local distribution and extend it to higher $\log R_{23}$ and $\log O_{32}$ values,
typical of SFGs at $z=2$--$3$ \citep[e.g.,][]{henry:2013, nakajima:2014, sanders:2015:ionization}.
Locally, the tail of the distribution consists of metal-poor galaxies,  typically with $\ohmetal \lesssim 8.5$. 
Since the majority of our sample also shows $\ohmetal \lesssim 8.5$ (See \autorefsec{sec:metal}), we argue that the 
ionization parameter could be similar at a given metallicity in both low and high redshift galaxies, 
consistent with  \citet{sanders:2015:ionization}.

\begin{figure*}
  \centerline{
    \includegraphics[width=\linewidth]{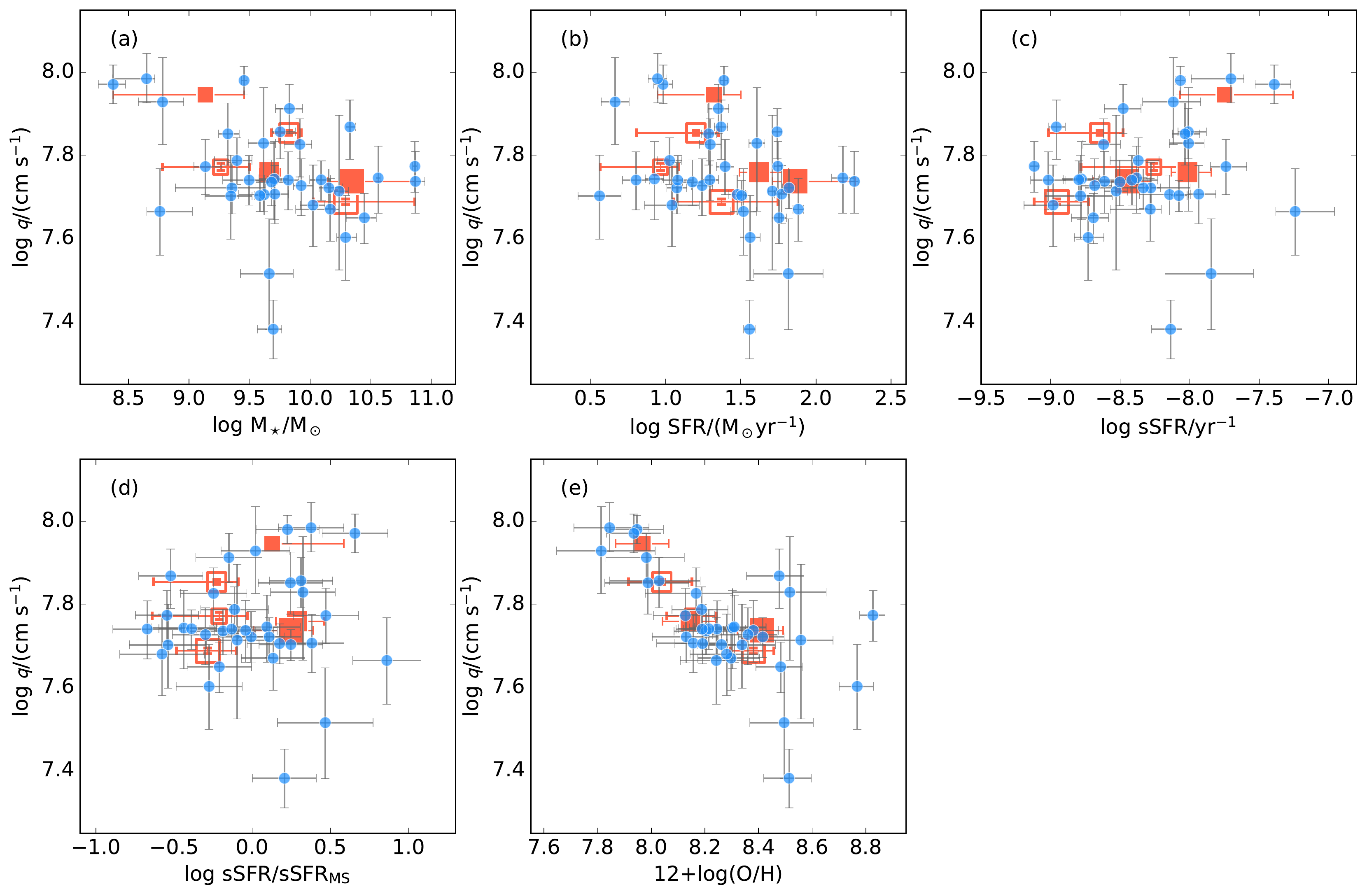}
  }
  \caption{
    Comparison of the ionization parameter of our sample with (a) stellar mass, (b) SFR, (c) sSFR,
    (d) $\Delta_\text{MS} \equiv \text{sSFR}/\text{sSFR}_\text{MS}$,
    and (e) gas-phase oxygen abundance \ohmetal.
    Symbols are same as \autoref{fig:o32r23}.
    \label{fig:logq_prop}
  }
\end{figure*}

\autoref{fig:logq_prop} compares various galaxy physical  quantities  with the ionization parameter. 
The strong correlation between $\log q$ and \ohmetal is trivial, 
as both \ohmetal and $\log q$ strongly depend on \oiii/[\ion{O}{2}]. 
Among the other parameters shown in \autoref{fig:logq_prop},
we do not find significant correlations between  any of them, with the exception of the SFR-log $q$ plot.  
The Spearman's correlation coefficient between SFR and $\log q$ is $-0.35$,
corresponding to a two-sided $p$-value of $0.04$,
when considering objects without upper or lower limit in $\log q$. 
A similar, but more significant correlation between SFR and $O_{32}$ 
is found by \citet{sanders:2015:ionization}
for $z\sim2.3$ SFGs,  with a two-sided $p$-value of 0.002.
This correlation can be understood as a result of a correlation between SFR and metallicity
and an anti-correlation between ionization parameter and metallicity \citep[][but see \citealt{dors:2011}]{perezmontero:2014}.
\citet{nakajima:2014} and \citet{sanders:2015:ionization} found correlations of $O_{32}$ 
with stellar mass and sSFR as well as with SFR.
While we do not see such correlations for individual objects,
stacked data points show some hints of a similar correlation for $\log q$ with stellar mass and sSFR. 

\begin{figure*}
  \centerline{
    \includegraphics[width=\linewidth]{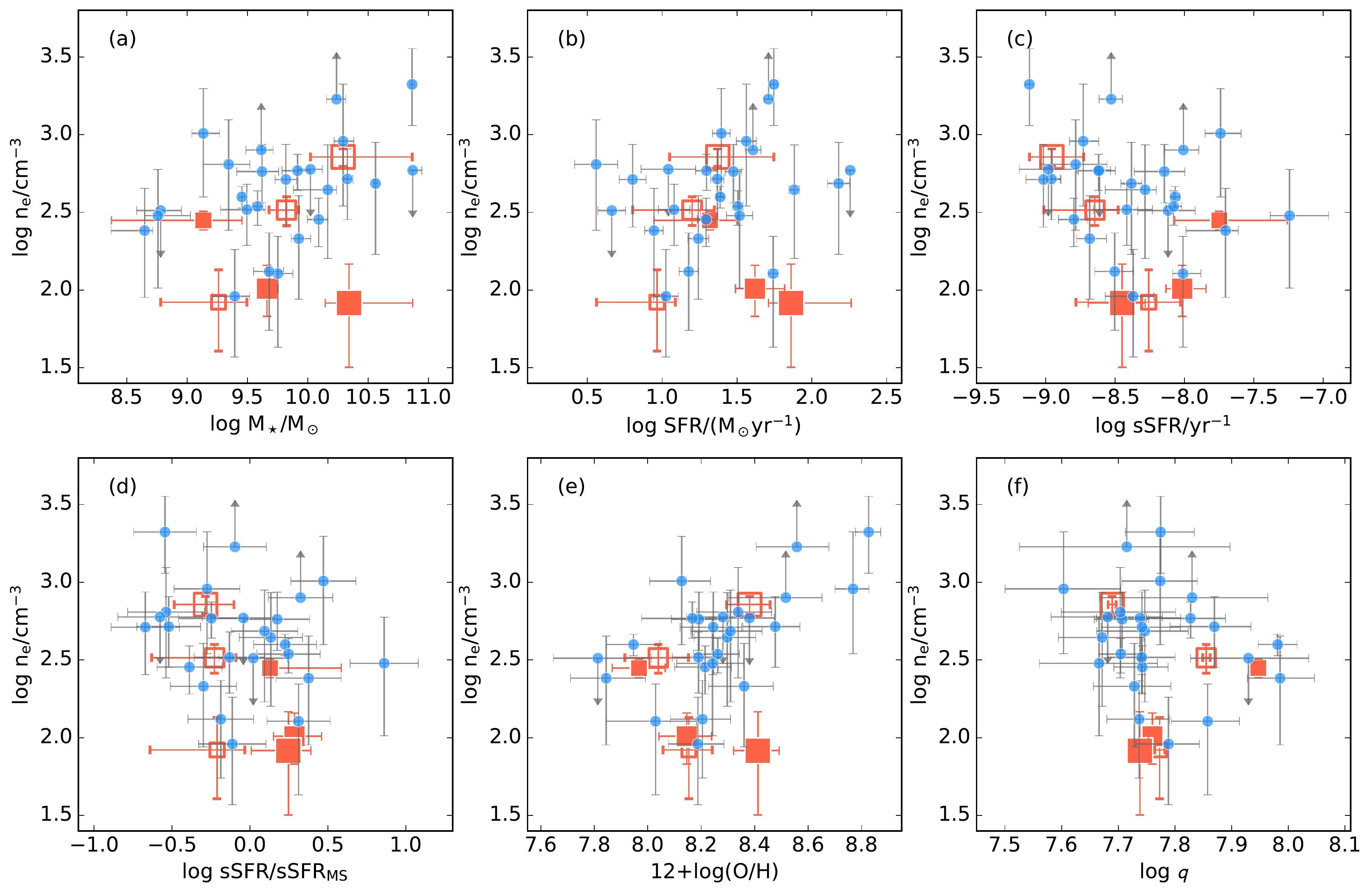}
  }
  \caption{
    Comparison of the electron density of our sample with (a) stellar mass, (b) SFR, (c) sSFR,
    (d) $\Delta_\text{MS} \equiv \text{sSFR}/\text{sSFR}_\text{MS}$, 
    (e) gas-phase oxygen abundance \ohmetal, and
    (f) ionization parameter $\log q$. 
    Symbols are same as \autoref{fig:o32r23}.
    \label{fig:ne_prop}
  }
\end{figure*}

In \autoref{fig:ne_prop} we also compare various physical parameters with the electron density
measured from the line ratio of \oiitot doublet. 
The measured electron densities of $\sim 100$--$1,000\,\text{cm}^{-3}$ are about one order of magnitude
higher than those of typical SFGs at $z=0$,  
roughly consistent with those reported for other high redshift galaxies
\citep[e.g.,][]{masters:2014,shirazi:2014,sanders:2015:ionization,shimakawa:2015}. In \autoref{fig:ne_prop}
there is no indication of strong correlations in any of the parameters with the electron density, 
which is also confirmed by the very low Spearman's rank correlation coefficients, indicating less than $1\sigma$ significance,
for individual objects  as well as for stacked points.

\citet{shimakawa:2015} measured the electron density of a sample of \halpha narrow-band-selected SFGs at $z=2.5$ through resolved \oii doublet.
They found a positive correlation with a $4\sigma$ significance between $n_e$ and sSFR. 
However, from  Panel (c) of \autoref{fig:ne_prop}, the correlation between sSFR and $n_e$ does not seem to be strong
in our  sample. If at all, there is a hint for an opposite trend with  higher sSFR galaxies with lower $n_e$.
Based on a larger sample at $z\sim2.3$, \citet{sanders:2015:ionization} also do not find any correlation
of electron density with stellar mass, SFR or sSFR.

\subsection{[\ion{O}{2}] luminosity as a SFR indicator}

\begin{figure}
  \centerline{
    \includegraphics[width=0.95\linewidth]{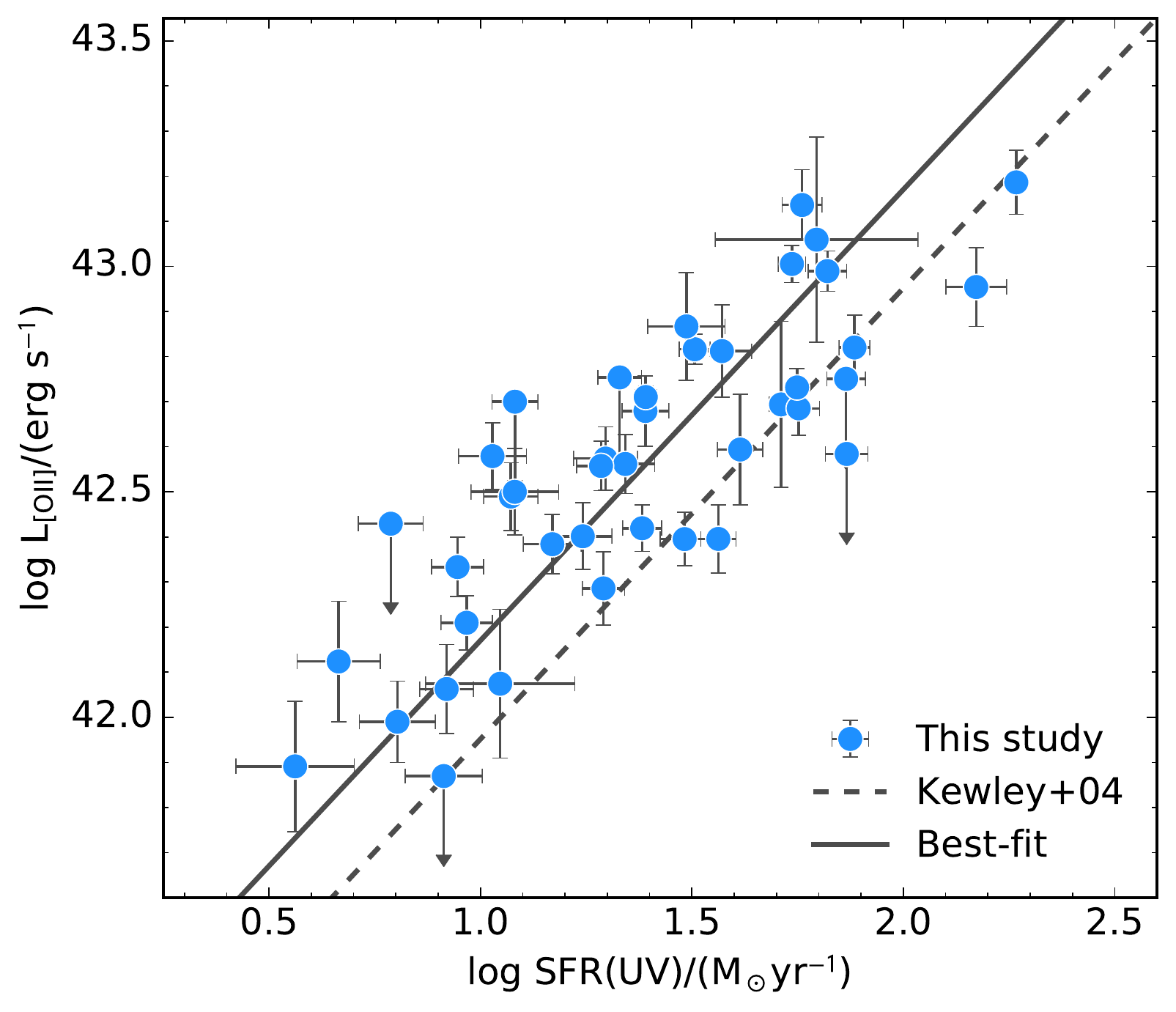}
  }
  \caption{
    \oii luminosity vs UV-based SFR.
    Solid line shows the best-fit linear regression with a slope of unity
    computed by using \texttt{LTS\_LINEFIT} \citep{cappellari:2013:atlas3d15}.
    Dashed line corresponds to Equation 4 in \citet{kewley:2004} converted from Salpeter to Chabrier IMF. 
  \label{fig:sfroii}}
\end{figure}

The \oii emission line luminosity has been used as an indicator of SFR \citep[e.g.,][]{kennicutt:1998},
though it depends not only on SFR, but also on metallicity and ionized gas properties. 
\autoref{fig:sfroii}  shows the relation between the [\ion{O}{2}] luminosity and the UV-based SFR.
Here, both quantities were extinction corrected using UV-based estimates and assuming $E(B-V)_\text{neb}=E(B-V)_\text{star}$ as before. 
The [\ion{O}{2}] luminosities of our sample do not seem to follow the calibration of
$\text{SFR}(\text{[\ion{O}{2}]})$ by \citet{kewley:2004},  shown as  the dashed line in \autoref{fig:sfroii}.
We derived the best-fit calibration of \oii SFR for our $z\gtrsim3$ main-sequence galaxies as
\begin{equation}
  \log \text{SFR}_\text{[O$\;$II]} (M_\odot\,\text{yr}^{-1}) = \log L_\text{[O$\;$II]} (\text{erg}\,\text{s}^{-1}) - 41.17,
\end{equation}
having  fixed the slope to unity.
Our best-fit gives 0.22 dex lower SFRs compared to  the \citet{kewley:2004} calibration.
  If the extinction towards \ion{H}{2} region were higher,  like the original Calzetti law,
  i.e., $E(B-V)_\text{gas}=E(B-V)_\text{star}/0.44$, then
  the discrepancy would become more prominent. 

The elevated [\ion{O}{2}] luminosity relative to the local relation could be due to
a change in the physical condition of star-forming regions as discussed above,
i.e., higher ionization parameter and electron density in high redshift galaxies.

\subsection{Mass--metallicity relation}
\label{sec:mzr}

\begin{figure*}
  \centerline{
    \includegraphics[width=0.48\linewidth]{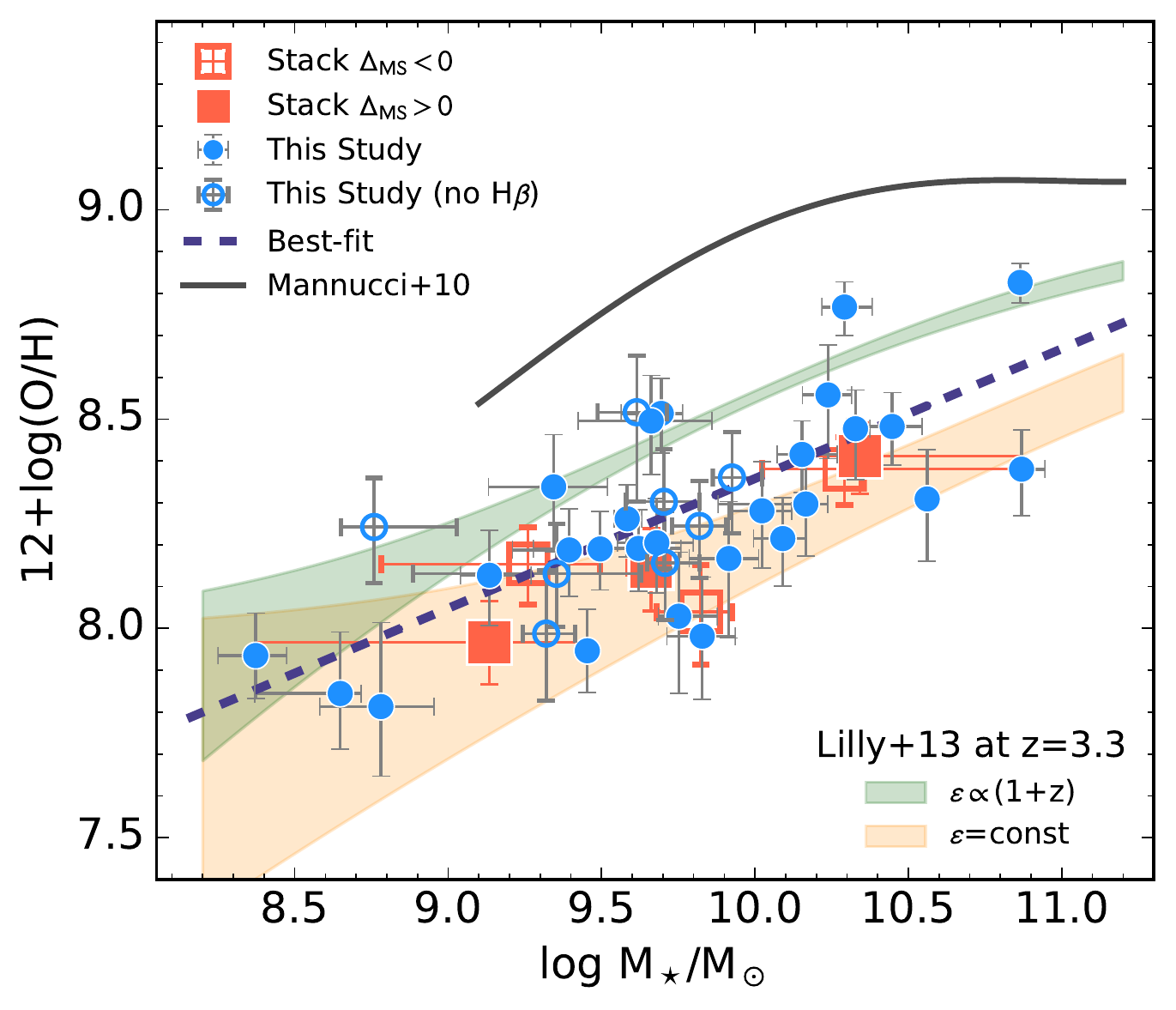}
    \includegraphics[width=0.48\linewidth]{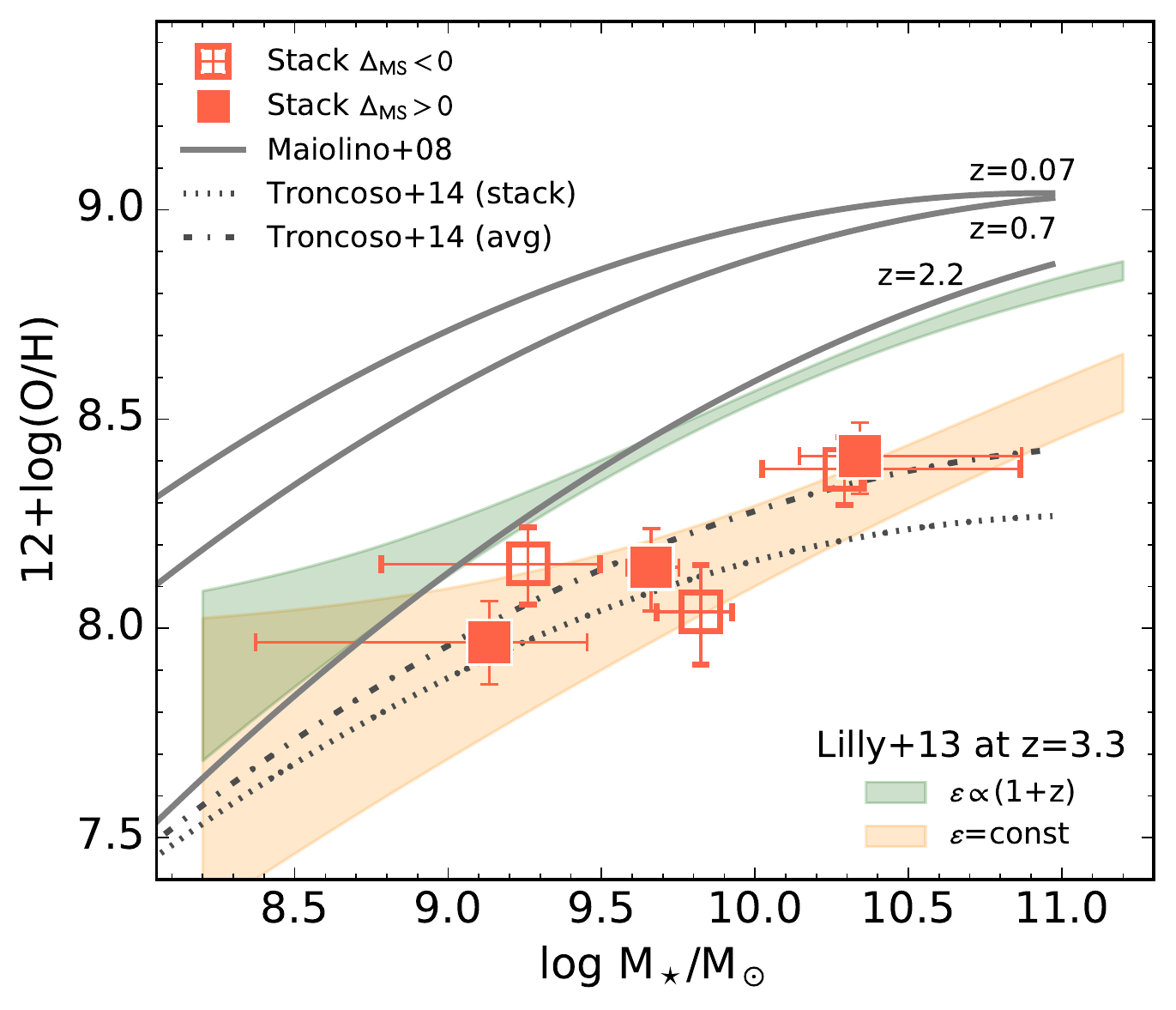}
  }
  \caption{
    (Left) Mass--metallicity relation of our sample is shown with circles.
    The filled and open circles are those with and without \hbeta detection, respectively.
    Squares represent the measurements on the stacked spectra in bins of stellar mass and SFR
    The bins above and below the best-fit MS are shown with filled and open symbols.
    The best-fit linear relation for our sample is shown with dashed line.
    Solid line is the $z=0$ relation \citep{mannucci:2010}.
    (Right) Our $z\sim3.3$ stacked data points (orange squares)
    shown in the left panel are compared with MZR at various redshifts.
    Solid lines are taken from \citet{maiolino:2008}
    for $z=0.07$, $0.72$, and $2.2$ relations
    by \citet{tremonti:2004}, \citet{savaglio:2005}, and \citet{erb:2006:metallicity}, respectively.
    Dotted and dot-dashed lines are mass--metallicity relations from \citet{troncoso:2014}
    defined based on the measurement on the stacked spectra 
    and the average of individual objects, respectively.
    Two filled regions are predictions at $z=3.3$ by a gas-regulation model by \citet{lilly:2013} shown in the left panel.
    In both panels, orange and green shaded areas are prediction from a gas-regulator model \citep{lilly:2013}
    fixing the star formation efficiency $\varepsilon$ to the locally calibrated value
    and allowing $\varepsilon$ to increase as $(1+z)$.
    The areas enclose the case for $0<Z_0/y<0.1$. 
    \label{fig:mzr}
  }
\end{figure*}

\autoref{fig:mzr} shows the MZR for our galaxies, both individually and for the stacked spectra, in comparison with the MZRs at lower redshifts.
The MZR at $z=0.07$, $0.7$, and $2.2$ are from \citet{tremonti:2004}, \citet{savaglio:2005}, and \citet{erb:2006:metallicity}, respectively,
converted to the same metallicity calibration, and parameterized by \citet{maiolino:2008}.
We also plot the MZR at $z\simeq3.4$ from  \citet{troncoso:2014} and that at $z=0$ taken from \citet{mannucci:2010}.
The majority of our sample follows the previously defined MZR at $z\simeq3.4$,
i.e., our MZR offsets by $\simeq 0.7$ dex and $\simeq 0.3$ dex from those at $z\simeq0$ and $z\simeq2$, respectively.
There are, however, a few objects showing higher metallicity, by up to $0.3$--$0.4$ dex compared to the stack points.
Due to a small sample size especially at $M_\star>10^{10.5}M_\odot$,
we do not attempt to constrain the turnover mass of MZR here.
Instead, we carried out a linear regression and found the best-fit MZR for the individual objects in our sample as
\begin{equation}
  \begin{split}
    & \ohmetal \\
    & = (8.36 \pm 0.03) + (0.31+0.05)\times(\log M_\star/M_\odot - 10)
  \label{eq:bestmzr}
  \end{split}
\end{equation}
which is shown in the left panel of \autoref{fig:mzr} with dashed line.

In \autoref{fig:mzr}
we plot ranges between minimum and maximum stellar masses in each bin as error bars for the stack points.
The metallicities from the stacked spectra appear to be lower than the average or median of the individual measurements
at a given stellar mass bin.
This may be due to the way employed for stacking: we normalized each spectrum by the total \oiiitwo flux and
then stacked with weights proportional to the inverse variance.
This procedure gives more weights for \oiii bright objects which tend to have lower metallicities.
Indeed, as shown in \autoref{fig:metal_lineflux} in \autoref{sec:misc_figures},
objects with higher S/N in \oiii tend to have lower metallicity. 
The same trend is also seen in \citet{troncoso:2014} for SFGs at $z\simeq3.4$.
Their best-fit MZR for the average of individual measurements 
shows higher metallicity than that of the measurement on the stacked spectra
(dot-dashed and dotted lines in the left panel of \autoref{fig:mzr}, respectively).
  We also tried stacking with the \oiii normalization but without weighing, and measuring
  metallicity in the same way,
  finding  about 0.1 dex higher \ohmetal compared with those presented above.
  However, our conclusion does not depend  on the choice of the stacking method
  since our analysis on the metallicity is based mainly on individual objects. 

\begin{figure}
  \centering
  \includegraphics[width=0.95\linewidth]{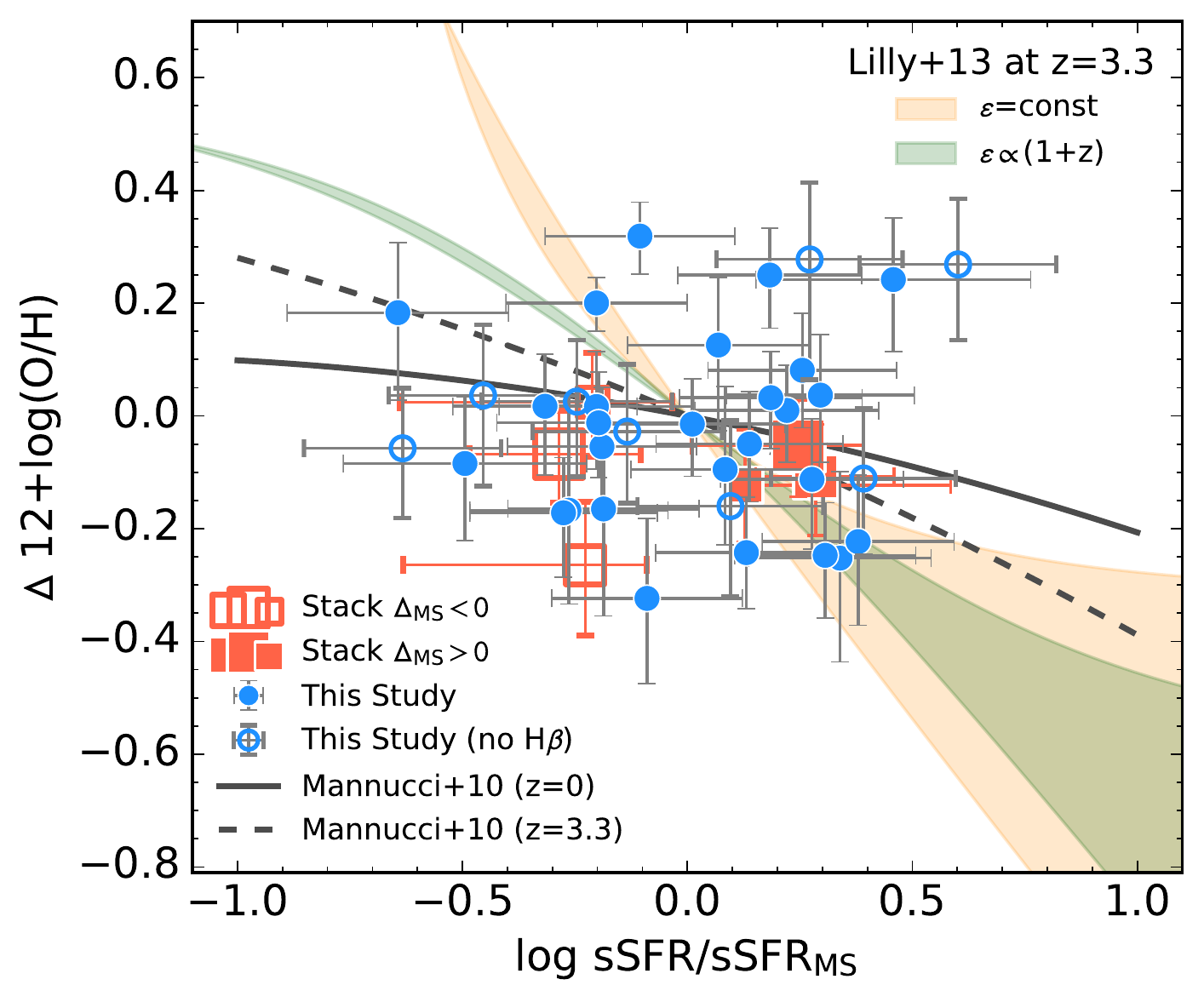}
  \caption{Difference of \ohmetal from the best-fit linear regression in the MZR for our sample,
    $\ohmetal = 8.358 + 0.31(\log M_\star/M_\odot - 10)$,
    as a function of $\Delta_\text{MS}$.
    Symbols of data points and two filled regions are same as \autoref{fig:o32r23} and \autoref{fig:mzr}, respectively,
    and solid and dashed lines represent the fundamental \zmsfr relation by \citet{mannucci:2010}
    adjusted to $z=0$ and $z=3.3$ MS, respectively.
    \citet{lilly:2013} models and \citet{mannucci:2010} relations are obtained for $M=10^{10}M_\odot$ galaxies.
    \label{fig:oh12dms}
  }
\end{figure}

\autoref{fig:oh12dms} shows that 
there appear to be no correlation between \ohmetal and SFR relative to the distances from the best-fit MZR and MS within the error bars
for both stack and individual points, respectively.
A  similar behavior has been reported for $z\sim2$ SFGs
\citep{steidel:2014,wuyts:2014,sanders:2015:metal},
while \citet{zahid:2014} found an anti-correlation between SFR and metallicity in a sample of SFGs at $z\sim 1.6$ \citep[see also][]{yabe:2015}.
This suggests that the role of SFR as a second parameter in the MZR
may be less important at $z \simeq 3.3$ compared to that in the local Universe \citep[e.g.,][]{mannucci:2010,andrews:2013}.
We shall return to  this issue in \autorefsec{sec:mzsfr}.

\begin{figure}
  \centerline{
    \includegraphics[width=0.95\linewidth]{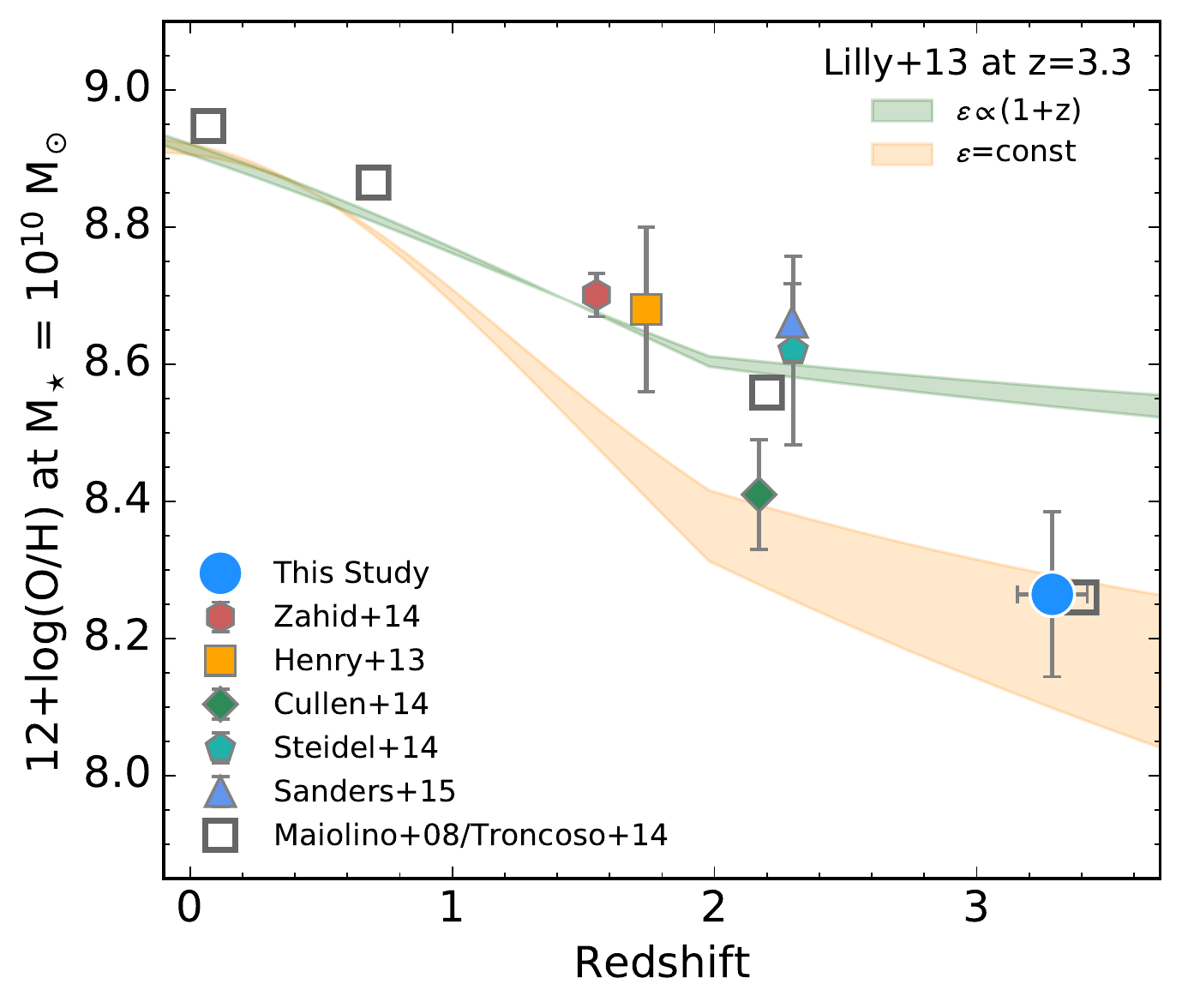}
  }
  \caption{
    Redshift evolution of MZR.
    Our $z\sim3.3$ data is shown with a filled circle.
    The data point and error bar indicate the average and standard deviation of the metallicity
    of the objects with $\log M_\star/M_\odot = 10 \pm 0.2$. 
    Open squares are adopted from Table 5 in \citet{maiolino:2008}
    for $z=0.07$, $0.72$, and $2.2$ relations
    by \citet{tremonti:2004}, \citet{savaglio:2005}, and \citet{erb:2006:metallicity}, respectively.
    An open square at $z=3.4$ is taken from \citet{troncoso:2014}.
    Filled symbols at $z<3$ are taken from the following literature:
    \citet{zahid:2014} for $z=1.55$ (hexagon),
    \citet{henry:2013} for $z=1.7$ (square),
    \citet{cullen:2014} for $z=2.2$ (diamond),
    \citet{steidel:2014} for $z=2.3$ (pentagon), and
    \citet{sanders:2015:metal} for $z=2.3$ (triangle).
    Two filled regions are predictions at $z=3.3$ by a gas-regulation model by \citet{lilly:2013} shown in \autoref{fig:mzr}.
    \label{fig:mzr_redshift}
  }
\end{figure}

The redshift evolution of metallicity at a stellar mass of $10^{10}M_\odot$ is illustrated in the right panel of \autoref{fig:mzr_redshift}.
For the comparison, data points at different redshifts are taken from \citet{tremonti:2004} for $z=0.07$, \citet{savaglio:2005} for $z=0.72$, and \citet{erb:2006:metallicity} for $z=2.2$
after corrected to the same metallicity calibration used here by \citet{maiolino:2008};
\citet{troncoso:2014} for $z=3.4$;
\citet{zahid:2014} for $z=1.55$; \citet{henry:2013} for $z=1.7$;
\citet{cullen:2014} for $z=2.2$;
\citet{steidel:2014} for $z=2.3$;
and \citet{sanders:2015:metal} for $z=2.3$. 
Note that all these  data  are converted to the metallicity calibration of \citet{maiolino:2008},  when needed.

\subsection{Mass--metallicity--SFR relation}
\label{sec:mzsfr}

\begin{figure}
  \centerline{
    \includegraphics[width=0.95\linewidth]{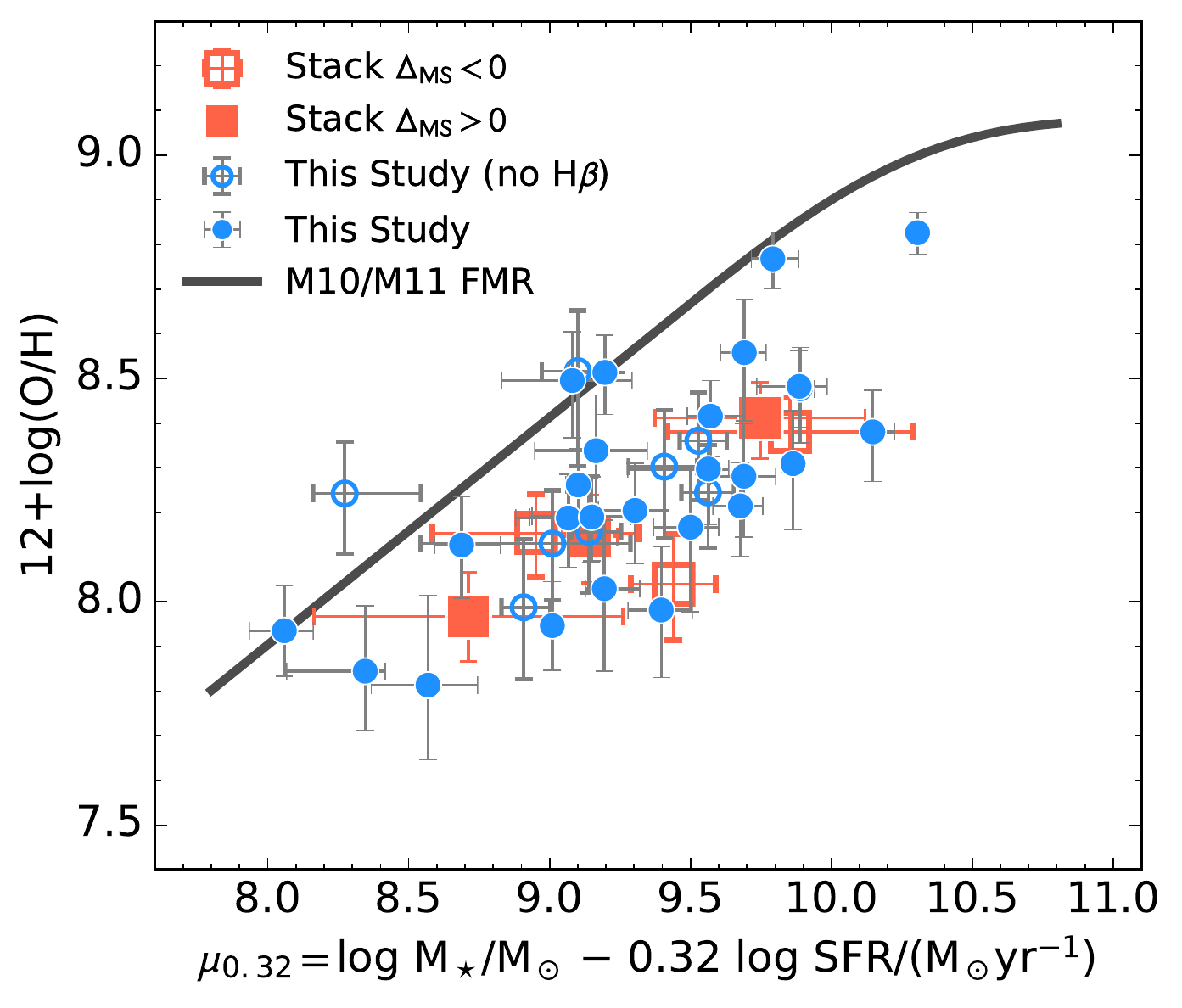}
  }
  \caption{
    Comparison of gas-phase oxygen abundance with $\mu_{0.32}\equiv \log(M_\star [M_\odot]) - 0.32 \log(\text{SFR} [M_\odot\,\text{yr}^{-1}])$.
    Fundamental metallicity relation proposed by \citet{mannucci:2010} is shown with the solid line extended to $\mu_{0.32}<9.5$
    following \citet{mannucci:2011}.
    \label{fig:fmr}
  }
\end{figure}

\citet{mannucci:2010} found that in the local Universe SFGs lie close to a 
three-dimensional surface in the space with SFR, stellar mass and metallicity as coordinates \citep[see also][]{laralopez:2010}. 
They also found that SFGs lie close to the same surface at least to $z\simeq2.5$,
suggesting the existence of the so called fundamental metallicity relation (FMR) and showed that 
a projection of the surface over the plane with coordinates \ohmetal and
$\mu_{0.32}\equiv \log M/M_\odot - 0.32\log\text{SFR}/(M_\odot\,\text{yr}^{-1})$ is able to minimize
the scatter about the surface itself.
\autoref{fig:fmr} shows such two-dimensional plane, where most of the objects in our sample
are offset from the locally defined FMR by $\simeq 0.3$~dex.

\begin{figure}
  \centerline{
    \includegraphics[width=0.9\linewidth]{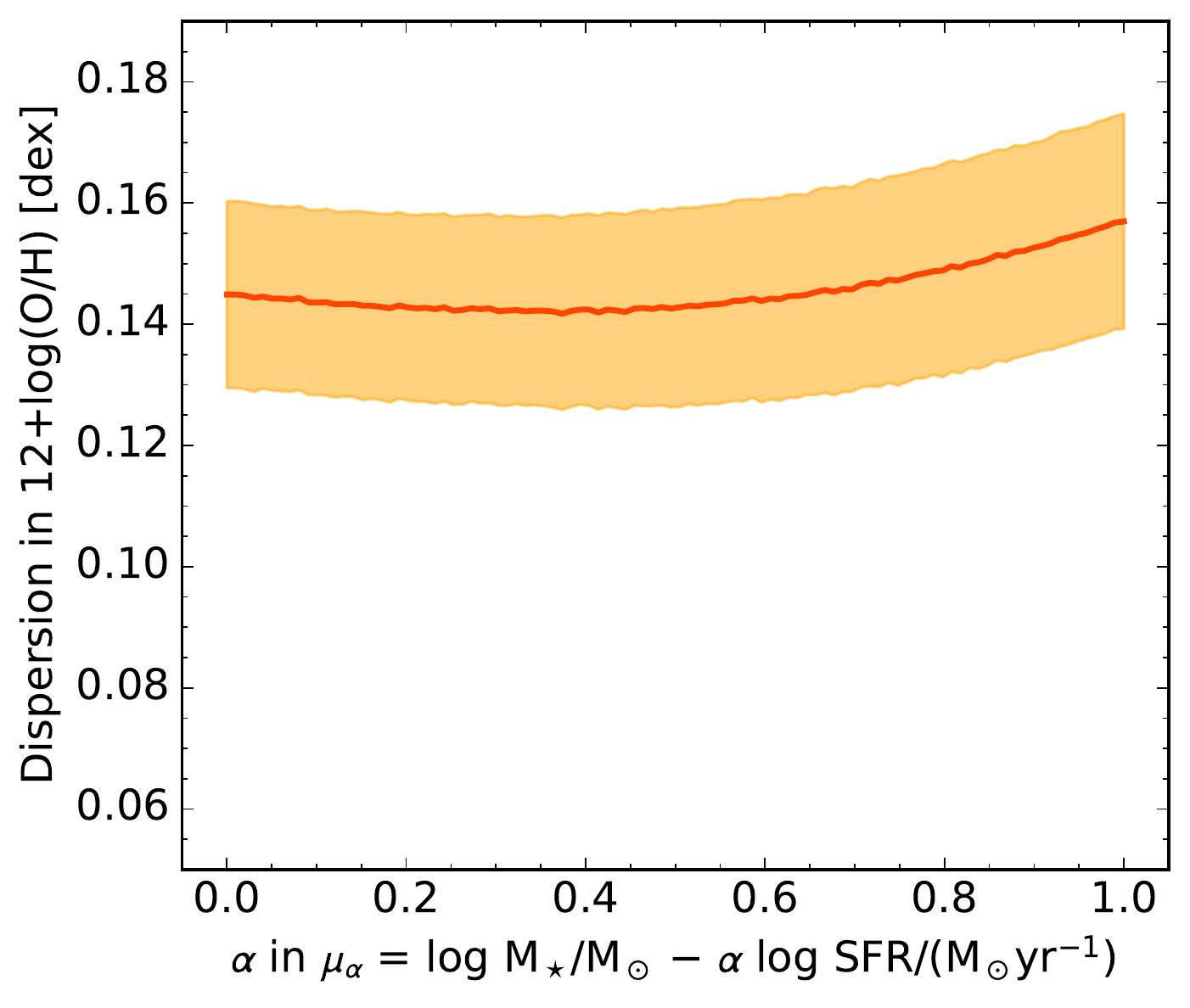}
  }
  \caption{Standard deviation around the best-fit linear regression in the $\ohmetal$--$\mu_\alpha$ relation
    as a function of $\alpha$, where $\mu_\alpha \equiv \log(M_\star)-\alpha\log(\text{SFR})$.
    Solid line and filled region are median and $1\sigma$ confidence interval based on $10^4$ bootstrap resampling. 
    \label{fig:fmr_sigma}}
\end{figure}

Comparing the left panel in \autoref{fig:mzr} and \autoref{fig:fmr}, 
it is not obvious whether in the case of our galaxies using $\mu_{0.32}$ as a coordinate reduces
the scatter in metallicity  in the MZR, as it does locally.
We left the $\alpha$ in $\mu_\alpha \equiv \log(M_\star)-\alpha\log(\text{SFR})$
as a free parameter, and computed the standard deviation around the best-fit linear regression
in the $\ohmetal$--$\mu_\alpha$ relation for our sample.
The scatter as a function of $\alpha$ is shown in \autoref{fig:fmr_sigma}.
\citet{mannucci:2010} claimed that  at $z=0$ with $\alpha=0.32$ the scatter decreases to
$\simeq0.02$ dex compared to $\simeq0.06$ dex in the case of the simple MZR (i.e., $\alpha=0$).
In contrast, varying $\alpha$ does not seem to reduce 
the observed scatter around the best-fit line for our $z=3.3$ galaxies,
but it remains essentially constant around 0.15~dex.
We should note that the typical uncertainties of metallicity from strong lines   are in general 
comparable to the scatter in the MZR,  in this study as well as  in the literature
\citep[e.g.,][]{marino:2013,steidel:2014},
which may explain  why by varying $\alpha$ the scatter does not change.
Indeed, the number of objects in our  sample is still too small to
overcome the dominance of the measurement error and
to draw  firm conclusions  on the intrinsic scatter of our $z\sim3.3$ MZR. 
To  push down the sampling error,
orders of magnitude larger samples would be required, 
e.g.,  $\gtrsim 10^{5}$ objects are used in the  study of \citealt{mannucci:2010} for the local galaxies,
whereas  $z \gtrsim 2$  samples include only $\gtrsim 10^{1\text{-}2}$ objects). 

\subsection{A comparison with a simple gas-regulator model}

Because of the redshift evolution in both  the MS and metallicity,
our sample spans a parameter space which is not well covered by the local galaxy population \citep{maier:2014}.
Therefore, an extrapolation of the \zmsfr relation  is
inevitable when trying to  compare the local relation with the high redshift one. 
A more physically motivated relation among  the three parameters, \zmsfr,
is proposed by \citet{lilly:2013}, which indeed can reproduce the FMR at $z=0$ and at $z\simeq 2.3$ 
with a sensible choice of parameters.
We then use the  \zmsfr relation in Equation (40) in \citet{lilly:2013}:
\begin{equation}
  \begin{split}
    & Z_\text{eq}(M_\star, \text{SFR}) = Z_{0}\, + \\
    & \frac{y}{1 + \lambda(1-R)^{-1} + \varepsilon^{-1}\left\{(1+\beta-b)\text{SFR}/M_\star - (1-R)^{-1}\frac{1.2}{t}\right\}},
  \end{split}
\end{equation}
where $Z_\text{eq}$ is the equilibrium metallicity,
$Z_0$ is the metallicity of the incoming gas,
$y$ is the chemical yield,
$\lambda$ is the mass-loading factor ($\equiv$ outflow rate/SFR),
$R$ is the fraction of mass return due to stellar evolution,
$\varepsilon$ is the star formation efficiency (SFE $\equiv$ SFR/$M_{\rm gas}$),
$\beta$ is the MS slope defined as $\text{SFR} \propto M_\star^{1+\beta}$,
and $t$ is the age of the Universe in units of Gyr.
In  \citet{lilly:2013} $\lambda$ and $\varepsilon$ are parameterized as follows:
\begin{gather}
  \lambda = \lambda_{10} m_{10}^a \\
  \varepsilon = \varepsilon_{10} m_{10}^b,
\end{gather}
where $m_{10}$ is the stellar mass in units of $10^{10}M_\odot$. 

Following  \citet{lilly:2013}, we fix the $\varepsilon$ and $\lambda$ parameters to the values that reproduce the local FMR of 
\citet{mannucci:2010}, and then use Equation (11) to derive the MZR  at  $z=3.3$ predicted by this regulator model.  
The result is shown in \autoref{fig:mzr}  with the orange shaded area encompassing the cases between $Z_0/y=0$ and $Z_0/y=0.1$.

Observations suggest that the SFE may be higher at higher redshift
\citep[e.g.][]{daddi:2010:co,genzel:2010}.
We also show with the green shaded area in \autoref{fig:mzr} the model predictions  
in the case for $\varepsilon \propto (1+z)$.
These two cases virtually enclose our $z\simeq 3.3$ sample, 
while the majority of them is consistent with the prediction with non-evolving $\varepsilon$.

In \autorefsec{sec:mzsfr} we have shown that the scatter of the MZR  does not change appreciably
by introducing the SFR as a second parameter. 
Together with the observed data, \autoref{fig:oh12dms} shows the corresponding relations
as predicted by the regulator models as well as  those observationally defined and extrapolated for the MS at $z=3.3$.
In general,  compared to our data a steeper dependence is predicted of the metallicity offset from  the MZR
as function of the SFR offset from the MS.  
Although this apparent mismatch may be due to  large  errors and small sample size,
it may give some insight into the functional form of the SFH of the galaxies  studied here. 
Indeed, the dependence of metallicity on SFR would disappear
if galaxies were evolving at constant  SFR, i.e., $d\text{SFR}/dt\simeq 0$
(see \autoref{sec:l13eq28} for the detail).

In \autoref{fig:mzr_redshift} we compare the redshift evolution of the MZR at $M_\star = 10^{10} M_\odot$ with
the model predictions described above. 
The evolving \zmsfr relation assuming $\varepsilon \propto (1+z)$ 
traces the observed trend up to $z \simeq 2.5$, with an exception for the data point of \citet[][]{cullen:2014}.
As mentioned above, our sample  appears to prefer a non-evolving star formation efficiency, as do the  data by \citet{maier:2014} at $z\sim 2.3$. 
However, the observational scatter of $\gtrsim 0.1$~dex for the high redshift measurements may
hamper to distinguish models at better than the $\simeq 2.5\sigma$ level.
On the other hand, the departure in metallicity from the original $z=0$ FMR by \citet{mannucci:2010},
as seen in \autoref{fig:fmr}, can be due the extrapolation of
the local FMR into a  parameter space which is basically unpopulated at $z=0$.
Perhaps a more suitable $\gamma$ would lie in-between the two cases above, 0 and 1. 
For instance, a recent study by \citet{genzel:2015} derived $\gamma = 0.34 \pm 0.15$,
based on a combined analysis of CO and dust scaling relations at $0 \lesssim z \lesssim 3$.
Of course, the other parameters of the regulator model
and their mass dependence may also evolve with redshift,
hence being different at $z\simeq 3.3$, but the current sample size is too small to explore the entire parameter space.

\section{Summary}
\label{sec:summary}

Using the rest-frame optical spectra of 43 normal star-forming galaxies lying close to  the star-forming main-sequence at $z\sim3.3$,
we carried out a study of the properties of the ionized gas and their relations with global galaxy parameters.
Our main results can be summarized as follows.

\begin{enumerate}
\item Strong optical emission lines contribute significantly  to the broad-band flux, especially in the \textit{K} band,
  with a median of 21\%, but up to $\sim 100\%$ in a few cases.
    These emission lines affect the estimate of stellar masses
    from SED fitting  by  0.13 dex (median), but the difference may exceed $\sim 0.5$ dex when the emission line contribution
    is $\gtrsim 40\%$.
\item A comparison between UV- and \hbeta-based SFRs  suggests  a lower additional extinction
  toward \ion{H}{2} regions at $z\sim3.3$, compared to the local calibration \citet{calzetti:2000}.
\item The ionization parameter appears to be systematically higher  than its average for  the local SFGs,  extending the local relationship
  in the $R_{23}$ vs. $O_{32}$ diagram.
  The ionization parameter derived from the $O_{32}$ indices does not show any correlation with galaxy global properties,
  except with the SFR.  However, the correlation between SFR and ionization parameter for our sample is not very significant ($\simeq 2\sigma$)
  compared with those in the literature \citep[e.g.,][]{sanders:2015:ionization}.
  Electron density derived from resolved \oii doublets also does not show any correlation with galaxy properties.
\item The MZR of our $z\sim3.3$ galaxies shows a $\sim 0.7$ dex offset from the local relation
  and a $\sim 0.3$ dex offset relative to  $z\sim2$, indicating very rapid evolution of gas-phase metallicity
  at $z\sim 3$.
\item Among our galaxies at $z \simeq 3.3$ we do not find any correlation of metallicity with SFR.
  If such a correlation exists it has to span a metallicity range 
  narrower than measurement errors.
\item Our $z\sim3.3$ sample does not follow the locally defined FMR,
  with metallicities being offset by $\sim 0.3$ dex compared to the value predicted by the FMR for $z=3.3$.
  This mismatch may result from the extrapolation of the \zmsfr relation empirically defined at $z=0$
  to $Z$, SFR and $M_\star$ combinations which are not well populated by $z=0$ galaxies, hence
  giving  an incorrect prediction for the evolution of MZR. 
\item For our sample, no projection of the \zmsfr relation to the $\mu_\alpha$ space is able to reduce the scatter in metallicity,
suggesting that the SFR may not play a significant role as a second parameter in the  MZR.
  However, the uncertainties in metallicity determination and the small sample size
  prevent us to draw a firm conclusion on this issue.
\item A comparison of the MZR of our galaxies and those at different redshifts
  with the prediction of a simple gas-regulator model suggests that a weakly evolving star formation efficiency could better account for 
   the observed redshift evolution of MZR.
\end{enumerate}

\acknowledgments
The data presented herein were obtained at the W.M. Keck Observatory,
which is operated as a scientific partnership among the California Institute of Technology,
the University of California and the National Aeronautics and Space Administration.
The Observatory was made possible by the generous financial support of the W.M. Keck Foundation.
The authors wish to recognize and acknowledge the very significant cultural role and reverence
that the summit of Mauna Kea has always had within the indigenous Hawaiian community.
We are most fortunate to have the opportunity to conduct observations from this mountain.
The observations were partly carried out within the framework of Subaru--Keck time exchange program.
This research made use of \texttt{Astropy}\footnote{\url{http://www.astropy.org}},
a community-developed core Python package for Astronomy \citep{astropy},
\texttt{APLpy}\footnote{\url{http://aplpy.github.com}}, an open-source plotting package for Python, 
and \texttt{matplotlib}\footnote{\url{http://matplotlib.org}}, a Python 2D plotting library \citep{matplotlib}.
We thank Scott Dahm, Marc Kassis, Jim Lyke, and Greg Wirth, and the rest of the staff at the Keck observatory
for supporting the observations,
Andreas Faisst, Nicholas Konidaris, Luca Rizzi, and Benny Trakhtenbrot for the assistance on the MOSFIRE data reduction, 
and Roberto Maiolino, Claudia Scarlata, and Maryam Shirazi for insightful discussions.
We thank the anonymous referee for providing constructive comments. 
AR is grateful to the Institute for Astronomy at ETH Zurich for its kind hospitality while working at this project.

{\it Facilities:} \facility{Keck I (MOSFIRE)}




\begin{thebibliography}{}
\providecommand\natexlab[1]{#1}
\providecommand\JournalTitle[1]{#1}

\bibitem[{{Andrews} \& {Martini}(2013)}]{andrews:2013}
{Andrews}, B.~H., \& {Martini}, P. 2013,
  \href{http://dx.doi.org/10.1088/0004-637X/765/2/140}{\JournalTitle{\apj},
  765, 140}

\bibitem[{{Astropy Collaboration} {et~al.}(2013){Astropy Collaboration},
  {Robitaille}, {Tollerud}, {Greenfield}, {Droettboom}, {Bray}, {Aldcroft},
  {Davis}, {Ginsburg}, {Price-Whelan}, {Kerzendorf}, {Conley}, {Crighton},
  {Barbary}, {Muna}, {Ferguson}, {Grollier}, {Parikh}, {Nair}, {Unther},
  {Deil}, {Woillez}, {Conseil}, {Kramer}, {Turner}, {Singer}, {Fox}, {Weaver},
  {Zabalza}, {Edwards}, {Azalee Bostroem}, {Burke}, {Casey}, {Crawford},
  {Dencheva}, {Ely}, {Jenness}, {Labrie}, {Lim}, {Pierfederici}, {Pontzen},
  {Ptak}, {Refsdal}, {Servillat}, \& {Streicher}}]{astropy}
{Astropy Collaboration}, {Robitaille}, T.~P., {Tollerud}, E.~J., {et~al.} 2013,
  \href{http://dx.doi.org/10.1051/0004-6361/201322068}{\JournalTitle{\aap},
  558, A33}

\bibitem[{{Baldwin} {et~al.}(1991){Baldwin}, {Ferland}, {Martin}, {Corbin},
  {Cota}, {Peterson}, \& {Slettebak}}]{baldwin:1991}
{Baldwin}, J.~A., {Ferland}, G.~J., {Martin}, P.~G., {et~al.} 1991,
  \href{http://dx.doi.org/10.1086/170146}{\JournalTitle{\apj}, 374, 580}

\bibitem[{{Barro} {et~al.}(2015){Barro}, {Faber}, {Dekel}, {Pacifici},
  {Perez-Gonzalez}, {Toloba}, {Koo}, {Trump}, {Inoue}, {Guo}, {Liu}, {Primack},
  {Koekemoer}, {Brammer}, {Cava}, {Cardiel}, {Ceverino}, {Eliche}, {Fang},
  {Finkelstein}, {Kocevski}, {Livermore}, \& {McGrath}}]{barro:2015}
{Barro}, G., {Faber}, S.~M., {Dekel}, A., {et~al.} 2015, \JournalTitle{ArXiv
  e-prints}, \href{http://arxiv.org/abs/1503.07164}{{\sffamily
  arXiv:1503.07164}}

\bibitem[{{Belli} {et~al.}(2013){Belli}, {Jones}, {Ellis}, \&
  {Richard}}]{belli:2013}
{Belli}, S., {Jones}, T., {Ellis}, R.~S., \& {Richard}, J. 2013,
  \href{http://dx.doi.org/10.1088/0004-637X/772/2/141}{\JournalTitle{\apj},
  772, 141}

\bibitem[{{Belli} {et~al.}(2015){Belli}, {Newman}, \& {Ellis}}]{belli:2015}
{Belli}, S., {Newman}, A.~B., \& {Ellis}, R.~S. 2015,
  \href{http://dx.doi.org/10.1088/0004-637X/799/2/206}{\JournalTitle{\apj},
  799, 206}

\bibitem[{{Bouch{\'e}} {et~al.}(2010){Bouch{\'e}}, {Dekel}, {Genzel}, {Genel},
  {Cresci}, {F{\"o}rster Schreiber}, {Shapiro}, {Davies}, \&
  {Tacconi}}]{bouche:2010}
{Bouch{\'e}}, N., {Dekel}, A., {Genzel}, R., {et~al.} 2010,
  \href{http://dx.doi.org/10.1088/0004-637X/718/2/1001}{\JournalTitle{\apj},
  718, 1001}

\bibitem[{{Brammer} {et~al.}(2012){Brammer}, {van Dokkum}, {Franx},
  {Fumagalli}, {Patel}, {Rix}, {Skelton}, {Kriek}, {Nelson}, {Schmidt},
  {Bezanson}, {da Cunha}, {Erb}, {Fan}, {F{\"o}rster Schreiber}, {Illingworth},
  {Labb{\'e}}, {Leja}, {Lundgren}, {Magee}, {Marchesini}, {McCarthy},
  {Momcheva}, {Muzzin}, {Quadri}, {Steidel}, {Tal}, {Wake}, {Whitaker}, \&
  {Williams}}]{brammer:2012}
{Brammer}, G.~B., {van Dokkum}, P.~G., {Franx}, M., {et~al.} 2012,
  \href{http://dx.doi.org/10.1088/0067-0049/200/2/13}{\JournalTitle{\apjs},
  200, 13}

\bibitem[{{Brinchmann} {et~al.}(2004){Brinchmann}, {Charlot}, {White},
  {Tremonti}, {Kauffmann}, {Heckman}, \& {Brinkmann}}]{brinchmann:2004}
{Brinchmann}, J., {Charlot}, S., {White}, S.~D.~M., {et~al.} 2004,
  \href{http://dx.doi.org/10.1111/j.1365-2966.2004.07881.x}{\JournalTitle{\mnras},
  351, 1151}

\bibitem[{{Bruzual} \& {Charlot}(2003)}]{bruzual:2003}
{Bruzual}, G., \& {Charlot}, S. 2003,
  \href{http://dx.doi.org/10.1046/j.1365-8711.2003.06897.x}{\JournalTitle{\mnras},
  344, 1000}

\bibitem[{{Calzetti} {et~al.}(2000){Calzetti}, {Armus}, {Bohlin}, {Kinney},
  {Koornneef}, \& {Storchi-Bergmann}}]{calzetti:2000}
{Calzetti}, D., {Armus}, L., {Bohlin}, R.~C., {et~al.} 2000,
  \href{http://dx.doi.org/10.1086/308692}{\JournalTitle{\apj}, 533, 682}

\bibitem[{{Capak} {et~al.}(2007){Capak}, {Aussel}, {Ajiki}, {McCracken},
  {Mobasher}, {Scoville}, {Shopbell}, {Taniguchi}, {Thompson}, {Tribiano},
  {Sasaki}, {Blain}, {Brusa}, {Carilli}, {Comastri}, {Carollo}, {Cassata},
  {Colbert}, {Ellis}, {Elvis}, {Giavalisco}, {Green}, {Guzzo}, {Hasinger},
  {Ilbert}, {Impey}, {Jahnke}, {Kartaltepe}, {Kneib}, {Koda}, {Koekemoer},
  {Komiyama}, {Leauthaud}, {Lefevre}, {Lilly}, {Liu}, {Massey}, {Miyazaki},
  {Murayama}, {Nagao}, {Peacock}, {Pickles}, {Porciani}, {Renzini}, {Rhodes},
  {Rich}, {Salvato}, {Sanders}, {Scarlata}, {Schiminovich}, {Schinnerer},
  {Scodeggio}, {Sheth}, {Shioya}, {Tasca}, {Taylor}, {Yan}, \&
  {Zamorani}}]{capak:2007}
{Capak}, P., {Aussel}, H., {Ajiki}, M., {et~al.} 2007,
  \href{http://dx.doi.org/10.1086/519081}{\JournalTitle{\apjs}, 172, 99}

\bibitem[{{Cappellari} {et~al.}(2013){Cappellari}, {Scott}, {Alatalo}, {Blitz},
  {Bois}, {Bournaud}, {Bureau}, {Crocker}, {Davies}, {Davis}, {de Zeeuw},
  {Duc}, {Emsellem}, {Khochfar}, {Krajnovi{\'c}}, {Kuntschner}, {McDermid},
  {Morganti}, {Naab}, {Oosterloo}, {Sarzi}, {Serra}, {Weijmans}, \&
  {Young}}]{cappellari:2013:atlas3d15}
{Cappellari}, M., {Scott}, N., {Alatalo}, K., {et~al.} 2013,
  \href{http://dx.doi.org/10.1093/mnras/stt562}{\JournalTitle{\mnras}, 432,
  1709}

\bibitem[{{Cardelli} {et~al.}(1989){Cardelli}, {Clayton}, \&
  {Mathis}}]{cardelli:1989}
{Cardelli}, J.~A., {Clayton}, G.~C., \& {Mathis}, J.~S. 1989,
  \href{http://dx.doi.org/10.1086/167900}{\JournalTitle{\apj}, 345, 245}

\bibitem[{{Carollo} \& {Lilly}(2001)}]{carollo:2001}
{Carollo}, C.~M., \& {Lilly}, S.~J. 2001,
  \href{http://dx.doi.org/10.1086/319104}{\JournalTitle{\apjl}, 548, L153}

\bibitem[{{Carollo} {et~al.}(2013){Carollo}, {Bschorr}, {Renzini}, {Lilly},
  {Capak}, {Cibinel}, {Ilbert}, {Onodera}, {Scoville}, {Cameron}, {Mobasher},
  {Sanders}, \& {Taniguchi}}]{carollo:2013}
{Carollo}, C.~M., {Bschorr}, T.~J., {Renzini}, A., {et~al.} 2013,
  \href{http://dx.doi.org/10.1088/0004-637X/773/2/112}{\JournalTitle{\apj},
  773, 112}

\bibitem[{{Carollo} {et~al.}(2014){Carollo}, {Cibinel}, {Lilly}, {Pipino},
  {Bonoli}, {Finoguenov}, {Miniati}, {Norberg}, \& {Silverman}}]{carollo:2014}
{Carollo}, C.~M., {Cibinel}, A., {Lilly}, S.~J., {et~al.} 2014,
  \JournalTitle{ArXiv e-prints},
  \href{http://arxiv.org/abs/1402.1172}{{\sffamily arXiv:1402.1172
  [astro-ph.CO]}}

\bibitem[{{Castellano} {et~al.}(2014){Castellano}, {Sommariva}, {Fontana},
  {Pentericci}, {Santini}, {Grazian}, {Amorin}, {Donley}, {Dunlop}, {Ferguson},
  {Fiore}, {Galametz}, {Giallongo}, {Guo}, {Huang}, {Koekemoer}, {Maiolino},
  {McLure}, {Paris}, {Schaerer}, {Troncoso}, \& {Vanzella}}]{castellano:2014}
{Castellano}, M., {Sommariva}, V., {Fontana}, A., {et~al.} 2014,
  \href{http://dx.doi.org/10.1051/0004-6361/201322704}{\JournalTitle{\aap},
  566, A19}

\bibitem[{{Chabrier}(2003)}]{chabrier:2003}
{Chabrier}, G. 2003,
  \href{http://dx.doi.org/10.1086/376392}{\JournalTitle{\pasp}, 115, 763}

\bibitem[{{Civano} {et~al.}(2011){Civano}, {Brusa}, {Comastri}, {Elvis},
  {Salvato}, {Zamorani}, {Capak}, {Fiore}, {Gilli}, {Hao}, {Ikeda}, {Kakazu},
  {Kartaltepe}, {Masters}, {Miyaji}, {Mignoli}, {Puccetti}, {Shankar},
  {Silverman}, {Vignali}, {Zezas}, \& {Koekemoer}}]{civano:2011}
{Civano}, F., {Brusa}, M., {Comastri}, A., {et~al.} 2011,
  \href{http://dx.doi.org/10.1088/0004-637X/741/2/91}{\JournalTitle{\apj}, 741,
  91}

\bibitem[{{Civano} {et~al.}(2016){Civano}, {Marchesi}, {Comastri}, {Urry},
  {Elvis}, {Cappelluti}, {Puccetti}, {Brusa}, {Zamorani}, {Hasinger},
  {Aldcroft}, {Alexander}, {Allevato}, {Brunner}, {Capak}, {Finoguenov},
  {Fiore}, {Fruscione}, {Gilli}, {Glotfelty}, {Griffiths}, {Hao}, {Harrison},
  {Jahnke}, {Kartaltepe}, {Karim}, {LaMassa}, {Lanzuisi}, {Miyaji}, {Ranalli},
  {Salvato}, {Sargent}, {Scoville}, {Schawinski}, {Schinnerer}, {Silverman},
  {Smolcic}, {Stern}, {Toft}, {Trakhenbrot}, {Treister}, \&
  {Vignali}}]{civano:2016}
{Civano}, F., {Marchesi}, S., {Comastri}, A., {et~al.} 2016,
  \JournalTitle{ArXiv e-prints},
  \href{http://arxiv.org/abs/1601.00941}{{\sffamily arXiv:1601.00941}}

\bibitem[{{Cullen} {et~al.}(2014){Cullen}, {Cirasuolo}, {McLure}, {Dunlop}, \&
  {Bowler}}]{cullen:2014}
{Cullen}, F., {Cirasuolo}, M., {McLure}, R.~J., {Dunlop}, J.~S., \& {Bowler},
  R.~A.~A. 2014,
  \href{http://dx.doi.org/10.1093/mnras/stu443}{\JournalTitle{\mnras}, 440,
  2300}

\bibitem[{{Daddi} {et~al.}(2004){Daddi}, {Cimatti}, {Renzini}, {Fontana},
  {Mignoli}, {Pozzetti}, {Tozzi}, \& {Zamorani}}]{daddi:2004:bzk}
{Daddi}, E., {Cimatti}, A., {Renzini}, A., {et~al.} 2004,
  \href{http://dx.doi.org/10.1086/425569}{\JournalTitle{\apj}, 617, 746}

\bibitem[{{Daddi} {et~al.}(2007){Daddi}, {Dickinson}, {Morrison}, {Chary},
  {Cimatti}, {Elbaz}, {Frayer}, {Renzini}, {Pope}, {Alexander}, {Bauer},
  {Giavalisco}, {Huynh}, {Kurk}, \& {Mignoli}}]{daddi:2007:sfr}
{Daddi}, E., {Dickinson}, M., {Morrison}, G., {et~al.} 2007,
  \href{http://dx.doi.org/10.1086/521818}{\JournalTitle{\apj}, 670, 156}

\bibitem[{{Daddi} {et~al.}(2010){Daddi}, {Bournaud}, {Walter}, {Dannerbauer},
  {Carilli}, {Dickinson}, {Elbaz}, {Morrison}, {Riechers}, {Onodera}, {Salmi},
  {Krips}, \& {Stern}}]{daddi:2010:co}
{Daddi}, E., {Bournaud}, F., {Walter}, F., {et~al.} 2010,
  \href{http://dx.doi.org/10.1088/0004-637X/713/1/686}{\JournalTitle{\apj},
  713, 686}

\bibitem[{{Dav{\'e}} {et~al.}(2012){Dav{\'e}}, {Finlator}, \&
  {Oppenheimer}}]{dave:2012}
{Dav{\'e}}, R., {Finlator}, K., \& {Oppenheimer}, B.~D. 2012,
  \href{http://dx.doi.org/10.1111/j.1365-2966.2011.20148.x}{\JournalTitle{\mnras},
  421, 98}

\bibitem[{{Dayal} {et~al.}(2013){Dayal}, {Ferrara}, \& {Dunlop}}]{dayal:2013}
{Dayal}, P., {Ferrara}, A., \& {Dunlop}, J.~S. 2013,
  \href{http://dx.doi.org/10.1093/mnras/stt083}{\JournalTitle{\mnras}, 430,
  2891}

\bibitem[{{Dekel} {et~al.}(2013){Dekel}, {Zolotov}, {Tweed}, {Cacciato},
  {Ceverino}, \& {Primack}}]{dekel:2013b}
{Dekel}, A., {Zolotov}, A., {Tweed}, D., {et~al.} 2013,
  \href{http://dx.doi.org/10.1093/mnras/stt1338}{\JournalTitle{\mnras}, 435,
  999}

\bibitem[{{Dopita} {et~al.}(2013){Dopita}, {Sutherland}, {Nicholls}, {Kewley},
  \& {Vogt}}]{dopita:2013}
{Dopita}, M.~A., {Sutherland}, R.~S., {Nicholls}, D.~C., {Kewley}, L.~J., \&
  {Vogt}, F.~P.~A. 2013,
  \href{http://dx.doi.org/10.1088/0067-0049/208/1/10}{\JournalTitle{\apjs},
  208, 10}

\bibitem[{{Dors} {et~al.}(2011){Dors}, {Krabbe}, {H{\"a}gele}, \&
  {P{\'e}rez-Montero}}]{dors:2011}
{Dors}, Jr., O.~L., {Krabbe}, A., {H{\"a}gele}, G.~F., \& {P{\'e}rez-Montero},
  E. 2011,
  \href{http://dx.doi.org/10.1111/j.1365-2966.2011.18978.x}{\JournalTitle{\mnras},
  415, 3616}

\bibitem[{{Dutton} {et~al.}(2010){Dutton}, {van den Bosch}, \&
  {Dekel}}]{dutton:2010}
{Dutton}, A.~A., {van den Bosch}, F.~C., \& {Dekel}, A. 2010,
  \href{http://dx.doi.org/10.1111/j.1365-2966.2010.16620.x}{\JournalTitle{\mnras},
  405, 1690}

\bibitem[{{Elbaz} {et~al.}(2007){Elbaz}, {Daddi}, {Le Borgne}, {Dickinson},
  {Alexander}, {Chary}, {Starck}, {Brandt}, {Kitzbichler}, {MacDonald},
  {Nonino}, {Popesso}, {Stern}, \& {Vanzella}}]{elbaz:2007}
{Elbaz}, D., {Daddi}, E., {Le Borgne}, D., {et~al.} 2007,
  \href{http://dx.doi.org/10.1051/0004-6361:20077525}{\JournalTitle{\aap}, 468,
  33}

\bibitem[{{Ellison} {et~al.}(2008){Ellison}, {Patton}, {Simard}, \&
  {McConnachie}}]{ellison:2008}
{Ellison}, S.~L., {Patton}, D.~R., {Simard}, L., \& {McConnachie}, A.~W. 2008,
  \href{http://dx.doi.org/10.1086/527296}{\JournalTitle{\apjl}, 672, L107}

\bibitem[{{Erb} {et~al.}(2006{\natexlab{a}}){Erb}, {Shapley}, {Pettini},
  {Steidel}, {Reddy}, \& {Adelberger}}]{erb:2006:metallicity}
{Erb}, D.~K., {Shapley}, A.~E., {Pettini}, M., {et~al.} 2006{\natexlab{a}},
  \href{http://dx.doi.org/10.1086/503623}{\JournalTitle{\apj}, 644, 813}

\bibitem[{{Erb} {et~al.}(2006{\natexlab{b}}){Erb}, {Steidel}, {Shapley},
  {Pettini}, {Reddy}, \& {Adelberger}}]{erb:2006:survey}
{Erb}, D.~K., {Steidel}, C.~C., {Shapley}, A.~E., {et~al.} 2006{\natexlab{b}},
  \href{http://dx.doi.org/10.1086/505341}{\JournalTitle{\apj}, 647, 128}

\bibitem[{{Erb} {et~al.}(2014){Erb}, {Steidel}, {Trainor}, {Bogosavljevi{\'c}},
  {Shapley}, {Nestor}, {Kulas}, {Law}, {Strom}, {Rudie}, {Reddy}, {Pettini},
  {Konidaris}, {Mace}, {Matthews}, \& {McLean}}]{erb:2014}
{Erb}, D.~K., {Steidel}, C.~C., {Trainor}, R.~F., {et~al.} 2014,
  \href{http://dx.doi.org/10.1088/0004-637X/795/1/33}{\JournalTitle{\apj}, 795,
  33}

\bibitem[{{Feldmann} {et~al.}(2006){Feldmann}, {Carollo}, {Porciani}, {Lilly},
  {Capak}, {Taniguchi}, {Le F{\`e}vre}, {Renzini}, {Scoville}, {Ajiki},
  {Aussel}, {Contini}, {McCracken}, {Mobasher}, {Murayama}, {Sanders},
  {Sasaki}, {Scarlata}, {Scodeggio}, {Shioya}, {Silverman}, {Takahashi},
  {Thompson}, \& {Zamorani}}]{feldmann:2006:zebra}
{Feldmann}, R., {Carollo}, C.~M., {Porciani}, C., {et~al.} 2006,
  \href{http://dx.doi.org/10.1111/j.1365-2966.2006.10930.x}{\JournalTitle{\mnras},
  372, 565}

\bibitem[{{Finkelstein} {et~al.}(2011){Finkelstein}, {Hill}, {Gebhardt},
  {Adams}, {Blanc}, {Papovich}, {Ciardullo}, {Drory}, {Gawiser}, {Gronwall},
  {Schneider}, \& {Tran}}]{finkelstein:2011}
{Finkelstein}, S.~L., {Hill}, G.~J., {Gebhardt}, K., {et~al.} 2011,
  \href{http://dx.doi.org/10.1088/0004-637X/729/2/140}{\JournalTitle{\apj},
  729, 140}

\bibitem[{{Fitzpatrick}(1999)}]{fitzpatrick:1999}
{Fitzpatrick}, E.~L. 1999,
  \href{http://dx.doi.org/10.1086/316293}{\JournalTitle{\pasp}, 111, 63}

\bibitem[{{Forbes} {et~al.}(2014){Forbes}, {Krumholz}, {Burkert}, \&
  {Dekel}}]{forbes:2014}
{Forbes}, J.~C., {Krumholz}, M.~R., {Burkert}, A., \& {Dekel}, A. 2014,
  \href{http://dx.doi.org/10.1093/mnras/stu1142}{\JournalTitle{\mnras}, 443,
  168}

\bibitem[{{Francis} {et~al.}(1991){Francis}, {Hewett}, {Foltz}, {Chaffee},
  {Weymann}, \& {Morris}}]{francis:1991}
{Francis}, P.~J., {Hewett}, P.~C., {Foltz}, C.~B., {et~al.} 1991,
  \href{http://dx.doi.org/10.1086/170066}{\JournalTitle{\apj}, 373, 465}

\bibitem[{{Genzel} {et~al.}(2010){Genzel}, {Tacconi}, {Gracia-Carpio},
  {Sternberg}, {Cooper}, {Shapiro}, {Bolatto}, {Bouch{\'e}}, {Bournaud},
  {Burkert}, {Combes}, {Comerford}, {Cox}, {Davis}, {Schreiber},
  {Garcia-Burillo}, {Lutz}, {Naab}, {Neri}, {Omont}, {Shapley}, \&
  {Weiner}}]{genzel:2010}
{Genzel}, R., {Tacconi}, L.~J., {Gracia-Carpio}, J., {et~al.} 2010,
  \href{http://dx.doi.org/10.1111/j.1365-2966.2010.16969.x}{\JournalTitle{\mnras},
  407, 2091}

\bibitem[{{Genzel} {et~al.}(2015){Genzel}, {Tacconi}, {Lutz}, {Saintonge},
  {Berta}, {Magnelli}, {Combes}, {Garc{\'{\i}}a-Burillo}, {Neri}, {Bolatto},
  {Contini}, {Lilly}, {Boissier}, {Boone}, {Bouch{\'e}}, {Bournaud}, {Burkert},
  {Carollo}, {Colina}, {Cooper}, {Cox}, {Feruglio}, {F{\"o}rster Schreiber},
  {Freundlich}, {Gracia-Carpio}, {Juneau}, {Kovac}, {Lippa}, {Naab}, {Salome},
  {Renzini}, {Sternberg}, {Walter}, {Weiner}, {Weiss}, \&
  {Wuyts}}]{genzel:2015}
{Genzel}, R., {Tacconi}, L.~J., {Lutz}, D., {et~al.} 2015,
  \href{http://dx.doi.org/10.1088/0004-637X/800/1/20}{\JournalTitle{\apj}, 800,
  20}

\bibitem[{{Gonz{\'a}lez} {et~al.}(2014){Gonz{\'a}lez}, {Bouwens},
  {Illingworth}, {Labb{\'e}}, {Oesch}, {Franx}, \& {Magee}}]{gonzalez:2014}
{Gonz{\'a}lez}, V., {Bouwens}, R., {Illingworth}, G., {et~al.} 2014,
  \href{http://dx.doi.org/10.1088/0004-637X/781/1/34}{\JournalTitle{\apj}, 781,
  34}

\bibitem[{{Gordon} {et~al.}(2003){Gordon}, {Clayton}, {Misselt}, {Landolt}, \&
  {Wolff}}]{gordon:2003}
{Gordon}, K.~D., {Clayton}, G.~C., {Misselt}, K.~A., {Landolt}, A.~U., \&
  {Wolff}, M.~J. 2003,
  \href{http://dx.doi.org/10.1086/376774}{\JournalTitle{\apj}, 594, 279}

\bibitem[{{Grogin} {et~al.}(2011){Grogin}, {Kocevski}, {Faber}, {Ferguson},
  {Koekemoer}, {Riess}, {Acquaviva}, {Alexander}, {Almaini}, {Ashby}, {Barden},
  {Bell}, {Bournaud}, {Brown}, {Caputi}, {Casertano}, {Cassata}, {Castellano},
  {Challis}, {Chary}, {Cheung}, {Cirasuolo}, {Conselice}, {Roshan Cooray},
  {Croton}, {Daddi}, {Dahlen}, {Dav{\'e}}, {de Mello}, {Dekel}, {Dickinson},
  {Dolch}, {Donley}, {Dunlop}, {Dutton}, {Elbaz}, {Fazio}, {Filippenko},
  {Finkelstein}, {Fontana}, {Gardner}, {Garnavich}, {Gawiser}, {Giavalisco},
  {Grazian}, {Guo}, {Hathi}, {H{\"a}ussler}, {Hopkins}, {Huang}, {Huang},
  {Jha}, {Kartaltepe}, {Kirshner}, {Koo}, {Lai}, {Lee}, {Li}, {Lotz}, {Lucas},
  {Madau}, {McCarthy}, {McGrath}, {McIntosh}, {McLure}, {Mobasher},
  {Moustakas}, {Mozena}, {Nandra}, {Newman}, {Niemi}, {Noeske}, {Papovich},
  {Pentericci}, {Pope}, {Primack}, {Rajan}, {Ravindranath}, {Reddy}, {Renzini},
  {Rix}, {Robaina}, {Rodney}, {Rosario}, {Rosati}, {Salimbeni}, {Scarlata},
  {Siana}, {Simard}, {Smidt}, {Somerville}, {Spinrad}, {Straughn}, {Strolger},
  {Telford}, {Teplitz}, {Trump}, {van der Wel}, {Villforth}, {Wechsler},
  {Weiner}, {Wiklind}, {Wild}, {Wilson}, {Wuyts}, {Yan}, \&
  {Yun}}]{grogin:2011}
{Grogin}, N.~A., {Kocevski}, D.~D., {Faber}, S.~M., {et~al.} 2011,
  \href{http://dx.doi.org/10.1088/0067-0049/197/2/35}{\JournalTitle{\apjs},
  197, 35}

\bibitem[{{Hashimoto} {et~al.}(2013){Hashimoto}, {Ouchi}, {Shimasaku}, {Ono},
  {Nakajima}, {Rauch}, {Lee}, \& {Okamura}}]{hashimoto:2013}
{Hashimoto}, T., {Ouchi}, M., {Shimasaku}, K., {et~al.} 2013,
  \href{http://dx.doi.org/10.1088/0004-637X/765/1/70}{\JournalTitle{\apj}, 765,
  70}

\bibitem[{{Hayashi} {et~al.}(2009){Hayashi}, {Motohara}, {Shimasaku},
  {Onodera}, {Uchimoto}, {Kashikawa}, {Yoshida}, {Okamura}, {Ly}, \&
  {Malkan}}]{hayashi:2009}
{Hayashi}, M., {Motohara}, K., {Shimasaku}, K., {et~al.} 2009,
  \href{http://dx.doi.org/10.1088/0004-637X/691/1/140}{\JournalTitle{\apj},
  691, 140}

\bibitem[{{Henry} {et~al.}(2013){Henry}, {Scarlata}, {Dom{\'{\i}}nguez},
  {Malkan}, {Martin}, {Siana}, {Atek}, {Bedregal}, {Colbert}, {Rafelski},
  {Ross}, {Teplitz}, {Bunker}, {Dressler}, {Hathi}, {Masters}, {McCarthy}, \&
  {Straughn}}]{henry:2013}
{Henry}, A., {Scarlata}, C., {Dom{\'{\i}}nguez}, A., {et~al.} 2013,
  \href{http://dx.doi.org/10.1088/2041-8205/776/2/L27}{\JournalTitle{\apjl},
  776, L27}

\bibitem[{Hunter(2007)}]{matplotlib}
Hunter, J.~D. 2007, \JournalTitle{Computing In Science \& Engineering}, 9, 90

\bibitem[{{Ilbert} {et~al.}(2009){Ilbert}, {Capak}, {Salvato}, {Aussel},
  {McCracken}, {Sanders}, {Scoville}, {Kartaltepe}, {Arnouts}, {Le Floc'h},
  {Mobasher}, {Taniguchi}, {Lamareille}, {Leauthaud}, {Sasaki}, {Thompson},
  {Zamojski}, {Zamorani}, {Bardelli}, {Bolzonella}, {Bongiorno}, {Brusa},
  {Caputi}, {Carollo}, {Contini}, {Cook}, {Coppa}, {Cucciati}, {de la Torre},
  {de Ravel}, {Franzetti}, {Garilli}, {Hasinger}, {Iovino}, {Kampczyk},
  {Kneib}, {Knobel}, {Kovac}, {Le Borgne}, {Le Brun}, {F{\`e}vre}, {Lilly},
  {Looper}, {Maier}, {Mainieri}, {Mellier}, {Mignoli}, {Murayama}, {Pell{\`o}},
  {Peng}, {P{\'e}rez-Montero}, {Renzini}, {Ricciardelli}, {Schiminovich},
  {Scodeggio}, {Shioya}, {Silverman}, {Surace}, {Tanaka}, {Tasca}, {Tresse},
  {Vergani}, \& {Zucca}}]{ilbert:2009}
{Ilbert}, O., {Capak}, P., {Salvato}, M., {et~al.} 2009,
  \href{http://dx.doi.org/10.1088/0004-637X/690/2/1236}{\JournalTitle{\apj},
  690, 1236}

\bibitem[{{Ilbert} {et~al.}(2013){Ilbert}, {McCracken}, {Le F{\`e}vre},
  {Capak}, {Dunlop}, {Karim}, {Renzini}, {Caputi}, {Boissier}, {Arnouts},
  {Aussel}, {Comparat}, {Guo}, {Hudelot}, {Kartaltepe}, {Kneib}, {Krogager},
  {Le Floc'h}, {Lilly}, {Mellier}, {Milvang-Jensen}, {Moutard}, {Onodera},
  {Richard}, {Salvato}, {Sanders}, {Scoville}, {Silverman}, {Taniguchi},
  {Tasca}, {Thomas}, {Toft}, {Tresse}, {Vergani}, {Wolk}, \&
  {Zirm}}]{ilbert:2013}
{Ilbert}, O., {McCracken}, H.~J., {Le F{\`e}vre}, O., {et~al.} 2013,
  \href{http://dx.doi.org/10.1051/0004-6361/201321100}{\JournalTitle{\aap},
  556, A55}

\bibitem[{{Izotov} {et~al.}(2006){Izotov}, {Stasi{\'n}ska}, {Meynet}, {Guseva},
  \& {Thuan}}]{izotov:2006}
{Izotov}, Y.~I., {Stasi{\'n}ska}, G., {Meynet}, G., {Guseva}, N.~G., \&
  {Thuan}, T.~X. 2006,
  \href{http://dx.doi.org/10.1051/0004-6361:20053763}{\JournalTitle{\aap}, 448,
  955}

\bibitem[{{Jones} {et~al.}(2015){Jones}, {Martin}, \& {Cooper}}]{jones:2015}
{Jones}, T., {Martin}, C., \& {Cooper}, M.~C. 2015,
  \href{http://dx.doi.org/10.1088/0004-637X/813/2/126}{\JournalTitle{\apj},
  813, 126}

\bibitem[{{Karim} {et~al.}(2011){Karim}, {Schinnerer},
  {Mart{\'{\i}}nez-Sansigre}, {Sargent}, {van der Wel}, {Rix}, {Ilbert},
  {Smol{\v c}i{\'c}}, {Carilli}, {Pannella}, {Koekemoer}, {Bell}, \&
  {Salvato}}]{karim:2011}
{Karim}, A., {Schinnerer}, E., {Mart{\'{\i}}nez-Sansigre}, A., {et~al.} 2011,
  \href{http://dx.doi.org/10.1088/0004-637X/730/2/61}{\JournalTitle{\apj}, 730,
  61}

\bibitem[{{Kashino} {et~al.}(2013){Kashino}, {Silverman}, {Rodighiero},
  {Renzini}, {Arimoto}, {Daddi}, {Lilly}, {Sanders}, {Kartaltepe}, {Zahid},
  {Nagao}, {Sugiyama}, {Capak}, {Carollo}, {Chu}, {Hasinger}, {Ilbert},
  {Kajisawa}, {Kewley}, {Koekemoer}, {Kova{\v c}}, {Le F{\`e}vre}, {Masters},
  {McCracken}, {Onodera}, {Scoville}, {Strazzullo}, {Symeonidis}, \&
  {Taniguchi}}]{kashino:2013}
{Kashino}, D., {Silverman}, J.~D., {Rodighiero}, G., {et~al.} 2013,
  \href{http://dx.doi.org/10.1088/2041-8205/777/1/L8}{\JournalTitle{\apjl},
  777, L8}

\bibitem[{{Kennicutt} \& {Evans}(2012)}]{kennicutt:2012}
{Kennicutt}, R.~C., \& {Evans}, N.~J. 2012,
  \href{http://dx.doi.org/10.1146/annurev-astro-081811-125610}{\JournalTitle{\araa},
  50, 531}

\bibitem[{{Kennicutt}(1998)}]{kennicutt:1998}
{Kennicutt}, Jr., R.~C. 1998,
  \href{http://dx.doi.org/10.1146/annurev.astro.36.1.189}{\JournalTitle{\araa},
  36, 189}

\bibitem[{{Kewley} \& {Dopita}(2002)}]{kewley:2002}
{Kewley}, L.~J., \& {Dopita}, M.~A. 2002,
  \href{http://dx.doi.org/10.1086/341326}{\JournalTitle{\apjs}, 142, 35}

\bibitem[{{Kewley} {et~al.}(2004){Kewley}, {Geller}, \& {Jansen}}]{kewley:2004}
{Kewley}, L.~J., {Geller}, M.~J., \& {Jansen}, R.~A. 2004,
  \href{http://dx.doi.org/10.1086/382723}{\JournalTitle{\aj}, 127, 2002}

\bibitem[{{Kewley} {et~al.}(2006){Kewley}, {Groves}, {Kauffmann}, \&
  {Heckman}}]{kewley:2006}
{Kewley}, L.~J., {Groves}, B., {Kauffmann}, G., \& {Heckman}, T. 2006,
  \href{http://dx.doi.org/10.1111/j.1365-2966.2006.10859.x}{\JournalTitle{\mnras},
  372, 961}

\bibitem[{{Kewley} {et~al.}(2013){Kewley}, {Maier}, {Yabe}, {Ohta}, {Akiyama},
  {Dopita}, \& {Yuan}}]{kewley:2013}
{Kewley}, L.~J., {Maier}, C., {Yabe}, K., {et~al.} 2013,
  \href{http://dx.doi.org/10.1088/2041-8205/774/1/L10}{\JournalTitle{\apjl},
  774, L10}

\bibitem[{{Kobulnicky} \& {Kewley}(2004)}]{kobulnicky:2004}
{Kobulnicky}, H.~A., \& {Kewley}, L.~J. 2004,
  \href{http://dx.doi.org/10.1086/425299}{\JournalTitle{\apj}, 617, 240}

\bibitem[{{Kobulnicky} \& {Phillips}(2003)}]{kobulnicky:2003}
{Kobulnicky}, H.~A., \& {Phillips}, A.~C. 2003,
  \href{http://dx.doi.org/10.1086/379361}{\JournalTitle{\apj}, 599, 1031}

\bibitem[{{Kodama} {et~al.}(1999){Kodama}, {Bell}, \& {Bower}}]{kodama:1999}
{Kodama}, T., {Bell}, E.~F., \& {Bower}, R.~G. 1999,
  \href{http://dx.doi.org/10.1046/j.1365-8711.1999.02184.x}{\JournalTitle{\mnras},
  302, 152}

\bibitem[{{Koekemoer} {et~al.}(2011){Koekemoer}, {Faber}, {Ferguson}, {Grogin},
  {Kocevski}, {Koo}, {Lai}, {Lotz}, {Lucas}, {McGrath}, {Ogaz}, {Rajan},
  {Riess}, {Rodney}, {Strolger}, {Casertano}, {Castellano}, {Dahlen},
  {Dickinson}, {Dolch}, {Fontana}, {Giavalisco}, {Grazian}, {Guo}, {Hathi},
  {Huang}, {van der Wel}, {Yan}, {Acquaviva}, {Alexander}, {Almaini}, {Ashby},
  {Barden}, {Bell}, {Bournaud}, {Brown}, {Caputi}, {Cassata}, {Challis},
  {Chary}, {Cheung}, {Cirasuolo}, {Conselice}, {Roshan Cooray}, {Croton},
  {Daddi}, {Dav{\'e}}, {de Mello}, {de Ravel}, {Dekel}, {Donley}, {Dunlop},
  {Dutton}, {Elbaz}, {Fazio}, {Filippenko}, {Finkelstein}, {Frazer}, {Gardner},
  {Garnavich}, {Gawiser}, {Gruetzbauch}, {Hartley}, {H{\"a}ussler},
  {Herrington}, {Hopkins}, {Huang}, {Jha}, {Johnson}, {Kartaltepe},
  {Khostovan}, {Kirshner}, {Lani}, {Lee}, {Li}, {Madau}, {McCarthy},
  {McIntosh}, {McLure}, {McPartland}, {Mobasher}, {Moreira}, {Mortlock},
  {Moustakas}, {Mozena}, {Nandra}, {Newman}, {Nielsen}, {Niemi}, {Noeske},
  {Papovich}, {Pentericci}, {Pope}, {Primack}, {Ravindranath}, {Reddy},
  {Renzini}, {Rix}, {Robaina}, {Rosario}, {Rosati}, {Salimbeni}, {Scarlata},
  {Siana}, {Simard}, {Smidt}, {Snyder}, {Somerville}, {Spinrad}, {Straughn},
  {Telford}, {Teplitz}, {Trump}, {Vargas}, {Villforth}, {Wagner}, {Wandro},
  {Wechsler}, {Weiner}, {Wiklind}, {Wild}, {Wilson}, {Wuyts}, \&
  {Yun}}]{koekemoer:2011}
{Koekemoer}, A.~M., {Faber}, S.~M., {Ferguson}, H.~C., {et~al.} 2011,
  \href{http://dx.doi.org/10.1088/0067-0049/197/2/36}{\JournalTitle{\apjs},
  197, 36}

\bibitem[{{Krumholz} \& {Dekel}(2012)}]{krumholz:2012}
{Krumholz}, M.~R., \& {Dekel}, A. 2012,
  \href{http://dx.doi.org/10.1088/0004-637X/753/1/16}{\JournalTitle{\apj}, 753,
  16}

\bibitem[{{Lara-L{\'o}pez} {et~al.}(2010){Lara-L{\'o}pez}, {Cepa},
  {Bongiovanni}, {P{\'e}rez Garc{\'{\i}}a}, {Ederoclite}, {Casta{\~n}eda},
  {Fern{\'a}ndez Lorenzo}, {Povi{\'c}}, \&
  {S{\'a}nchez-Portal}}]{laralopez:2010}
{Lara-L{\'o}pez}, M.~A., {Cepa}, J., {Bongiovanni}, A., {et~al.} 2010,
  \href{http://dx.doi.org/10.1051/0004-6361/201014803}{\JournalTitle{\aap},
  521, L53}

\bibitem[{{Lequeux} {et~al.}(1979){Lequeux}, {Peimbert}, {Rayo}, {Serrano}, \&
  {Torres-Peimbert}}]{lequeux:1979}
{Lequeux}, J., {Peimbert}, M., {Rayo}, J.~F., {Serrano}, A., \&
  {Torres-Peimbert}, S. 1979, \JournalTitle{\aap}, 80, 155

\bibitem[{{Lilly} {et~al.}(2013){Lilly}, {Carollo}, {Pipino}, {Renzini}, \&
  {Peng}}]{lilly:2013}
{Lilly}, S.~J., {Carollo}, C.~M., {Pipino}, A., {Renzini}, A., \& {Peng}, Y.
  2013,
  \href{http://dx.doi.org/10.1088/0004-637X/772/2/119}{\JournalTitle{\apj},
  772, 119}

\bibitem[{{Lilly} {et~al.}(2003){Lilly}, {Carollo}, \& {Stockton}}]{lilly:2003}
{Lilly}, S.~J., {Carollo}, C.~M., \& {Stockton}, A.~N. 2003,
  \href{http://dx.doi.org/10.1086/378389}{\JournalTitle{\apj}, 597, 730}

\bibitem[{{Lilly} {et~al.}(2007){Lilly}, {Le F{\`e}vre}, {Renzini}, {Zamorani},
  {Scodeggio}, {Contini}, {Carollo}, {Hasinger}, {Kneib}, {Iovino}, {Le Brun},
  {Maier}, {Mainieri}, {Mignoli}, {Silverman}, {Tasca}, {Bolzonella},
  {Bongiorno}, {Bottini}, {Capak}, {Caputi}, {Cimatti}, {Cucciati}, {Daddi},
  {Feldmann}, {Franzetti}, {Garilli}, {Guzzo}, {Ilbert}, {Kampczyk}, {Kovac},
  {Lamareille}, {Leauthaud}, {Borgne}, {McCracken}, {Marinoni}, {Pello},
  {Ricciardelli}, {Scarlata}, {Vergani}, {Sanders}, {Schinnerer}, {Scoville},
  {Taniguchi}, {Arnouts}, {Aussel}, {Bardelli}, {Brusa}, {Cappi}, {Ciliegi},
  {Finoguenov}, {Foucaud}, {Franceschini}, {Halliday}, {Impey}, {Knobel},
  {Koekemoer}, {Kurk}, {Maccagni}, {Maddox}, {Marano}, {Marconi}, {Meneux},
  {Mobasher}, {Moreau}, {Peacock}, {Porciani}, {Pozzetti}, {Scaramella},
  {Schiminovich}, {Shopbell}, {Smail}, {Thompson}, {Tresse}, {Vettolani},
  {Zanichelli}, \& {Zucca}}]{lilly:2007}
{Lilly}, S.~J., {Le F{\`e}vre}, O., {Renzini}, A., {et~al.} 2007,
  \href{http://dx.doi.org/10.1086/516589}{\JournalTitle{\apjs}, 172, 70}

\bibitem[{{Lilly} {et~al.}(2009){Lilly}, {Le Brun}, {Maier}, {Mainieri},
  {Mignoli}, {Scodeggio}, {Zamorani}, {Carollo}, {Contini}, {Kneib}, {Le
  F{\`e}vre}, {Renzini}, {Bardelli}, {Bolzonella}, {Bongiorno}, {Caputi},
  {Coppa}, {Cucciati}, {de la Torre}, {de Ravel}, {Franzetti}, {Garilli},
  {Iovino}, {Kampczyk}, {Kovac}, {Knobel}, {Lamareille}, {Le Borgne}, {Pello},
  {Peng}, {P{\'e}rez-Montero}, {Ricciardelli}, {Silverman}, {Tanaka}, {Tasca},
  {Tresse}, {Vergani}, {Zucca}, {Ilbert}, {Salvato}, {Oesch}, {Abbas},
  {Bottini}, {Capak}, {Cappi}, {Cassata}, {Cimatti}, {Elvis}, {Fumana},
  {Guzzo}, {Hasinger}, {Koekemoer}, {Leauthaud}, {Maccagni}, {Marinoni},
  {McCracken}, {Memeo}, {Meneux}, {Porciani}, {Pozzetti}, {Sanders},
  {Scaramella}, {Scarlata}, {Scoville}, {Shopbell}, \&
  {Taniguchi}}]{lilly:2009}
{Lilly}, S.~J., {Le Brun}, V., {Maier}, C., {et~al.} 2009,
  \href{http://dx.doi.org/10.1088/0067-0049/184/2/218}{\JournalTitle{\apjs},
  184, 218}

\bibitem[{{Luridiana} {et~al.}(2015){Luridiana}, {Morisset}, \& {Shaw}}]{pyneb}
{Luridiana}, V., {Morisset}, C., \& {Shaw}, R.~A. 2015,
  \href{http://dx.doi.org/10.1051/0004-6361/201323152}{\JournalTitle{\aap},
  573, A42}

\bibitem[{{Ly} {et~al.}(2014){Ly}, {Malkan}, {Nagao}, {Kashikawa}, {Shimasaku},
  \& {Hayashi}}]{ly:2014}
{Ly}, C., {Malkan}, M.~A., {Nagao}, T., {et~al.} 2014,
  \href{http://dx.doi.org/10.1088/0004-637X/780/2/122}{\JournalTitle{\apj},
  780, 122}

\bibitem[{{Magdis} {et~al.}(2010){Magdis}, {Rigopoulou}, {Huang}, \&
  {Fazio}}]{magdis:2010}
{Magdis}, G.~E., {Rigopoulou}, D., {Huang}, J.-S., \& {Fazio}, G.~G. 2010,
  \href{http://dx.doi.org/10.1111/j.1365-2966.2009.15779.x}{\JournalTitle{\mnras},
  401, 1521}

\bibitem[{{Maier} {et~al.}(2014){Maier}, {Lilly}, {Ziegler}, {Contini},
  {P{\'e}rez Montero}, {Peng}, \& {Balestra}}]{maier:2014}
{Maier}, C., {Lilly}, S.~J., {Ziegler}, B.~L., {et~al.} 2014,
  \href{http://dx.doi.org/10.1088/0004-637X/792/1/3}{\JournalTitle{\apj}, 792,
  3}

\bibitem[{{Maier} {et~al.}(2015){Maier}, {Ziegler}, {Lilly}, {Contini},
  {P{\'e}rez-Montero}, {Lamareille}, {Bolzonella}, \& {Le Floc'h}}]{maier:2015}
{Maier}, C., {Ziegler}, B.~L., {Lilly}, S.~J., {et~al.} 2015,
  \href{http://dx.doi.org/10.1051/0004-6361/201425224}{\JournalTitle{\aap},
  577, A14}

\bibitem[{{Maiolino} {et~al.}(2008){Maiolino}, {Nagao}, {Grazian}, {Cocchia},
  {Marconi}, {Mannucci}, {Cimatti}, {Pipino}, {Ballero}, {Calura}, {Chiappini},
  {Fontana}, {Granato}, {Matteucci}, {Pastorini}, {Pentericci}, {Risaliti},
  {Salvati}, \& {Silva}}]{maiolino:2008}
{Maiolino}, R., {Nagao}, T., {Grazian}, A., {et~al.} 2008,
  \href{http://dx.doi.org/10.1051/0004-6361:200809678}{\JournalTitle{\aap},
  488, 463}

\bibitem[{{Mannucci} {et~al.}(2010){Mannucci}, {Cresci}, {Maiolino}, {Marconi},
  \& {Gnerucci}}]{mannucci:2010}
{Mannucci}, F., {Cresci}, G., {Maiolino}, R., {Marconi}, A., \& {Gnerucci}, A.
  2010,
  \href{http://dx.doi.org/10.1111/j.1365-2966.2010.17291.x}{\JournalTitle{\mnras},
  408, 2115}

\bibitem[{{Mannucci} {et~al.}(2011){Mannucci}, {Salvaterra}, \&
  {Campisi}}]{mannucci:2011}
{Mannucci}, F., {Salvaterra}, R., \& {Campisi}, M.~A. 2011,
  \href{http://dx.doi.org/10.1111/j.1365-2966.2011.18459.x}{\JournalTitle{\mnras},
  414, 1263}

\bibitem[{{Mannucci} {et~al.}(2009){Mannucci}, {Cresci}, {Maiolino}, {Marconi},
  {Pastorini}, {Pozzetti}, {Gnerucci}, {Risaliti}, {Schneider}, {Lehnert}, \&
  {Salvati}}]{mannucci:2009}
{Mannucci}, F., {Cresci}, G., {Maiolino}, R., {et~al.} 2009,
  \href{http://dx.doi.org/10.1111/j.1365-2966.2009.15185.x}{\JournalTitle{\mnras},
  398, 1915}

\bibitem[{{Maraston} {et~al.}(2010){Maraston}, {Pforr}, {Renzini}, {Daddi},
  {Dickinson}, {Cimatti}, \& {Tonini}}]{maraston:2010}
{Maraston}, C., {Pforr}, J., {Renzini}, A., {et~al.} 2010,
  \href{http://dx.doi.org/10.1111/j.1365-2966.2010.16973.x}{\JournalTitle{\mnras},
  407, 830}

\bibitem[{{Marchesi} {et~al.}(2016){Marchesi}, {Civano}, {Elvis}, {Salvato},
  {Brusa}, {Comastri}, {Gilli}, {Hasinger}, {Lanzuisi}, {Miyaji}, {Treister},
  {Urry}, {Vignali}, {Zamorani}, {Allevato}, {Cappelluti}, {Cardamone},
  {Finoguenov}, {Griffiths}, {Karim}, {Laigle}, {LaMassa}, {Jahnke}, {Ranalli},
  {Schawinski}, {Schinnerer}, {Silverman}, {Smolcic}, {Suh}, \&
  {Trakhtenbrot}}]{marchesi:2016}
{Marchesi}, S., {Civano}, F., {Elvis}, M., {et~al.} 2016,
  \href{http://dx.doi.org/10.3847/0004-637X/817/1/34}{\JournalTitle{\apj}, 817,
  34}

\bibitem[{{Marino} {et~al.}(2013){Marino}, {Rosales-Ortega}, {S{\'a}nchez},
  {Gil de Paz}, {V{\'{\i}}lchez}, {Miralles-Caballero}, {Kehrig},
  {P{\'e}rez-Montero}, {Stanishev}, {Iglesias-P{\'a}ramo}, {D{\'{\i}}az},
  {Castillo-Morales}, {Kennicutt}, {L{\'o}pez-S{\'a}nchez}, {Galbany},
  {Garc{\'{\i}}a-Benito}, {Mast}, {Mendez-Abreu}, {Monreal-Ibero}, {Husemann},
  {Walcher}, {Garc{\'{\i}}a-Lorenzo}, {Masegosa}, {Del Olmo Orozco},
  {Mour{\~a}o}, {Ziegler}, {Moll{\'a}}, {Papaderos},
  {S{\'a}nchez-Bl{\'a}zquez}, {Gonz{\'a}lez Delgado}, {Falc{\'o}n-Barroso},
  {Roth}, {van de Ven}, \& {Califa Team}}]{marino:2013}
{Marino}, R.~A., {Rosales-Ortega}, F.~F., {S{\'a}nchez}, S.~F., {et~al.} 2013,
  \href{http://dx.doi.org/10.1051/0004-6361/201321956}{\JournalTitle{\aap},
  559, A114}

\bibitem[{{Markwardt}(2009)}]{markwardt:2009}
{Markwardt}, C.~B. 2009, in Astronomical Society of the Pacific Conference
  Series, Vol. 411, Astronomical Data Analysis Software and Systems XVIII, ed.
  D.~A. {Bohlender}, D.~{Durand}, \& P.~{Dowler}, 251

\bibitem[{{Masters} {et~al.}(2014){Masters}, {McCarthy}, {Siana}, {Malkan},
  {Mobasher}, {Atek}, {Henry}, {Martin}, {Rafelski}, {Hathi}, {Scarlata},
  {Ross}, {Bunker}, {Blanc}, {Bedregal}, {Dom{\'{\i}}nguez}, {Colbert},
  {Teplitz}, \& {Dressler}}]{masters:2014}
{Masters}, D., {McCarthy}, P., {Siana}, B., {et~al.} 2014,
  \href{http://dx.doi.org/10.1088/0004-637X/785/2/153}{\JournalTitle{\apj},
  785, 153}

\bibitem[{{McCracken} {et~al.}(2010){McCracken}, {Capak}, {Salvato}, {Aussel},
  {Thompson}, {Daddi}, {Sanders}, {Kneib}, {Willott}, {Mancini}, {Renzini},
  {Cook}, {Le F{\`e}vre}, {Ilbert}, {Kartaltepe}, {Koekemoer}, {Mellier},
  {Murayama}, {Scoville}, {Shioya}, \& {Tanaguchi}}]{mccracken:2010}
{McCracken}, H.~J., {Capak}, P., {Salvato}, M., {et~al.} 2010,
  \href{http://dx.doi.org/10.1088/0004-637X/708/1/202}{\JournalTitle{\apj},
  708, 202}

\bibitem[{{McCracken} {et~al.}(2012){McCracken}, {Milvang-Jensen}, {Dunlop},
  {Franx}, {Fynbo}, {Le F{\`e}vre}, {Holt}, {Caputi}, {Goranova}, {Buitrago},
  {Emerson}, {Freudling}, {Hudelot}, {L{\'o}pez-Sanjuan}, {Magnard}, {Mellier},
  {M{\o}ller}, {Nilsson}, {Sutherland}, {Tasca}, \& {Zabl}}]{mccracken:2012}
{McCracken}, H.~J., {Milvang-Jensen}, B., {Dunlop}, J., {et~al.} 2012,
  \href{http://dx.doi.org/10.1051/0004-6361/201219507}{\JournalTitle{\aap},
  544, A156}

\bibitem[{{McDermid} {et~al.}(2015){McDermid}, {Alatalo}, {Blitz}, {Bournaud},
  {Bureau}, {Cappellari}, {Crocker}, {Davies}, {Davis}, {de Zeeuw}, {Duc},
  {Emsellem}, {Khochfar}, {Krajnovi{\'c}}, {Kuntschner}, {Morganti}, {Naab},
  {Oosterloo}, {Sarzi}, {Scott}, {Serra}, {Weijmans}, \&
  {Young}}]{mcdermid:2015}
{McDermid}, R.~M., {Alatalo}, K., {Blitz}, L., {et~al.} 2015,
  \href{http://dx.doi.org/10.1093/mnras/stv105}{\JournalTitle{\mnras}, 448,
  3484}

\bibitem[{{McGaugh}(1991)}]{mcgaugh:1991}
{McGaugh}, S.~S. 1991,
  \href{http://dx.doi.org/10.1086/170569}{\JournalTitle{\apj}, 380, 140}

\bibitem[{{McLean} {et~al.}(2010){McLean}, {Steidel}, {Epps}, {Matthews},
  {Adkins}, {Konidaris}, {Weber}, {Aliado}, {Brims}, {Canfield}, {Cromer},
  {Fucik}, {Kulas}, {Mace}, {Magnone}, {Rodriguez}, {Wang}, \&
  {Weiss}}]{mclean:2010:mosfire}
{McLean}, I.~S., {Steidel}, C.~C., {Epps}, H., {et~al.} 2010,
  \href{http://dx.doi.org/10.1117/12.856715}{in Society of Photo-Optical
  Instrumentation Engineers (SPIE) Conference Series, Vol. 7735, Society of
  Photo-Optical Instrumentation Engineers (SPIE) Conference Series}, 1

\bibitem[{{McLean} {et~al.}(2012){McLean}, {Steidel}, {Epps}, {Konidaris},
  {Matthews}, {Adkins}, {Aliado}, {Brims}, {Canfield}, {Cromer}, {Fucik},
  {Kulas}, {Mace}, {Magnone}, {Rodriguez}, {Rudie}, {Trainor}, {Wang}, {Weber},
  \& {Weiss}}]{mclean:2012:mosfire}
{McLean}, I.~S., {Steidel}, C.~C., {Epps}, H.~W., {et~al.} 2012,
  \href{http://dx.doi.org/10.1117/12.924794}{in Society of Photo-Optical
  Instrumentation Engineers (SPIE) Conference Series, Vol. 8446, Society of
  Photo-Optical Instrumentation Engineers (SPIE) Conference Series}, 0

\bibitem[{{McLinden} {et~al.}(2011){McLinden}, {Finkelstein}, {Rhoads},
  {Malhotra}, {Hibon}, {Richardson}, {Cresci}, {Quirrenbach}, {Pasquali},
  {Bian}, {Fan}, \& {Woodward}}]{mclinden:2011}
{McLinden}, E.~M., {Finkelstein}, S.~L., {Rhoads}, J.~E., {et~al.} 2011,
  \href{http://dx.doi.org/10.1088/0004-637X/730/2/136}{\JournalTitle{\apj},
  730, 136}

\bibitem[{{Meurer} {et~al.}(1999){Meurer}, {Heckman}, \&
  {Calzetti}}]{meurer:1999}
{Meurer}, G.~R., {Heckman}, T.~M., \& {Calzetti}, D. 1999,
  \href{http://dx.doi.org/10.1086/307523}{\JournalTitle{\apj}, 521, 64}

\bibitem[{{Micha{\l}owski} {et~al.}(2014){Micha{\l}owski}, {Hayward}, {Dunlop},
  {Bruce}, {Cirasuolo}, {Cullen}, \& {Hernquist}}]{michalowski:2014}
{Micha{\l}owski}, M.~J., {Hayward}, C.~C., {Dunlop}, J.~S., {et~al.} 2014,
  \href{http://dx.doi.org/10.1051/0004-6361/201424174}{\JournalTitle{\aap},
  571, A75}

\bibitem[{{Morishita} {et~al.}(2015){Morishita}, {Ichikawa}, {Noguchi},
  {Akiyama}, {Patel}, {Kajisawa}, \& {Obata}}]{morishita:2015}
{Morishita}, T., {Ichikawa}, T., {Noguchi}, M., {et~al.} 2015,
  \href{http://dx.doi.org/10.1088/0004-637X/805/1/34}{\JournalTitle{\apj}, 805,
  34}

\bibitem[{{Nagao} {et~al.}(2006){Nagao}, {Maiolino}, \& {Marconi}}]{nagao:2006}
{Nagao}, T., {Maiolino}, R., \& {Marconi}, A. 2006,
  \href{http://dx.doi.org/10.1051/0004-6361:20065216}{\JournalTitle{\aap}, 459,
  85}

\bibitem[{{Nakajima} \& {Ouchi}(2014)}]{nakajima:2014}
{Nakajima}, K., \& {Ouchi}, M. 2014,
  \href{http://dx.doi.org/10.1093/mnras/stu902}{\JournalTitle{\mnras}, 442,
  900}

\bibitem[{{Nakamura} {et~al.}(2004){Nakamura}, {Fukugita}, {Brinkmann}, \&
  {Schneider}}]{nakamura:2004}
{Nakamura}, O., {Fukugita}, M., {Brinkmann}, J., \& {Schneider}, D.~P. 2004,
  \href{http://dx.doi.org/10.1086/386350}{\JournalTitle{\aj}, 127, 2511}

\bibitem[{{Nelson} {et~al.}(2015){Nelson}, {van Dokkum}, {F{\"o}rster
  Schreiber}, {Franx}, {Brammer}, {Momcheva}, {Wuyts}, {Whitaker}, {Skelton},
  {Fumagalli}, {Kriek}, {Labb{\'e}}, {Leja}, {Rix}, {Tacconi}, {van der Wel},
  {van den Bosch}, {Oesch}, {Dickey}, \& {Ulf Lange}}]{nelson:2015}
{Nelson}, E.~J., {van Dokkum}, P.~G., {F{\"o}rster Schreiber}, N.~M., {et~al.}
  2015, \JournalTitle{ArXiv e-prints},
  \href{http://arxiv.org/abs/1507.03999}{{\sffamily arXiv:1507.03999}}

\bibitem[{{Noeske} {et~al.}(2007){Noeske}, {Weiner}, {Faber}, {Papovich},
  {Koo}, {Somerville}, {Bundy}, {Conselice}, {Newman}, {Schiminovich}, {Le
  Floc'h}, {Coil}, {Rieke}, {Lotz}, {Primack}, {Barmby}, {Cooper}, {Davis},
  {Ellis}, {Fazio}, {Guhathakurta}, {Huang}, {Kassin}, {Martin}, {Phillips},
  {Rich}, {Small}, {Willmer}, \& {Wilson}}]{noeske:2007:ms}
{Noeske}, K.~G., {Weiner}, B.~J., {Faber}, S.~M., {et~al.} 2007,
  \href{http://dx.doi.org/10.1086/517926}{\JournalTitle{\apjl}, 660, L43}

\bibitem[{{Nordon} {et~al.}(2013){Nordon}, {Lutz}, {Saintonge}, {Berta},
  {Wuyts}, {F{\"o}rster Schreiber}, {Genzel}, {Magnelli}, {Poglitsch},
  {Popesso}, {Rosario}, {Sturm}, \& {Tacconi}}]{nordon:2013}
{Nordon}, R., {Lutz}, D., {Saintonge}, A., {et~al.} 2013,
  \href{http://dx.doi.org/10.1088/0004-637X/762/2/125}{\JournalTitle{\apj},
  762, 125}

\bibitem[{{Oh} {et~al.}(2011){Oh}, {Sarzi}, {Schawinski}, \& {Yi}}]{oh:2011}
{Oh}, K., {Sarzi}, M., {Schawinski}, K., \& {Yi}, S.~K. 2011,
  \href{http://dx.doi.org/10.1088/0067-0049/195/2/13}{\JournalTitle{\apjs},
  195, 13}

\bibitem[{{Oke} \& {Gunn}(1983)}]{oke:1983}
{Oke}, J.~B., \& {Gunn}, J.~E. 1983,
  \href{http://dx.doi.org/10.1086/160817}{\JournalTitle{\apj}, 266, 713}

\bibitem[{{Onodera} {et~al.}(2010{\natexlab{a}}){Onodera}, {Arimoto}, {Daddi},
  {Renzini}, {Kong}, {Cimatti}, {Broadhurst}, \&
  {Alexander}}]{onodera:2010:sbzk}
{Onodera}, M., {Arimoto}, N., {Daddi}, E., {et~al.} 2010{\natexlab{a}},
  \href{http://dx.doi.org/10.1088/0004-637X/715/1/385}{\JournalTitle{\apj},
  715, 385}

\bibitem[{{Onodera} {et~al.}(2010{\natexlab{b}}){Onodera}, {Daddi}, {Gobat},
  {Cappellari}, {Arimoto}, {Renzini}, {Yamada}, {McCracken}, {Mancini},
  {Capak}, {Carollo}, {Cimatti}, {Giavalisco}, {Ilbert}, {Kong}, {Lilly},
  {Motohara}, {Ohta}, {Sanders}, {Scoville}, {Tamura}, \&
  {Taniguchi}}]{onodera:2010:pbzk}
{Onodera}, M., {Daddi}, E., {Gobat}, R., {et~al.} 2010{\natexlab{b}},
  \href{http://dx.doi.org/10.1088/2041-8205/715/1/L6}{\JournalTitle{\apjl},
  715, L6}

\bibitem[{{Onodera} {et~al.}(2012){Onodera}, {Renzini}, {Carollo},
  {Cappellari}, {Mancini}, {Strazzullo}, {Daddi}, {Arimoto}, {Gobat}, {Yamada},
  {McCracken}, {Ilbert}, {Capak}, {Cimatti}, {Giavalisco}, {Koekemoer}, {Kong},
  {Lilly}, {Motohara}, {Ohta}, {Sanders}, {Scoville}, {Tamura}, \&
  {Taniguchi}}]{onodera:2012}
{Onodera}, M., {Renzini}, A., {Carollo}, M., {et~al.} 2012,
  \href{http://dx.doi.org/10.1088/0004-637X/755/1/26}{\JournalTitle{\apj}, 755,
  26}

\bibitem[{{Onodera} {et~al.}(2015){Onodera}, {Carollo}, {Renzini},
  {Cappellari}, {Mancini}, {Arimoto}, {Daddi}, {Gobat}, {Strazzullo},
  {Tacchella}, \& {Yamada}}]{onodera:2015}
{Onodera}, M., {Carollo}, C.~M., {Renzini}, A., {et~al.} 2015,
  \href{http://dx.doi.org/10.1088/0004-637X/808/2/161}{\JournalTitle{\apj},
  808, 161}

\bibitem[{{Osterbrock} \& {Ferland}(2006)}]{osterbrock:agnagn}
{Osterbrock}, D.~E., \& {Ferland}, G.~J. 2006, {Astrophysics of gaseous nebulae
  and active galactic nuclei}

\bibitem[{{Pagel} {et~al.}(1979){Pagel}, {Edmunds}, {Blackwell}, {Chun}, \&
  {Smith}}]{pagel:1979}
{Pagel}, B.~E.~J., {Edmunds}, M.~G., {Blackwell}, D.~E., {Chun}, M.~S., \&
  {Smith}, G. 1979, \JournalTitle{\mnras}, 189, 95

\bibitem[{{Pannella} {et~al.}(2009){Pannella}, {Carilli}, {Daddi}, {McCracken},
  {Owen}, {Renzini}, {Strazzullo}, {Civano}, {Koekemoer}, {Schinnerer},
  {Scoville}, {Smol{\v c}i{\'c}}, {Taniguchi}, {Aussel}, {Kneib}, {Ilbert},
  {Mellier}, {Salvato}, {Thompson}, \& {Willott}}]{pannella:2009}
{Pannella}, M., {Carilli}, C.~L., {Daddi}, E., {et~al.} 2009,
  \href{http://dx.doi.org/10.1088/0004-637X/698/2/L116}{\JournalTitle{\apjl},
  698, L116}

\bibitem[{{Pannella} {et~al.}(2015){Pannella}, {Elbaz}, {Daddi}, {Dickinson},
  {Hwang}, {Schreiber}, {Strazzullo}, {Aussel}, {Bethermin}, {Buat},
  {Charmandaris}, {Cibinel}, {Juneau}, {Ivison}, {Le Borgne}, {Le Floc'h},
  {Leiton}, {Lin}, {Magdis}, {Morrison}, {Mullaney}, {Onodera}, {Renzini},
  {Salim}, {Sargent}, {Scott}, {Shu}, \& {Wang}}]{pannella:2015}
{Pannella}, M., {Elbaz}, D., {Daddi}, E., {et~al.} 2015,
  \href{http://dx.doi.org/10.1088/0004-637X/807/2/141}{\JournalTitle{\apj},
  807, 141}

\bibitem[{{Peng} {et~al.}(2010){Peng}, {Lilly}, {Kova{\v c}}, {Bolzonella},
  {Pozzetti}, {Renzini}, {Zamorani}, {Ilbert}, {Knobel}, {Iovino}, {Maier},
  {Cucciati}, {Tasca}, {Carollo}, {Silverman}, {Kampczyk}, {de Ravel},
  {Sanders}, {Scoville}, {Contini}, {Mainieri}, {Scodeggio}, {Kneib}, {Le
  F{\`e}vre}, {Bardelli}, {Bongiorno}, {Caputi}, {Coppa}, {de la Torre},
  {Franzetti}, {Garilli}, {Lamareille}, {Le Borgne}, {Le Brun}, {Mignoli},
  {Perez Montero}, {Pello}, {Ricciardelli}, {Tanaka}, {Tresse}, {Vergani},
  {Welikala}, {Zucca}, {Oesch}, {Abbas}, {Barnes}, {Bordoloi}, {Bottini},
  {Cappi}, {Cassata}, {Cimatti}, {Fumana}, {Hasinger}, {Koekemoer},
  {Leauthaud}, {Maccagni}, {Marinoni}, {McCracken}, {Memeo}, {Meneux}, {Nair},
  {Porciani}, {Presotto}, \& {Scaramella}}]{peng:2010}
{Peng}, Y.-j., {Lilly}, S.~J., {Kova{\v c}}, K., {et~al.} 2010,
  \href{http://dx.doi.org/10.1088/0004-637X/721/1/193}{\JournalTitle{\apj},
  721, 193}

\bibitem[{{P{\'e}rez-Montero}(2014)}]{perezmontero:2014}
{P{\'e}rez-Montero}, E. 2014,
  \href{http://dx.doi.org/10.1093/mnras/stu753}{\JournalTitle{\mnras}, 441,
  2663}

\bibitem[{{Puglisi} {et~al.}(2016){Puglisi}, {Rodighiero}, {Franceschini},
  {Talia}, {Cimatti}, {Baronchelli}, {Daddi}, {Renzini}, {Schawinski},
  {Mancini}, {Silverman}, {Gruppioni}, {Lutz}, {Berta}, \&
  {Oliver}}]{puglisi:2016}
{Puglisi}, A., {Rodighiero}, G., {Franceschini}, A., {et~al.} 2016,
  \href{http://dx.doi.org/10.1051/0004-6361/201526782}{\JournalTitle{\aap},
  586, A83}

\bibitem[{{Reddy} {et~al.}(2010){Reddy}, {Erb}, {Pettini}, {Steidel}, \&
  {Shapley}}]{reddy:2010}
{Reddy}, N.~A., {Erb}, D.~K., {Pettini}, M., {Steidel}, C.~C., \& {Shapley},
  A.~E. 2010,
  \href{http://dx.doi.org/10.1088/0004-637X/712/2/1070}{\JournalTitle{\apj},
  712, 1070}

\bibitem[{{Reddy} {et~al.}(2012){Reddy}, {Pettini}, {Steidel}, {Shapley},
  {Erb}, \& {Law}}]{reddy:2012}
{Reddy}, N.~A., {Pettini}, M., {Steidel}, C.~C., {et~al.} 2012,
  \href{http://dx.doi.org/10.1088/0004-637X/754/1/25}{\JournalTitle{\apj}, 754,
  25}

\bibitem[{{Reddy} {et~al.}(2015){Reddy}, {Kriek}, {Shapley}, {Freeman},
  {Siana}, {Coil}, {Mobasher}, {Price}, {Sanders}, \& {Shivaei}}]{reddy:2015}
{Reddy}, N.~A., {Kriek}, M., {Shapley}, A.~E., {et~al.} 2015,
  \href{http://dx.doi.org/10.1088/0004-637X/806/2/259}{\JournalTitle{\apj},
  806, 259}

\bibitem[{{Renzini}(2009)}]{renzini:2009}
{Renzini}, A. 2009,
  \href{http://dx.doi.org/10.1111/j.1745-3933.2009.00710.x}{\JournalTitle{\mnras},
  398, L58}

\bibitem[{{Renzini} \& {Peng}(2015)}]{renzini:2015:ms}
{Renzini}, A., \& {Peng}, Y.-j. 2015,
  \href{http://dx.doi.org/10.1088/2041-8205/801/2/L29}{\JournalTitle{\apjl},
  801, L29}

\bibitem[{{Rodighiero} {et~al.}(2011){Rodighiero}, {Daddi}, {Baronchelli},
  {Cimatti}, {Renzini}, {Aussel}, {Popesso}, {Lutz}, {Andreani}, {Berta},
  {Cava}, {Elbaz}, {Feltre}, {Fontana}, {F{\"o}rster Schreiber},
  {Franceschini}, {Genzel}, {Grazian}, {Gruppioni}, {Ilbert}, {Le Floch},
  {Magdis}, {Magliocchetti}, {Magnelli}, {Maiolino}, {McCracken}, {Nordon},
  {Poglitsch}, {Santini}, {Pozzi}, {Riguccini}, {Tacconi}, {Wuyts}, \&
  {Zamorani}}]{rodighiero:2011}
{Rodighiero}, G., {Daddi}, E., {Baronchelli}, I., {et~al.} 2011,
  \href{http://dx.doi.org/10.1088/2041-8205/739/2/L40}{\JournalTitle{\apjl},
  739, L40}

\bibitem[{{Rodighiero} {et~al.}(2014){Rodighiero}, {Renzini}, {Daddi},
  {Baronchelli}, {Berta}, {Cresci}, {Franceschini}, {Gruppioni}, {Lutz},
  {Mancini}, {Santini}, {Zamorani}, {Silverman}, {Kashino}, {Andreani},
  {Cimatti}, {S{\'a}nchez}, {Le Floch}, {Magnelli}, {Popesso}, \&
  {Pozzi}}]{rodighiero:2014}
{Rodighiero}, G., {Renzini}, A., {Daddi}, E., {et~al.} 2014,
  \href{http://dx.doi.org/10.1093/mnras/stu1110}{\JournalTitle{\mnras}, 443,
  19}

\bibitem[{{Salmi} {et~al.}(2012){Salmi}, {Daddi}, {Elbaz}, {Sargent},
  {Dickinson}, {Renzini}, {Bethermin}, \& {Le Borgne}}]{salmi:2012}
{Salmi}, F., {Daddi}, E., {Elbaz}, D., {et~al.} 2012,
  \href{http://dx.doi.org/10.1088/2041-8205/754/1/L14}{\JournalTitle{\apjl},
  754, L14}

\bibitem[{{Sanders} {et~al.}(2007){Sanders}, {Salvato}, {Aussel}, {Ilbert},
  {Scoville}, {Surace}, {Frayer}, {Sheth}, {Helou}, {Brooke}, {Bhattacharya},
  {Yan}, {Kartaltepe}, {Barnes}, {Blain}, {Calzetti}, {Capak}, {Carilli},
  {Carollo}, {Comastri}, {Daddi}, {Ellis}, {Elvis}, {Fall}, {Franceschini},
  {Giavalisco}, {Hasinger}, {Impey}, {Koekemoer}, {Le F{\`e}vre}, {Lilly},
  {Liu}, {McCracken}, {Mobasher}, {Renzini}, {Rich}, {Schinnerer}, {Shopbell},
  {Taniguchi}, {Thompson}, {Urry}, \& {Williams}}]{sanders:2007}
{Sanders}, D.~B., {Salvato}, M., {Aussel}, H., {et~al.} 2007,
  \href{http://dx.doi.org/10.1086/517885}{\JournalTitle{\apjs}, 172, 86}

\bibitem[{{Sanders} {et~al.}(2015){Sanders}, {Shapley}, {Kriek}, {Reddy},
  {Freeman}, {Coil}, {Siana}, {Mobasher}, {Shivaei}, {Price}, \& {de
  Groot}}]{sanders:2015:metal}
{Sanders}, R.~L., {Shapley}, A.~E., {Kriek}, M., {et~al.} 2015,
  \href{http://dx.doi.org/10.1088/0004-637X/799/2/138}{\JournalTitle{\apj},
  799, 138}

\bibitem[{{Sanders} {et~al.}(2016){Sanders}, {Shapley}, {Kriek}, {Reddy},
  {Freeman}, {Coil}, {Siana}, {Mobasher}, {Shivaei}, {Price}, \& {de
  Groot}}]{sanders:2015:ionization}
---. 2016,
  \href{http://dx.doi.org/10.3847/0004-637X/816/1/23}{\JournalTitle{\apj}, 816,
  23}

\bibitem[{{Savaglio} {et~al.}(2005){Savaglio}, {Glazebrook}, {Le Borgne},
  {Juneau}, {Abraham}, {Chen}, {Crampton}, {McCarthy}, {Carlberg}, {Marzke},
  {Roth}, {J{\o}rgensen}, \& {Murowinski}}]{savaglio:2005}
{Savaglio}, S., {Glazebrook}, K., {Le Borgne}, D., {et~al.} 2005,
  \href{http://dx.doi.org/10.1086/497331}{\JournalTitle{\apj}, 635, 260}

\bibitem[{{Schaerer} {et~al.}(2013){Schaerer}, {de Barros}, \&
  {Sklias}}]{schaerer:2013}
{Schaerer}, D., {de Barros}, S., \& {Sklias}, P. 2013,
  \href{http://dx.doi.org/10.1051/0004-6361/201220002}{\JournalTitle{\aap},
  549, A4}

\bibitem[{{Schlafly} \& {Finkbeiner}(2011)}]{schalafly:2011}
{Schlafly}, E.~F., \& {Finkbeiner}, D.~P. 2011,
  \href{http://dx.doi.org/10.1088/0004-637X/737/2/103}{\JournalTitle{\apj},
  737, 103}

\bibitem[{{Schreiber} {et~al.}(2015){Schreiber}, {Pannella}, {Elbaz},
  {B{\'e}thermin}, {Inami}, {Dickinson}, {Magnelli}, {Wang}, {Aussel}, {Daddi},
  {Juneau}, {Shu}, {Sargent}, {Buat}, {Faber}, {Ferguson}, {Giavalisco},
  {Koekemoer}, {Magdis}, {Morrison}, {Papovich}, {Santini}, \&
  {Scott}}]{schreiber:2015}
{Schreiber}, C., {Pannella}, M., {Elbaz}, D., {et~al.} 2015,
  \href{http://dx.doi.org/10.1051/0004-6361/201425017}{\JournalTitle{\aap},
  575, A74}

\bibitem[{{Shimakawa} {et~al.}(2015){Shimakawa}, {Kodama}, {Steidel}, {Tadaki},
  {Tanaka}, {Strom}, {Hayashi}, {Koyama}, {Suzuki}, \&
  {Yamamoto}}]{shimakawa:2015}
{Shimakawa}, R., {Kodama}, T., {Steidel}, C.~C., {et~al.} 2015,
  \href{http://dx.doi.org/10.1093/mnras/stv915}{\JournalTitle{\mnras}, 451,
  1284}

\bibitem[{{Shirazi} \& {Brinchmann}(2012)}]{shirazi:2012}
{Shirazi}, M., \& {Brinchmann}, J. 2012,
  \href{http://dx.doi.org/10.1111/j.1365-2966.2012.20439.x}{\JournalTitle{\mnras},
  421, 1043}

\bibitem[{{Shirazi} {et~al.}(2014){Shirazi}, {Brinchmann}, \&
  {Rahmati}}]{shirazi:2014}
{Shirazi}, M., {Brinchmann}, J., \& {Rahmati}, A. 2014,
  \href{http://dx.doi.org/10.1088/0004-637X/787/2/120}{\JournalTitle{\apj},
  787, 120}

\bibitem[{{Skelton} {et~al.}(2014){Skelton}, {Whitaker}, {Momcheva}, {Brammer},
  {van Dokkum}, {Labb{\'e}}, {Franx}, {van der Wel}, {Bezanson}, {Da Cunha},
  {Fumagalli}, {F{\"o}rster Schreiber}, {Kriek}, {Leja}, {Lundgren}, {Magee},
  {Marchesini}, {Maseda}, {Nelson}, {Oesch}, {Pacifici}, {Patel}, {Price},
  {Rix}, {Tal}, {Wake}, \& {Wuyts}}]{skelton:2014}
{Skelton}, R.~E., {Whitaker}, K.~E., {Momcheva}, I.~G., {et~al.} 2014,
  \href{http://dx.doi.org/10.1088/0067-0049/214/2/24}{\JournalTitle{\apjs},
  214, 24}

\bibitem[{{Sparre} {et~al.}(2015){Sparre}, {Hayward}, {Springel},
  {Vogelsberger}, {Genel}, {Torrey}, {Nelson}, {Sijacki}, \&
  {Hernquist}}]{sparre:2015}
{Sparre}, M., {Hayward}, C.~C., {Springel}, V., {et~al.} 2015,
  \href{http://dx.doi.org/10.1093/mnras/stu2713}{\JournalTitle{\mnras}, 447,
  3548}

\bibitem[{{Speagle} {et~al.}(2014){Speagle}, {Steinhardt}, {Capak}, \&
  {Silverman}}]{speagle:2014}
{Speagle}, J.~S., {Steinhardt}, C.~L., {Capak}, P.~L., \& {Silverman}, J.~D.
  2014,
  \href{http://dx.doi.org/10.1088/0067-0049/214/2/15}{\JournalTitle{\apjs},
  214, 15}

\bibitem[{{Stark} {et~al.}(2013){Stark}, {Schenker}, {Ellis}, {Robertson},
  {McLure}, \& {Dunlop}}]{stark:2013}
{Stark}, D.~P., {Schenker}, M.~A., {Ellis}, R., {et~al.} 2013,
  \href{http://dx.doi.org/10.1088/0004-637X/763/2/129}{\JournalTitle{\apj},
  763, 129}

\bibitem[{{Steidel} {et~al.}(2014){Steidel}, {Rudie}, {Strom}, {Pettini},
  {Reddy}, {Shapley}, {Trainor}, {Erb}, {Turner}, {Konidaris}, {Kulas}, {Mace},
  {Matthews}, \& {McLean}}]{steidel:2014}
{Steidel}, C.~C., {Rudie}, G.~C., {Strom}, A.~L., {et~al.} 2014,
  \href{http://dx.doi.org/10.1088/0004-637X/795/2/165}{\JournalTitle{\apj},
  795, 165}

\bibitem[{{Steinhardt} {et~al.}(2014){Steinhardt}, {Speagle}, {Capak},
  {Silverman}, {Carollo}, {Dunlop}, {Hashimoto}, {Hsieh}, {Ilbert}, {Le Fevre},
  {Le Floc'h}, {Lee}, {Lin}, {Lin}, {Masters}, {McCracken}, {Nagao}, {Petric},
  {Salvato}, {Sanders}, {Scoville}, {Sheth}, {Strauss}, \&
  {Taniguchi}}]{steinhardt:2014}
{Steinhardt}, C.~L., {Speagle}, J.~S., {Capak}, P., {et~al.} 2014,
  \href{http://dx.doi.org/10.1088/2041-8205/791/2/L25}{\JournalTitle{\apjl},
  791, L25}

\bibitem[{{Tacchella} {et~al.}(2015{\natexlab{a}}){Tacchella}, {Dekel},
  {Carollo}, {Ceverino}, {DeGraf}, {Lapiner}, {Mandelker}, \&
  {Primack}}]{tacchella:2015:profile}
{Tacchella}, S., {Dekel}, A., {Carollo}, C.~M., {et~al.} 2015{\natexlab{a}},
  \JournalTitle{ArXiv e-prints},
  \href{http://arxiv.org/abs/1509.00017}{{\sffamily arXiv:1509.00017}}

\bibitem[{{Tacchella} {et~al.}(2015{\natexlab{b}}){Tacchella}, {Dekel},
  {Carollo}, {Ceverino}, {DeGraf}, {Lapiner}, {Mandelker}, \&
  {Primack}}]{tacchella:2015:ms}
---. 2015{\natexlab{b}}, \JournalTitle{ArXiv e-prints},
  \href{http://arxiv.org/abs/1509.02529}{{\sffamily arXiv:1509.02529}}

\bibitem[{{Tacchella} {et~al.}(2015{\natexlab{c}}){Tacchella}, {Carollo},
  {Renzini}, {Schreiber}, {Lang}, {Wuyts}, {Cresci}, {Dekel}, {Genzel},
  {Lilly}, {Mancini}, {Newman}, {Onodera}, {Shapley}, {Tacconi}, {Woo}, \&
  {Zamorani}}]{tacchella:2015:science}
{Tacchella}, S., {Carollo}, C.~M., {Renzini}, A., {et~al.} 2015{\natexlab{c}},
  \href{http://dx.doi.org/10.1126/science.1261094}{\JournalTitle{Science}, 348,
  314}

\bibitem[{{Tacchella} {et~al.}(2015{\natexlab{d}}){Tacchella}, {Lang},
  {Carollo}, {F{\"o}rster Schreiber}, {Renzini}, {Shapley}, {Wuyts}, {Cresci},
  {Genzel}, {Lilly}, {Mancini}, {Newman}, {Tacconi}, {Zamorani}, {Davies},
  {Kurk}, \& {Pozzetti}}]{tacchella:2015:data}
{Tacchella}, S., {Lang}, P., {Carollo}, C.~M., {et~al.} 2015{\natexlab{d}},
  \href{http://dx.doi.org/10.1088/0004-637X/802/2/101}{\JournalTitle{\apj},
  802, 101}

\bibitem[{{Thomas} {et~al.}(2005){Thomas}, {Maraston}, {Bender}, \& {Mendes de
  Oliveira}}]{thomas:2005}
{Thomas}, D., {Maraston}, C., {Bender}, R., \& {Mendes de Oliveira}, C. 2005,
  \href{http://dx.doi.org/10.1086/426932}{\JournalTitle{\apj}, 621, 673}

\bibitem[{{Thomas} {et~al.}(2010){Thomas}, {Maraston}, {Schawinski}, {Sarzi},
  \& {Silk}}]{thomas:2010}
{Thomas}, D., {Maraston}, C., {Schawinski}, K., {Sarzi}, M., \& {Silk}, J.
  2010,
  \href{http://dx.doi.org/10.1111/j.1365-2966.2010.16427.x}{\JournalTitle{\mnras},
  404, 1775}

\bibitem[{{Tremonti} {et~al.}(2004){Tremonti}, {Heckman}, {Kauffmann},
  {Brinchmann}, {Charlot}, {White}, {Seibert}, {Peng}, {Schlegel}, {Uomoto},
  {Fukugita}, \& {Brinkmann}}]{tremonti:2004}
{Tremonti}, C.~A., {Heckman}, T.~M., {Kauffmann}, G., {et~al.} 2004,
  \href{http://dx.doi.org/10.1086/423264}{\JournalTitle{\apj}, 613, 898}

\bibitem[{{Troncoso} {et~al.}(2014){Troncoso}, {Maiolino}, {Sommariva},
  {Cresci}, {Mannucci}, {Marconi}, {Meneghetti}, {Grazian}, {Cimatti},
  {Fontana}, {Nagao}, \& {Pentericci}}]{troncoso:2014}
{Troncoso}, P., {Maiolino}, R., {Sommariva}, V., {et~al.} 2014,
  \href{http://dx.doi.org/10.1051/0004-6361/201322099}{\JournalTitle{\aap},
  563, A58}

\bibitem[{{van de Sande} {et~al.}(2013){van de Sande}, {Kriek}, {Franx}, {van
  Dokkum}, {Bezanson}, {Bouwens}, {Quadri}, {Rix}, \&
  {Skelton}}]{vandesande:2013}
{van de Sande}, J., {Kriek}, M., {Franx}, M., {et~al.} 2013,
  \href{http://dx.doi.org/10.1088/0004-637X/771/2/85}{\JournalTitle{\apj}, 771,
  85}

\bibitem[{{Vanden Berk} {et~al.}(2001){Vanden Berk}, {Richards}, {Bauer},
  {Strauss}, {Schneider}, {Heckman}, {York}, {Hall}, {Fan}, {Knapp},
  {Anderson}, {Annis}, {Bahcall}, {Bernardi}, {Briggs}, {Brinkmann}, {Brunner},
  {Burles}, {Carey}, {Castander}, {Connolly}, {Crocker}, {Csabai}, {Doi},
  {Finkbeiner}, {Friedman}, {Frieman}, {Fukugita}, {Gunn}, {Hennessy},
  {Ivezi{\'c}}, {Kent}, {Kunszt}, {Lamb}, {Leger}, {Long}, {Loveday}, {Lupton},
  {Meiksin}, {Merelli}, {Munn}, {Newberg}, {Newcomb}, {Nichol}, {Owen}, {Pier},
  {Pope}, {Rockosi}, {Schlegel}, {Siegmund}, {Smee}, {Snir}, {Stoughton},
  {Stubbs}, {SubbaRao}, {Szalay}, {Szokoly}, {Tremonti}, {Uomoto}, {Waddell},
  {Yanny}, \& {Zheng}}]{vandenberk:2001}
{Vanden Berk}, D.~E., {Richards}, G.~T., {Bauer}, A., {et~al.} 2001,
  \href{http://dx.doi.org/10.1086/321167}{\JournalTitle{\aj}, 122, 549}

\bibitem[{{Whitaker} {et~al.}(2012){Whitaker}, {van Dokkum}, {Brammer}, \&
  {Franx}}]{whitaker:2012:ms}
{Whitaker}, K.~E., {van Dokkum}, P.~G., {Brammer}, G., \& {Franx}, M. 2012,
  \href{http://dx.doi.org/10.1088/2041-8205/754/2/L29}{\JournalTitle{\apjl},
  754, L29}

\bibitem[{{Whitaker} {et~al.}(2014){Whitaker}, {Franx}, {Leja}, {van Dokkum},
  {Henry}, {Skelton}, {Fumagalli}, {Momcheva}, {Brammer}, {Labb{\'e}},
  {Nelson}, \& {Rigby}}]{whitaker:2014:ms}
{Whitaker}, K.~E., {Franx}, M., {Leja}, J., {et~al.} 2014,
  \href{http://dx.doi.org/10.1088/0004-637X/795/2/104}{\JournalTitle{\apj},
  795, 104}

\bibitem[{{Wilkins} {et~al.}(2011){Wilkins}, {Bunker}, {Stanway}, {Lorenzoni},
  \& {Caruana}}]{wilkins:2011}
{Wilkins}, S.~M., {Bunker}, A.~J., {Stanway}, E., {Lorenzoni}, S., \&
  {Caruana}, J. 2011,
  \href{http://dx.doi.org/10.1111/j.1365-2966.2011.19315.x}{\JournalTitle{\mnras},
  417, 717}

\bibitem[{{Wuyts} {et~al.}(2014){Wuyts}, {Kurk}, {F{\"o}rster Schreiber},
  {Genzel}, {Wisnioski}, {Bandara}, {Wuyts}, {Beifiori}, {Bender}, {Brammer},
  {Burkert}, {Buschkamp}, {Carollo}, {Chan}, {Davies}, {Eisenhauer}, {Fossati},
  {Kulkarni}, {Lang}, {Lilly}, {Lutz}, {Mancini}, {Mendel}, {Momcheva}, {Naab},
  {Nelson}, {Renzini}, {Rosario}, {Saglia}, {Seitz}, {Sharples}, {Sternberg},
  {Tacchella}, {Tacconi}, {van Dokkum}, \& {Wilman}}]{wuyts:2014}
{Wuyts}, E., {Kurk}, J., {F{\"o}rster Schreiber}, N.~M., {et~al.} 2014,
  \href{http://dx.doi.org/10.1088/2041-8205/789/2/L40}{\JournalTitle{\apjl},
  789, L40}

\bibitem[{{Yabe} {et~al.}(2012){Yabe}, {Ohta}, {Iwamuro}, {Yuma}, {Akiyama},
  {Tamura}, {Kimura}, {Takato}, {Moritani}, {Sumiyoshi}, {Maihara},
  {Silverman}, {Dalton}, {Lewis}, {Bonfield}, {Lee}, {Curtis Lake}, {Macaulay},
  \& {Clarke}}]{yabe:2012}
{Yabe}, K., {Ohta}, K., {Iwamuro}, F., {et~al.} 2012,
  \href{http://dx.doi.org/10.1093/pasj/64.3.60}{\JournalTitle{\pasj}, 64, 60}

\bibitem[{{Yabe} {et~al.}(2014){Yabe}, {Ohta}, {Iwamuro}, {Akiyama}, {Tamura},
  {Yuma}, {Kimura}, {Takato}, {Moritani}, {Sumiyoshi}, {Maihara}, {Silverman},
  {Dalton}, {Lewis}, {Bonfield}, {Lee}, {Curtis-Lake}, {Macaulay}, \&
  {Clarke}}]{yabe:2014}
---. 2014,
  \href{http://dx.doi.org/10.1093/mnras/stt2185}{\JournalTitle{\mnras}, 437,
  3647}

\bibitem[{{Yabe} {et~al.}(2015){Yabe}, {Ohta}, {Akiyama}, {Bunker}, {Dalton},
  {Ellis}, {Glazebrook}, {Goto}, {Imanishi}, {Iwamuro}, {Okada}, {Shimizu},
  {Takato}, {Tamura}, {Tonegawa}, \& {Totani}}]{yabe:2015}
{Yabe}, K., {Ohta}, K., {Akiyama}, M., {et~al.} 2015,
  \href{http://dx.doi.org/10.1093/pasj/psv079}{\JournalTitle{\pasj}, 67, 102}

\bibitem[{{Yoshikawa} {et~al.}(2010){Yoshikawa}, {Akiyama}, {Kajisawa},
  {Alexander}, {Ohta}, {Suzuki}, {Tokoku}, {Uchimoto}, {Konishi}, {Yamada},
  {Tanaka}, {Omata}, {Nishimura}, {Koekemoer}, {Brandt}, \&
  {Ichikawa}}]{yoshikawa:2010}
{Yoshikawa}, T., {Akiyama}, M., {Kajisawa}, M., {et~al.} 2010,
  \href{http://dx.doi.org/10.1088/0004-637X/718/1/112}{\JournalTitle{\apj},
  718, 112}

\bibitem[{{Zahid} {et~al.}(2014){Zahid}, {Kashino}, {Silverman}, {Kewley},
  {Daddi}, {Renzini}, {Rodighiero}, {Nagao}, {Arimoto}, {Sanders},
  {Kartaltepe}, {Lilly}, {Maier}, {Geller}, {Capak}, {Carollo}, {Chu},
  {Hasinger}, {Ilbert}, {Kajisawa}, {Koekemoer}, {Kovac{\#728}}, {Le
  F{\`e}vre}, {Masters}, {McCracken}, {Onodera}, {Scoville}, {Strazzullo},
  {Sugiyama}, {Taniguchi}, \& {The COSMOS Team}}]{zahid:2014}
{Zahid}, H.~J., {Kashino}, D., {Silverman}, J.~D., {et~al.} 2014,
  \href{http://dx.doi.org/10.1088/0004-637X/792/1/75}{\JournalTitle{\apj}, 792,
  75}

\end{thebibliography}

\clearpage

\appendix

\section{MOSFIRE spectra}
\label{sec:spectra}
\begin{figure*}[h]
  \centering
  \includegraphics[page=1,width=0.95\linewidth]{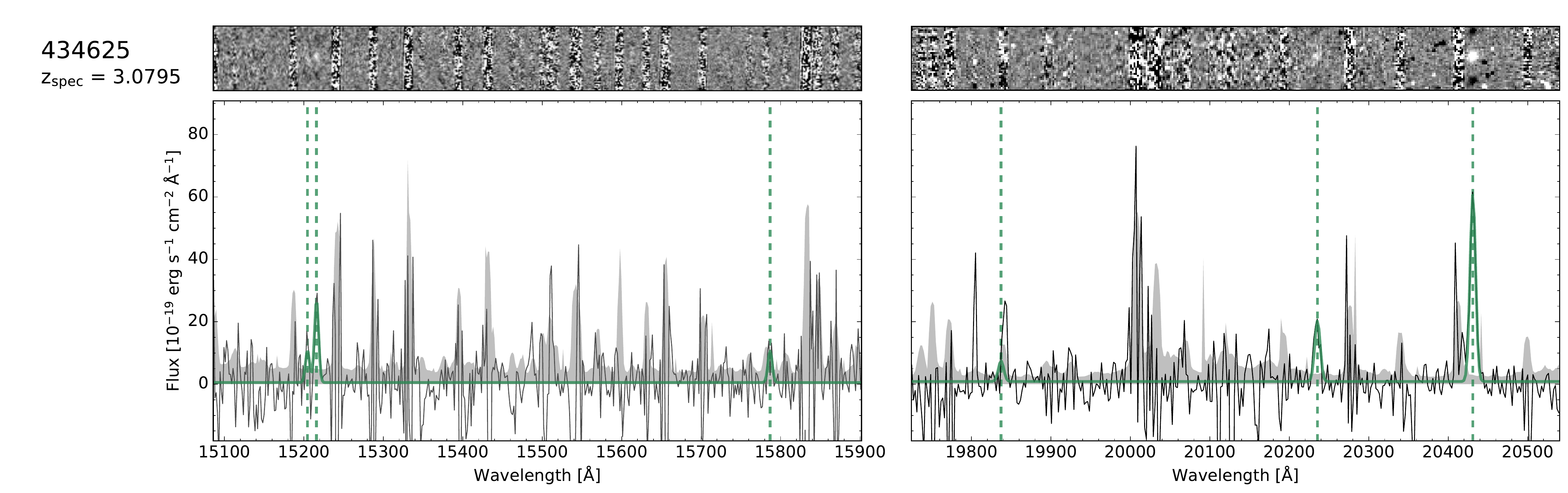}
  \includegraphics[page=2,width=0.95\linewidth]{figB.pdf}
  \includegraphics[page=3,width=0.95\linewidth]{figB.pdf}
  \includegraphics[page=4,width=0.95\linewidth]{figB.pdf}
  \caption{MOSFIRE spectra of each object in \textit{H}- (left) and \textit{K}-band (right).
    (Top) 2-dimensional spectra. (Bottom) 1-dimensional spectra for objects (solid line), $1\sigma$ noise (gray filled area),
    and the best-fit (green solid line). From left to right, dashed lines indicate the location of \oiitot, \neiiione,
    \hbeta, and \oiiitot.
    \label{fig:spec_begin}}
\end{figure*}
\addtocounter{figure}{-1}
\begin{figure*}
  \centering
  \includegraphics[page=5,width=0.95\linewidth]{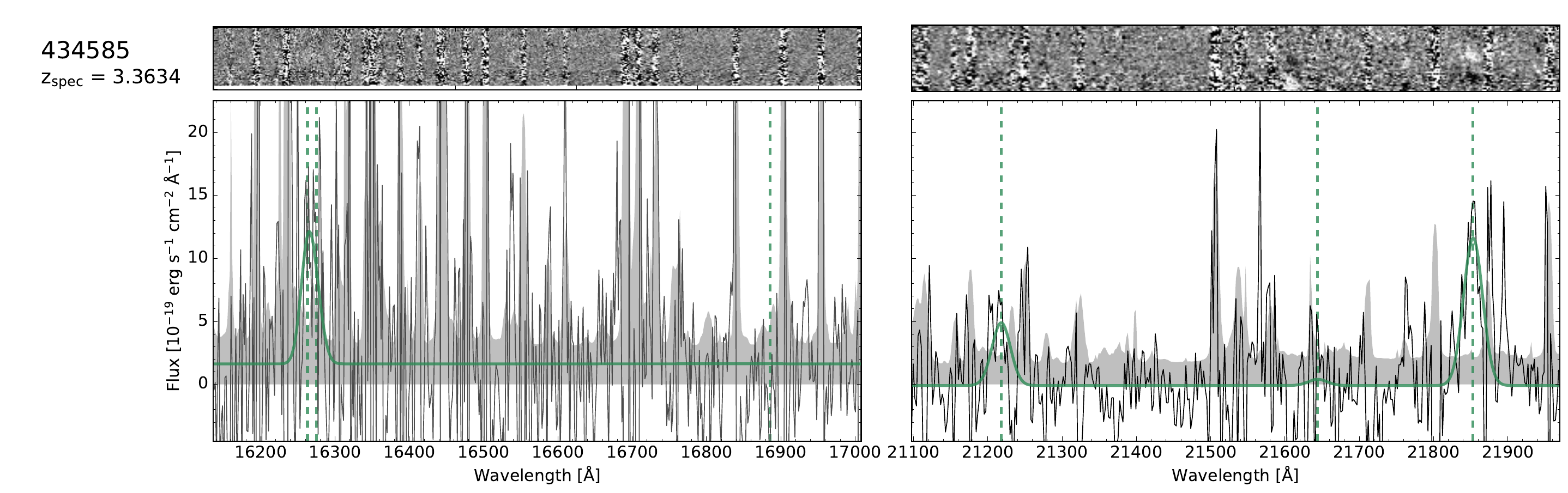}
  \includegraphics[page=6,width=0.95\linewidth]{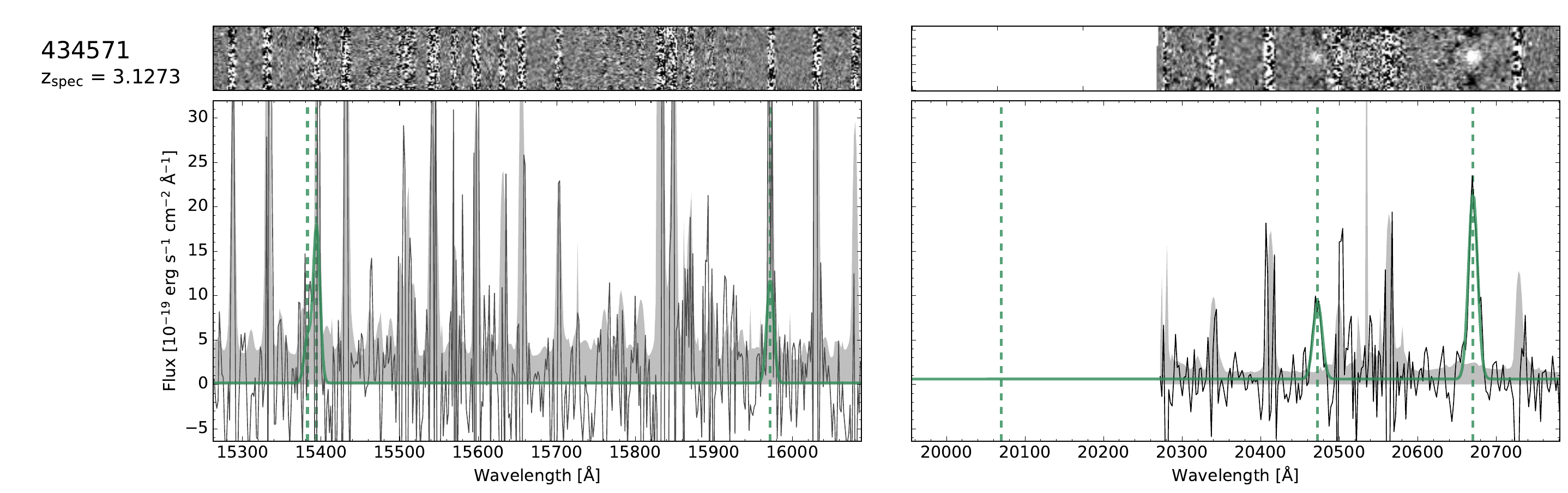}
  \includegraphics[page=7,width=0.95\linewidth]{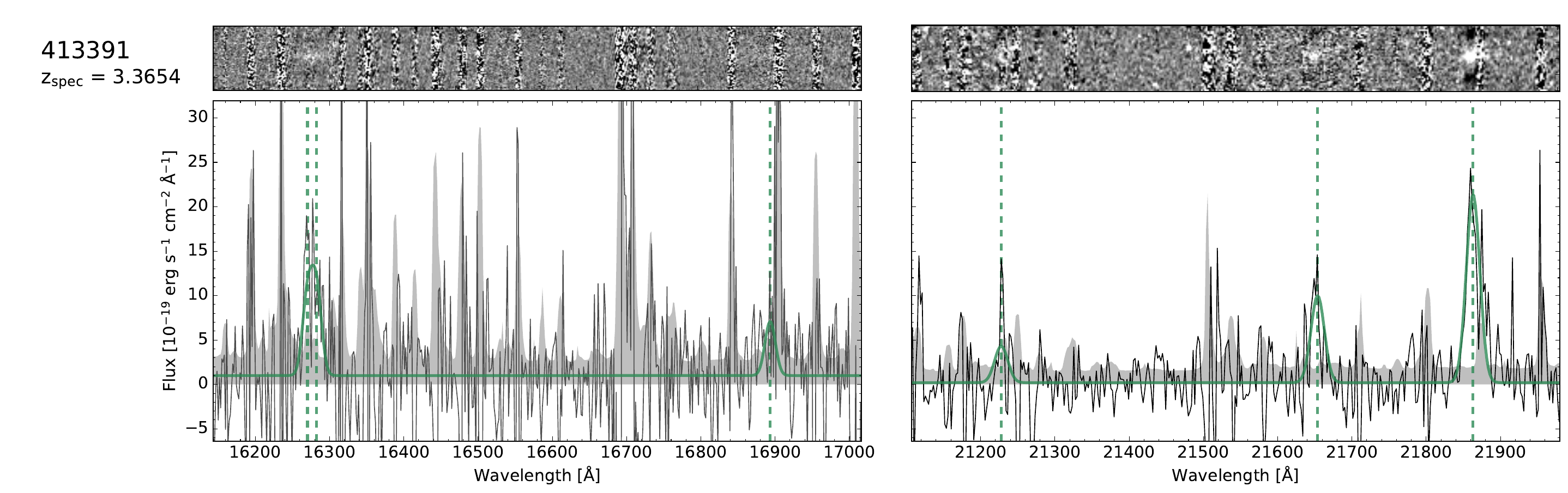}
  \includegraphics[page=8,width=0.95\linewidth]{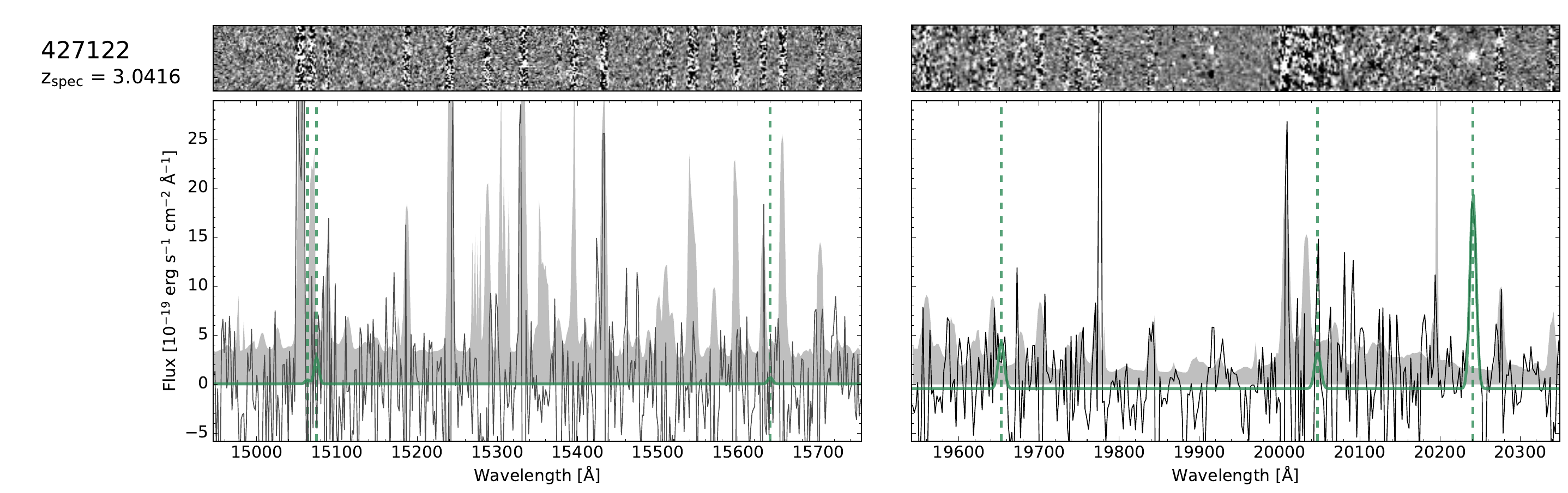}
  \caption{\textit{continued.}}
\end{figure*}
\addtocounter{figure}{-1}
\begin{figure*}
  \centering
  \includegraphics[page=9,width=0.95\linewidth]{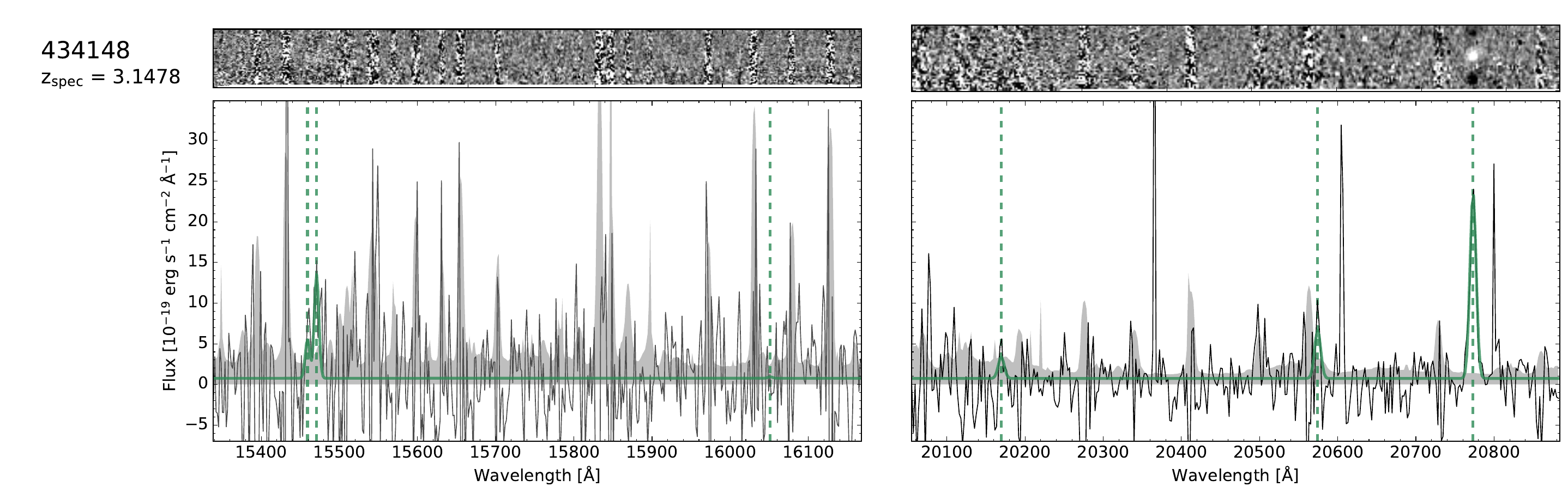}
  \includegraphics[page=10,width=0.95\linewidth]{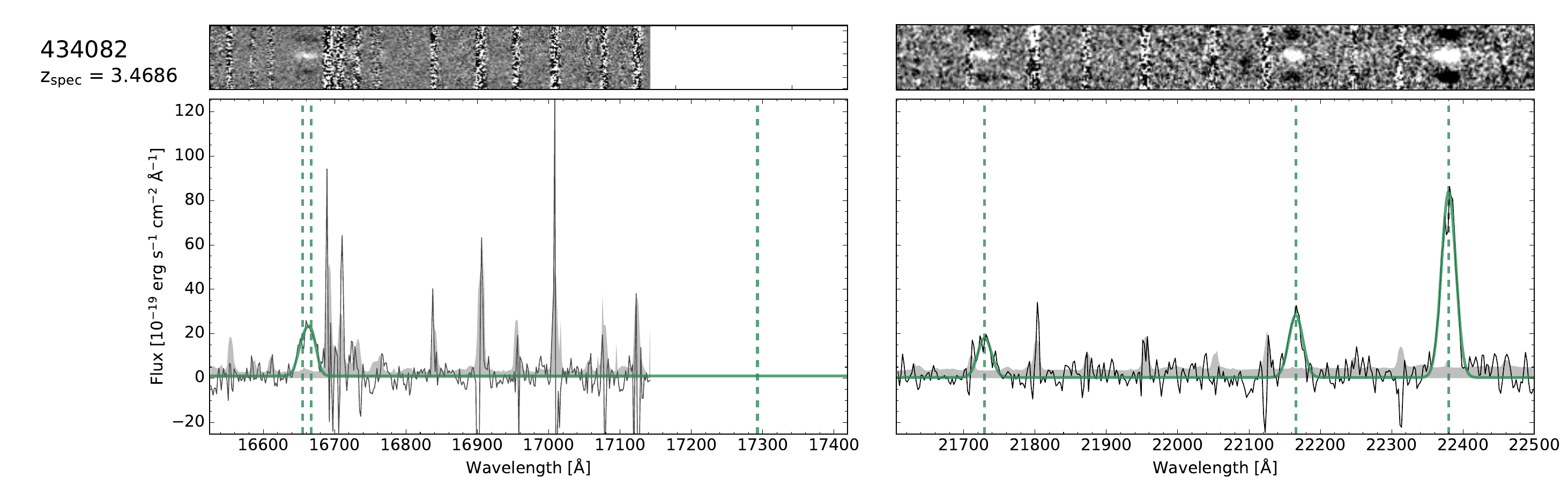}
  \includegraphics[page=11,width=0.95\linewidth]{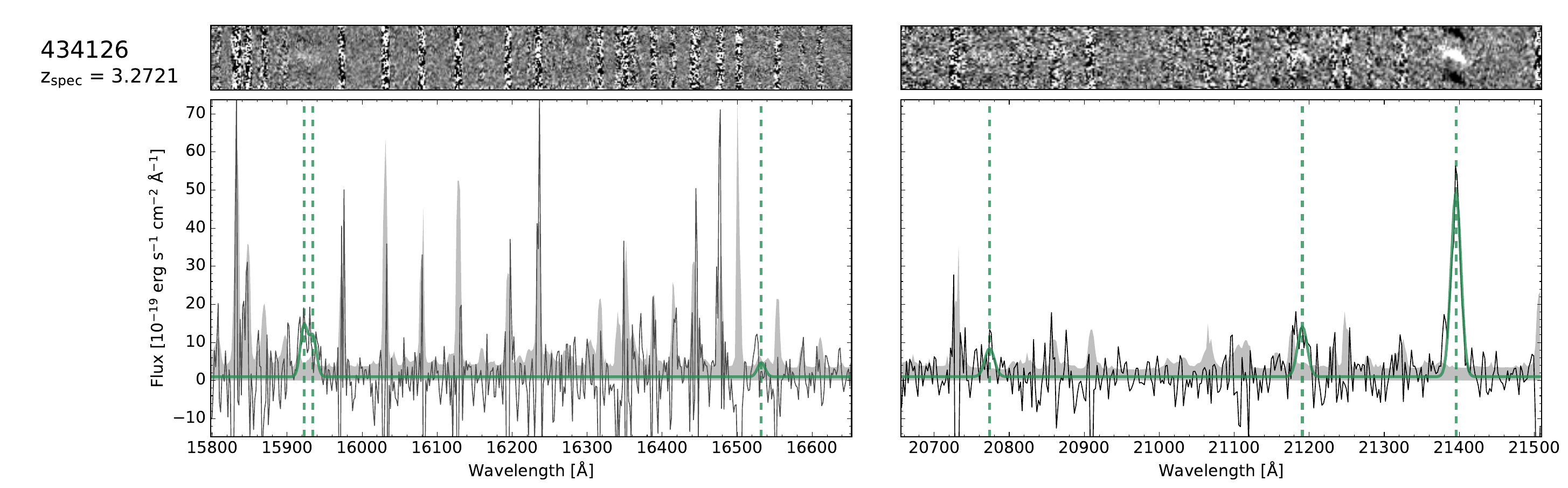}
  \includegraphics[page=12,width=0.95\linewidth]{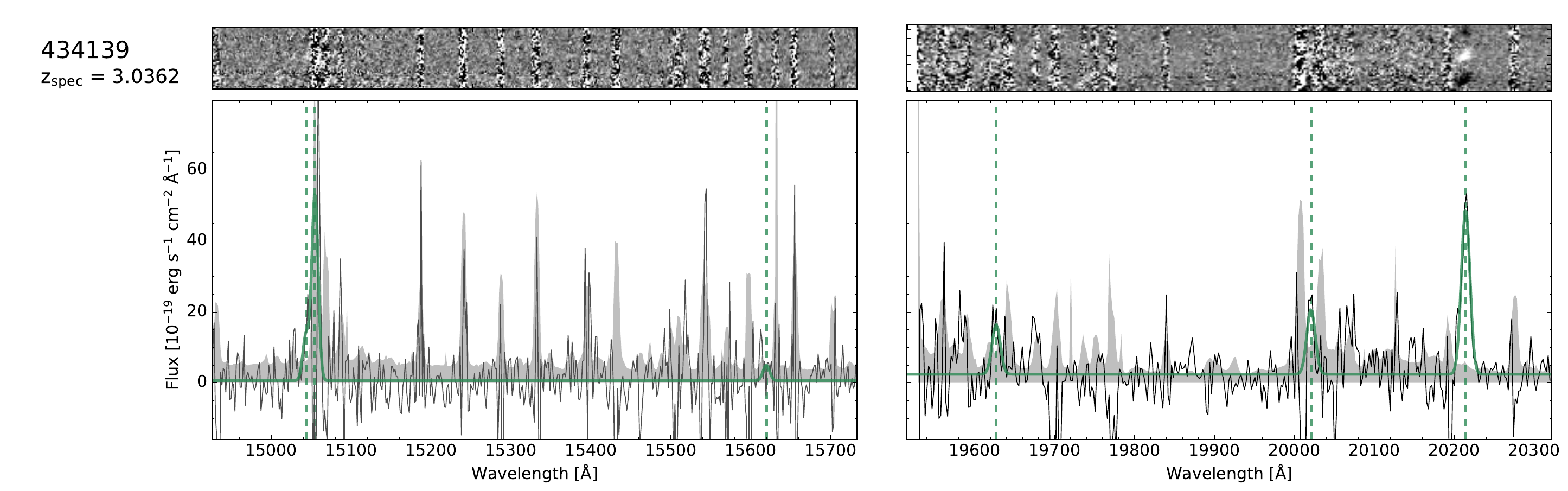}
  \caption{\textit{continued.}}
\end{figure*}
\addtocounter{figure}{-1}
\begin{figure*}
  \centering
  \includegraphics[page=13,width=0.95\linewidth]{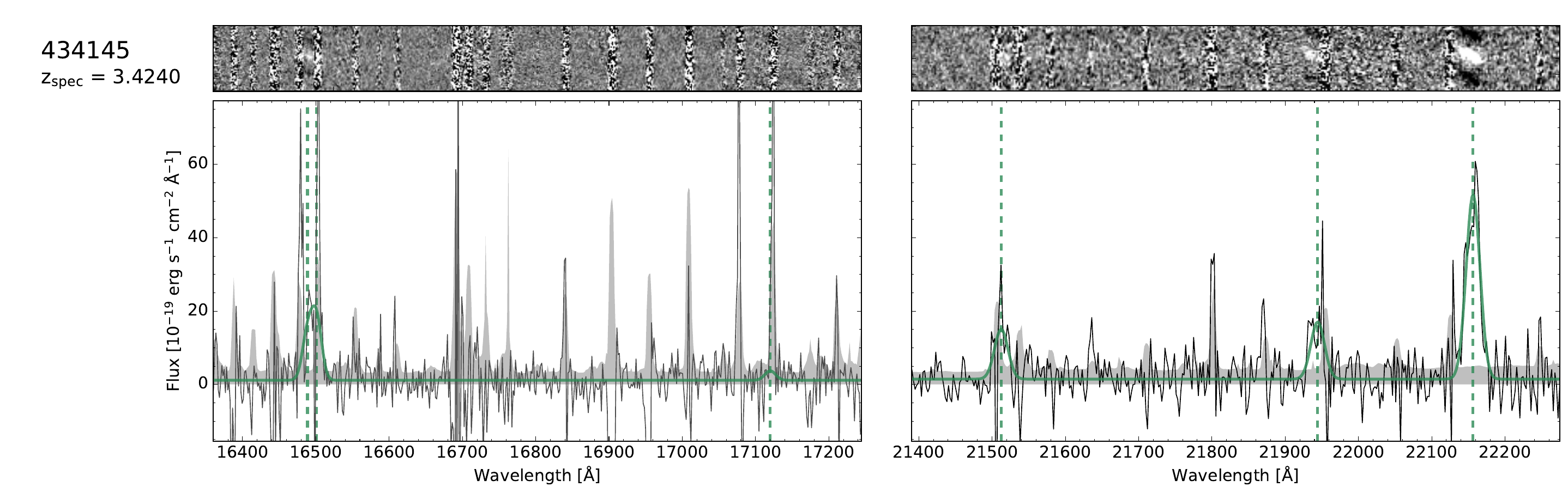}
  \includegraphics[page=14,width=0.95\linewidth]{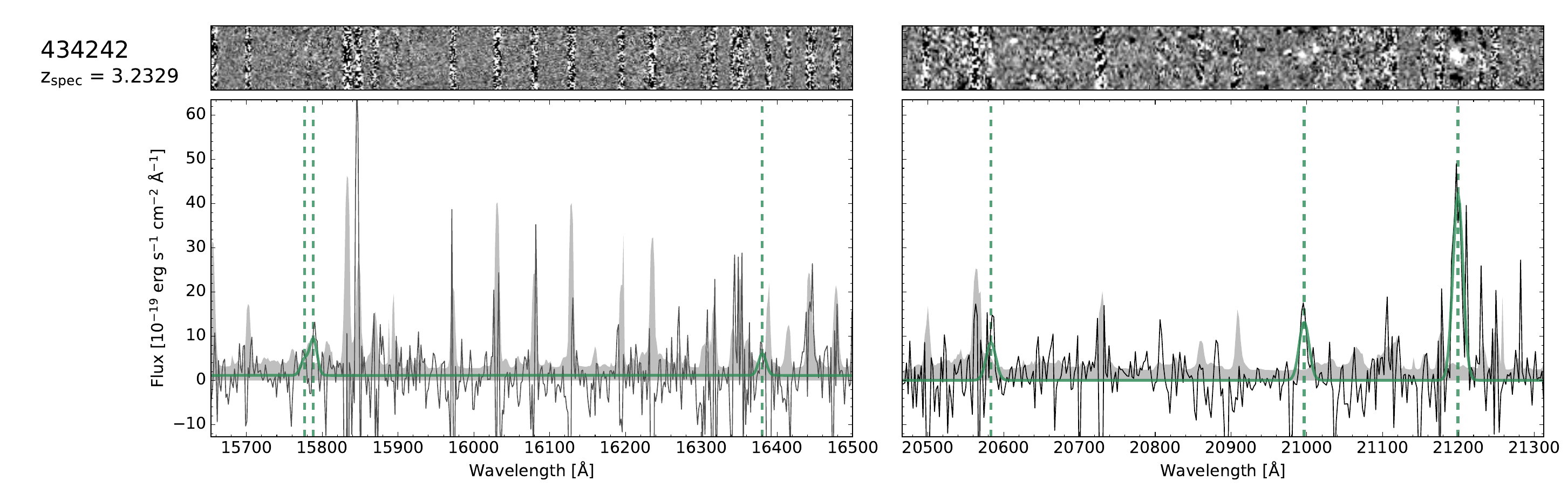}
  \includegraphics[page=15,width=0.95\linewidth]{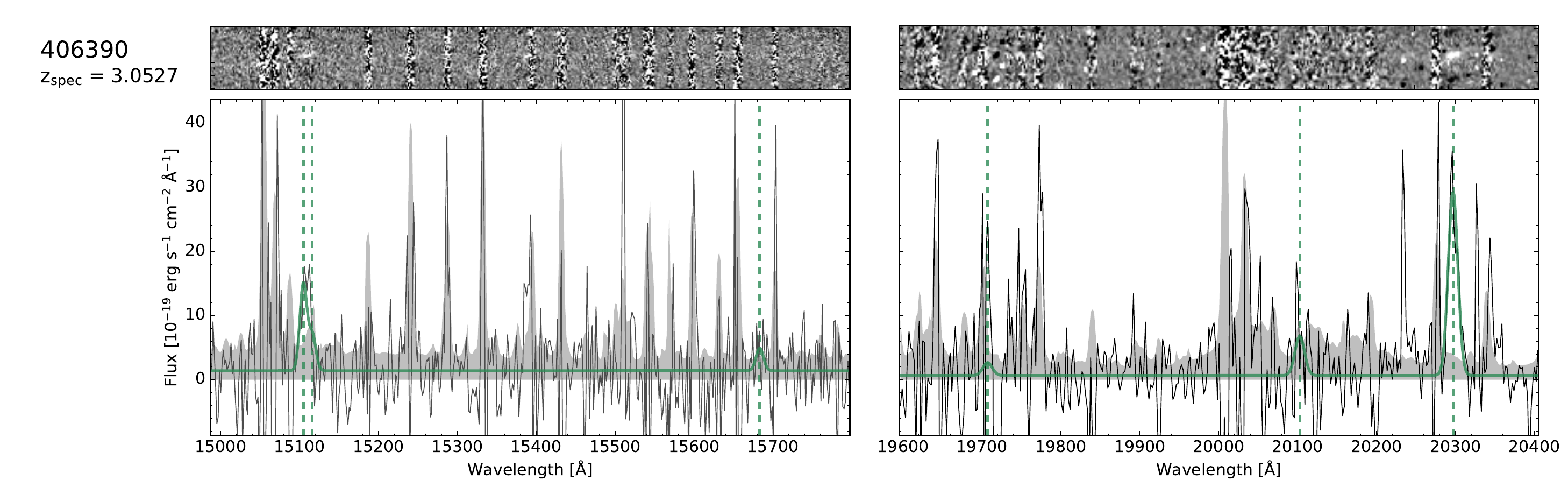}
  \includegraphics[page=16,width=0.95\linewidth]{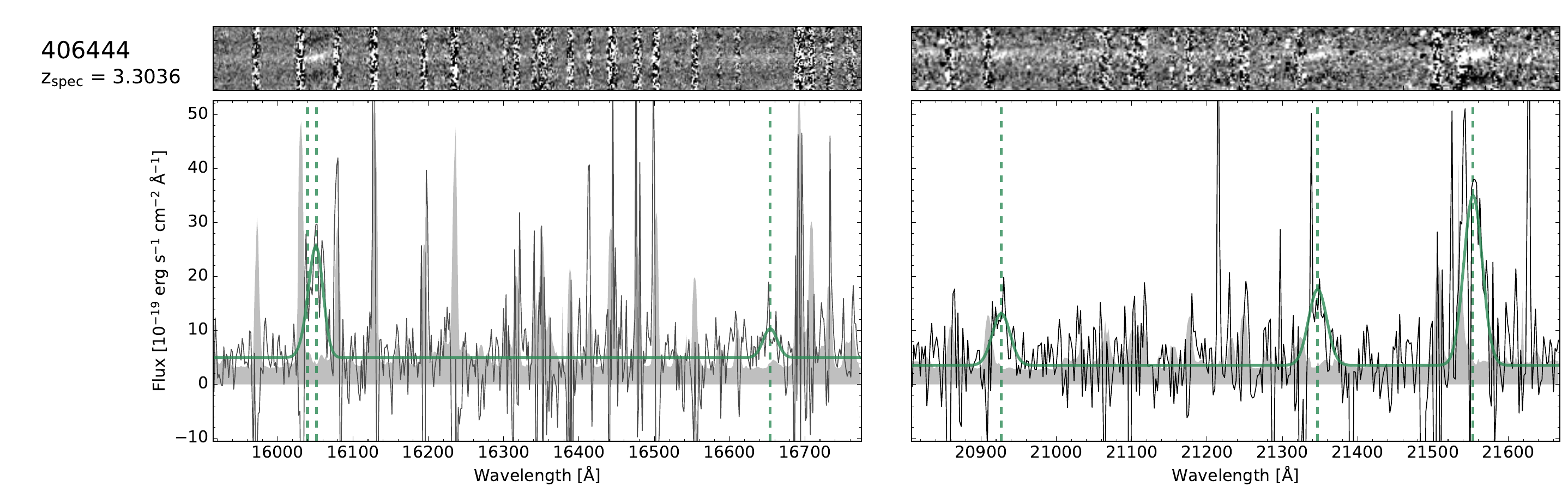}
  \caption{\textit{continued.}}
\end{figure*}
\addtocounter{figure}{-1}
\begin{figure*}
  \centering
  \includegraphics[page=17,width=0.95\linewidth]{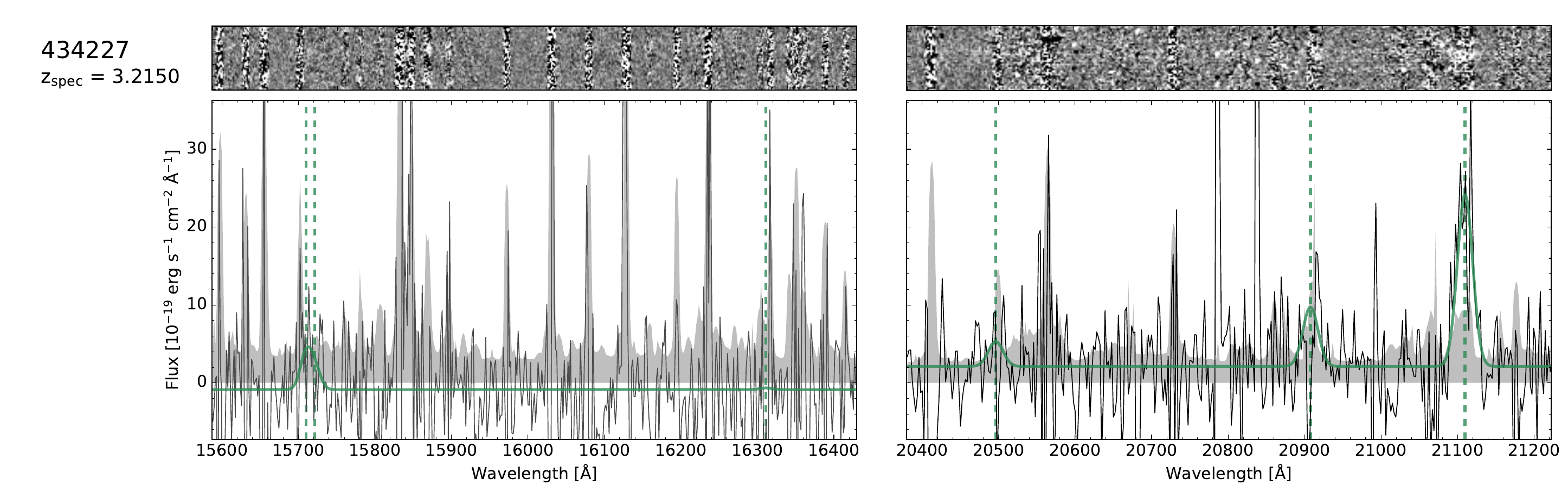}
  \includegraphics[page=18,width=0.95\linewidth]{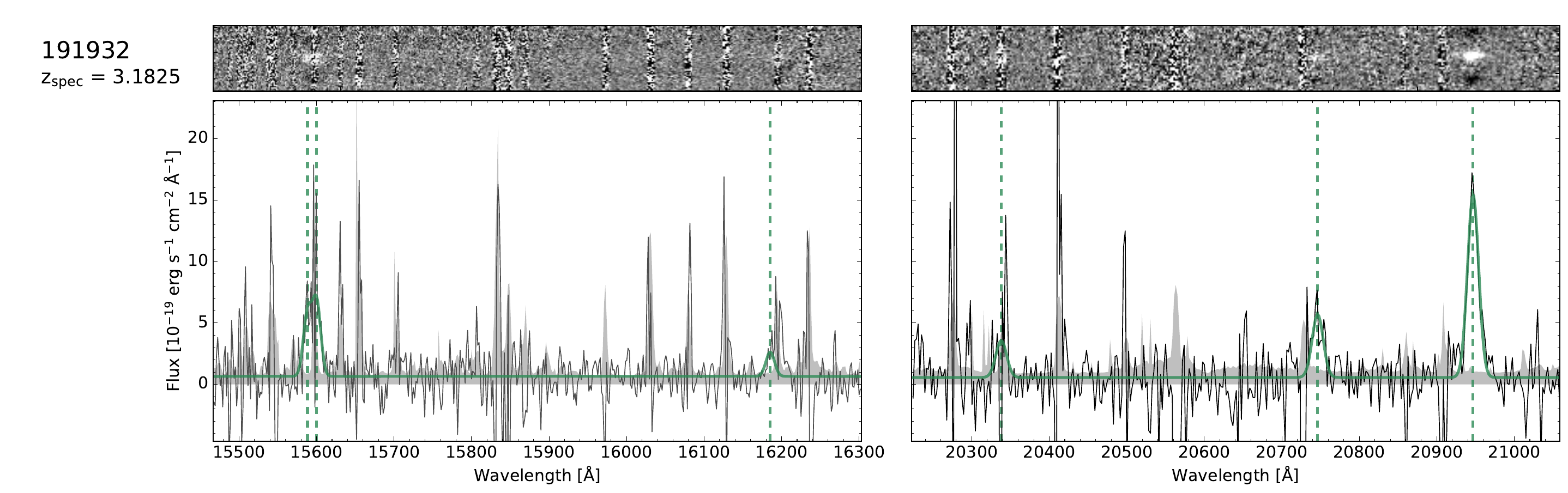}
  \includegraphics[page=19,width=0.95\linewidth]{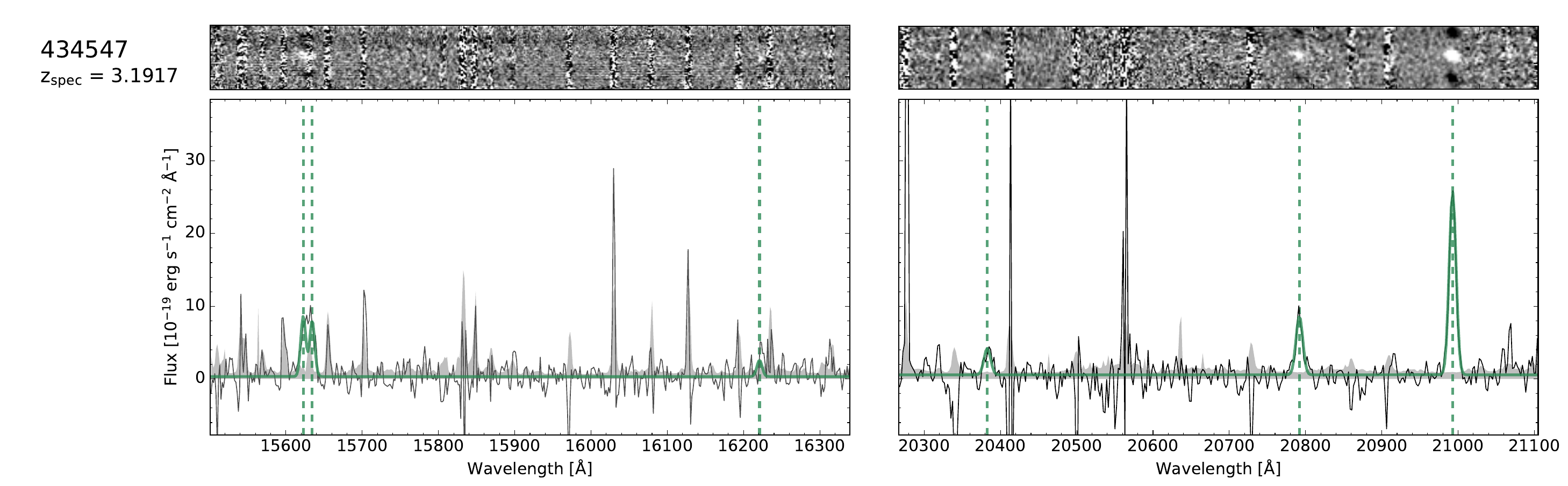}
  \includegraphics[page=20,width=0.95\linewidth]{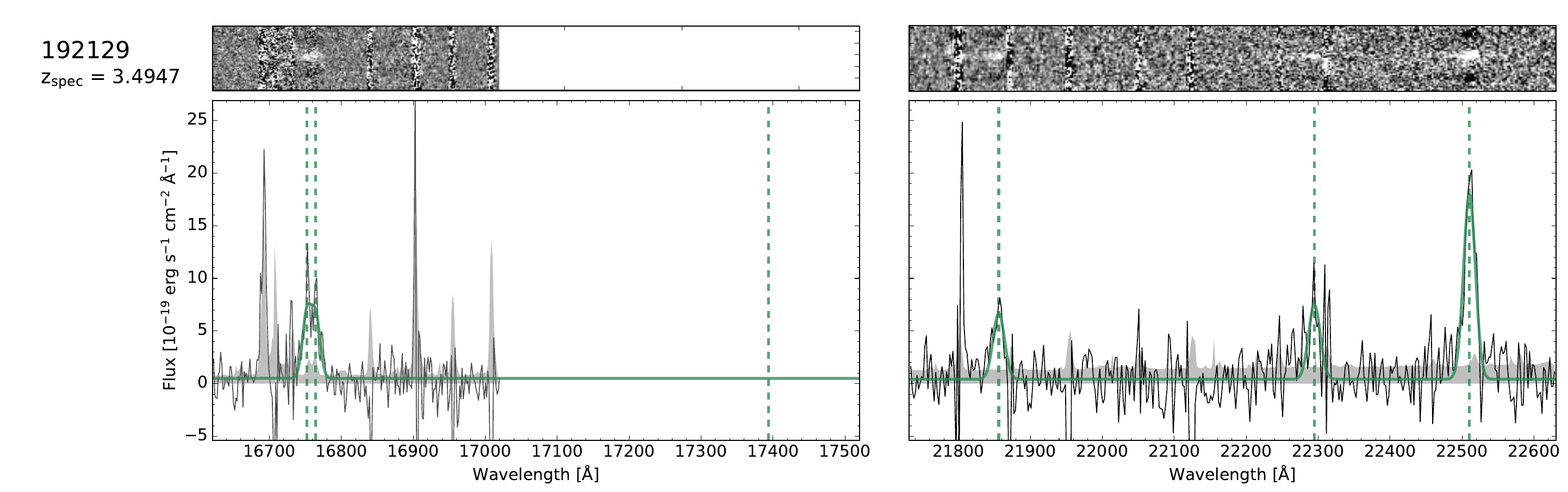}
  \caption{\textit{continued.}}
\end{figure*}
\addtocounter{figure}{-1}
\begin{figure*}
  \centering
  \includegraphics[page=21,width=0.95\linewidth]{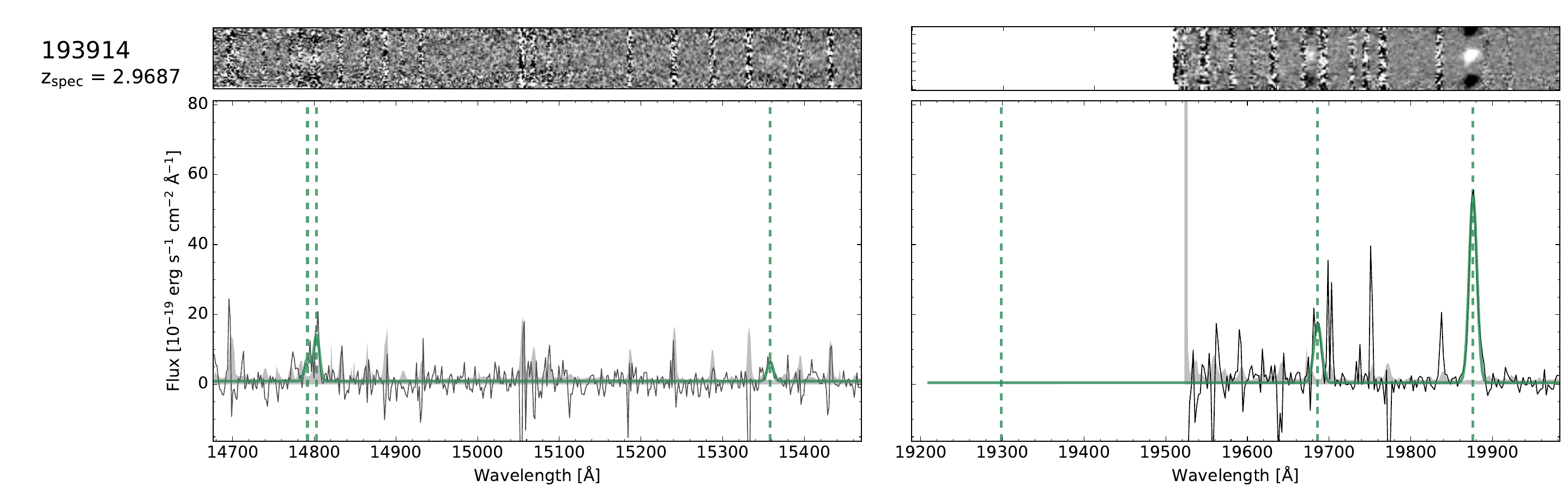}
  \includegraphics[page=22,width=0.95\linewidth]{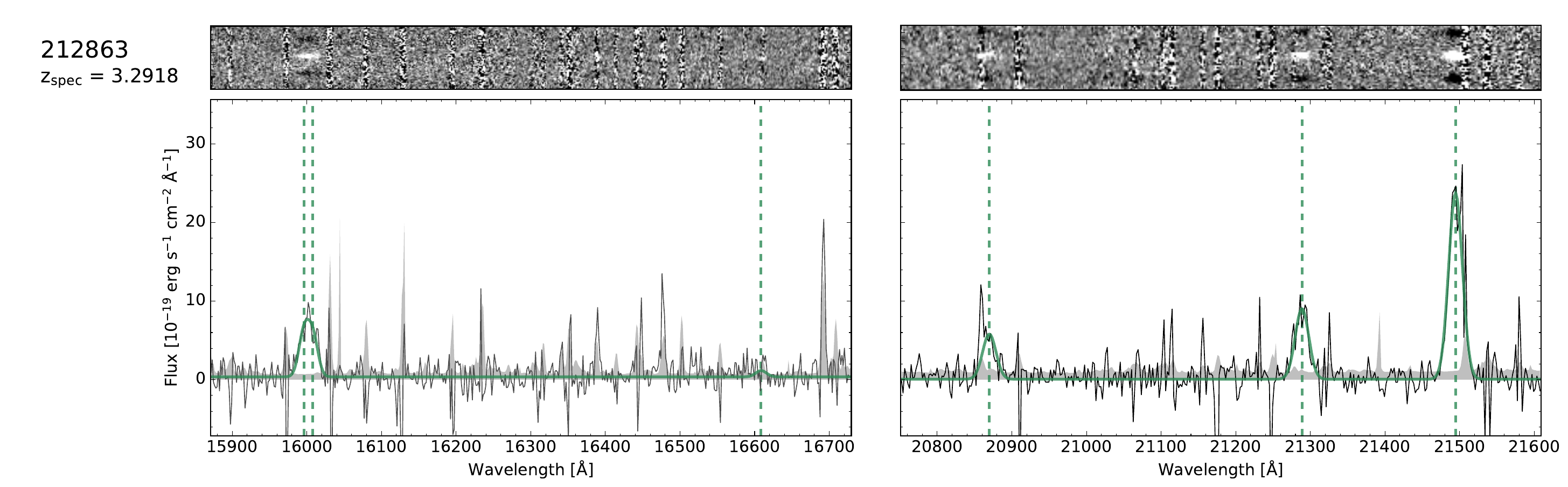}
  \includegraphics[page=23,width=0.95\linewidth]{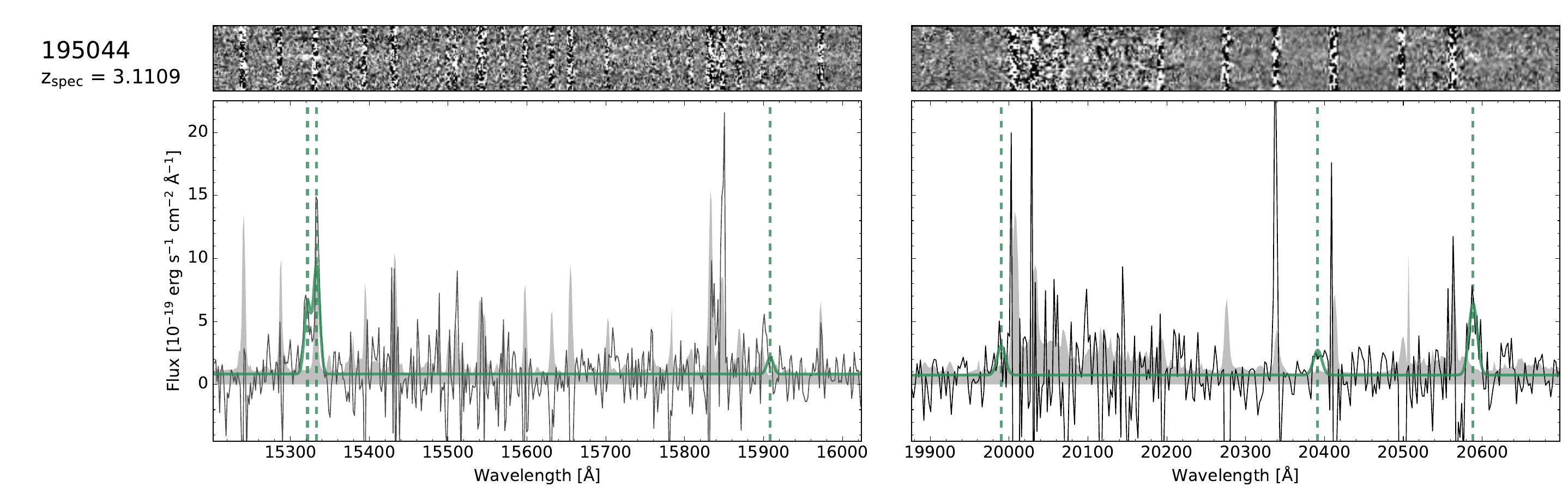}
  \includegraphics[page=24,width=0.95\linewidth]{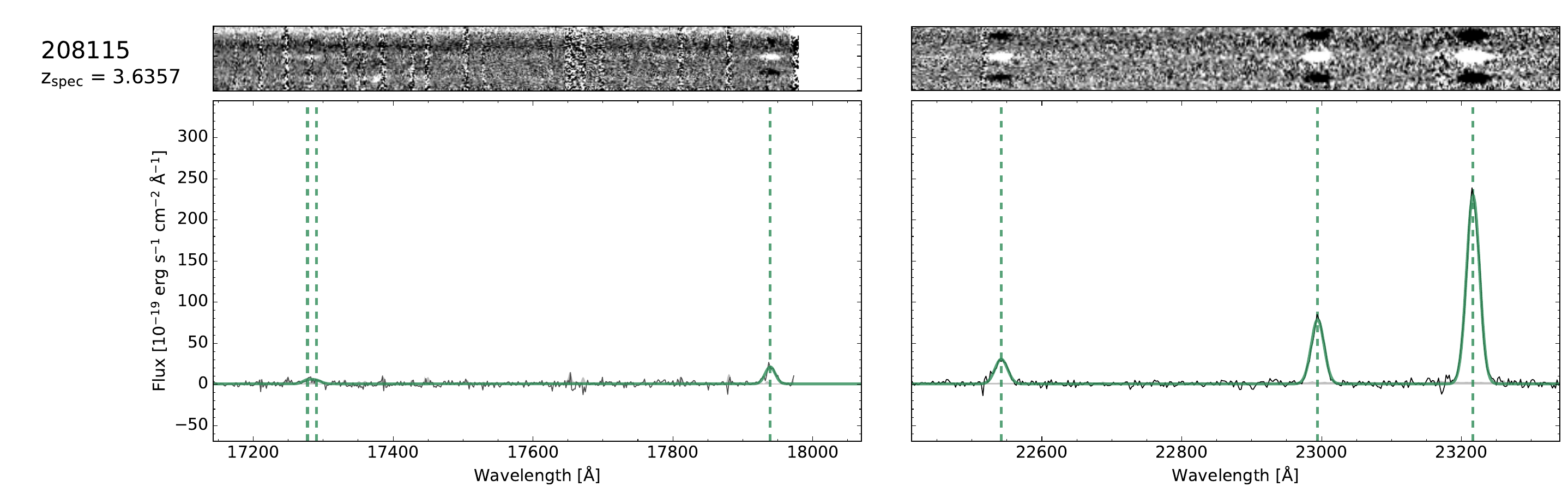}
  \caption{\textit{continued.}}
\end{figure*}
\addtocounter{figure}{-1}
\begin{figure*}
  \centering
  \includegraphics[page=25,width=0.95\linewidth]{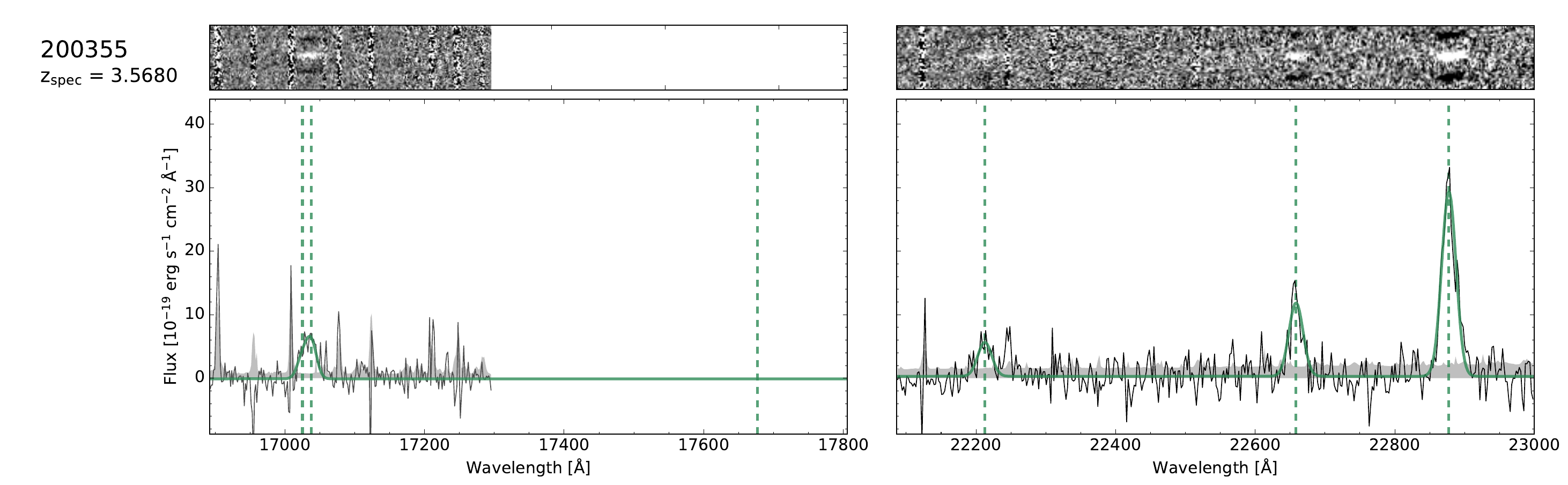}
  \includegraphics[page=26,width=0.95\linewidth]{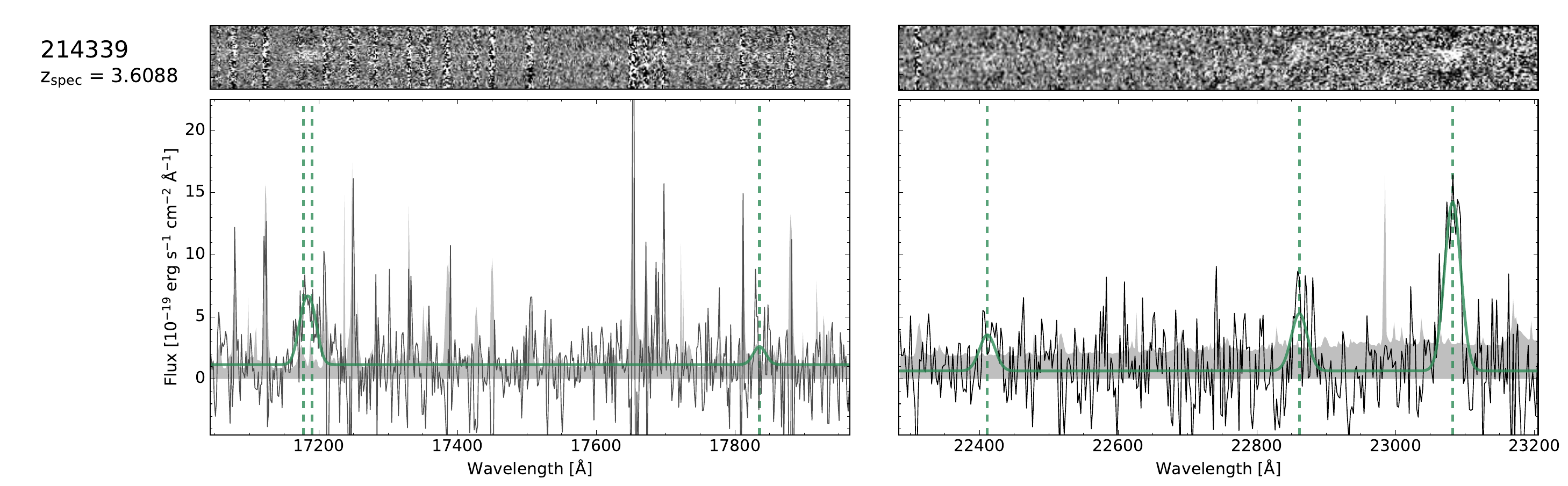}
  \includegraphics[page=27,width=0.95\linewidth]{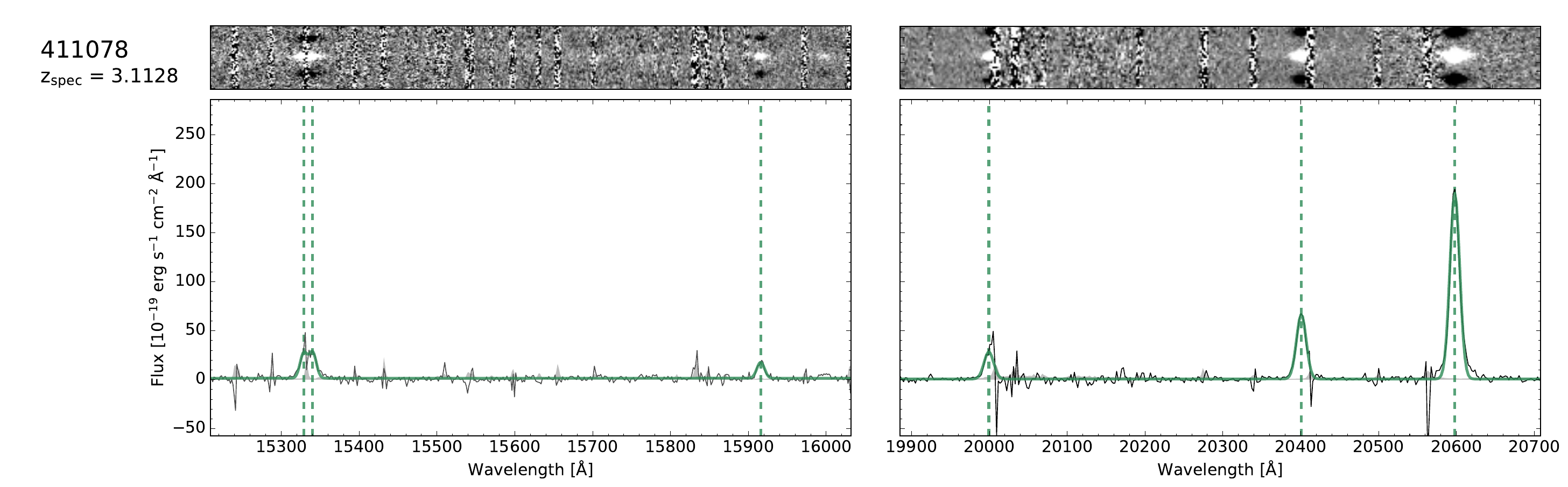}
  \includegraphics[page=28,width=0.95\linewidth]{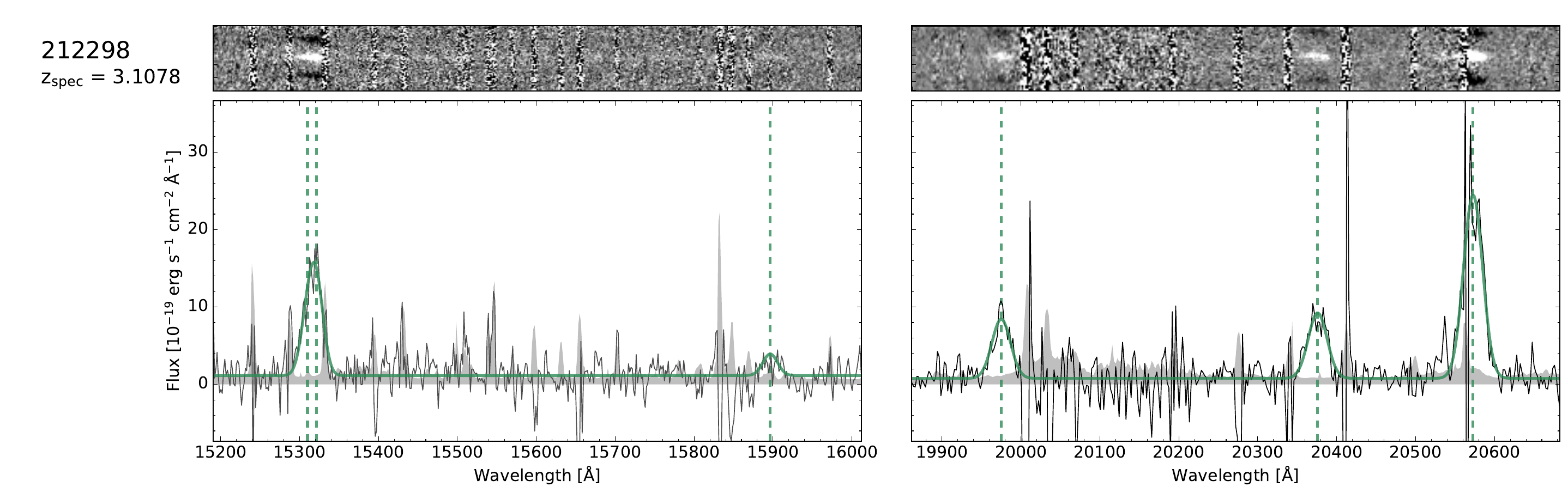}
  \caption{\textit{continued.}}
\end{figure*}
\addtocounter{figure}{-1}
\begin{figure*}
  \centering
  \includegraphics[page=29,width=0.95\linewidth]{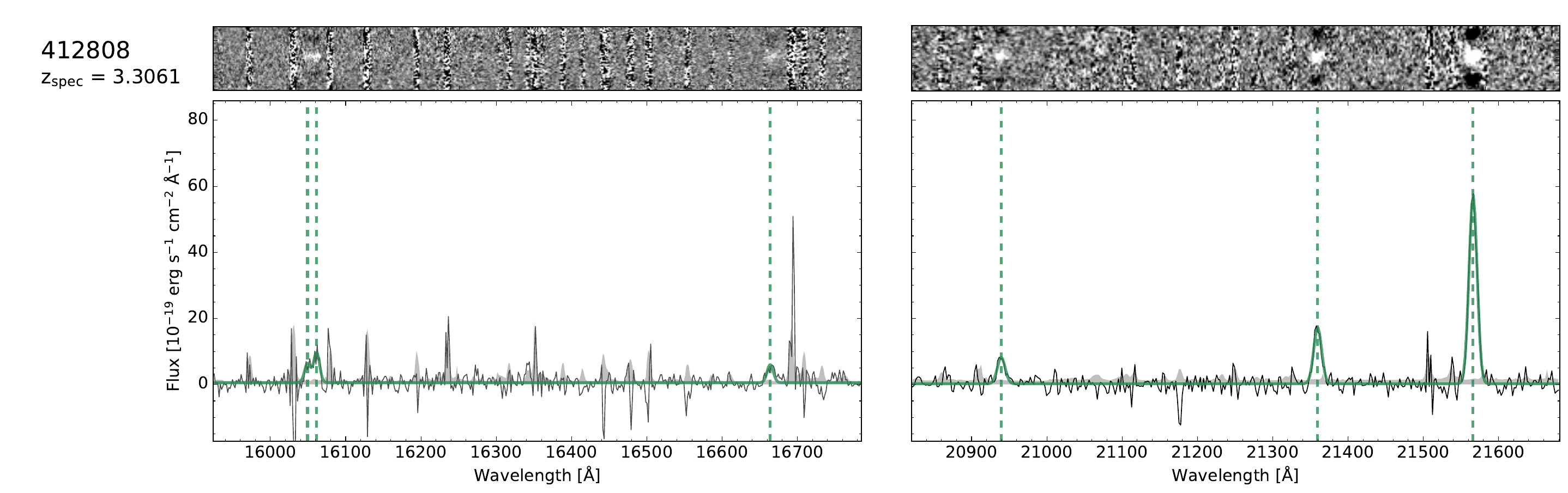}
  \includegraphics[page=30,width=0.95\linewidth]{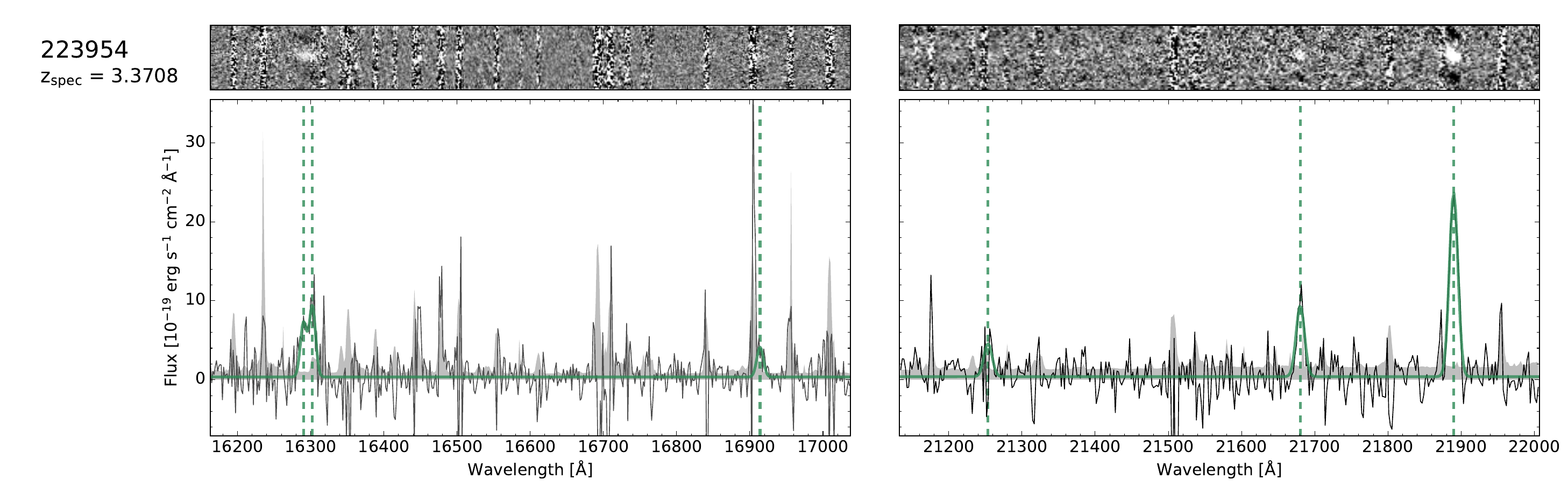}
  \includegraphics[page=31,width=0.95\linewidth]{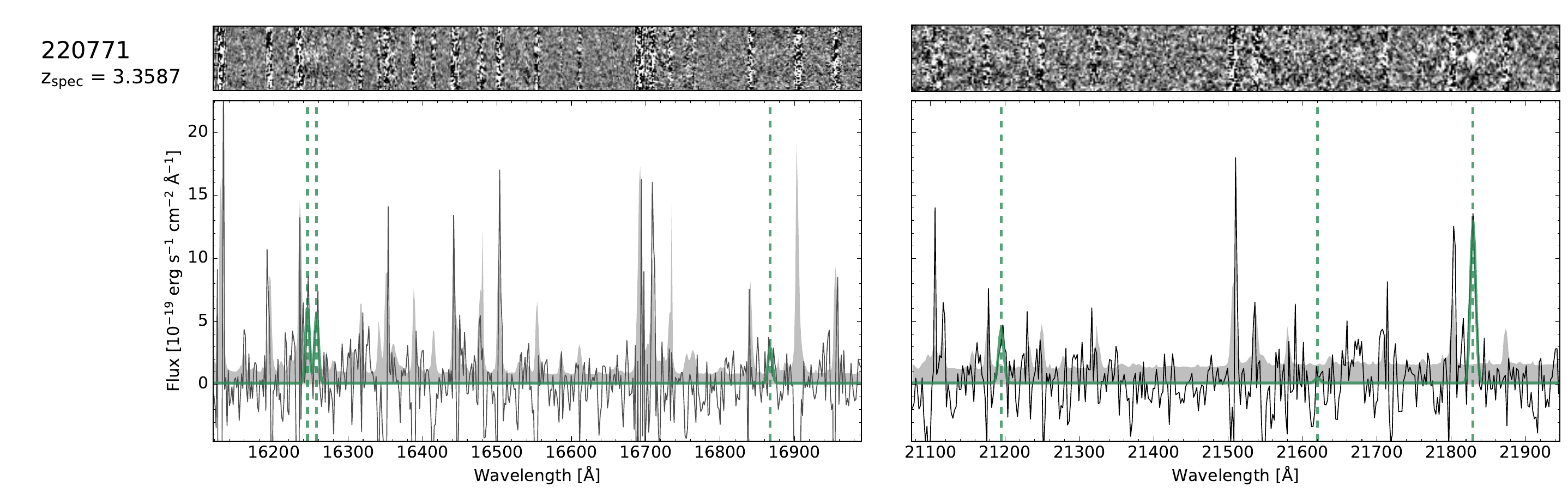}
  \includegraphics[page=32,width=0.95\linewidth]{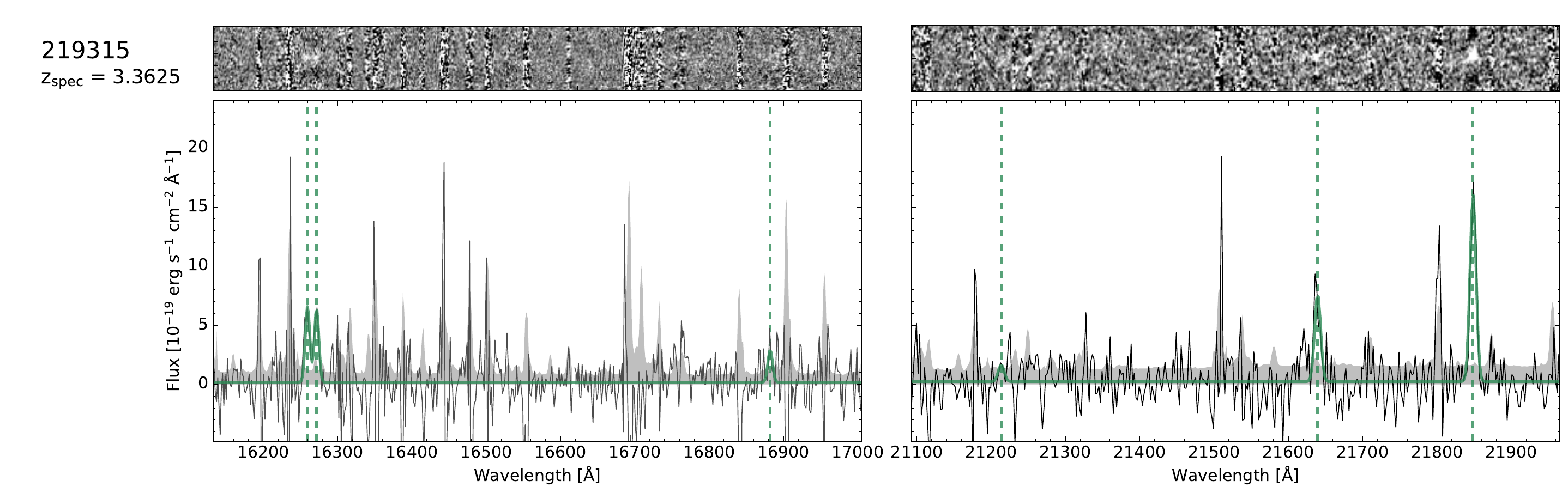}
  \caption{\textit{continued.}}
\end{figure*}
\addtocounter{figure}{-1}
\begin{figure*}
  \centering
  \includegraphics[page=33,width=0.95\linewidth]{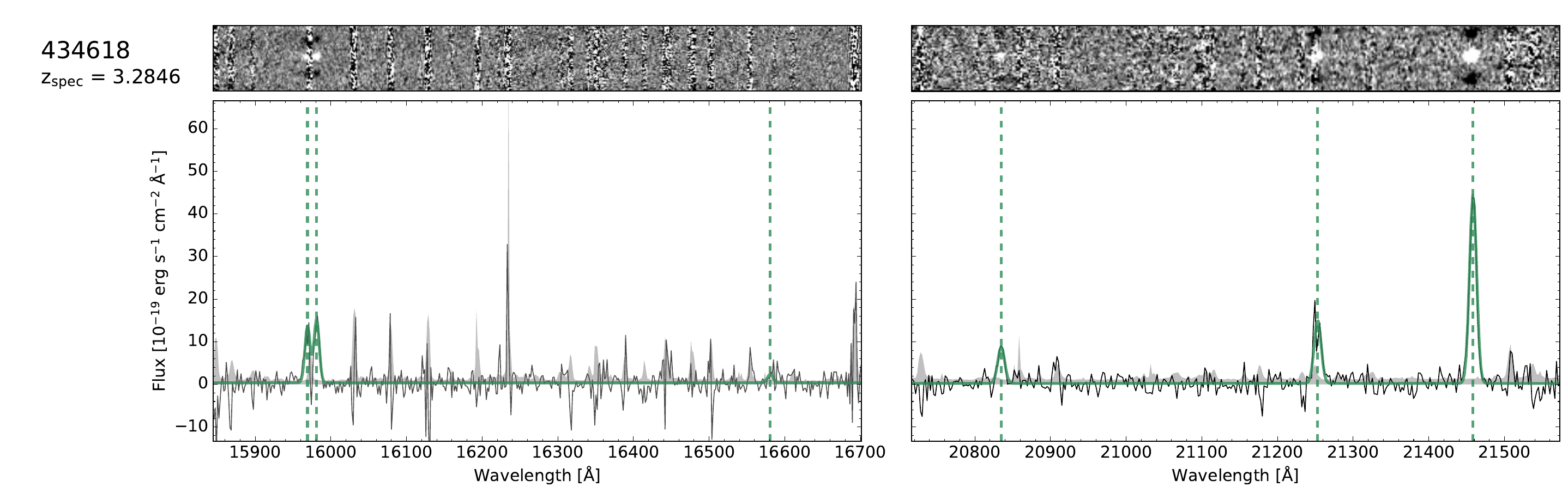}
  \includegraphics[page=34,width=0.95\linewidth]{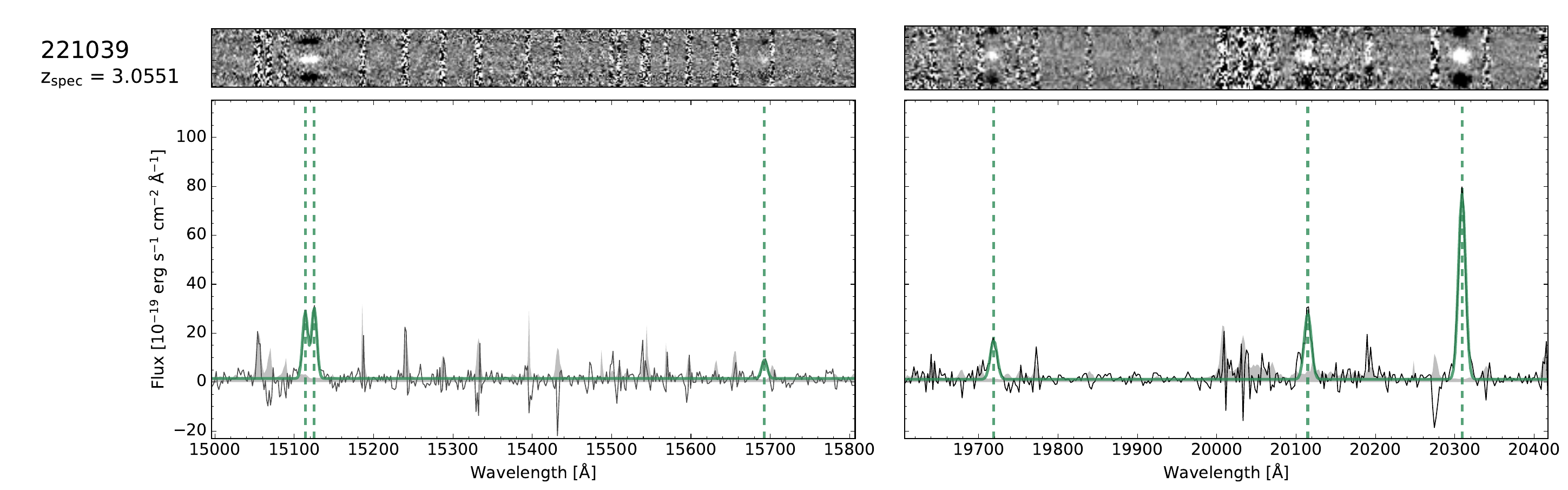}
  \includegraphics[page=35,width=0.95\linewidth]{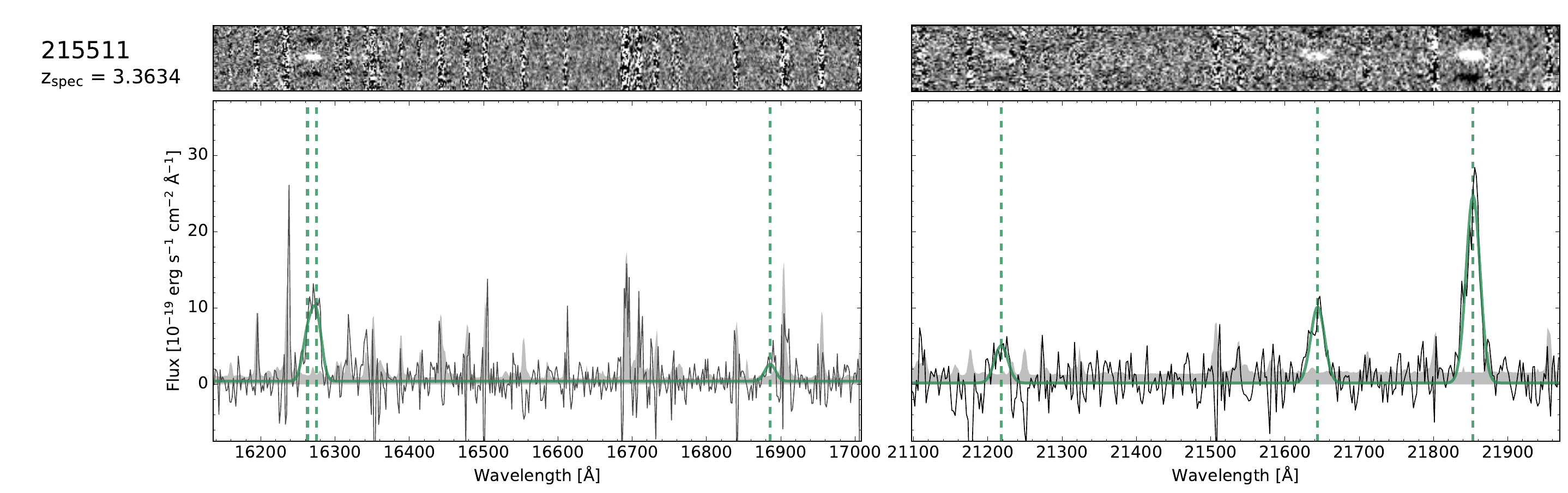}
  \includegraphics[page=36,width=0.95\linewidth]{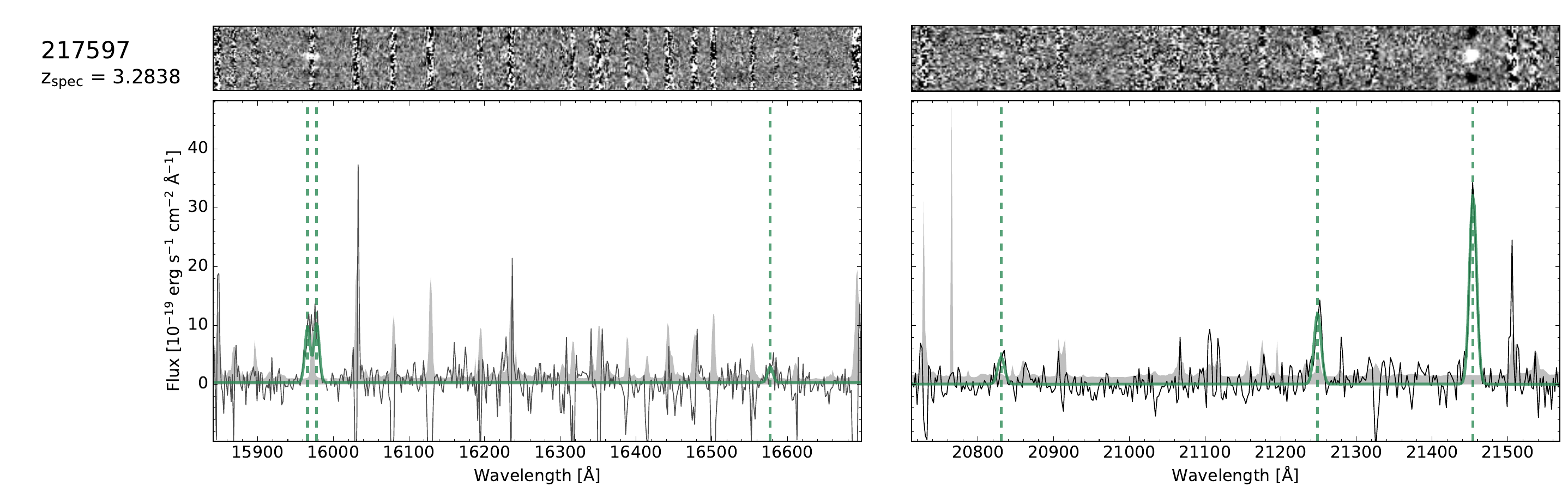}
  \caption{\textit{continued.}}
\end{figure*}
\addtocounter{figure}{-1}
\begin{figure*}
  \centering
  \includegraphics[page=37,width=0.95\linewidth]{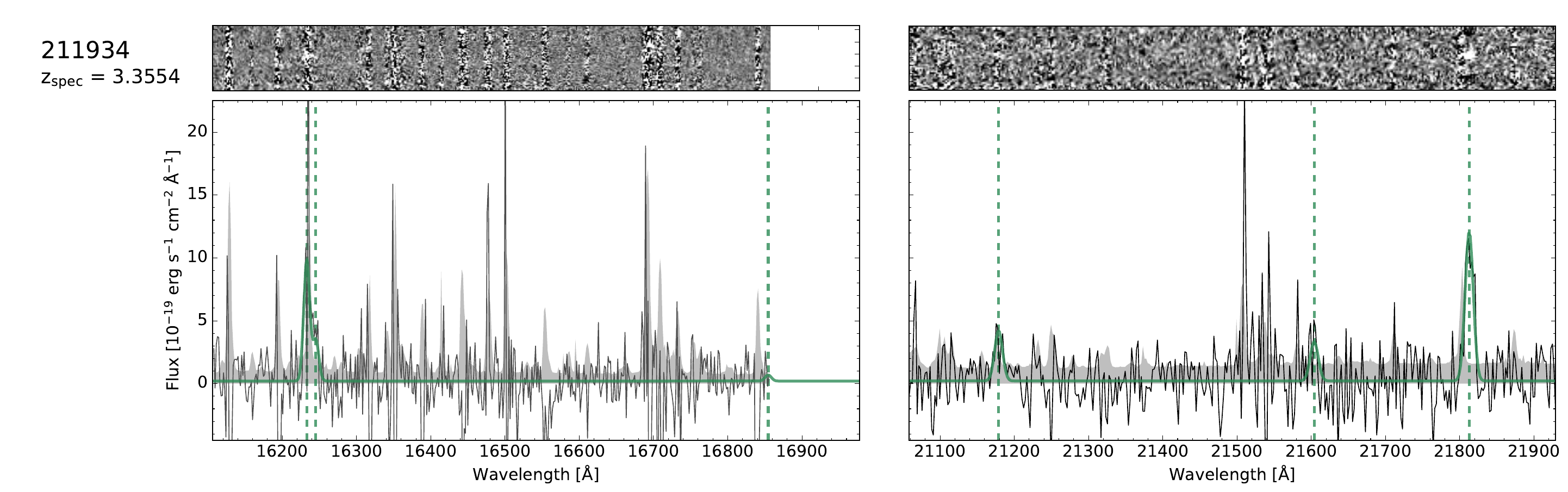}
  \includegraphics[page=38,width=0.95\linewidth]{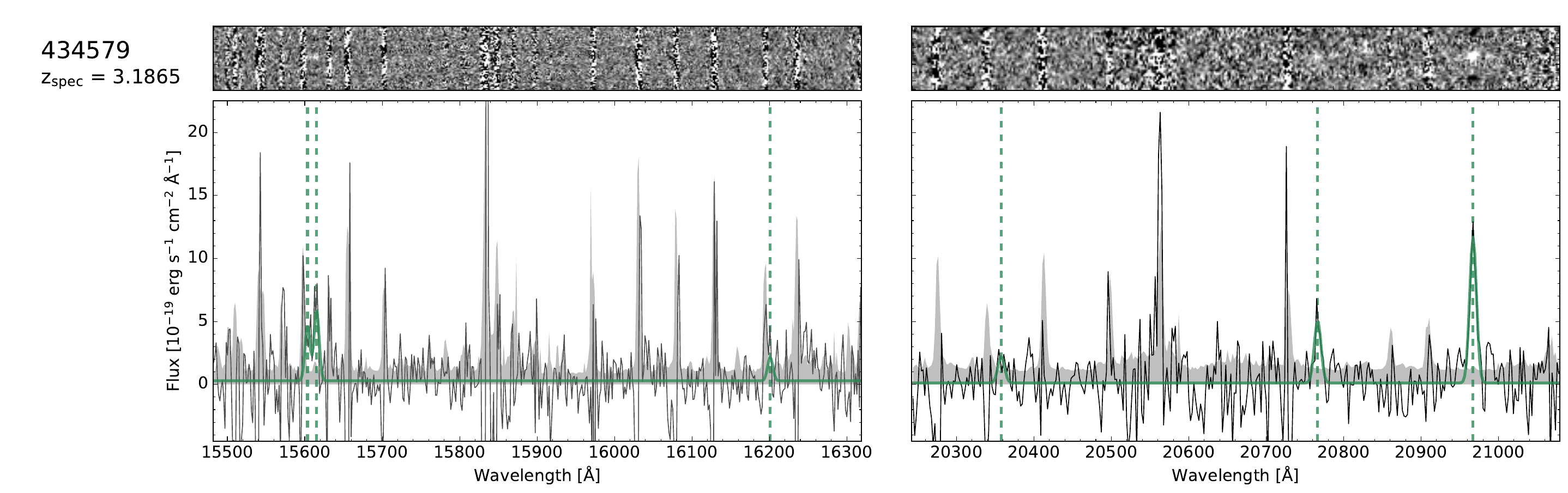}
  \includegraphics[page=39,width=0.95\linewidth]{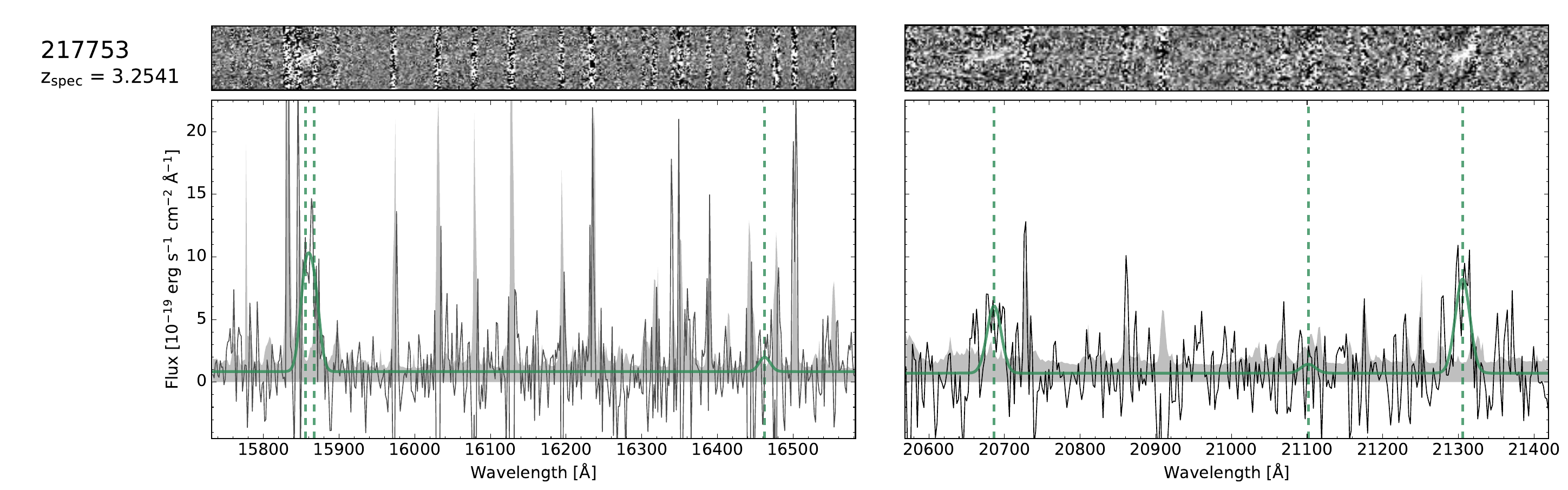}
  \includegraphics[page=40,width=0.95\linewidth]{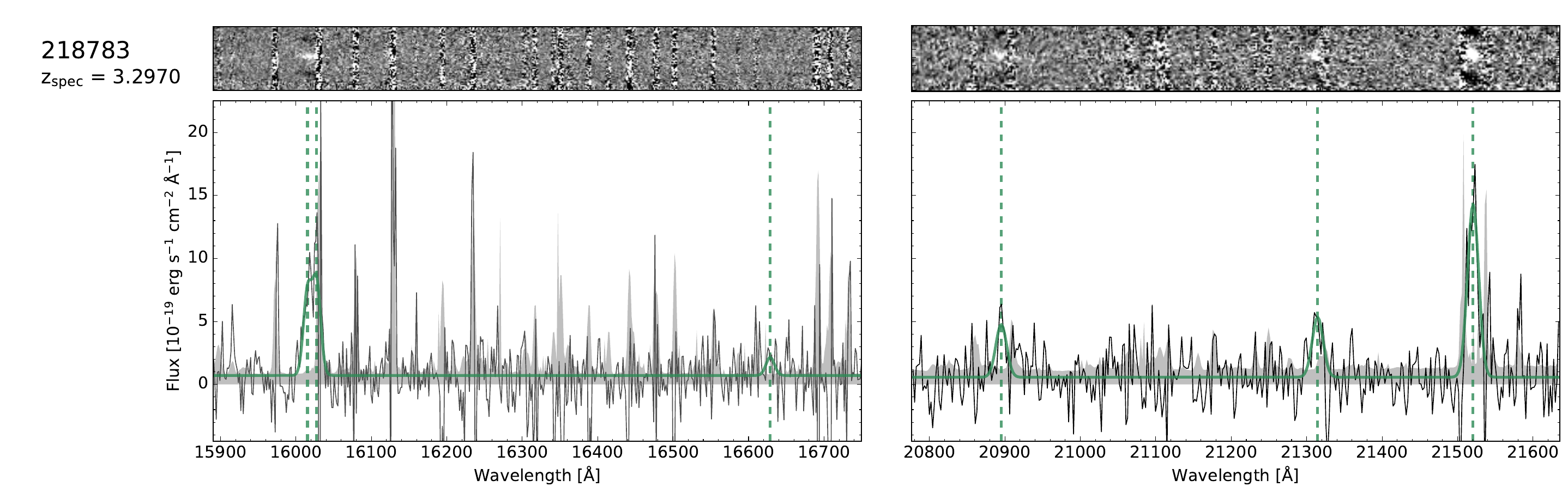}
  \caption{\textit{continued.}}
\end{figure*}
\addtocounter{figure}{-1}
\begin{figure*}
  \centering
  \includegraphics[page=41,width=0.95\linewidth]{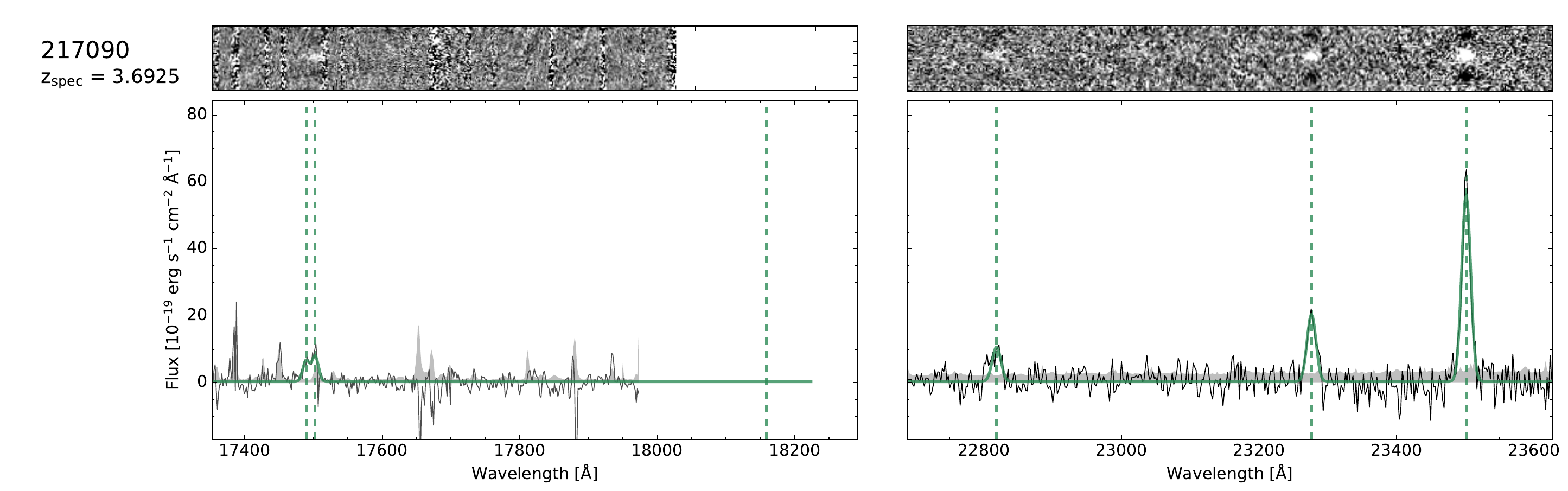}
  \includegraphics[page=42,width=0.95\linewidth]{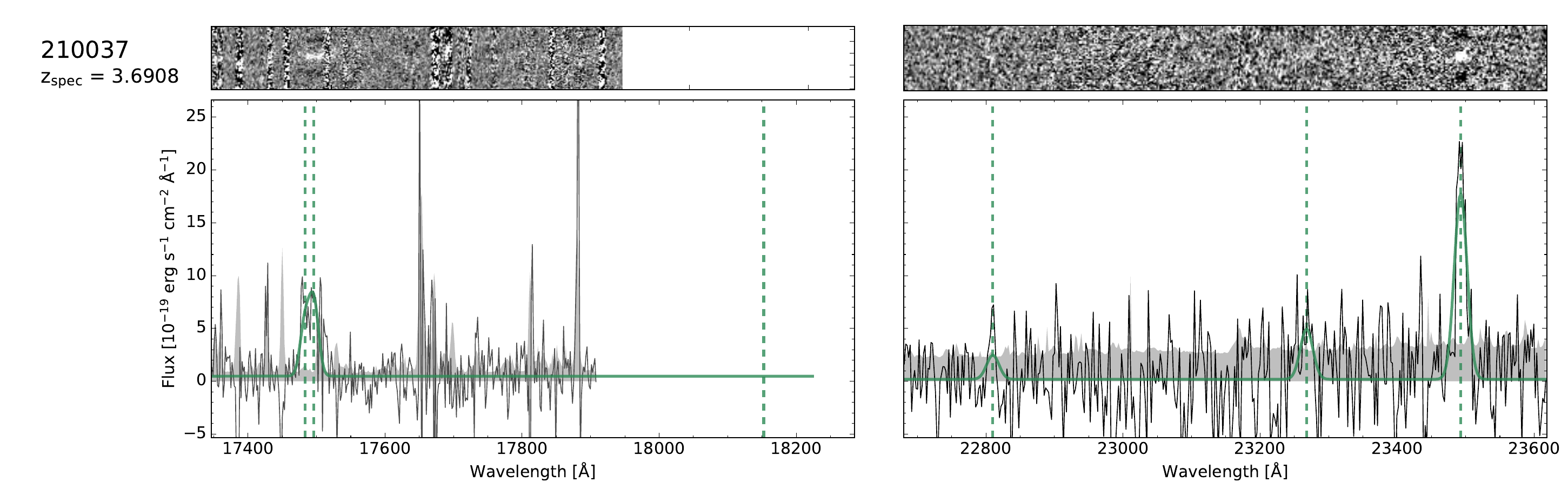}
  \includegraphics[page=43,width=0.95\linewidth]{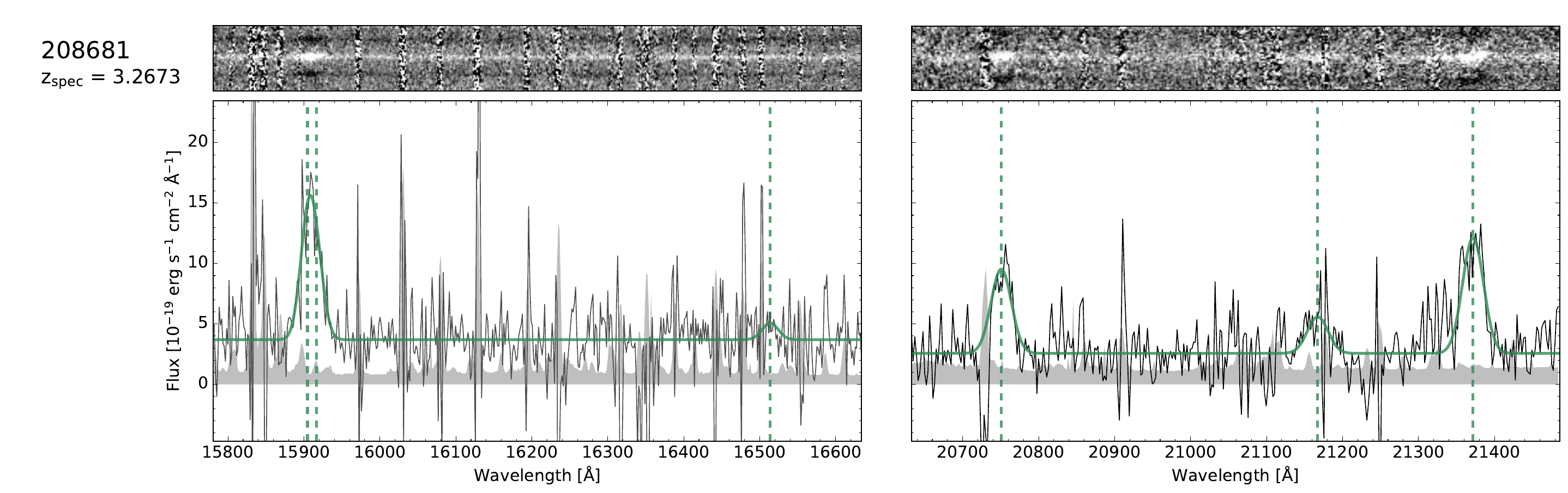}
  \caption{\textit{continued.}
    \label{fig:spec_end}}
\end{figure*}

\clearpage

\section{HST image stamps}
\begin{figure*}
  \centerline{
    \includegraphics[width=\linewidth]{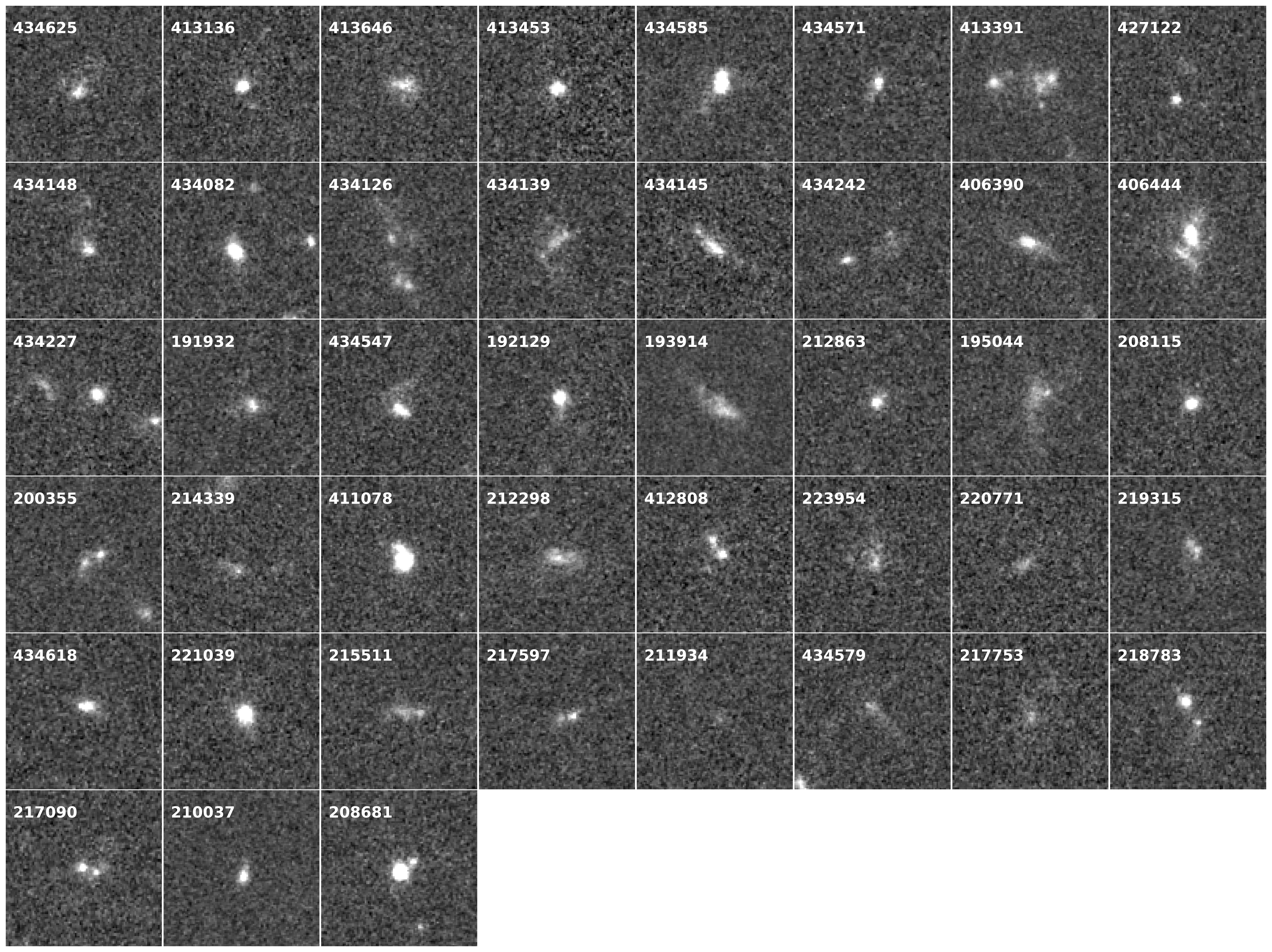}
  }
  \caption{
    \textit{HST}/ACS F814W stamps of the sample.  North is up and the size of each stamp is $3\times3$ arcsec$^2$.
  \label{fig:hstimg}}
\end{figure*}

\begin{figure*}
  \centerline{
    \includegraphics[width=0.7\linewidth]{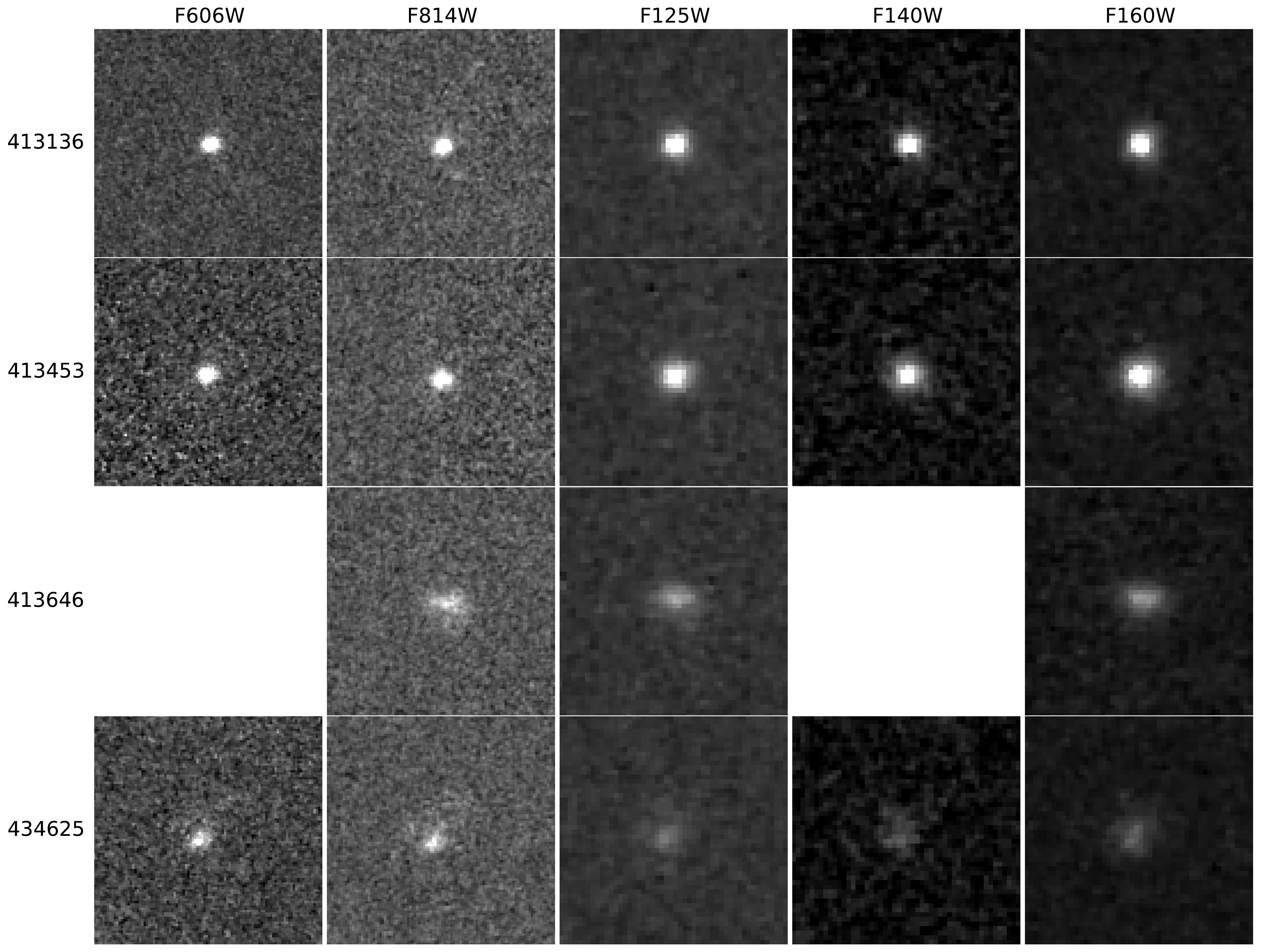}
  }
  \caption{
    \textit{HST} CANDELS stamps of the sample.  North is up and the size of each stamp is $3\times3$ arcsec$^2$.
    ACS/F814W image is the same as in \autoref{fig:hstimg} and WFC3/F140W image is from 3D-HST. 
  \label{fig:candelsimg}}
\end{figure*}

\autoref{fig:hstimg} shows \textit{Hubble Space Telescope} (\textit{HST})
Advance Camera for Surveys (ACS) F814W stamps of those with robust spectroscopic redshifts. 
For those observed with the CANDELS and 3D-HST surveys
\citep{grogin:2011, koekemoer:2011, brammer:2012, skelton:2014},
ACS F606W, Wide Field Camera 3 (WFC3) F125W, F140W, and F160W images
are shown in \autoref{fig:candelsimg}.

\clearpage

\section{Broad-band SED}
\label{sec:bestsed}
\begin{figure*}[h]
  \centerline{
    \includegraphics[page=1, width=\linewidth]{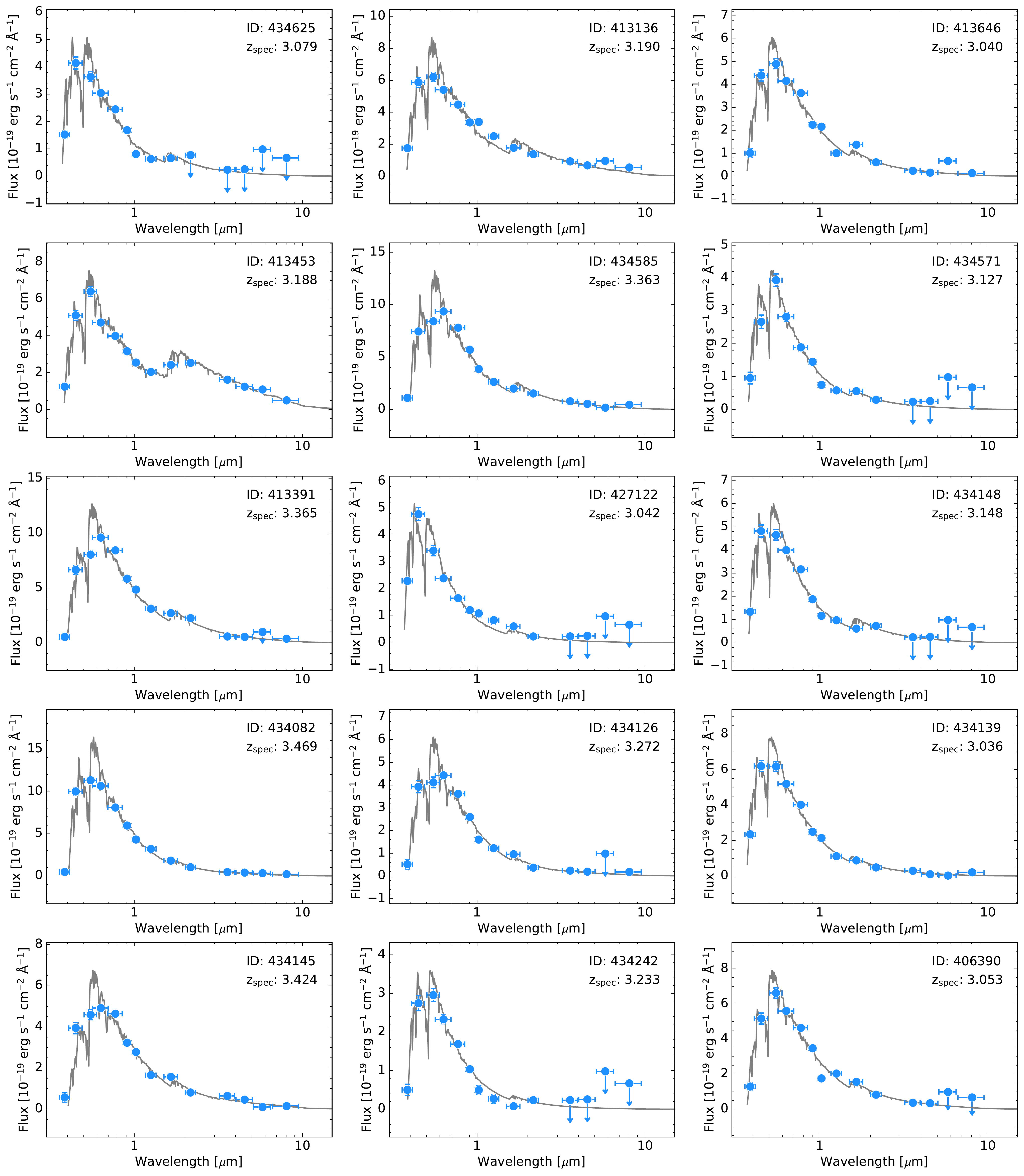}
  }
  \caption{Best-fit SED.
    \label{fig:bestsed_begin}
  }
\end{figure*}
\addtocounter{figure}{-1}
\begin{figure*}
  \centerline{
    \includegraphics[page=2, width=\linewidth]{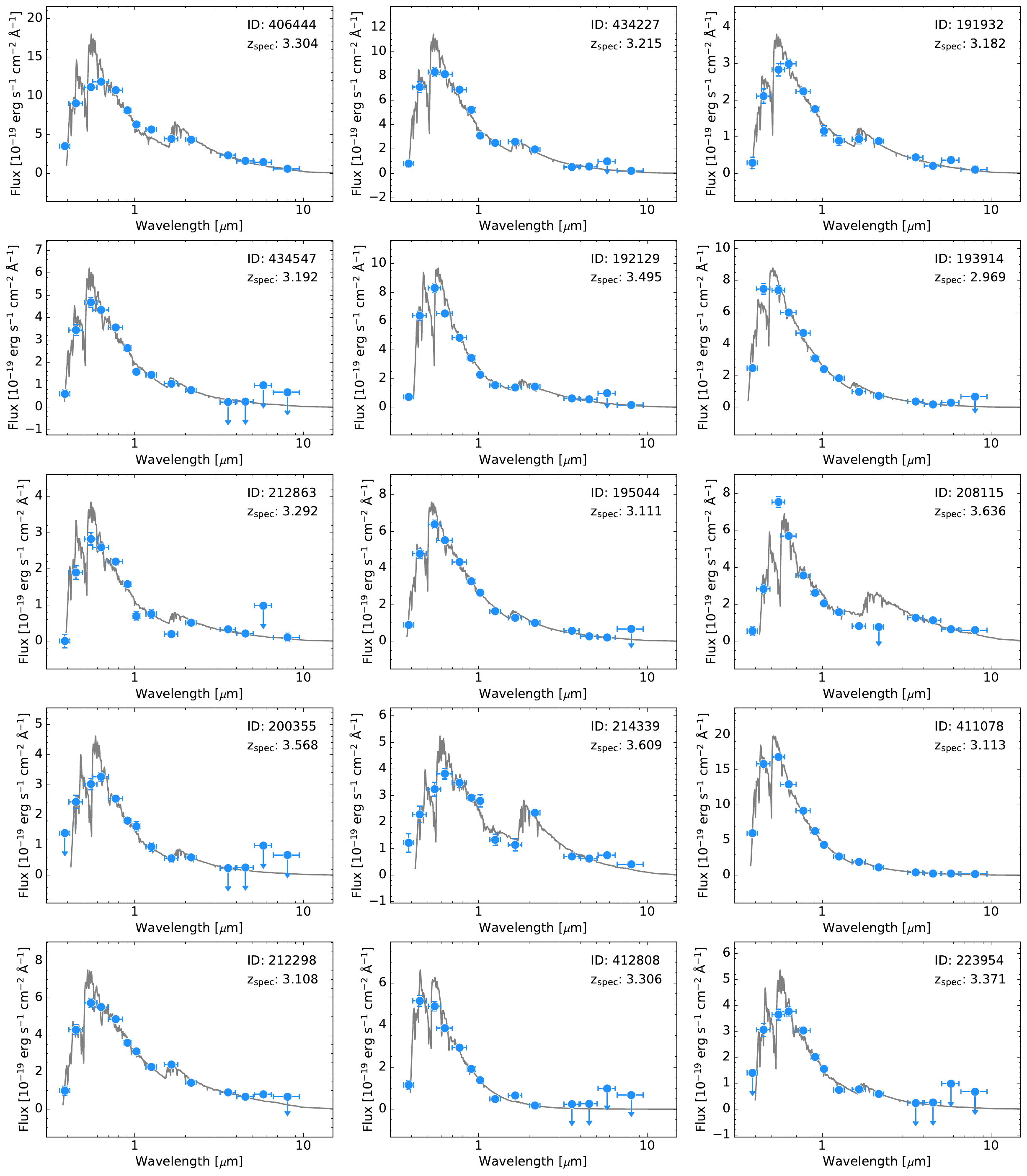}
  }
  \caption{\textit{continued.}
  }
\end{figure*}
\addtocounter{figure}{-1}
\begin{figure*}
  \centerline{
    \includegraphics[page=3, width=\linewidth]{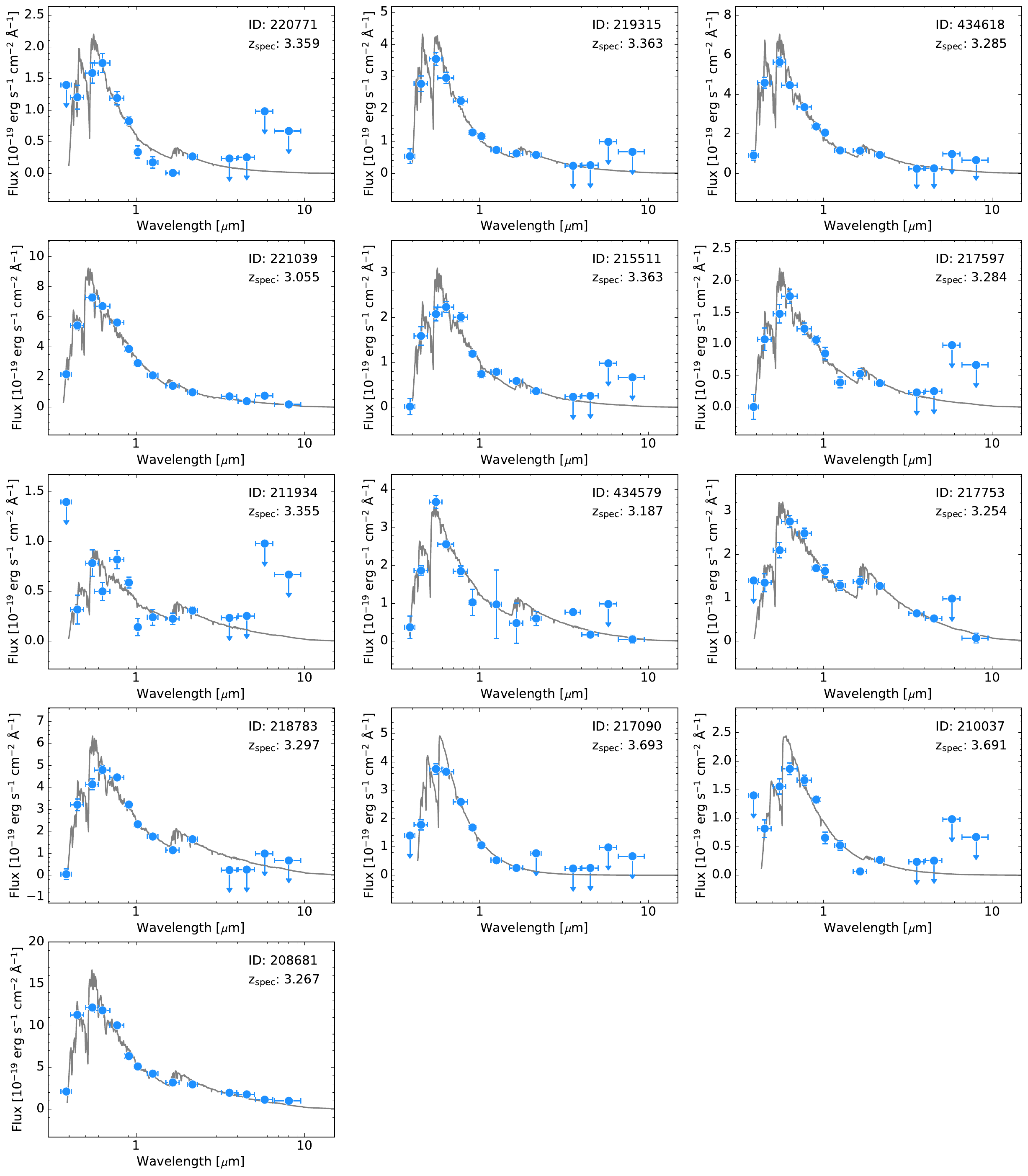}
  }
  \caption{\textit{continued.}
    \label{fig:bestsed_end}
  }
\end{figure*}

\clearpage

\section{Who does contribute the most in the stacking?}
\label{sec:misc_figures}
\begin{figure}[h]
  \centerline{
    \includegraphics[width=0.85\linewidth]{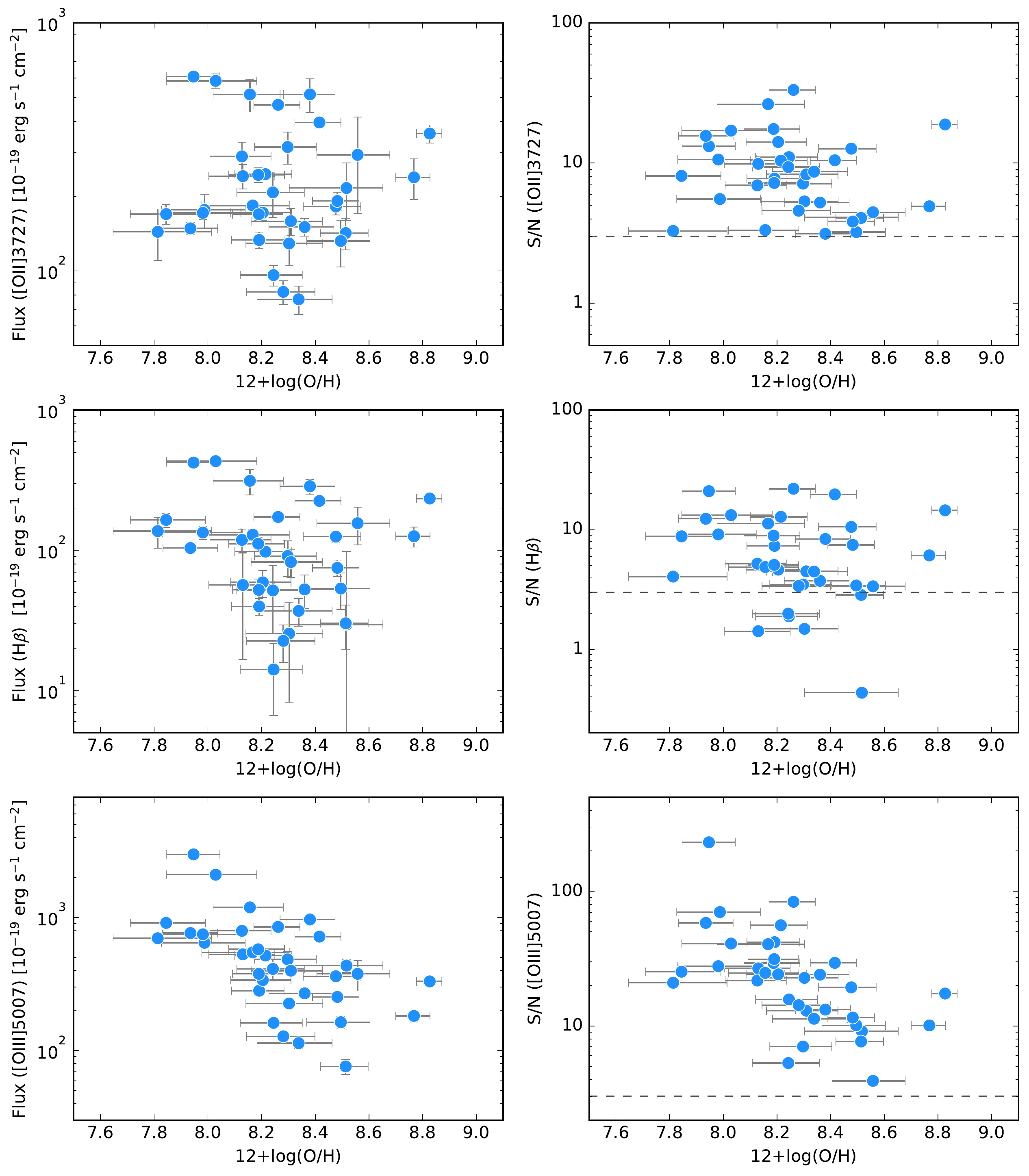}
  }
  \caption{
    \textit{Left}: Fluxes of \oii (\textit{Top}), \hbeta (\textit{Middle}), and \oiiione (\textit{Bottom}) as a function of
    gas-phase oxygen abundance. \textit{Right}: Signal-to-noise ratios for the same emission lines as the left panels
    as a function of \ohmetal. Dashed lines indicates $\text{S/N}=3$. 
  \label{fig:metal_lineflux}}
\end{figure}

\autoref{fig:metalmeasure} shows each line ratio as a function of \ohmetal and distribution of \ohmetal.
Stacked spectrum shows lower metallicity than the median metallicity of all objects.
The same trend was seen in a previous work by \citet{troncoso:2014}
in which they found slightly lower metallicity at a given stellar mass
for the stacked measurement than the average of individual measurement.
At least in our case, this discrepancy can be explained by the way we stacked spectra.
We normalized each spectrum by the \oiiitwo luminosity and stacked them
weighted by the inverse variance at each wavelength pixel.
Therefore, the resulting emission lines of stacked spectrum are dominated by
objects with higher S/N and less OH sky line contamination.
In \autoref{fig:metal_lineflux}, we show absolute fluxes and S/N ratios of
strong emission lines as a function of gas-phase oxygen abundance
for individual galaxies. 
As seen from the figure, \oiii is indeed dominated by
metal-poor objects, while \hbeta and \oii emission lines
have more contribution from more metal-rich objects. 
However, in the case that only \oiii has enhanced flux with respect to the other lines,
the line ratios used to investigate the metallicity tend to favor low metallicity,
in particular for the \oiii/\oii ratio.

\clearpage

\section{On the SFR dependence in metallicity in the regulator model of Lilly et al. (2013)}
\label{sec:l13eq28}

The general from of gas-phase metallicity in a quasi-steady state of the gas regulator model by \citet{lilly:2013} is their Equation (28) as follows.
\begin{equation}
  Z_\text{eq}(M_\star, \text{SFR}) = Z_{0}\, + \frac{y}{1 + \lambda(1-R)^{-1} + \varepsilon^{-1}\left\{M_\star^{-1}\cdot\text{SFR} + (1-R)^{-1}\frac{d\ln\mu}{dt}\right\}}.
  \label{eq:eq28l13}
\end{equation}

Since $\mu = \varepsilon^{-1}\cdot\text{sSFR}$, $d\ln\mu/dt$ is calculated as 
\begin{equation}
  \begin{split}
    \frac{d\ln\mu}{dt}
    & = -\varepsilon^{-1}\frac{d\varepsilon}{dt} + \text{SFR}^{-1}\frac{d\text{SFR}}{dt} - M_\star^{-1}\frac{dM_\star}{dt} \\
    & = -\varepsilon^{-1}\frac{d\varepsilon}{dt} + \text{SFR}^{-1}\frac{d\text{SFR}}{dt} - (1-R)\text{sSFR},
  \end{split} \label{eq:dmudt}
\end{equation}
where $dM_\star/dt$ is substituted by $(1-R)\text{SFR}$ as this term describes the build-up of long-lived stars in the system \citep[see][]{lilly:2013}.

Substituting the denominator of the second term in \autoref{eq:eq28l13} by \autoref{eq:dmudt} yields
\begin{equation}
  \begin{split}
    & 1 + \lambda(1-R)^{-1} + \varepsilon^{-1}\left\{M_\star^{-1}\cdot\text{SFR} + (1-R)^{-1}\frac{d\ln\mu}{dt}\right\}\\
    & = 1 + \lambda(1-R)^{-1} + \varepsilon^{-1}(1-R)^{-1}\left\{\text{SFR}^{-1}\frac{d\text{SFR}}{dt} - \varepsilon^{-1}\frac{d\varepsilon}{dt}\right\}.
  \end{split}
\end{equation}

Therefore, \autoref{eq:eq28l13} can be written as
\begin{equation}
  Z_\text{eq}(M_\star, \text{SFR}) = Z_{0}\, + \frac{y}{1 + \lambda(1-R)^{-1} + \varepsilon^{-1}(1-R)^{-1}\left\{\text{SFR}^{-1}\frac{d\text{SFR}}{dt} - \varepsilon^{-1}\frac{d\varepsilon}{dt}\right\}}.
\end{equation}

If the SFR of the system is constant, obviously there is no SFR dependence in metallicity as $d\text{SFR}/dt=0$.

\end{document}